\documentclass[lineo]{jfm}
\usepackage{amssymb,amsmath,mathrsfs,bm}
\usepackage{color,tikz,breakcites}
\usepackage{psfrag,overpic}
\usepackage{graphicx}
\usepackage{newtxtext}
\usepackage{newtxmath}
\usepackage{natbib}
\usepackage[breaklinks=true]{hyperref}
\usepackage[export]{adjustbox}
\usepackage[mathlines,pagewise]{lineno}
\usepackage{multirow}

\makeatletter
\renewcommand*{\p@section}{\S\,}
\renewcommand*{\p@subsection}{\S\S\,}
\makeatother

\definecolor{bblue}{RGB}{52,152,219}
\hypersetup{
    colorlinks = true,
    urlcolor   = black,
    citecolor  = black,
    linkcolor  = black,
}

\newcommand{\RomanNumeralCaps}[1]
\linenumbers

\shorttitle{Inertial particle transport in compressible wall turbulence}
\shortauthor{M. Yu, L.H. Zhao, X.X. Yuan, C.X. Xu}

\title{Transport of inertial spherical particles in compressible turbulent boundary layers}
\author{Ming Yu \aff{1},
Lihao Zhao \aff{2},
Xianxu Yuan \aff{1}\corresp{\email{yuanxianxu@cardc.cn}},
Chunxiao Xu \aff{2}\corresp{\email{xucx@tsinghua.edu.cn}}}

\affiliation{
\aff{1} 
State Key Laboratory of Aerodynamics, China Aerodynamics R\&D Center, Mianyang 621000, China
\aff{2} Key Laboratory of Applied Mechanics, Ministry of Education, Institute of 
Fluid Mechanics, Department of Engineering Mechanics, Tsinghua University, Beijing 100084, China}

\begin{document}

\maketitle

\begin{abstract}
In the present study, we perform direct numerical simulations of compressible turbulent boundary
layers at the free stream Mach number of $2 \sim 6$ laden with dilute phase of spherical particles
to investigate the Mach number effects on particle transport and dynamics.
Most of the phenomena observed and well-recognized for inertia particles in incompressible 
wall-bounded turbulent flows, such as the near-wall preferential accumulation and clustering
beneath the low-speed streaks, the flatter mean velocity profiles and the trend variation of the
particle velocity fluctuations, are identified in the compressible turbulent boundary layer as well.
However, we find that the compressibility effects are significant for large inertia particles. 
As the Mach number increases, the near-wall accumulation and the small-scale clustering
are alleviated, which is probably caused by the variation of the fluid density and viscosity
that are crucial to particle dynamics.
This can be affected by the fact that the forces acting on the particles with 
the viscous Stokes number greater than 500 are modulated by the comparatively 
high particle Mach numbers in the near-wall region.
This is also the reason for the abatement of the streamwise particle velocity fluctuation 
intensities with the Mach numbers.
\end{abstract}

\section{Introduction} \label{sec:intro}

Particle-laden turbulent boundary layer flows are ubiquitous in nature and wide realms of
engineering applications~\citep{rudinger2012fundamentals}, to name but a few, 
the formation and evolution of the sandstorm~\citep{cheng2012stochastic,liu2021large},
aircraft in extreme weather conditions~\citep{cao2014effects}, chemical industries
~\citep{baltussen2018multiscale} and supersonic combustors~\citep{feng2023numerical,
feng2023numerical2}.
Under the conditions of low volume fraction and low mass loadings, the dispersed particles smaller
than the Kolmogorov length scales can be regarded as dilute suspensions of point particles, 
passively transported by the turbulent flows, 
which is usually referred to as the `one-way coupling', and can be simulated 
by the point-particle approach under the Eulerian-Lagragian framework
~\citep{elghobashi1994predicting,balachandar2009scaling,balachandar2010turbulent,m2016point}.
Depending on their densities and the diameters, the particles with different
inertia respond to the multi-scale turbulent motions to form various particle clusters
~\citep{crowe1996numerical,balkovsky2001intermittent,salazar2008experimental},
which, in isotropic turbulence, can be characterized by the Stokes number based on
the Kolmogorov scale $St_K$, namely the ratio between the particle response time $\tau_p$ and 
the Kolmogorov time scale $\tau_K$~\citep{eaton1994preferential,goto2008sweep,bragg2014new}.

In incompressible canonical wall turbulence, such as the turbulent channels, pipes 
and boundary layers,
the non-homogeneity and anisotropy of the turbulent fluctuations due to the restriction of the wall
further complicate the motions and distributions of the particles.
Early-year experimental and numerical investigations on the one-way coupling between the 
wall-bounded turbulence and particles have found that the departure of the particles from 
the fluid with their increasing inertia can be illustrated by the statistics that 
the particle mean streamwise velocity is higher than that of the fluid in the near-wall region 
while lower in the outer region, leading to the much flatter profiles
~\citep{rashidi1990particle,eaton1994preferential}.
The cross-stream particle velocity fluctuations are correspondingly weaker, 
indicating their incapability of following the motions of the fluid~\citep{zhao2012stokes}.
From the perspective of their accumulation, it is found that the particles incline to move
toward the near-wall region and form streaky structures that resemble
the low-speed streaks in the buffer region~\citep{soldati2005particles,
soldati2009physics,balachandar2010turbulent,brandt2022particle}, 
the degree of which is the highest for particles with $St^+ \approx 10 \sim 80$
~\citep{soldati2009physics}, with $St^+$ the Stokes number under viscous scales.
~\citet{marchioli2002mechanisms} revealed that the vertical transport of the particles is highly
correlated with the sweeping and ejection events in the near-wall region.
The near-wall accumulation of the comparatively large particles should be attributed to 
the hindering of the rear ends of the quasi-streamwise vortices from the particles being 
transported away from the wall.
~\citet{picciotto2005characterization} confirmed that the particle Stokes number determines
the near-wall accumulation and clustering, and proposed that the streamwise 
and spanwise wall shear stress can be used to control the particle distribution.
~\citet{vinkovic2011direct} showed that only when the instantaneous Reynolds shear stress
exceeds a threshold, which scales approximately with the square root of the Stokes number, 
are the ejection events capable of bringing the particle upwards.
~\citet{sardina2012wall} found that the near-wall accumulation and the clustering
are closely linked with each other and that they are, in fact, the two aspects of the same processes.
They pointed out that the movement of particles toward the wall due to the turbophoretic drift
should be balanced by their accumulating within the regions of ejection to remain in 
the statistical steady state, forming the directional clusters along the streamwise direction.
~\citet{mortimer2019near} analyzed the particle dynamics utilizing the probability density function
and found that the particles with large Stokes numbers entering the viscous sublayer 
from the buffer region tend to retain their streamwise velocity and spatial organization,
leading to the higher streamwise velocity fluctuations of particles than those of the fluid.
In the outer region, however, the correlation between the particle concentration and 
the flow topology is comparatively weak~\citep{rouson2001preferential}.

The above-mentioned studies mainly concern the particle tranport by low Reynolds number 
channel flows.
The ever-advancing computational resources enable the investigation of the transport of 
the inertia particles at moderate to high Reynolds numbers, though still for incompressible flows.
It has been revealed that the particles at different inertia respond effectively to flow structures
with similar eddy turn-over time, `filtering' the smaller-scale turbulent motions.
This is obvious in turbulent plane Couette flows~\citep{bernardini2013effect}, 
which contain large-scale streamwise rollers even when at low Reynolds numbers.
Besides the highest level of particle clustering at $St^+ \approx 25$
that respond the most efficiently to the near-wall cycles, 
another mode of concentration emerges in the plane Couette flow with the spanwise scales 
identical to those of the large-scale rollers.
~\citet{bernardini2014reynolds} investigated the particle distribution with different inertia
at the friction Reynolds number up to 1000. Although the wall-normal concentration, 
and hence the turbophoretic drift, are independent of the Reynolds numbers, the deposition rates 
are higher with the increasing Reynolds number at the same Stokes number.
~\citet{jie2022existence} and \citet{motoori2022role} further confirmed that the particles with 
different inertia respond the most significantly to the flow structures with similar turn-over time
in wall turbulence.
Therefore, the multi-scale clustering of the particles can be characterized by the structure-based
Stokes number.
~\citet{berk2020transport} detected in high Reynolds number turbulent boundary layers that
the particles tend to accumulate inside the upward moving ejection events for the particles 
with a wide range of Stokes numbers, suggesting that
the clustering is probably a multi-scale phenomenon.
In the core region of the channel, \citet{jie2021effects} demonstrated that the inertial particles
accumulate more preferably within the high-speed regions in the quiescent core
~\citep{kwon2014quiescent} 
but avoid the vortical structures due to the centrifugal mechanism, whose boundaries function 
as barriers, hindering the transport of particles.

Compressible turbulent boundary layers laden with dilute phase of particles
can be encountered in such engineering applications as high-speed vehicles travelling through
the rain, ice crystals and other types of particles suspending in the atmosphere,
the ablation of the fuselage materials generating small particles transported downstream
by the high shear rate, and so on.
However, related studies are comparatively scanty.
In compressible isotropic turbulence, the particles are found to concentrate within the regions 
of low vorticity and high density which, due to the increasing compressibility effects, 
are weakened by the stronger shocklets~\citep{yang2014interactions,zhang2016preferential,
dai2017direct}.
The shocklets are also found to modify the probability density function of
the particle accelerations and lead to differences of statistics between the traces and bubbles
~\citep{wang2022acceleration}.
Similar phenomena are observed in compressible mixing layers~\citep{dai2018direct,dai2019direct},
but due to the existence of the mean shear, the particles also cluster within 
the low- or high-speed streaks, depending on their appearance on either the high- or low-speed side.
~\citet{xiao2020eulerian} investigated the particle behaviour in a spatially developed turbulent
boundary layer at the Mach number of $2$, where the near-wall accumulation and 
the clustering of the particles with the velocity streaks are also observed.
By analyzing the equation of the dilatation of particles, they found that the small particles
accumulate within the low-density regions, but the large particles within the low-density
regions close to the wall and high-density regions in the wake region, which is attributed to
the different centrifugal effects and the variation of the fluid density.
~\citet{buchta2019sound} revealed that the particle-turbulence interactions alter the local
pressure intensities, which are stronger with the mass loading near the subsonic region but weaker
near the supersonic regions. The former was attributed to the increasing time-rate-of-change of
fluid dilatation and the latter to the weaker turbulent kinetic energy.
~\citet{li2023particle} further studied the particle-laden compressible turbulent channel flows
incorporating the effects of gravity. It is found that the mean and fluctuating particle velocities
in the streamwise direction are increased but those in the cross-stream directions are decreased
due to the compressibility effects. Moreover, the particles are more likely to be clustered
within the ejections and sweeping events compared with those in incompressible flows.
\citet{capecelatro2024gas} reviewed the development of the force models of single
and multiple particles in compressible flows and the flow modulations due to the presence of 
solid particles.
This review concerned mostly the dilute suspensions of finite-sized particles using 
the particle-resolved DNS, especially the shock-particle interactions.

Although much is learned on the effects of particle inertia and Reynolds numbers in incompressible
wall turbulence and those of compressibility on the particle concentration in compressible
isotropic turbulence, there lacks a systematic study of the particle transport 
in compressible turbulent boundary layers at various Mach numbers.
This serves as the motivation of the present study. In this work, 
we perform direct numerical simulations (DNSs) of compressible turbulent boundary layers
laden with inertial particles at the free stream Mach number ranging from 2 to 6
to explore the effects of the Mach number on the behaviour of particles,
encompassing the near-wall accumulation, clustering with the velocity streaks, 
statistics and dynamics.
The conclusions will benefit the modelling of particle motions in engineering applications
of compressible turbulence.

The remainder of this paper is organized as follows.
The physical model and numerical methods utilized to perform the DNS are
introduced in section~\ref{sec:model}.
The features of the particle distribution, including the instantaneous distribution,
near-wall accumulation and clustering behaviour, are discussed in section~\ref{sec:dist}.
The mean and fluctuating velocity and acceleration are presented in section~\ref{sec:pardyn}.
Lastly, the conclusions are summarized in section~\ref{sec:con}.

\section{Physical model and numerical methods} \label{sec:model}

In the present study, we perform DNS of the compressible turbulent boundary layers 
with particles utilizing the Eulerian-Lagrangian point particle method.
The physical model and numerical methods for the fluid and particles will be introduced
in the following subsections.

\subsection{Simulation of supersonic turbulent boundary layers}

We firstly introduce the governing equations and the numerical settings for the fluid phase.
The compressible turbulent boundary layer flows are governed by the three-dimensional 
Navier-Stokes equations for Newtonian perfect gases, cast as
\begin{equation}
\frac{\partial \rho}{\partial t} +\frac{\partial (\rho u_i)}{\partial x_i} = 0,
\end{equation}
\begin{equation}
\frac{\partial (\rho u_i)}{\partial t} +\frac{\partial (\rho u_i u_j)}{\partial x_j} = 
-\frac{\partial p}{\partial x_i} + \frac{\partial \tau_{ij}}{\partial x_j},
\end{equation}
\begin{equation}
\frac{\partial (\rho E)}{\partial t} +\frac{\partial (\rho E u_j)}{\partial x_j} = 
-\frac{\partial (p u_j)}{\partial x_j} + \frac{\partial (\tau_{ij} u_i)}{\partial x_j}
-\frac{\partial q_j}{\partial x_j},
\end{equation}
where $u_i$ is the velocity component of the fluid in the $x_i$ direction,
with $i=1$,2,3 (also $x$, $y$ and $z$) denoting the streamwise, wall-normal and spanwise directions.
The $\rho$ is density, $p$ is pressure and $E$ is total energy, 
related by the following state equations of perfect gases
\begin{equation}
p=\rho R T,~~ E=C_v T + \frac{1}{2} u_i u_i,
\end{equation}
where $T$ is temperature, $R$ is the gas constant and $C_v$ the constant volume specific heat.
The viscous stresses and the molecular heat conductivity are determined by the following
constitutive equation for Newtonian fluids and Fourier's law, respectively,
\begin{equation}
\tau_{ij}= \mu \left( \frac{\partial u_i}{\partial x_j}+\frac{\partial u_j}{\partial x_i} \right)
-\frac{2}{3} \mu \frac{\partial u_k}{\partial x_k} \delta_{ij},~~
q_j = - \kappa \frac{\partial T}{\partial x_j}
\end{equation}
with $\mu$ the viscosity,  calculated by Sutherland's law,
and $\kappa$ the molecular heat conductivity, determined as $\kappa = C_p \mu /Pr$.

DNS is performed utilizing the open-source code `STREAMs' developed by~\citet{bernardini2021streams},
using the finite difference method to solve the governing equations. 
The convective terms are approximated by the sixth-order kinetic energy preserving scheme
~\citep{kennedy2008reduced,pirozzoli2010generalized}
in the smooth region and the fifth-order weighted-essentially-non-oscillation (WENO) scheme
~\citep{shu1988efficient} when the flow discontinuity is detected by the Ducro's sensor
~\citep{ducros1999large}.
The viscous terms are approximated by the sixth-order central scheme.
The low-storage third-order Runge-Kutta scheme is adopted for time advancement
~\citep{wray1990minimal}.

The boundary conditions are given as follows.
The turbulent inlet is composed of the mean flow obtained by empirical formulas
~\citep{musker1979explicit} and the synthetic turbulent fluctuations using the
digital filtering method~\citep{klein2003digital}.
The no-reflecting conditions are enforced at the upper and the outlet boundaries.
Periodic conditions are adopted in the spanwise direction.
The no-slip condition for velocity and the isothermal condition for temperature are given 
at the lower wall.

Herein, the free stream flow parameters are denoted by the subscript $\infty$.
The free stream Mach number is defined as the ratio between the free stream velocity and
the sound speed $M_\infty = U_\infty / \sqrt{\gamma R T_\infty}$,
and the free stream Reynolds number is defined as 
$Re_\infty = \rho_\infty U_\infty \delta_0/\mu_\infty$,
with $\delta_0$ the nominal boundary layer thickness at the turbulent inlet.
The ensemble average of a generic flow quantity $\varphi$ is marked as $\bar \varphi$
and the corresponding fluctuations as $\varphi'$, and the density-weighted average (Favre average)
as $\tilde \varphi$ and the corresponding fluctuations as $\varphi''$.
The viscous scales are defined based on the mean shear stress $\tau_w$, viscosity $\mu_w$, 
density $\rho_w$ at the wall, resulting in the friction velocity $u_\tau =\sqrt{\tau_w/\rho_w}$,
viscous length scale $\delta_\nu = \mu_w / (\rho_w u_\tau)$
and the friction Reynolds number $Re_\tau = \rho_w u_\tau \delta/\mu_w$ 
($\delta$ the nominal boundary layer thickness at a given streamwise station).
The flow quantities normalized by these viscous scales are denoted by the superscript $+$.
Local viscous scales are defined accordingly by substituting the wall density $\rho_w$ and
wall viscosity $\mu_w$ with their local mean values, and the flow quantities normalized by
these local viscous scales are marked by the superscript $*$.

The streamwise, wall-normal and spanwise sizes of the computational domain are set as
$L_1=80 \delta_0$, $L_2 = 9 \delta_0$ and $L_3 = 8 \delta_0$, discretized by $2400$, $280$ 
and $280$ grids, respectively.
The grids are uniformly distributed in the streamwise and spanwise directions and stretched
by hyperbolic sine function in the wall-normal direction with $240$ grids clustered below
$y= 2.5 \delta_0$.
At the inlet, the friction Reynolds number is set as $Re_{\tau 0} = 200$, according to which
the free stream Reynolds number $Re_\infty$ is calculated.
The streamwise and spanwise grid intervals defined based on the viscous scales at the turbulent 
inlet are $\Delta x^+ = 6.67$ and $\Delta z^+ = 5.71$. The first grid off the wall is located at 
$\Delta y^+_w=0.5$ and the grid interval in the free stream is $\Delta y^+=7.06$.
Such grid intervals are sufficient with the presently used low-dissipative numerical schemes
~\citep{pirozzoli2011numerical}.

The free stream Mach numbers of the turbulent boundary layers are set as $M_\infty = 2$, 4 and 6.
The wall temperatures $T_w$ are set to be constant values equal to the recovery temperature 
$T_r = T_\infty (1+(\gamma-1) r M^2_\infty)/2$, the mean wall temperature with adiabatic conditions
at the given Mach number, with $\gamma=1.4$ the specific heat ratio, 
$r=Pr^{1/2}$ the recovery factor, and $Pr=0.71$ the Prandtl number.
The streamwise variation of the nominal ($Re_\delta$), displacement ($Re_{\delta^*}$), 
momentum ($Re_\theta$) and the friction Reynolds numbers ($Re_\tau$) in the statistically
equilibrium regions are listed in Table~\ref{tab:flowparam}.
Although the ranges of $Re_\delta$, $Re_{\delta^*}$ and $Re_\theta$ are different at various
Mach numbers, the friction Reynolds numbers $Re_\tau$ are of the same level,
which is the most important parameter in wall-bounded turbulence.
Hereinafter, the results reported are obtained within the streamwise range of 
$(60 \sim 70) \delta_0$ and the time span of $240 \delta_0/U_\infty$ to obtain converged statistics.
The data are not collected until the simulations have been run for $1000 \delta_0/U_\infty$, 
during which the turbulent flows are fully developed and the distributions of particles 
in the wall-normal direction reach steady states, as we will demonstrate subsequently.

\begin{table}
\centering
\begin{tabular}{ccccccc}
Flow Case & $M_\infty$  & $T_w/T_\infty$ & $Re_\delta$  &  $Re_{\delta^*}$  &  $Re_\theta$  &  $Re_\tau$  \\ 
M2   &  $2.0$  & $1.712$ & $13098 \sim 24879$ & $4460 \sim 8162$ & $1543 \sim 2947$ & $285 \sim 504$ \\
M4   &  $4.0$  & $3.848$ & $56805 \sim 87778$ & $28655 \sim 43108$ & $5795 \sim 8977$ & $322 \sim 499$ \\
M6   &  $6.0$  & $7.408$ & $153112 \sim 240250$ & $94000 \sim 128679$ & $12800 \sim 18712$ & $316 \sim 462$ \\ 
\end{tabular}
\caption{Flow parameters. Here, $Re_\delta$, $Re_\delta^*$ and $Re_\theta$ are the Reynolds numbers
defined by the nominal ($\delta$), displacement ($\delta^*$) and momentum ($\theta$) 
boundary layer thicknesses.}
\label{tab:flowparam}
\end{table}

\begin{figure}
\centering
\begin{overpic}[width=0.5\textwidth]{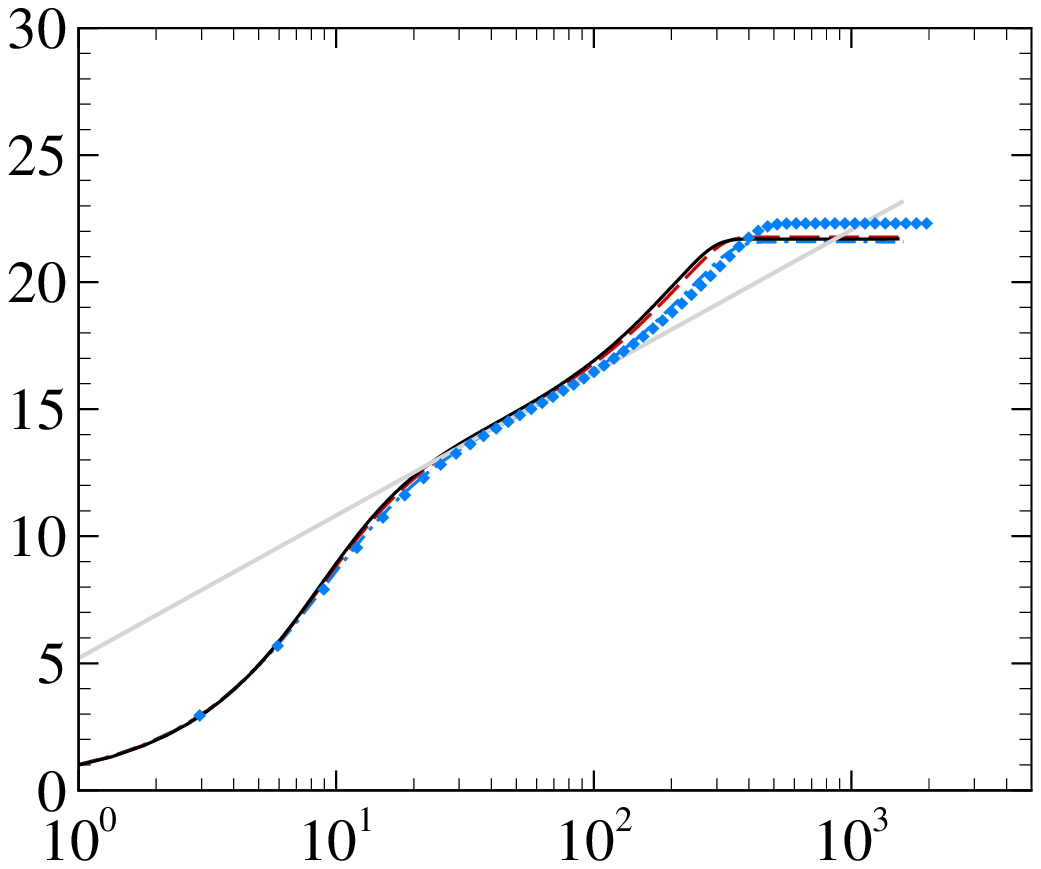}
\put(0,70){(a)}
\put(50,0){$y^+$}
\put(-1,35){\rotatebox{90}{$u^+_{1,VD}$}}
\end{overpic}~
\begin{overpic}[width=0.5\textwidth]{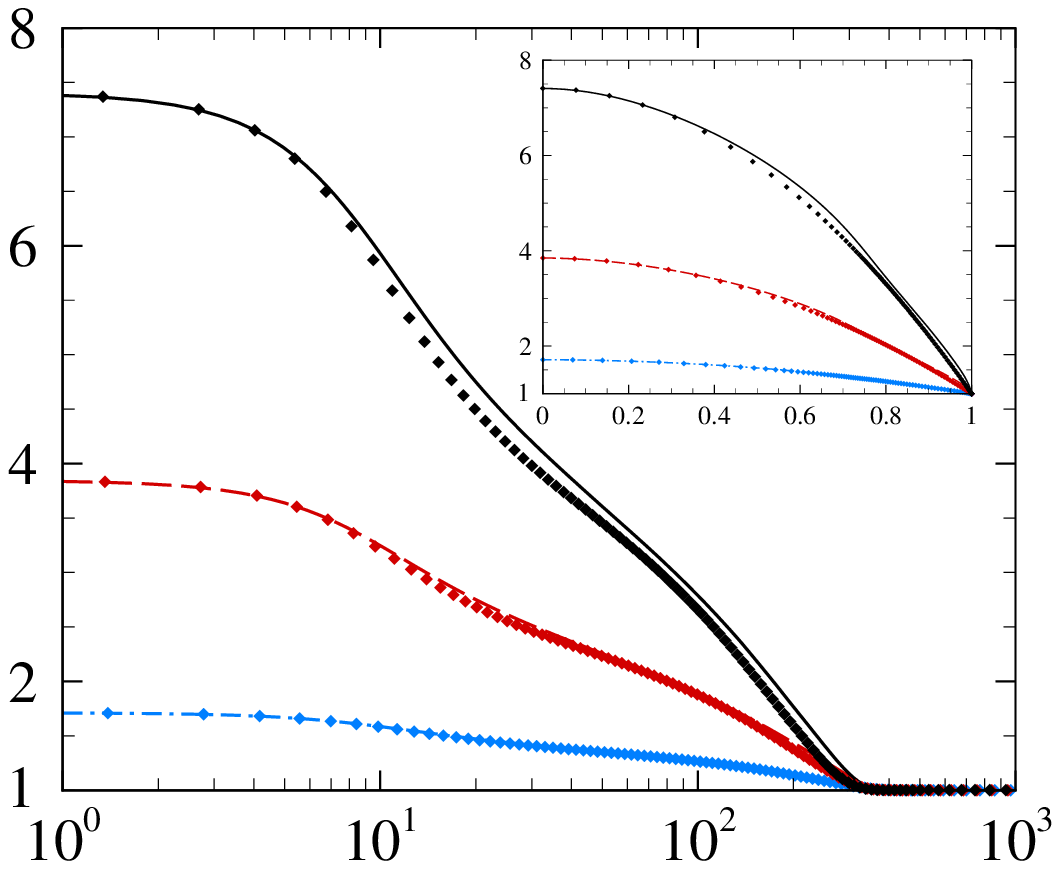}
\put(0,70){(b)}
\put(50,0){$y^+$}
\put(0,35){\rotatebox{90}{$\bar T/\bar T_\infty$}}
\put(65,38){\tiny $\bar u/U_\infty$}
\put(45,50){\tiny \rotatebox{90}{$\bar T/\bar T_\infty$}}
\end{overpic}\\[1.0ex]
\begin{overpic}[width=0.5\textwidth]{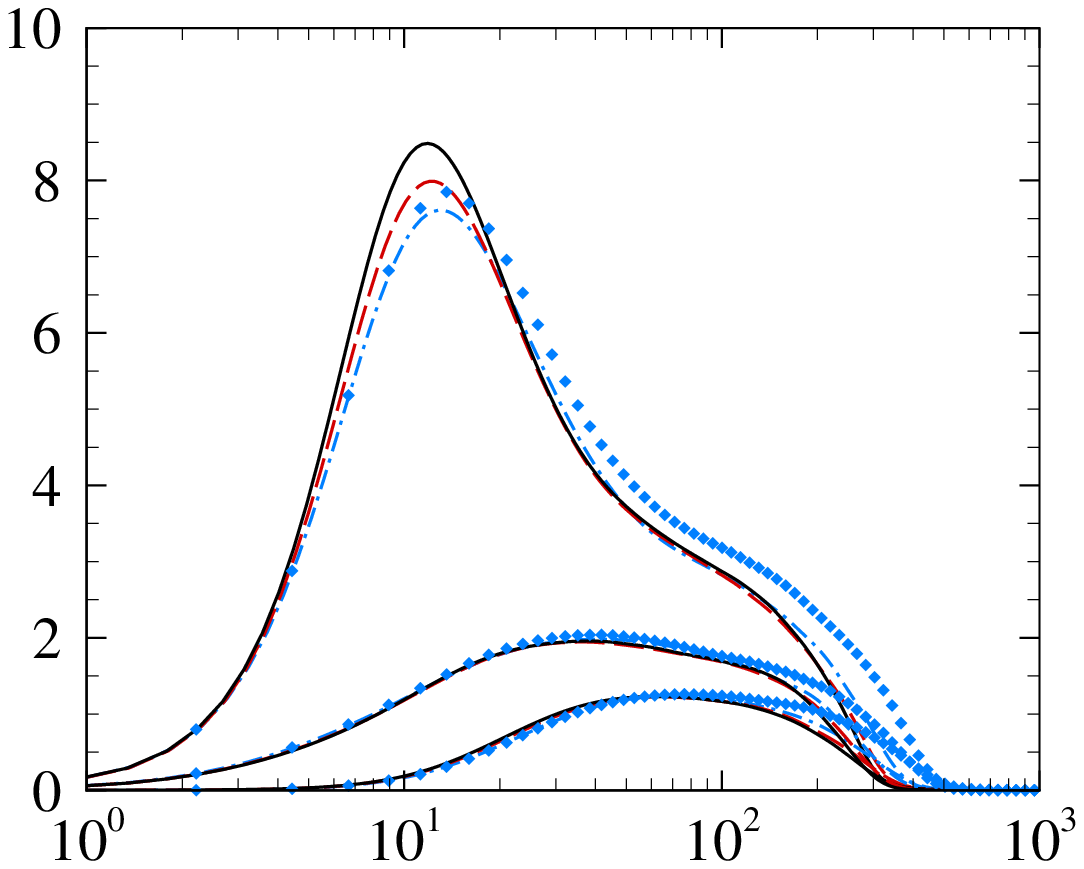}
\put(0,70){(c)}
\put(50,0){$y^+$}
\put(0,35){\rotatebox{90}{$R^+_{ij}$}}
\put(25,58){\small $R^+_{11}$}
\put(30,20){\small $R^+_{33}$}
\put(55,13){\small $R^+_{22}$}
\end{overpic}~
\begin{overpic}[width=0.5\textwidth]{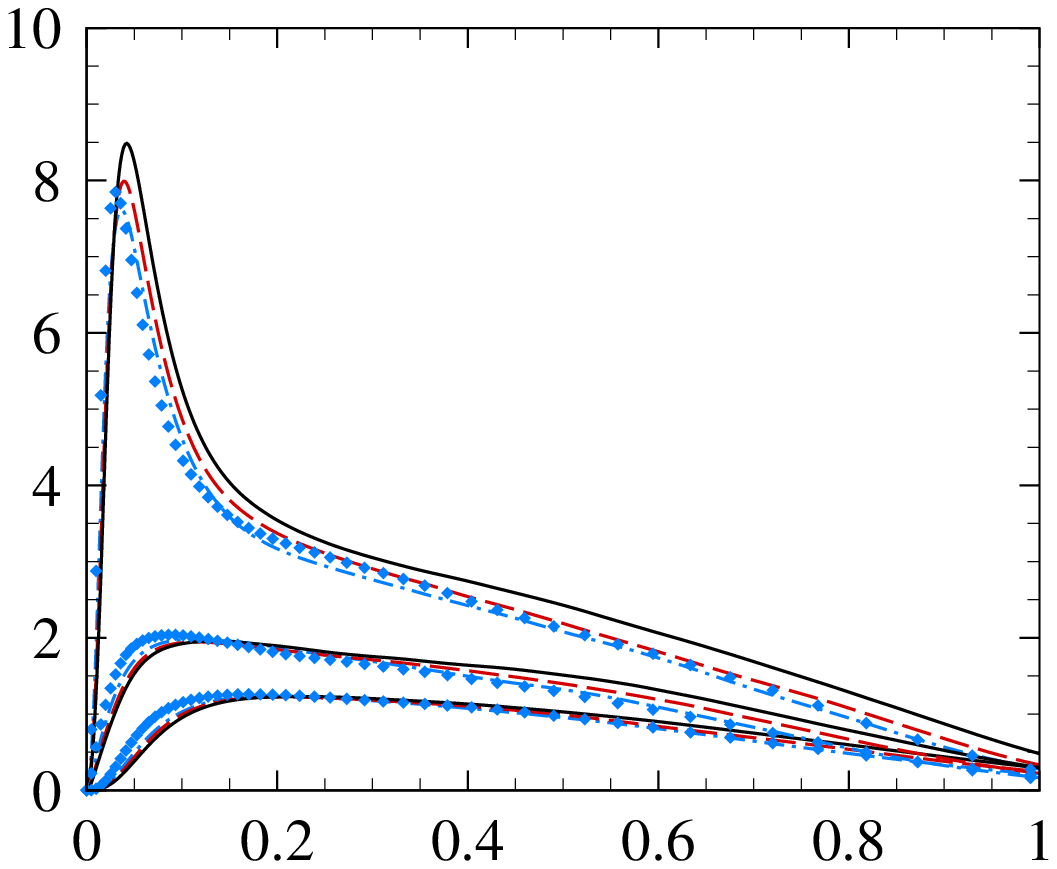}
\put(0,70){(d)}
\put(50,0){$y/\delta$}
\put(0,35){\rotatebox{90}{$R^+_{ij}$}}
\put(23,58){\small $R^+_{11}$}
\put(20,25){\small $R^+_{33}$}
\put(30,13){\small $R^+_{22}$}
\end{overpic}\\
\caption{Wall-normal distributions of (a) van Driest transformed mean velocity $u^+_{1,VD}$, 
(b) mean temperature $\bar T/T_\infty$, (c,d) Reynolds stresses $R^+_{ij}$.
Blue dash-dotted lines: case M2, red dashed lines: case M4, black solid lines: case M6.
Grey lines in (a): logarithmic law $2.44 \ln(y^+) + 5.2$.
Symbols in (a,c,d): reference data from~\citet{bernardini2011wall} 
at $M_\infty=2$ at $Re_\tau \approx 450$, symbols in (b): mean temperature obtained by GRA.}
\label{fig:meanprof}
\end{figure}

In figure~\ref{fig:meanprof}(a) we present the van Driest transformed mean velocity obtained by
the following integration
\begin{equation}
u^+_{1,VD} = \frac{1}{u_\tau} \int^{\bar u_1}_0 \left( \frac{\bar \rho}{\bar \rho_w} \right)^{1/2} 
{\rm d} u_1,
\end{equation}
along with those reported by~\citet{bernardini2011wall} at $M_\infty =2$ and $Re_\tau \approx 450$
for comparison.
For all the cases considered, the van Driest transformed mean velocity profiles 
are well-collapsed and consistent with the reference data below $y^+=100$, obeying the linear law
in the viscous sublayer and the logarithmic law in the log region (if any).
The disparity in the wake region can be ascribed to the slightly lower friction Reynolds number
$Re_\tau$ in the present study.
The mean temperatures $\bar T$ are shown in figure~\ref{fig:meanprof}(b), manifesting zero gradients
close to the wall due to the quasi-adiabatic thermal condition and monotonic decrement as 
it approaches the outer edge of the boundary layer.
The mean temperature profiles are well-collapsed with the generalized Reynolds analogy (GRA)
proposed by~\citet{zhang2014generalized} that relates the mean temperature and the mean velocity,
\begin{equation}
\frac{\bar T}{T_\infty} = \frac{T_w}{T_\infty} + \frac{T_{rg}-T_w}{T_\infty} \frac{\bar u}{U_\infty}
+\frac{T_\infty-T_{rg}}{T_\infty} \left( \frac{\bar u}{U_\infty} \right)^2,
\label{eqn:gra}
\end{equation}
where $T_{rg}$ is the generalized total temperature
\begin{equation}
T_{rg} = T_\infty + r_g \frac{U^2_\infty}{2C_p},
\end{equation}
with $r_g$ the generalized recovery coefficient.

Figures~\ref{fig:meanprof}(c,d) display the distribution of the Reynolds stresses
$R^+_{ij} = \overline{\rho u''_i u''_j}/\tau_w$ for the three cases,
amongst which those of case M2 show agreement with the reference data reported by
~\citet{bernardini2011wall} in the near-wall region when plotted against the viscous coordinate 
(figure~\ref{fig:meanprof}(c)) and also in the outer region when plotted against 
the global coordinate (figure~\ref{fig:meanprof}(d)), except for the slightly lower peaks of 
$R^+_{11}$ due to the lower Reynolds number.
The Reynolds stresses are weakly dependent on the Mach number, showing a slight increment of
the inner peaks with the Mach number.
This is consistent with the previous studies of~\citet{cogo2022direct} and~\citet{yu2019genuine}.

\subsection{Particle simulations in Lagragian framework}

The dispersed phase considered herein is composed of heavy spherical particles with infinitesimal
volume and mass fractions, thereby disregarding the inter-particle collisions and 
the feedback effects from the particles to the fluid~\citep{m2016point}.
In the one-way coupling approximation,
the trajectories and the motions of the spherical particles are solved by the following equations
\begin{equation}
\frac{{\rm d} r_{p,i}}{{\rm d}t} = v_i, ~~
\frac{{\rm d} v_{i}}{{\rm d}t} = a_i,
\end{equation}
with $r_{p,i}$, $v_i$ and $a_i$ denoting the particle position, velocity and acceleration.
We consider merely the Stokes drag force induced by the slip velocity while neglecting the other
components such as the lift force, Basset history force and virtual mass force due to the large
density of particles compared with that of the fluid
~\citep{maxey1983equation,armenio2001importance,mortimer2019near}, cast as
\begin{equation}
a_i = \frac{f_{D}}{\tau_p}(u_i-v_i)
\label{eqn:pforce}
\end{equation}
where the $\tau_p = \rho_p d^2_p/18 \mu$ is the particle relaxation time, 
$d_p$ is the particle diameter, $\rho_p$ is the particle density.
The coefficient $f_D$ is estimated as
\begin{equation}
f_D = (1+0.15 Re^{0.687}_p) H_M,
\label{eqn:cd}
\end{equation}
\begin{equation}
\begin{aligned}
H_M =
\begin{cases}
      0.0239 M^3_p + 0.212 M^2_p - 0.074 M_p + 1,~~~ & M_p \le 1 \\
      0.93 + \frac{1}{3.5 + M^5_p}, ~~~ & M_p >1   
\end{cases}
\end{aligned}
\label{eqn:hm}
\end{equation}
incorporating the compressibility effects, as suggested by~\citet{loth2021supersonic},
with $Re_p = \rho |u_i - v_i| d_p/\mu$ and $M_p = |u_i - v_i|/\sqrt{\gamma R T}$
being the particle Reynolds and Mach numbers, respectively.
Note that we have neglected the rarefied effects,
for according to a preliminary estimation and the direct numerical simulation results,
particle Knudsen numbers rarely reach higher than 0.01, the criterion where the rarefied effects
should be taken into account.
The high Reynolds number modification incorporating the effects of the fully turbulent wake
~\citep{clift1971motion}
has also been disregarded, for the particle Reynolds numbers are always lower than $100$.

We adopt the same strategy in the time advancement of the particle equations as the fluid phase
using the third-order low-storage Runge-Kutta scheme.
Trilinear interpolation is used to obtain the information of the fluid at the particle position
~\citep{eaton2009two,bernardini2014reynolds}.
We performed DNS of particles in turbulent channel flows for validation,
for details please refer to Appendix~\ref{sec:app}.
The initial positions of the particles in the simulation are randomly distributed within 
$y=2 \delta_0$, with their initial velocities set to be the same as those of the fluid.
The perfectly elastic collisions are assumed when the particles hit the wall.
Periodic conditions are imposed as in the fluid phase in the spanwise direction.
When the particles pass through the upper boundary and the flow outlet, they are recycled
to the flow inlet with a random wall-normal and spanwise position below $y=\delta_0$ 
and the same velocity
as that of the fluid at that location so as to retain the total number of particles.
As we have mentioned in the last subsection, the flow statistics reported hereinafter
are obtained within the range of $(60 \sim 70) \delta_0$, 
where the particle fields are considered to be fully developed, with the validation 
presented in Appendix~\ref{sec:stma}.

\begin{table}
\centering
\small
\begin{tabular}{ccc|ccc|ccc|ccc}
\multirow{2}*{Type} & \multirow{2}*{$d_p/\delta_0$}  &  \multirow{2}*{$\rho_p/\rho_\infty$} &
\multicolumn{3}{c|}{$St_\infty$}  &
\multicolumn{3}{c|}{$St^+$}  &  \multicolumn{3}{c}{$St_{K,w}$} \\ 
&  &  & M2 & M4 & M6 & M2 & M4 & M6 & M2 & M4 & M6 \\  \hline 
P1 & 0.001 & 20   &  0.011 & 0.041 & 0.11  & 0.065 & 0.14 & 0.27 &  0.031 & 0.068 & 0.12 \\
P2 & 0.001 & 100  &  0.056 & 0.20  & 0.55  & 0.33 & 0.70 & 1.33 & 0.15 & 0.34 & 0.63 \\
P3 & 0.001 & 500  &  0.28  & 1.01  & 2.74  & 1.63 & 3.52 & 6.63 & 0.77 & 1.70 & 3.17 \\
P4 & 0.001 & 2500 &  1.40  & 5.07  & 13.70 & 8.15 & 17.60 & 33.15 & 3.86 & 8.50 & 15.85 \\
P5 & 0.001 & 10000 & 5.60  & 20.28 & 54.78 & 32.59 & 70.44 & 132.59 & 15.46 & 34.00 & 63.41 \\
P6 & 0.001 & 20000 & 11.21 & 40.56 & 109.56 & 65.18 & 140.81 & 265.19 & 30.91 & 67.99 & 126.83 \\
P7 & 0.001 & 50000 & 28.02 & 101.39 & 273.90 & 162.94 & 352.21 & 662.97 & 77.28 & 169.98 & 317.07 \\
\end{tabular}
\caption{Particle diameters, density ratio and Stokes numbers.}
\label{tab:partparam}
\end{table}

\begin{figure}
\centering
\begin{overpic}[width=0.5\textwidth]{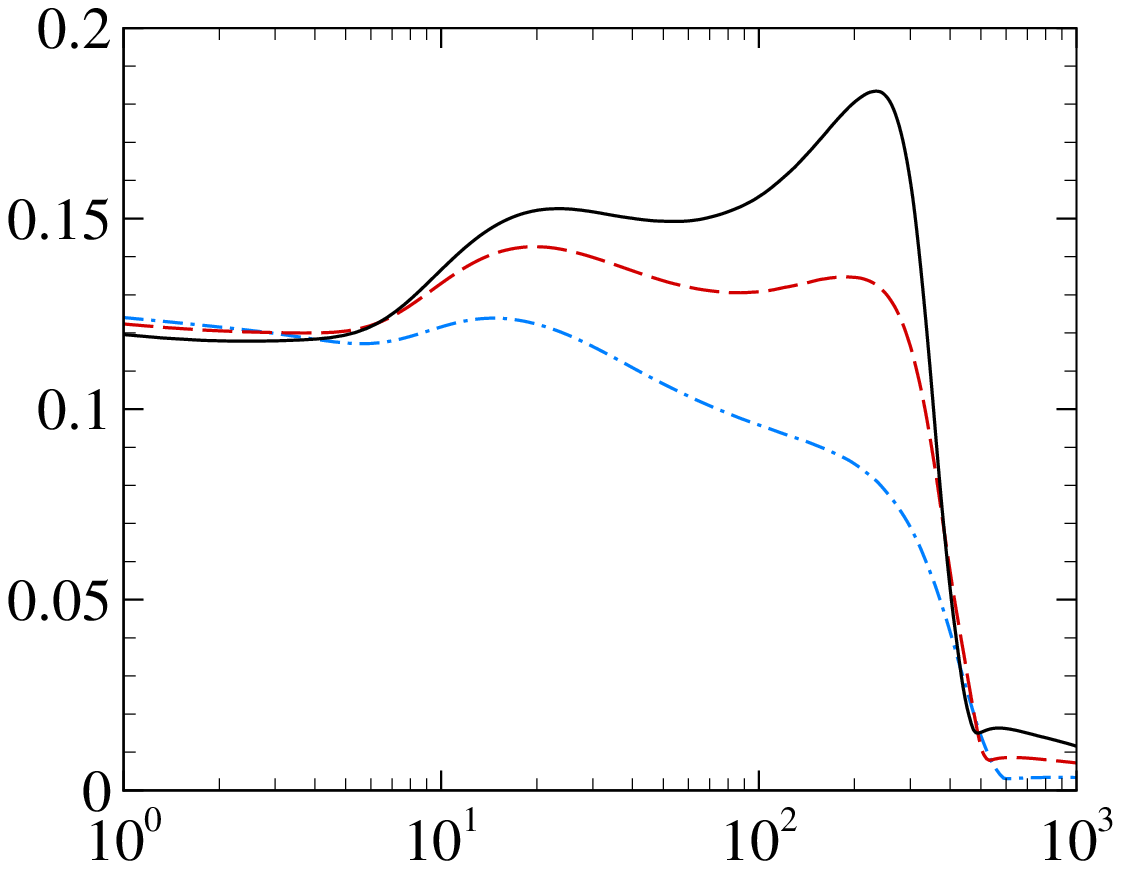}
\put(0,70){(a)}
\put(50,0.5){$y^+$}
\put(0,35){\rotatebox{90}{$d_p/\eta$}}
\end{overpic}~
\begin{overpic}[width=0.5\textwidth]{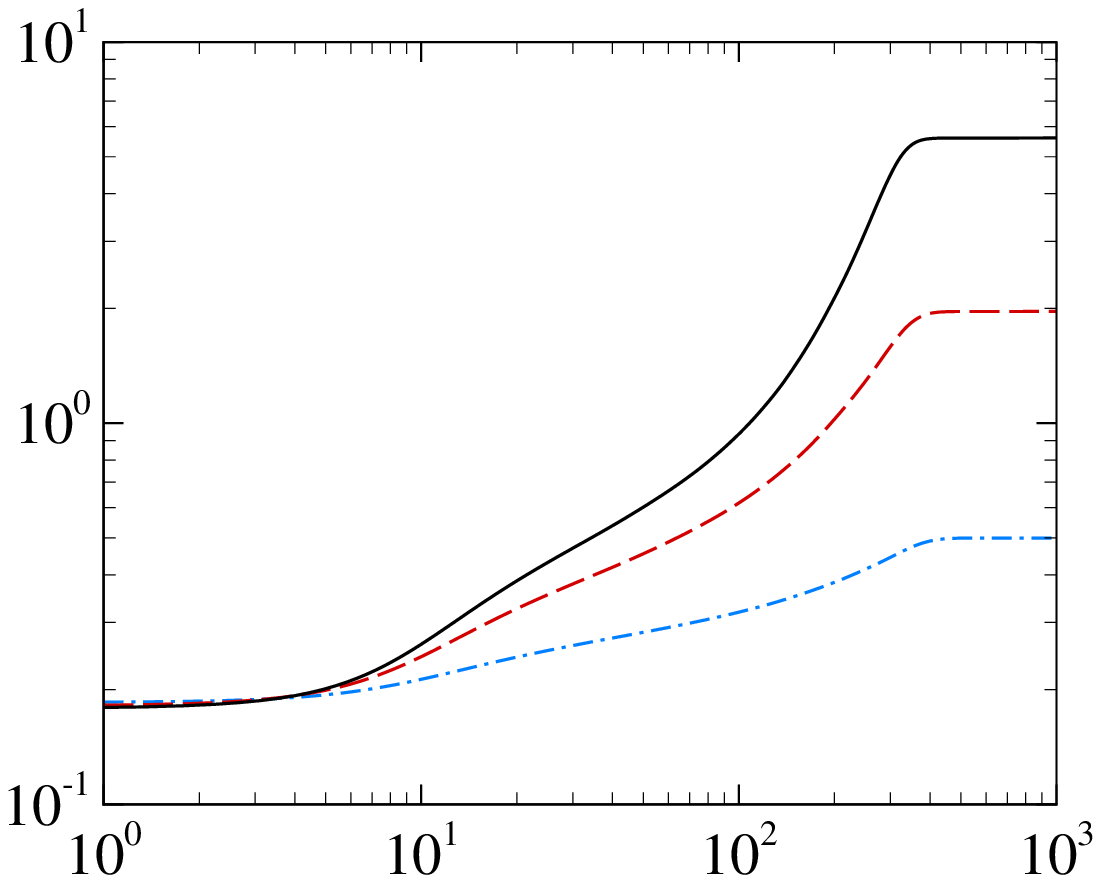}
\put(0,70){(b)}
\put(50,0.5){$y^+$}
\put(0,35){\rotatebox{90}{$d^*_p$}}
\end{overpic}\\
\begin{overpic}[width=0.5\textwidth]{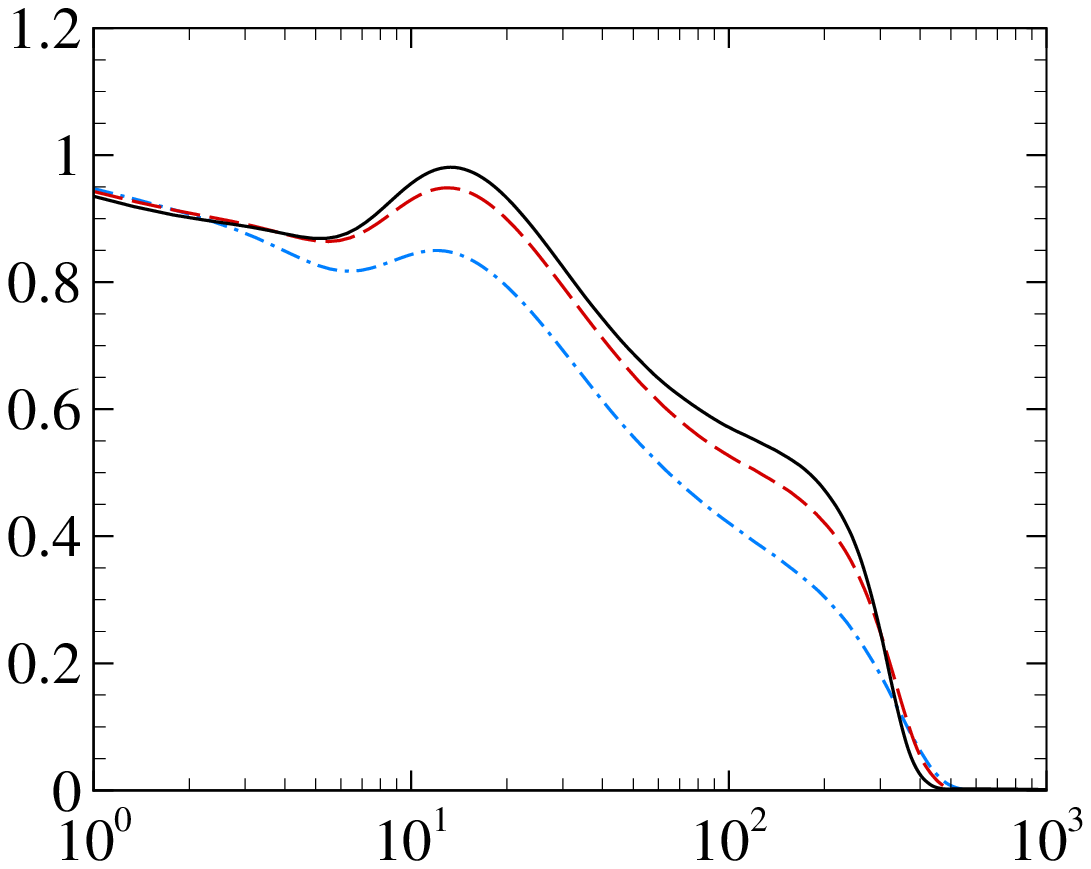}
\put(0,70){(c)}
\put(50,0.5){$y^+$}
\put(-1,30){\rotatebox{90}{$St_K/St_{Kw}$}}
\end{overpic}~
\begin{overpic}[width=0.5\textwidth]{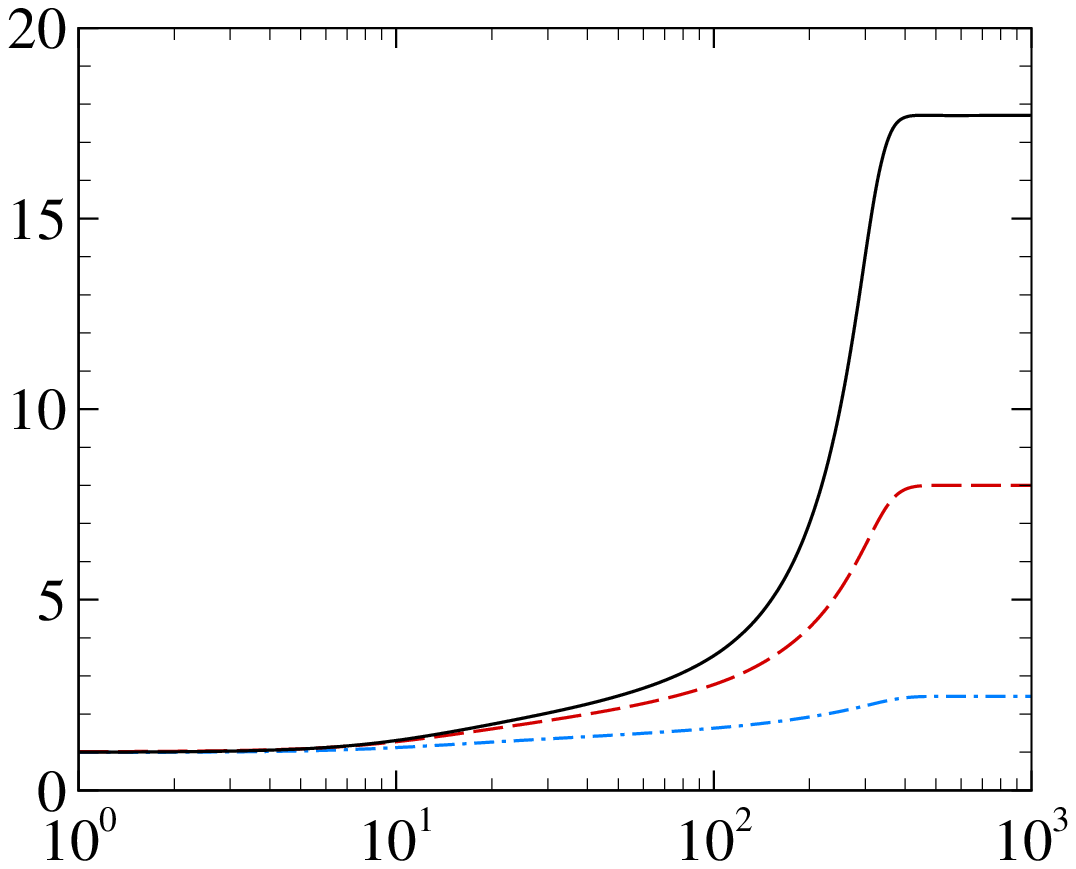}
\put(0,70){(d)}
\put(50,0.5){$y^+$}
\put(0,35){\rotatebox{90}{$St^*/St^+$}}
\end{overpic}\\
\caption{Wall-normal distribution of the particle diameter (a) $d_p/\eta$ and (b) $d^*_p$,
and the relative particle Stokes number (c) $St_K/St_{Kw}$ and (d) $St^*/St^+$.
Blue dash-dotted lines: case M2, red dashed lines: case M4, black solid lines: case M6.}
\label{fig:psize}
\end{figure}

The dispersed phase contains seven particle populations with the same particle numbers $N_p = 10^6$.
As reported in Table~\ref{tab:partparam}, these particle populations share the identical diameter
of $d_p=0.001 \delta_0$ and various density ratios $\rho_p / \rho_\infty$ ranging from 20 to 50000.
Within the nominal boundary layer thickness, the volume fraction of the dispersed particle phase 
is approximately $ 4\times 10^{-7}$, and the mass fraction ranges from 
$8 \times 10^{-6}$ to $0.02$, suggesting the appropriateness of the one-way coupling approximation.
In figure~\ref{fig:psize}(a) we present the ratios between the particle diameter and 
the Kolmogorov scale $d_p/\eta$, which are lower than $1.0$ across the boundary layer,
satisfying the requirements of the point-particle approach~\citep{maxey1983equation,m2016point}.
When normalized by the local viscous length scale, as shown in figure~\ref{fig:psize}(b),
the particle diameters $d^*_p$ are approximately $0.185$ at the wall and increase monotonically
toward the edge of the boundary layer, with the highest value of $4.734$ at $M_\infty=6$.

The different particle density leads to the disparity in the particle relaxation time $\tau_p$.
In Table~\ref{tab:partparam} we list three sets of particle Stokes numbers, 
\begin{equation}
St_\infty = \tau_p / \tau_\infty, ~~ St^+ = \tau_p / \tau_\nu, ~~ St_{K,w} = \tau_p / \tau_{K,w}
\end{equation}
representing the ratio between the particle relaxation time $\tau_p$ and the characteristic 
time scale of the flow under the outer scale $\tau_\infty = \delta_0 / U_\infty$, 
under the viscous scale $\tau_\nu = \delta_\nu / u_\tau$ and under the Kolmogorov scale at the wall
$\tau_{K,w} = (\mu_w/\varepsilon_w)^{1/2}$.
Amongst these parameters, the $St_{K,w}$ is a crucial parameter in isotropic turbulence,
determining the features of the particle dynamics and accumulation~\citep{balachandar2010turbulent,
brandt2022particle}, whereas the $St^+$ is commonly used in wall turbulence.
For particles in incompressible wall turbulence without viscosity stratification, 
both $\tau_p$ and $\tau_\nu$ are constant.
However, this is not the case for the presently considered flow, for the viscosity $\mu$ is 
a function of temperature, complicating the evaluation of the particle inertia.
In figure~\ref{fig:psize}(c,d) we present the distribution of $St_K$ and $St^+$ compared with 
their values at the wall.
Due to the lower dissipation, the $St_K$ decreases monotonically away from the wall,
showing weak Mach number dependence.
The $St^*$ values, on the other hand, increase with the wall-normal coordinate, 
and escalate with the Mach number, indicating that the response of the particle to 
the turbulent fluctuation of the fluid flow is varying across the boundary layer
due to the stratification of the mean density and viscosity, especially near the edge of
the boundary layer.

\section{Instantaneous and statistical particle distributions}  \label{sec:dist}

\subsection{Instantaneous particle distribution}

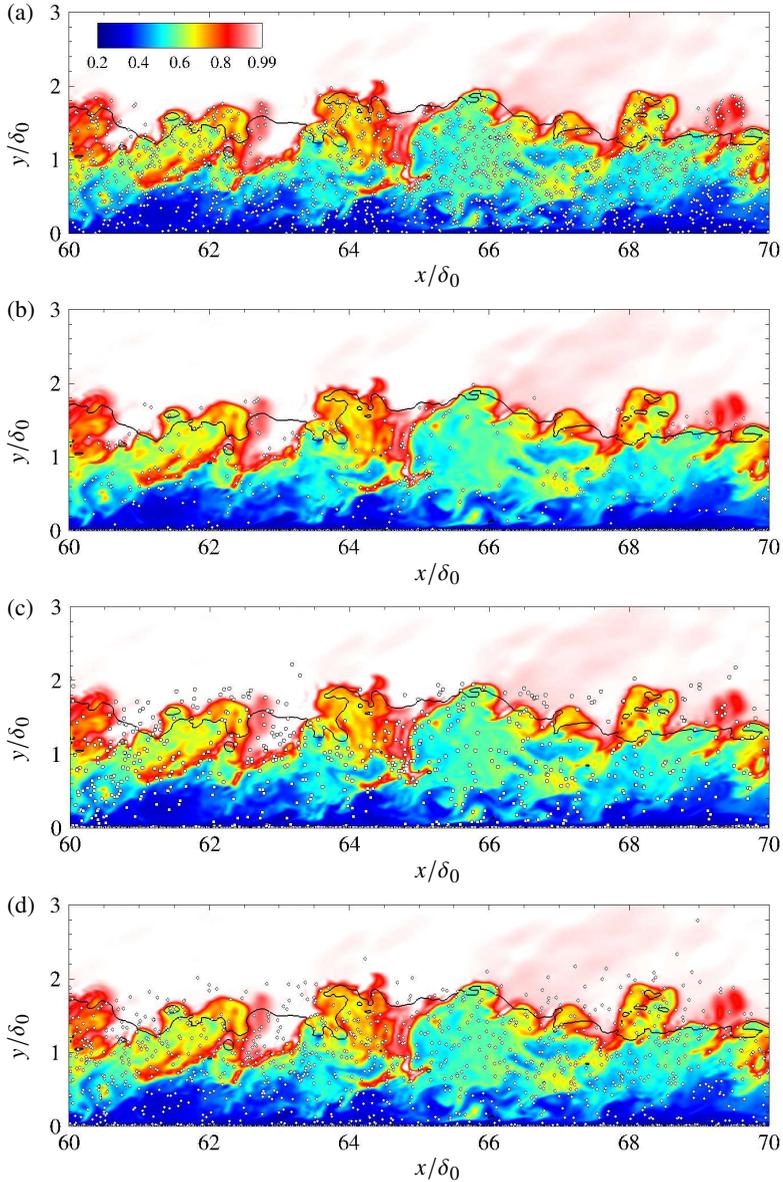
\begin{figure}
\centering
\begin{overpic}[width=0.8\textwidth,trim={0.2cm 0.2cm 0.2cm 0.2cm},clip]{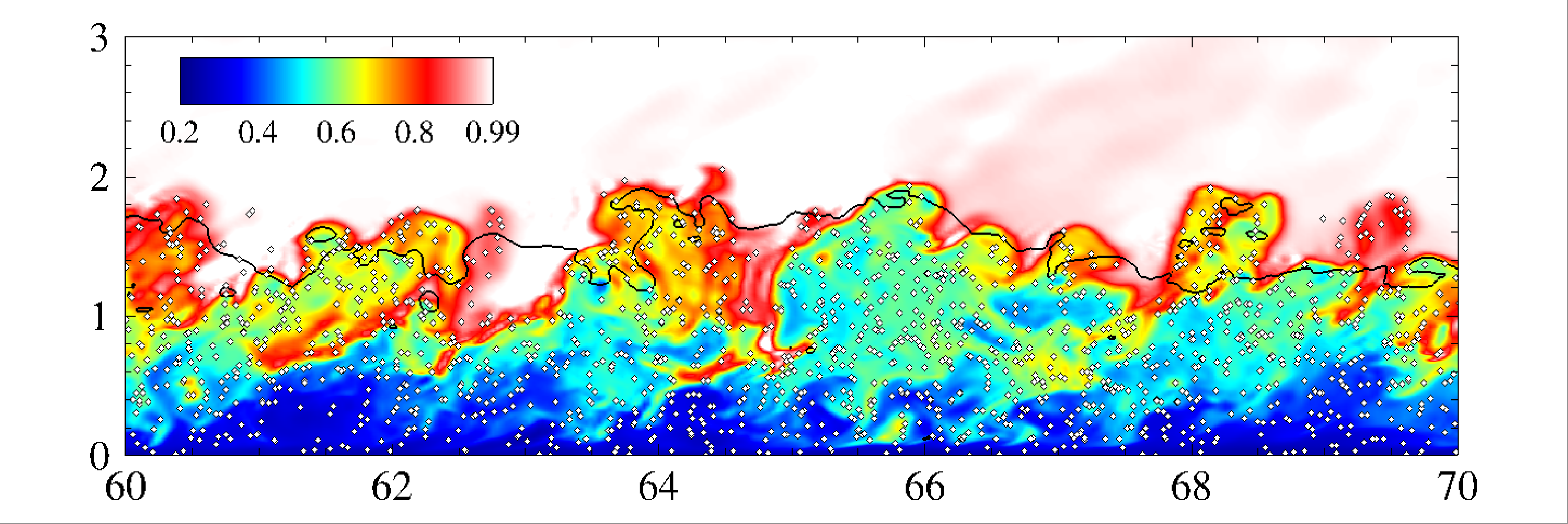}
\put(0,30){(a)}
\put(50,-2){$x/\delta_0$}
\put(0,12){\rotatebox{90}{$y/\delta_0$}}
\end{overpic}\\[3.0ex]
\begin{overpic}[width=0.8\textwidth,trim={0.2cm 0.2cm 0.2cm 0.2cm},clip]{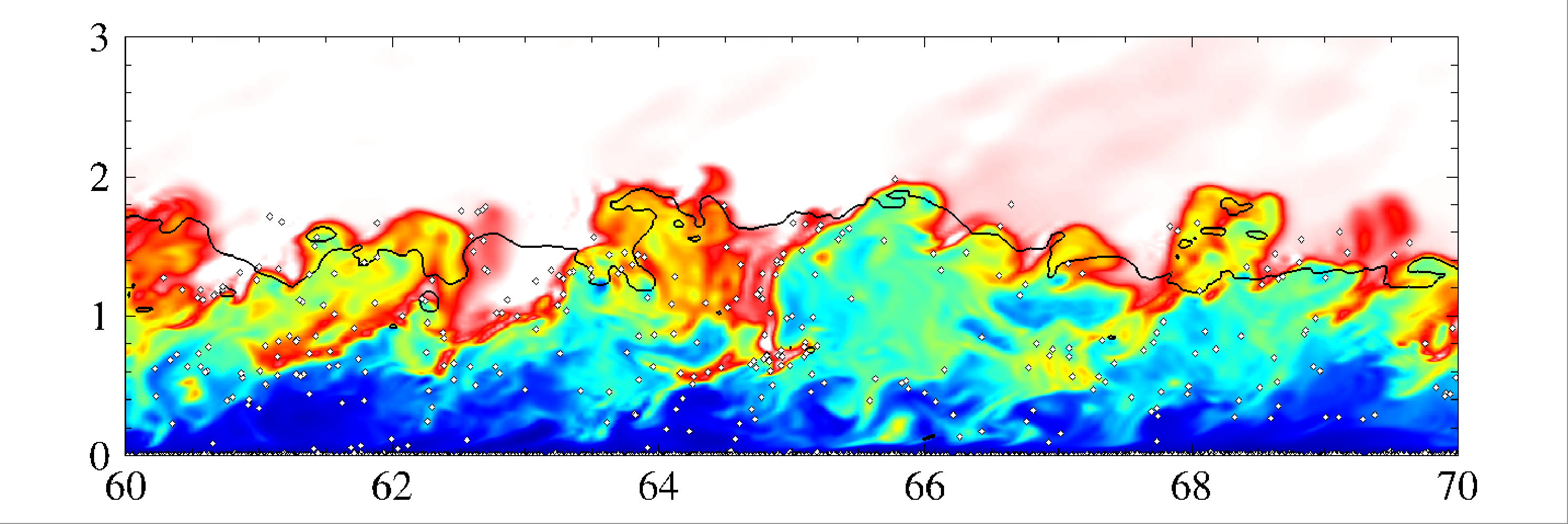}
\put(0,30){(b)}
\put(50,-2){$x/\delta_0$}
\put(0,12){\rotatebox{90}{$y/\delta_0$}}
\end{overpic}\\[3.0ex]
\begin{overpic}[width=0.8\textwidth,trim={0.2cm 0.2cm 0.2cm 0.2cm},clip]{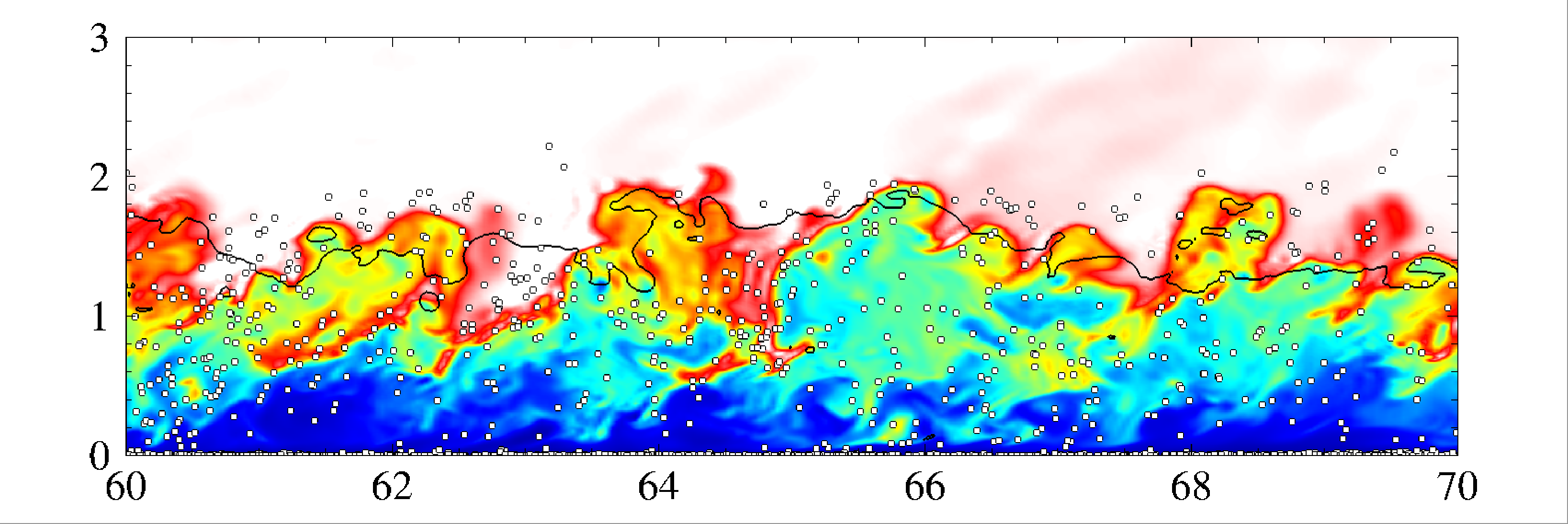}
\put(0,30){(c)}
\put(50,-2){$x/\delta_0$}
\put(0,12){\rotatebox{90}{$y/\delta_0$}}
\end{overpic}\\[3.0ex]
\begin{overpic}[width=0.8\textwidth,trim={0.2cm 0.2cm 0.2cm 0.2cm},clip]{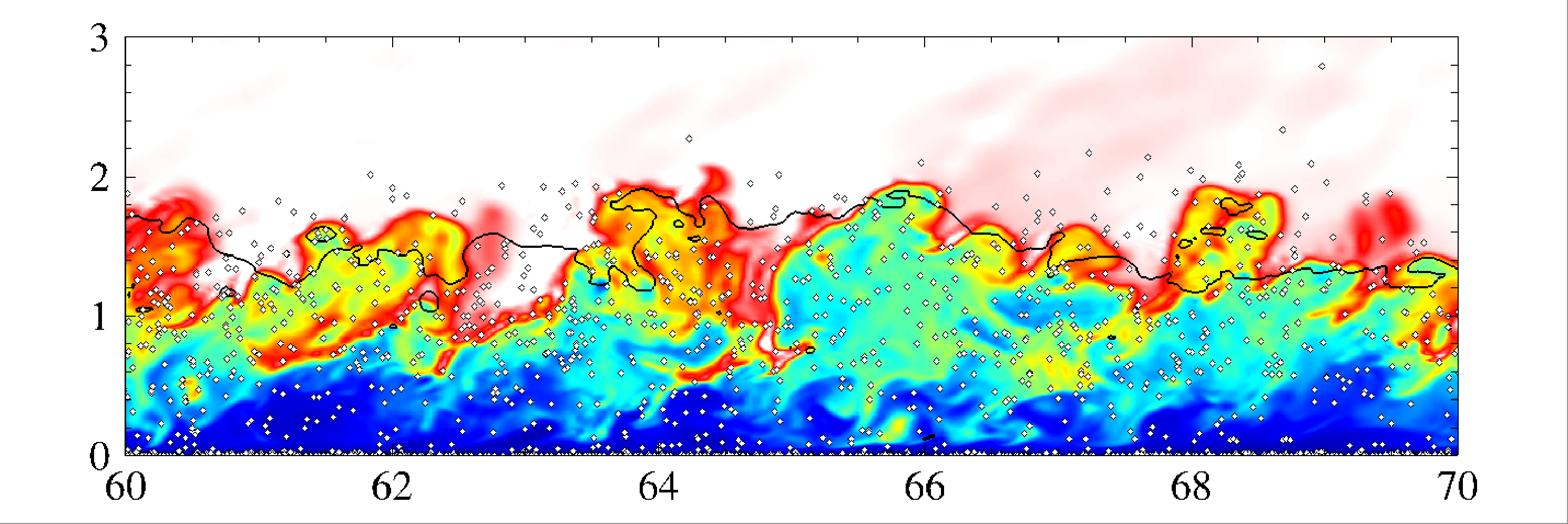}
\put(0,30){(d)}
\put(50,-2){$x/\delta_0$}
\put(0,12){\rotatebox{90}{$y/\delta_0$}}
\end{overpic}\\[0.0ex]
\caption{Instantaneous density distribution at $z=0$ (flooded) and particles within 
$z=0 \sim 0.01 \delta_0$ in case M4, particle populations (a) P1, (b) P4, (c) P5, (d) P6. 
Black solid lines: $u_1 = 0.99 U_\infty$.}
\label{fig:instxy4}
\end{figure}

We first present in figure~\ref{fig:instxy4} the instantaneous distribution of the particle 
populations P1, P4, P5 and P6 and the fluid density $\rho$ in case M4, along with the edge of 
the boundary layer marked by $u_1 = 0.99 U_\infty$.
For population P1 with the lowest particle density $\rho_p$ and Stokes number $St^+$, 
the particles appear to be uniformly distributed within the region where the fluid density 
is lower than that of the free stream value, even inside the small 'blobs' that are spatially 
separated from the turbulent boundary layer, 
but fail to fill in the regions between these `blobs' and the primary turbulent
regions of the boundary layers, the region where the density is the free-stream values 
but the momentum is lower.
As the Stokes number increases to $St^+ \approx 17.6$ (P4) and $70.44$ (P5), 
the particles accumulate close to the wall, thereby leading to their lower 
concentration at higher locations.
Due to their comparatively larger inertia away from the wall (recall figure~\ref{fig:psize}(d)),
some of the particles with high vertical velocities are capable of escaping from 
the low-fluid-density regions and reaching the free stream.
This is especially the case for population P6 ($St^+ \approx 140.81$) with an even larger particle
density, a considerable portion of which can be ejected to the free stream,
while the tendency of wall accumulation is weakened.
The variation of the particle distributions with the Stokes numbers in other flow cases
is similar to this case, showing non-monotonic behaviour of near-wall accumulation.
This is consistent with the previous studies in incompressible 
wall turbulence~\citep{marchioli2002mechanisms,bernardini2014reynolds}.

We would like to remark on the resemblance in the instantaneous distributions of the 
low Stokes number particles and the fluid density.
It has been shown that
in the limit of small volumetric loading, the concentration field $c$
(number of particles per volume) of the particles can be approximated as
~\citep{ferry2001fast,ferry2002equilibrium}
\begin{equation}
\frac{\partial c}{\partial t} + \frac{\partial c v_j}{\partial x_j} = 
c \frac{\partial u_i}{\partial x_i}.
\label{eqn:concen}
\end{equation}
Under the condition of $\rho_p / \rho_f >> 1$ and $St^+ <<1$, the velocities of the fluid
and particles are approximately the same $v_i \approx u_i$, with the deviation being the
order of $O(\tau_p)$ in outer scales.
For the presently considered compressible flow over a quasi-adiabatic flat plate, 
the flow dilatation can be disregarded in comparison with the vortical and shear
~\citep{yu2019genuine,yu2021compressibility},
so the right-hand-side of equation~\eqref{eqn:concen} can be neglected,
leading to the identical expression of equation~\eqref{eqn:concen} and the continuity equation.
It is, therefore, reasonable that the density and the low Stokes number particle
concentration fields
are similar, except that there is mean gradients in the fluid density 
due to the restriction of the state equation of the perfect gas and the nonuniformity of the mean
temperature in the wall-normal direction.

\begin{figure}
\centering
\scriptsize
\begin{overpic}[width=0.5\textwidth,trim={0.1cm 0.1cm 0.1cm 0.1cm},clip]{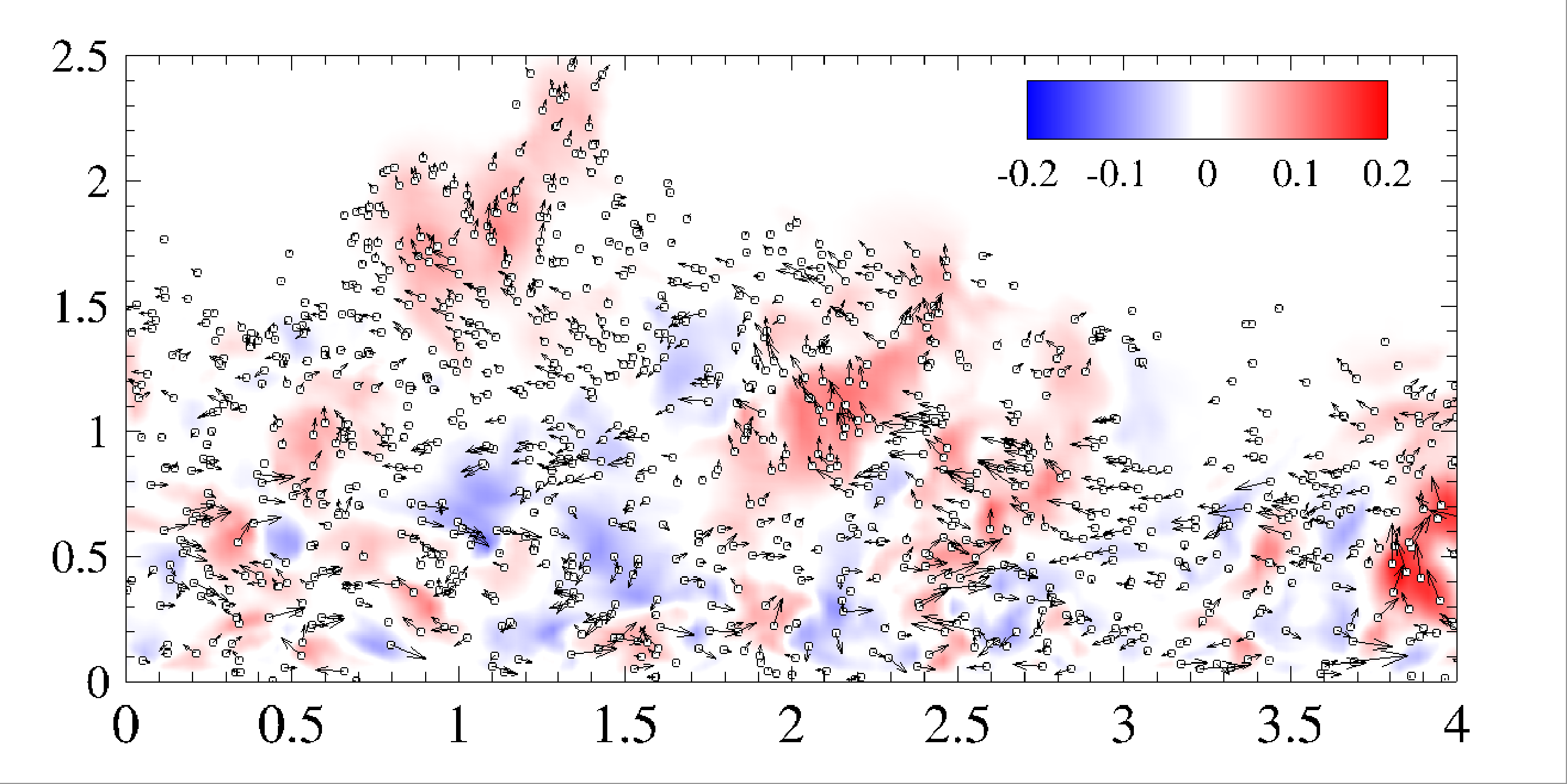}
\put(-3,45){(a)}
\put(50,-2){$z/\delta_0$}
\put(-2,22){\rotatebox{90}{$y/\delta_0$}}
\end{overpic}~
\begin{overpic}[width=0.5\textwidth,trim={0.1cm 0.1cm 0.1cm 0.1cm},clip]{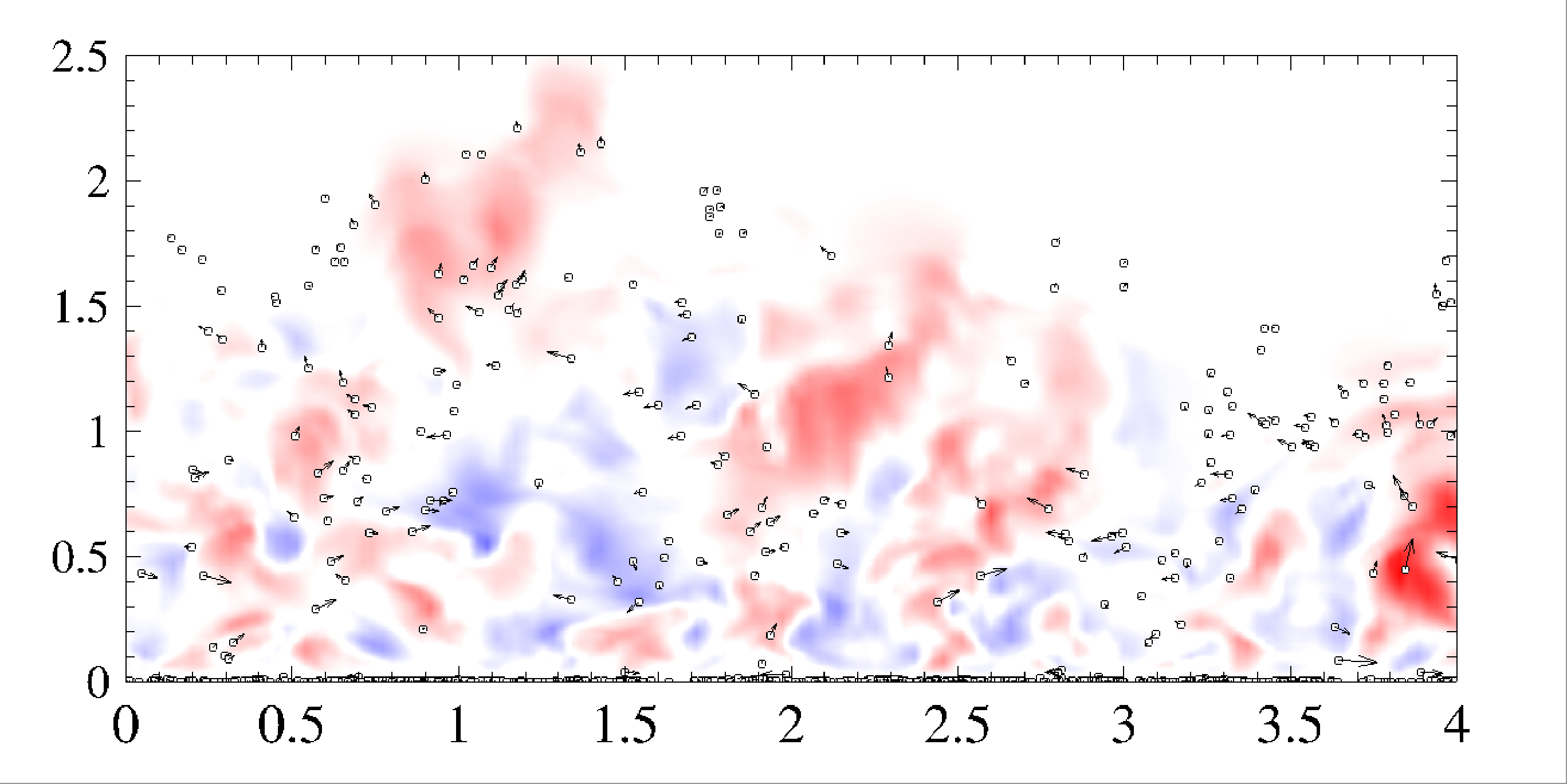}
\put(-3,45){(b)}
\put(50,-2){$z/\delta_0$}
\put(-2,22){\rotatebox{90}{$y/\delta_0$}}
\end{overpic}\\[2.0ex]
\begin{overpic}[width=0.5\textwidth,trim={0.1cm 0.1cm 0.1cm 0.1cm},clip]{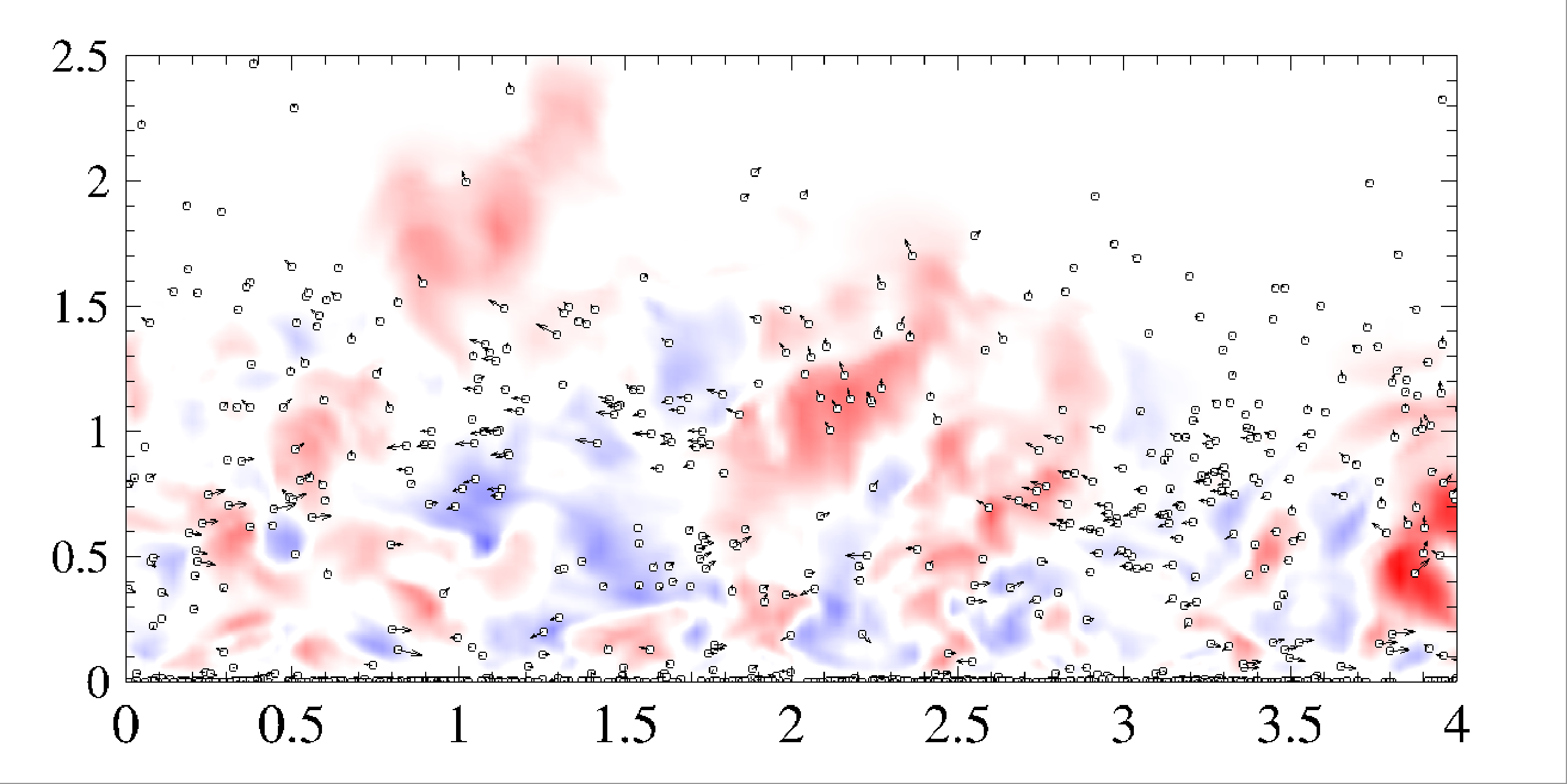}
\put(-3,45){(c)}
\put(50,-2){$z/\delta_0$}
\put(-2,22){\rotatebox{90}{$y/\delta_0$}}
\end{overpic}~
\begin{overpic}[width=0.5\textwidth,trim={0.1cm 0.1cm 0.1cm 0.1cm},clip]{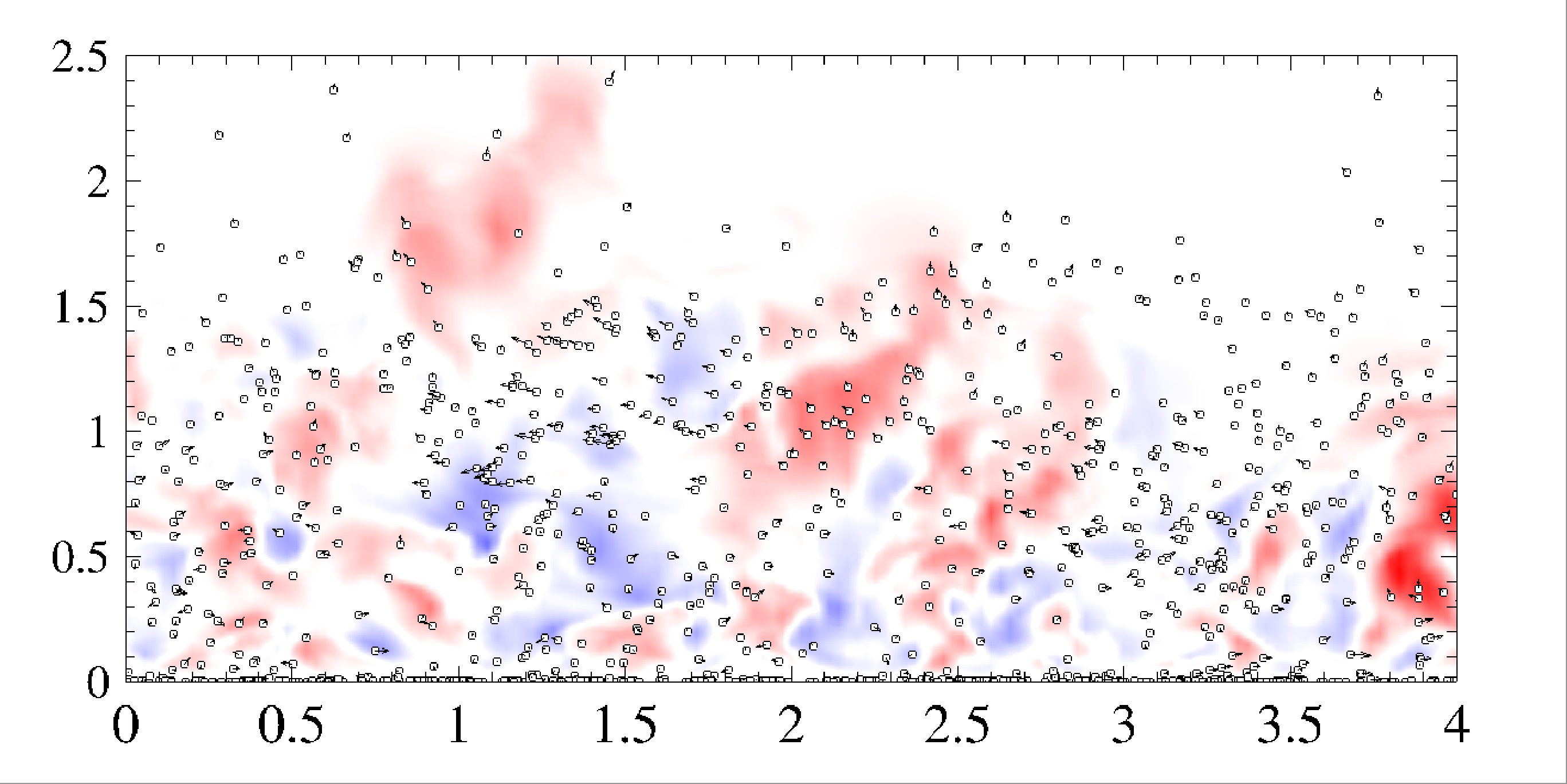}
\put(-3,45){(d)}
\put(50,-2){$z/\delta_0$}
\put(-2,22){\rotatebox{90}{$y/\delta_0$}}
\end{overpic}\\[0.0ex]
\caption{Instantaneous wall-normal velocity of the fluid $u_2$ (flooded) at $x=60 \delta_0$
and particle distribution within $x=(60 \sim 60.01) \delta_0$ in case M4,
(a) P1, (b) P4, (c) P5, (d) P6. Arrows: in-plane particle velocity vector.}
\label{fig:instyz2}
\end{figure}
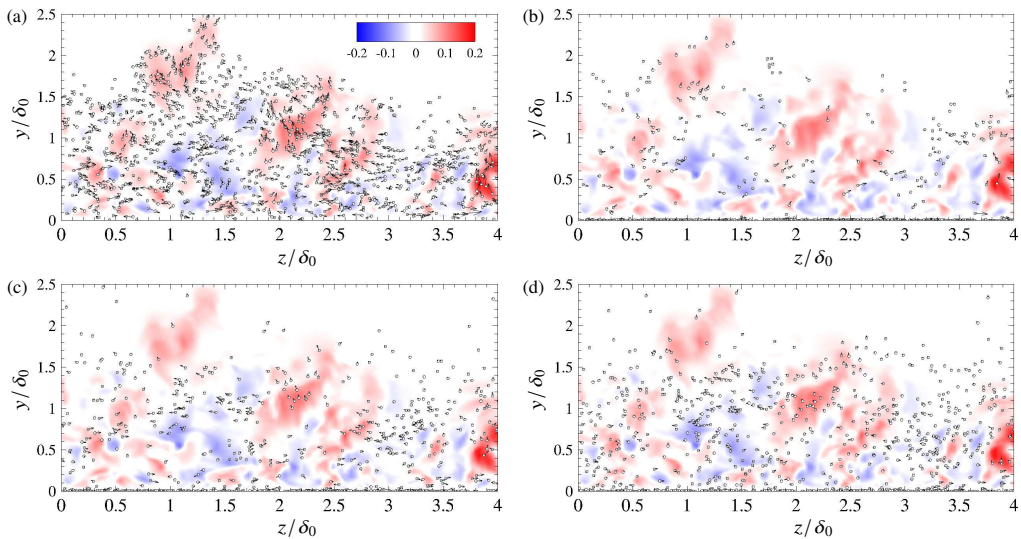

To further characterize the wall-normal transport of the particles, in figure~\ref{fig:instyz2}
we present the wall-normal velocity of the fluid and the in-plane particle velocity vectors.
For population P1, the wall-normal velocities of these low Stokes number particles
follow almost exactly that 
of the fluid, in that the particles are moving upwards in the regions of ejections and downwards
in the regions of sweeps within the boundary layer.
However, near the edge of the boundary layer, the particles prefer to accumulate by the
strong ejections, as suggested by higher particle concentration within the regions of 
positive wall-normal velocity $u_2>0$ than those of the negative wall-normal velocity $u_2<0$.
This has been examined via probability density distributions, as demonstrated in 
Appendix~\ref{sec:pdf}.
A simple reason is that there is no particle outside of the boundary layers, 
so no particle can be brought downwards into the turbulent boundary layers from the free stream flow.
This will be reflected in the statistics of particles as well, as it will be shown subsequently.
As for the other particle populations with larger Stokes numbers $St^+$,
their wall-normal transport processes are still roughly following those of the fluid,
while the magnitudes are reduced due to the larger particle inertia.
In spite of this, the particles are more inclined to move outwards to the free stream, 
even inside $u_2<0$ regions, 
which is probably the remnant of the strong ejection events or equivalently, 
the historical effects of the large inertia particles~\citep{soldati2005particles}.
However, it should be noted that the number of particles escaping the turbulent boundary layers 
does not increase monotonically with the Stokes number.
On the one hand, it can be derived that in the comparatively quiescent flow, 
the initial slip velocity required for the particles to reach the free stream is proportional to 
the reciprocal of the Stokes number. 
Henceforth, the high Stokes number particles are capable of escaping the boundary layers even though
their wall-normal velocities are generally small. 
On the other hand, the bursting events endowed with high-intensity vertical momentum
are usually short-lived~\citep{tardu1995characteristics,jimenez2013near,jimenez2018coherent},
so it is unlikely that the extremely high Stokes number particles can be accelerated to the threshold of escaping
the boundary layer.
Subject to these counteracting factors, the number of particles reaching the free stream
should be first increasing then diminished and, when the Stokes numbers are sufficiently large,
recovers to zero, for the particles will remain an almost straight trajectory with 
slight influences by the turbulent motions.

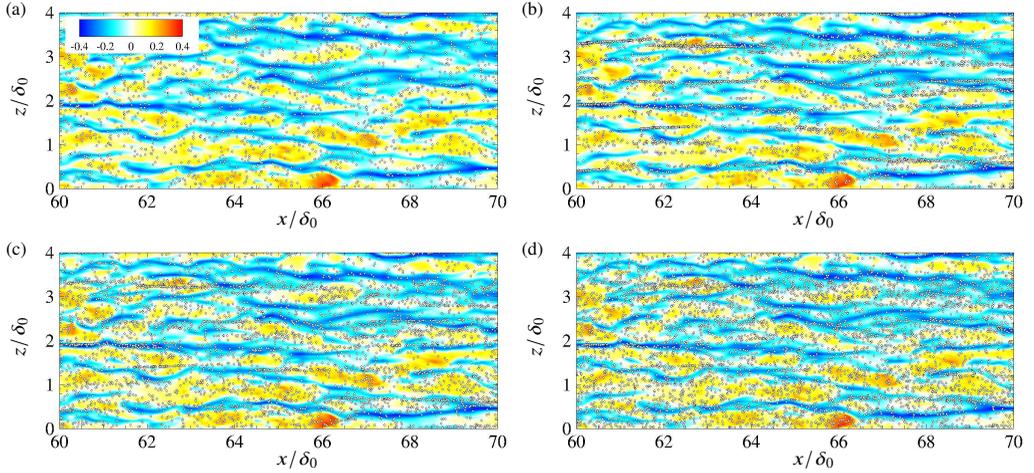
\begin{figure}
\centering
\scriptsize
\begin{overpic}[width=0.5\textwidth,trim={0.2cm 0.2cm 0.2cm 0.2cm},clip]{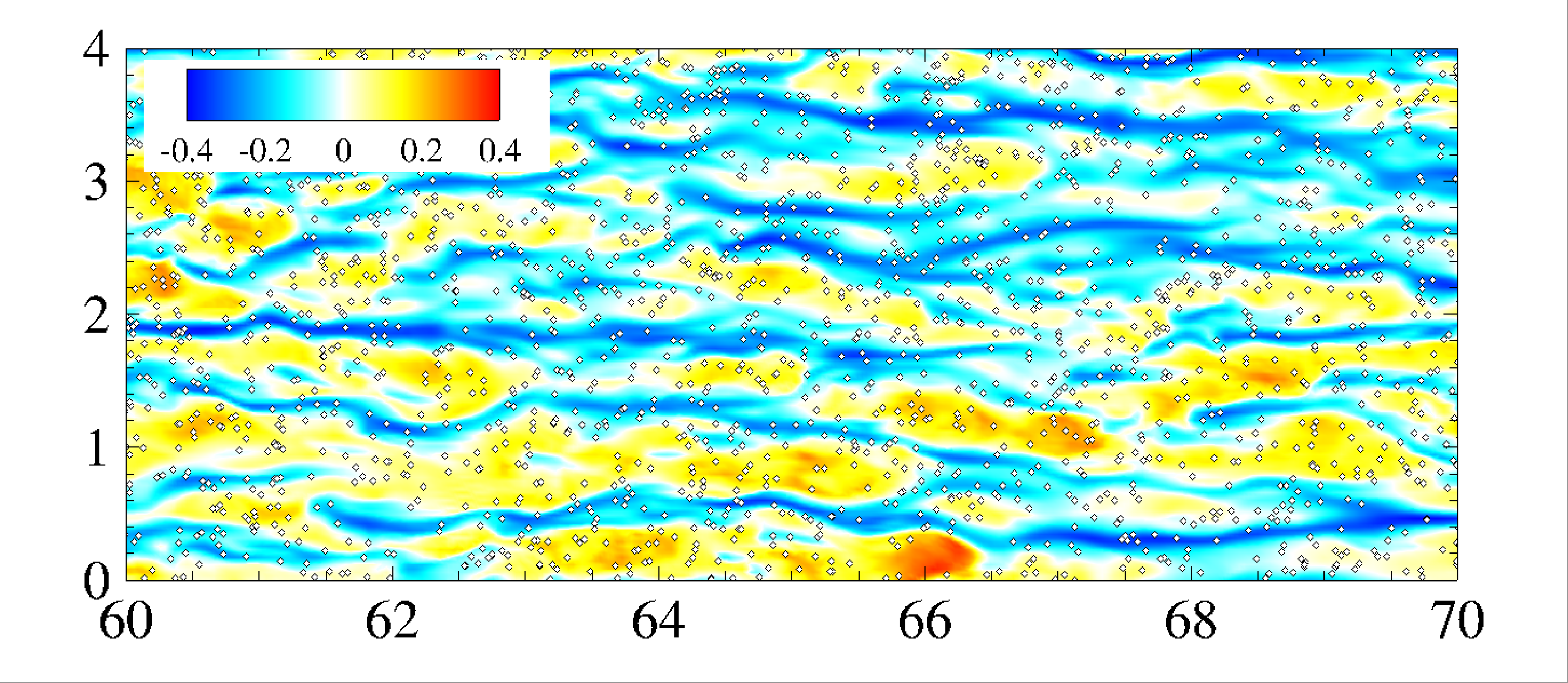}
\put(-3,40){(a)}
\put(50,-1){$x/\delta_0$}
\put(-2,20){\rotatebox{90}{$z/\delta_0$}}
\end{overpic}~
\begin{overpic}[width=0.5\textwidth,trim={0.2cm 0.2cm 0.2cm 0.2cm},clip]{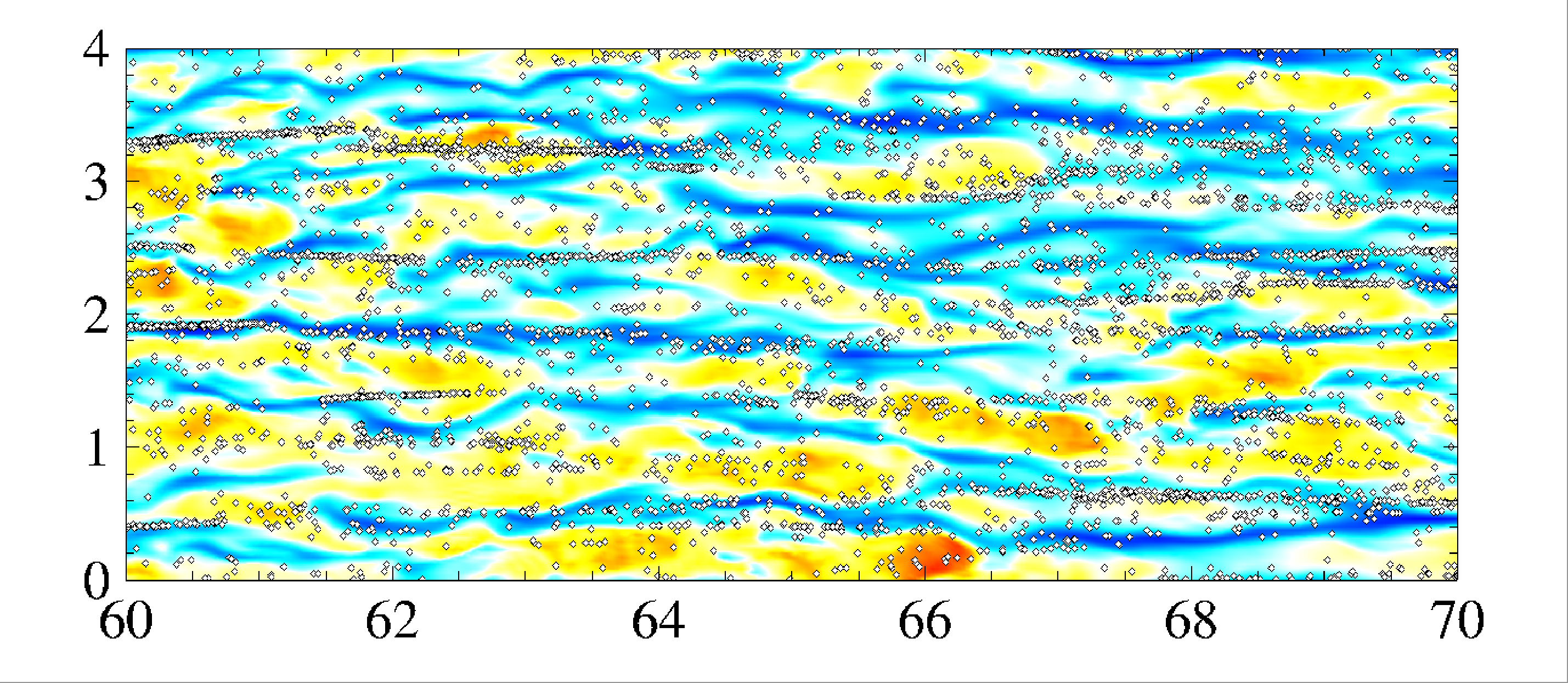}
\put(-3,40){(b)}
\put(50,-1){$x/\delta_0$}
\put(-2,20){\rotatebox{90}{$z/\delta_0$}}
\end{overpic}\\[2.0ex]
\begin{overpic}[width=0.5\textwidth,trim={0.2cm 0.2cm 0.2cm 0.2cm},clip]{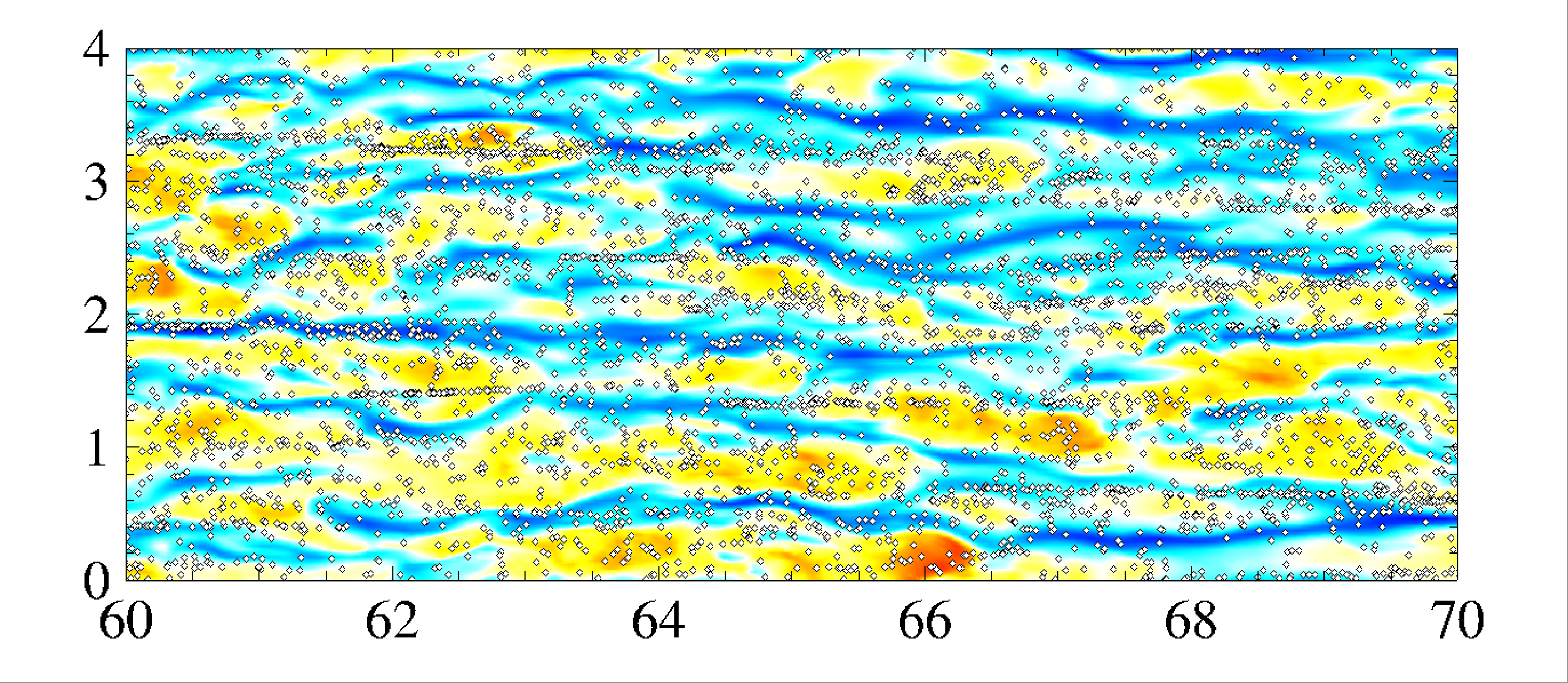}
\put(-3,40){(c)}
\put(50,-1){$x/\delta_0$}
\put(-2,20){\rotatebox{90}{$z/\delta_0$}}
\end{overpic}~
\begin{overpic}[width=0.5\textwidth,trim={0.2cm 0.2cm 0.2cm 0.2cm},clip]{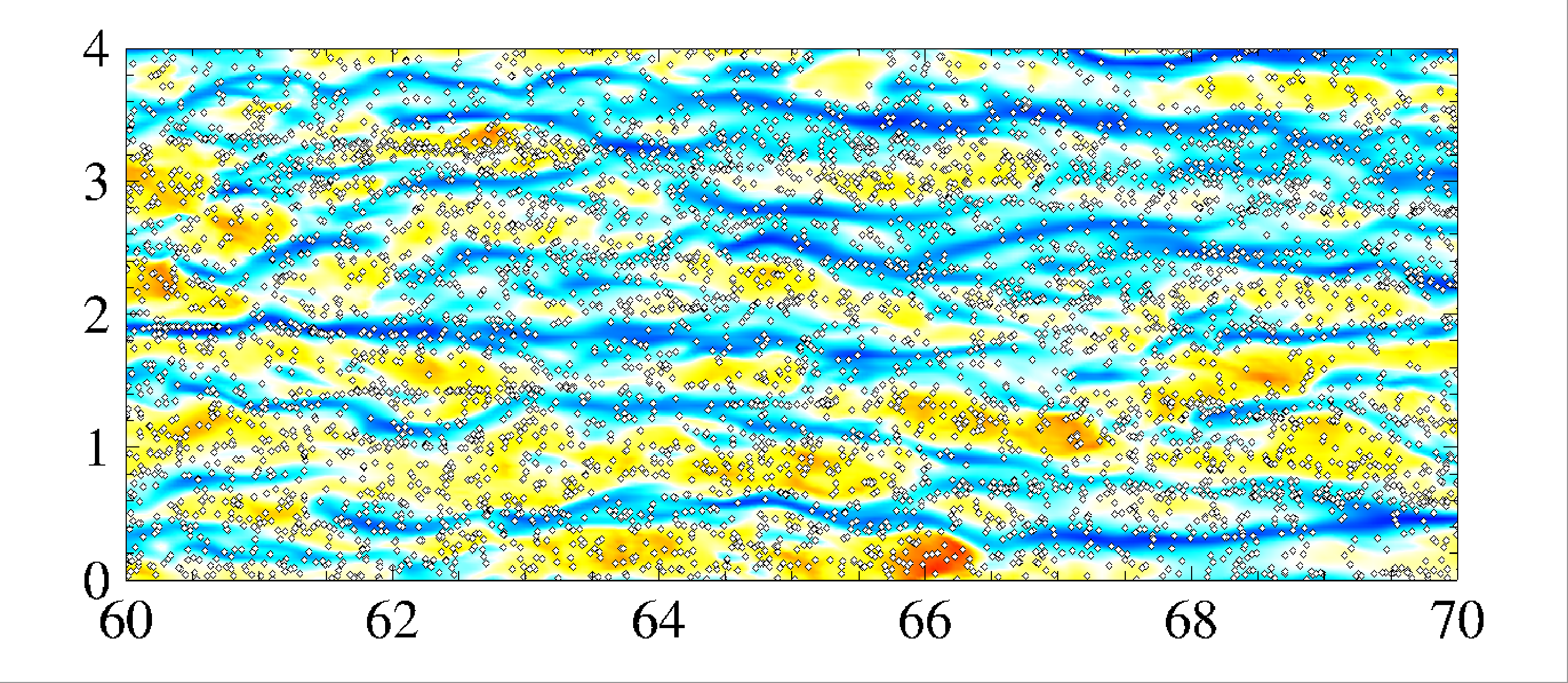}
\put(-3,40){(d)}
\put(50,-1){$x/\delta_0$}
\put(-2,20){\rotatebox{90}{$z/\delta_0$}}
\end{overpic}\\[0.0ex]
\caption{Instantaneous streamwise velocity fluctuation $u'_1$ at $y^+=15$ (flooded) and 
particle distribution within $y^+=3 \sim 15$ in case M4, (a) P1, (b) P4, (c) P5, (d) P6.}
\label{fig:instxz4}
\end{figure}

In figure~\ref{fig:instxz4} we present the distribution of the streamwise velocity fluctuation 
$u'_1$ at $y^+=15$ and the particles within the range of $y^+=3 \sim 15$.
As is commonly observed in dilute multiphase wall turbulence~\citep{rouson2001preferential,
sardina2012wall,bernardini2013effect}, the particles with low $St^+$ are evenly distributed 
in the wall-parallel plane, while their clustering with the low-speed streaks can be 
spotted for moderate $St^+$ particles (population P4 and P5), which is induced by the near-wall
quasi-streamwise vortices that transport the particles toward the viscous sublayer and deposited
by the spanwise velocity below the low-speed streaks where they are either trapped or brought
upwards when encountered with the strong ejections~\citep{marchioli2002mechanisms,
soldati2009physics}.
At larger $St^+$ in type P6, the `particle streaks' as weakened, showing a more uniform 
distribution, which is caused by the highly different particle response time and characteristic 
time scale of the near-wall self-sustaining cycle.
Much longer particle streaks with wider spanwise intervals will be observed in high Reynolds number 
flows in which the very-large-scale motions are becoming more and more manifest and 
contribute significantly to the Reynolds stress~\citep{jie2022existence,motoori2022role}.

\subsection{Near-wall accumulation and clustering} \label{subsec:dist}

\begin{figure}
\centering
\begin{overpic}[width=0.5\textwidth]{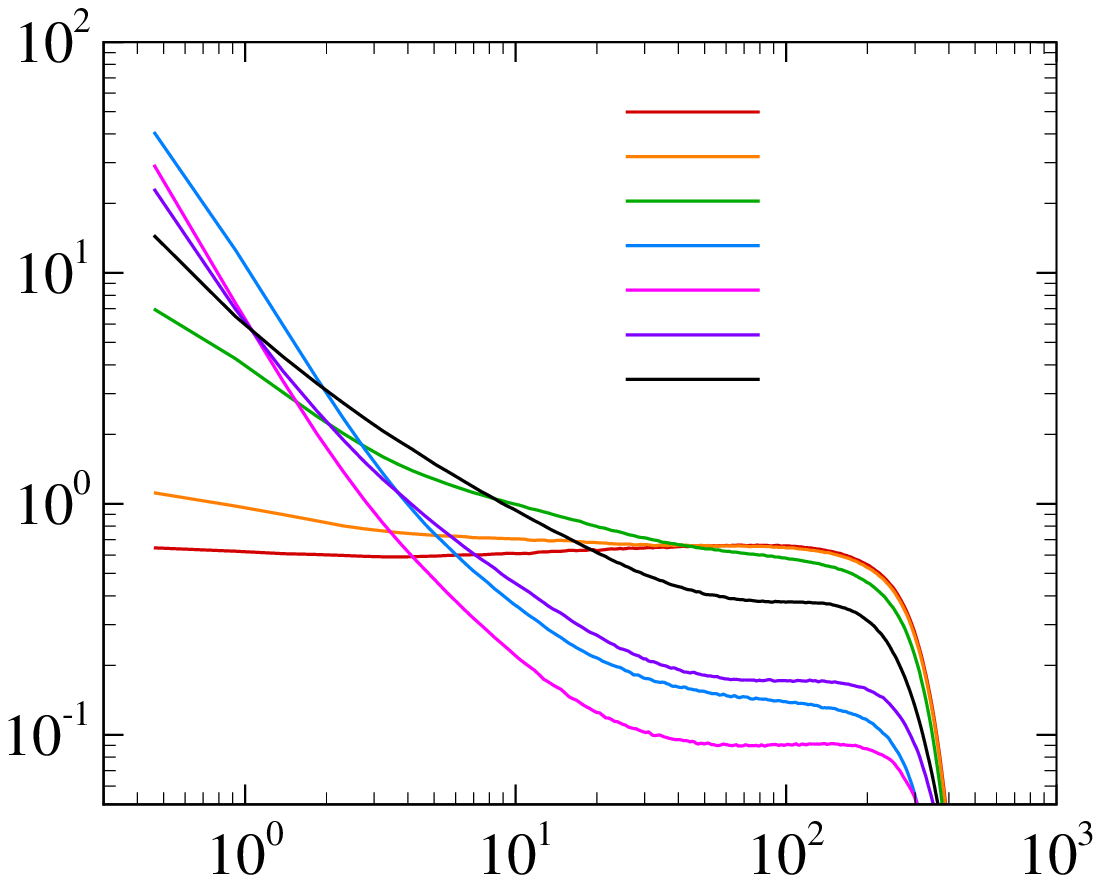}
\put(0,70){(a)}
\put(50,0){$y^+$}
\put(0,35){\rotatebox{90}{$\bar c(y)/c_0$}}
\put(68,65.5){\scriptsize P1}
\put(68,61.5){\scriptsize P2}
\put(68,58){\scriptsize P3}
\put(68,54.8){\scriptsize P4}
\put(68,51.5){\scriptsize P5}
\put(68,47.5){\scriptsize P6}
\put(68,44){\scriptsize P7}
\end{overpic}~
\begin{overpic}[width=0.5\textwidth]{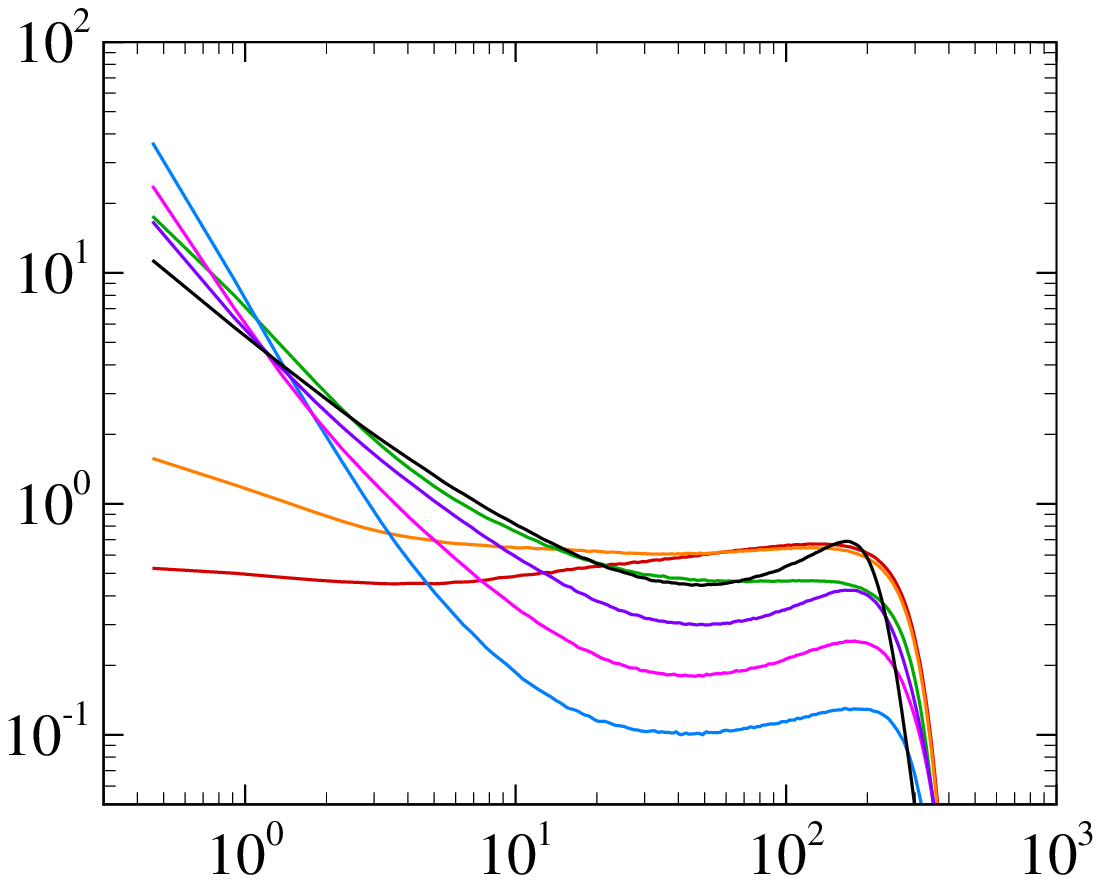}
\put(0,70){(b)}
\put(50,0){$y^+$}
\put(0,35){\rotatebox{90}{$\bar c(y)/c_0$}}
\end{overpic}\\[2.0ex]
\begin{overpic}[width=0.5\textwidth]{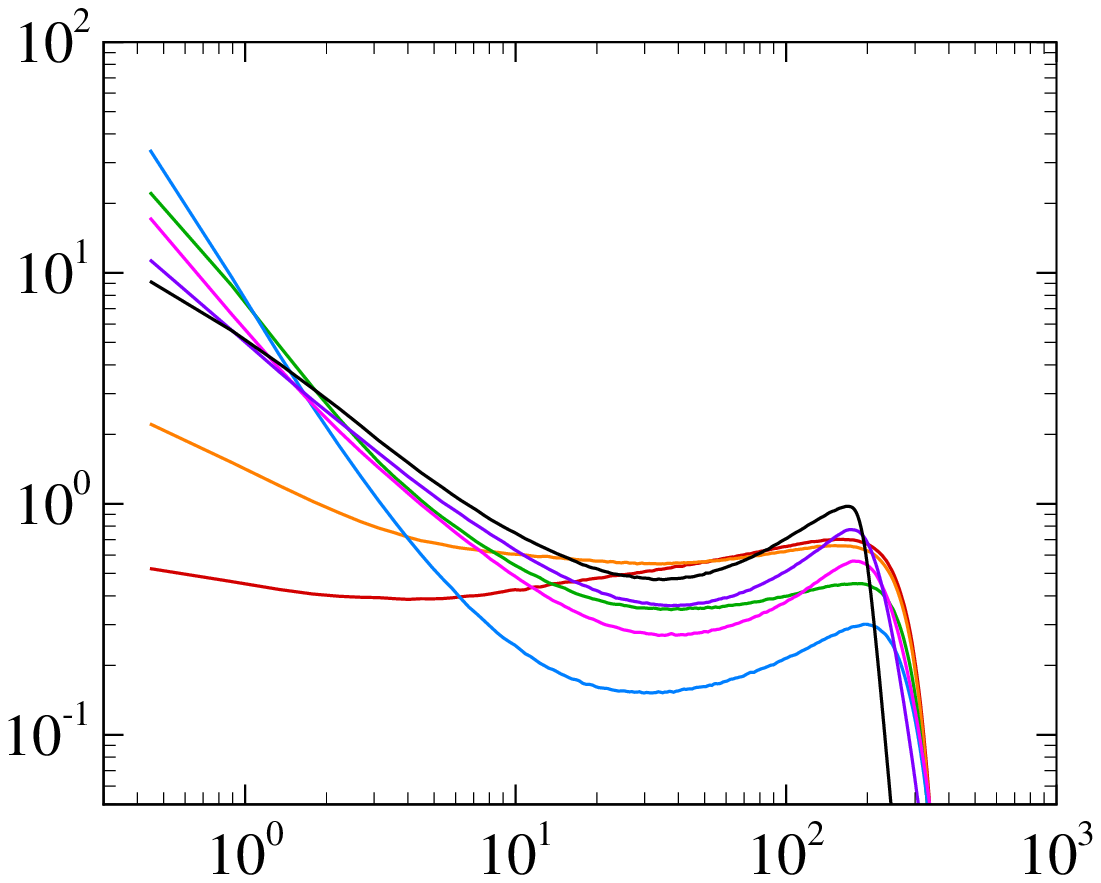}
\put(0,70){(c)}
\put(50,0){$y^+$}
\put(0,35){\rotatebox{90}{$\bar c(y)/c_0$}}
\end{overpic}~
\begin{overpic}[width=0.5\textwidth]{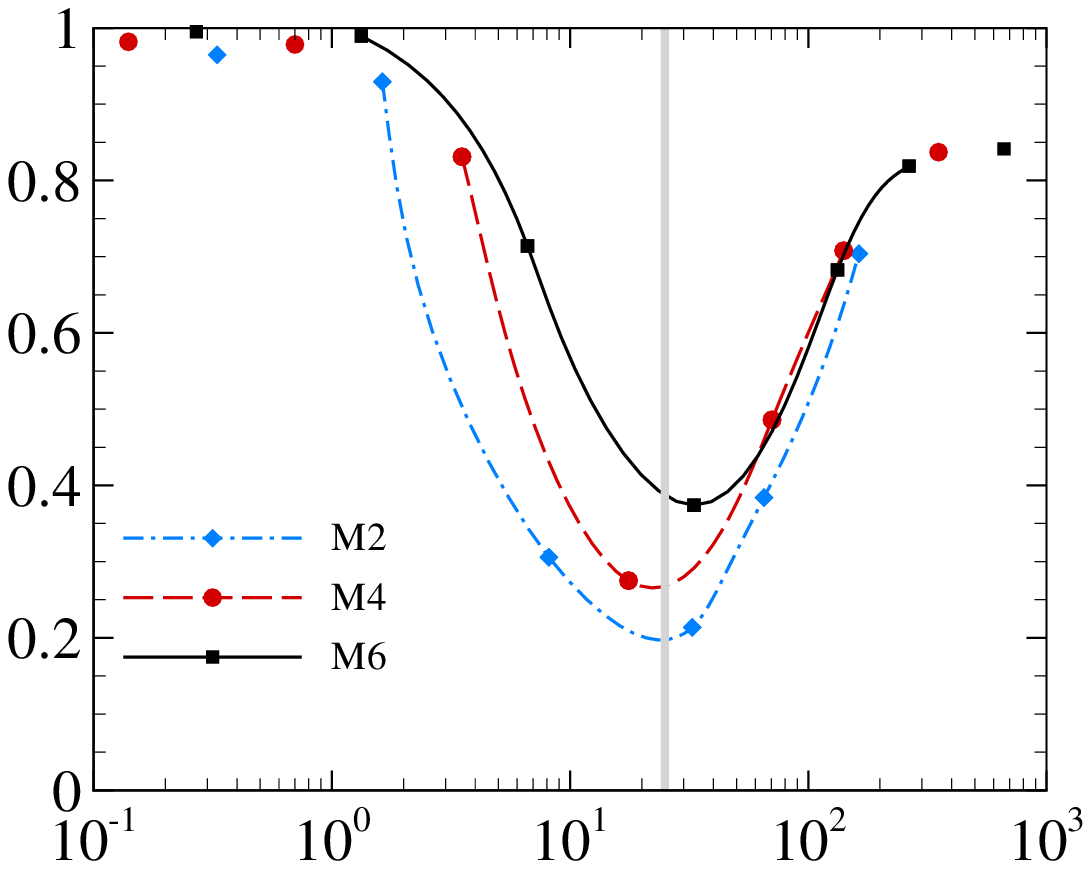}
\put(0,70){(d)}
\put(50,0){$St^+$}
\put(0,38){\rotatebox{90}{$S_p$}}
\put(60,65){\small $St^+=25$}
\end{overpic}\\
\caption{Wall-normal distribution of the particle concentration in cases (a) M2, (b) M4 and (c) M6,
and (d) the Shannon entropy at different Stokes numbers $St^+$ at $t=515 \delta_0/U_\infty$.
Lines in (d) are the cubic splines using the data within $St^+ = 1 \sim 300$.}
\label{fig:ccyy}
\end{figure}

The non-uniform wall-normal distribution can be quantitatively characterized by the mean
particle concentration, calculated based on the grid center of the Eulerian frame as
\begin{equation}
\displaystyle
c(y_j) = \frac{N_p(y_j)}{ \sum_{j} N_p(y_j)}
\frac{\sum_{j} \Delta h_j}{\Delta h_j},~~ j=1, 2, ..., N_y-1
\end{equation}
where $N_p (y_j)$ is the particle number located between the $j$-th and $(j+1)$-th grid point,
and $\Delta h_j$ the grid interval.
The distributions of the mean particle concentration $\bar c(y)/c_0$ are displayed 
in figure~\ref{fig:ccyy}(a-c), with $c_0$ the particle concentration
in the case of perfect uniform wall-normal distribution.
In general, the particles are almost evenly distributed across the boundary layer for
the populations with the lowest Stokes number $St^+$, and the phenomenon of the near-wall 
accumulation, namely the turbophoresis, becomes more evident with the increasing $St^+$.
The maximal near-wall particle number density is attained for the particle type P4, 
beyond which it is gradually diminished.
This is consistent with the conclusions in low-speed turbulent channel flows
~\citep{sardina2012wall,bernardini2014reynolds}. 
Comparing these cases, we found that the particle concentration $\bar c(y)$ within the range of
$y^+ = 50 \sim 200$ remain almost constant in case M2, whereas those in cases M4 and M6
slightly increase and manifest secondary peaks in the outer region, especially for the large $St^+$
particle populations.
Such non-monotonic variations should probably be caused by the evident variation of 
the particle Stokes number $St^*$ at high Mach numbers when evaluated by 
the local density and viscosity of the fluid.

A commonly used indicator of the non-uniform wall-normal distribution is the Shannon entropy
~\citep{picano2009spatial}.
It is defined by the ratio of two entropy parameters $S_p = \mathscr{S}/\mathscr{S}_{max}$,
in which $\mathscr{S}$ is calculated in equidistant slabs in the wall-normal direction
within $1.2 \delta$ 
\begin{equation}
\mathscr{S} = - \sum^{N_{yu}}_{j=1} P_j \ln (P_j)
\end{equation}
with $N_{yu}=200$ and the $P_j$ probability of finding a particle in the $j-$th slab.
In the case of evenly distributed particles, the $\mathscr{S}$ attains maximum 
$\mathscr{S}_{max}=\ln N_{yu}$.
The Shannon entropy $S_p$, therefore, ranges from 0 to 1.
In figure~\ref{fig:ccyy}(d) we plot the values of $S_p$ of each particle population in
all three cases against the $St^+$.
Expectedly, the values of $S_p$ are close to unity for low Stokes number particles, 
corresponding to 
the uniform distribution across the boundary layer, and decrease rapidly and attain minimum at 
$St^+ \approx 30$ for all the cases considered, irrelevant of the free stream Mach numbers, 
suggesting the highest level of near-wall accumulation.
This is consistent with the previous studies on the particles in incompressible turbulent
channels~\citep{marchioli2002mechanisms,bernardini2014reynolds}
For particles with $St^+$ greater than $300$, the $S_p$ values tend to reach an asymptotic value of 
$0.85$, implying that the tendency of near-wall accumulation of particles remains even when
the particles response to the turbulent motions rather slowly, which is probably caused by 
the integral effects in their remaining inside the turbulent boundary layer.
Comparing the cases at different Mach numbers, we further found that the higher Mach numbers 
tend to alleviate the turbophoresis phenomenon.
This can be inferred from the larger particle number density near the edge of the boundary layer.
It is reminiscent of the higher particle concentration near the channel center
within the quiescent cores~\citep{jie2021effects} and the free surface of the open channel
~\citep{gao2023direct,yu2017effects}, which was attributed to the corresponding
lower turbulent intensities~\citep{mortimer2020density}.
Such a phenomenon is found to be more evident for larger Stokes number particles.
Therefore, the more intense particle concentration near the edge of the boundary layer 
with the increasing Mach numbers should probably be attributed to the larger $St^*$, 
as reported in figure~\ref{fig:psize}(d).

\begin{figure}
\centering
\begin{overpic}[width=0.5\textwidth]{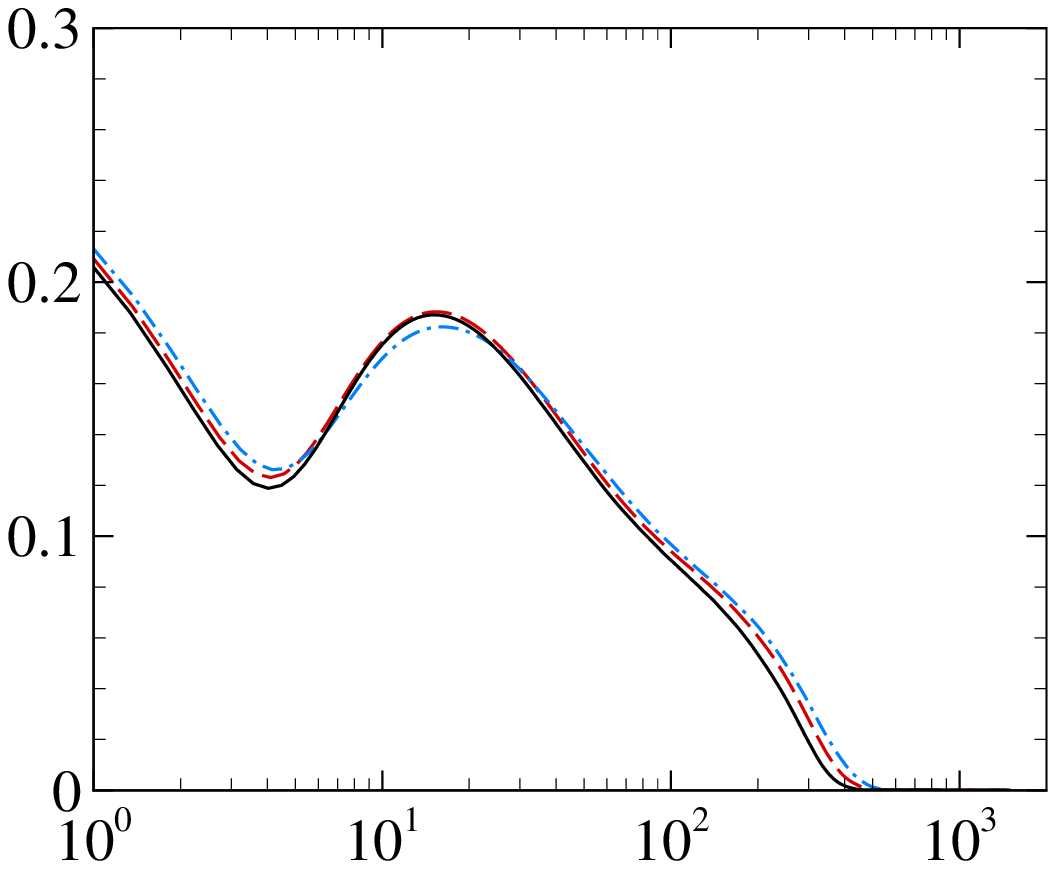}
\put(0,70){(a)}
\put(48,0){$y^+$}
\put(0,35){\rotatebox{90}{$\bar \omega'^+_x$}}
\end{overpic}~
\begin{overpic}[width=0.5\textwidth]{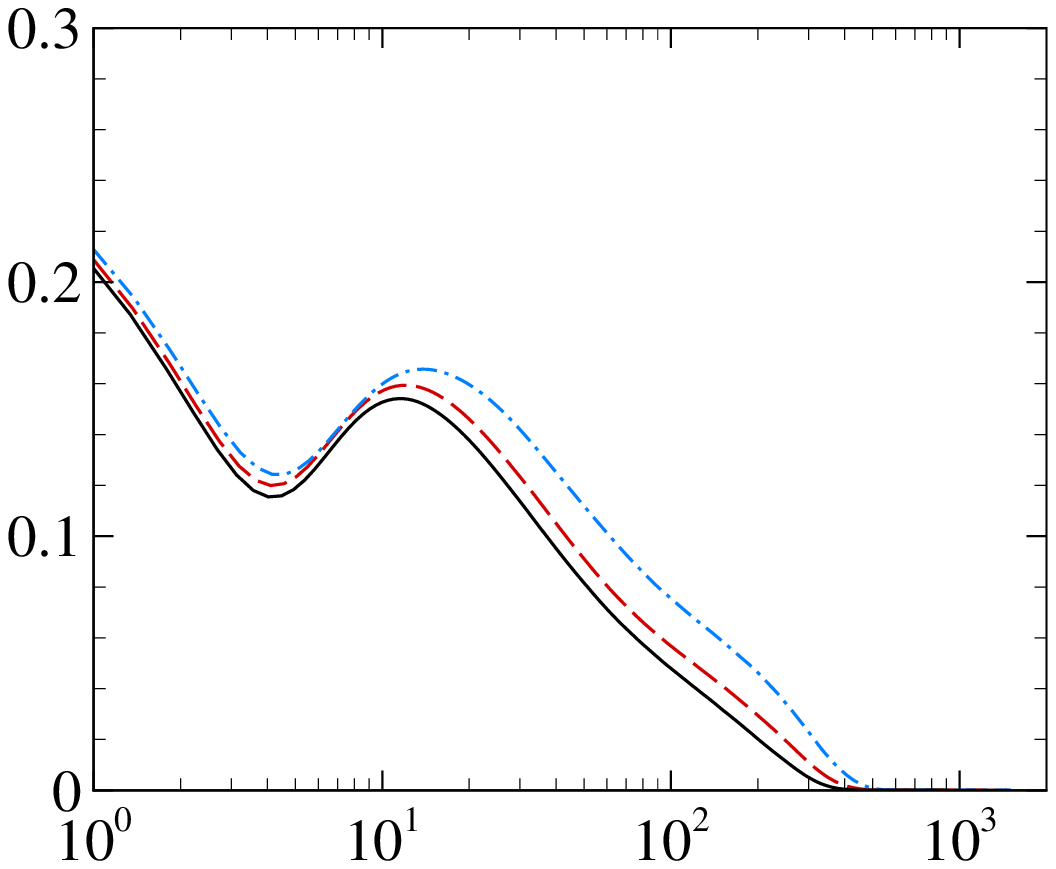}
\put(0,70){(b)}
\put(48,0){$y^+$}
\put(0,35){\rotatebox{90}{$\bar \omega'^*_x$}}
\end{overpic}\\
\caption{Wall-normal distribution of the streamwise vorticity fluctuation RMS, normalized by 
(a) viscous scale $\delta_\nu/u_\tau$, $\bar \omega'^+_x$,
(b) local viscous scales $\delta^*_\nu/u^*_\tau$, $\bar \omega'^*_x$.
Blue dash-dotted lines: case M2, red dashed lines: case M4, black solid lines: case M6.}
\label{fig:oxrms}
\end{figure}

Since the near-wall accumulation is mainly related to the ejection and sweeping events
induced by the quasi-streamwise vortices~\citep{marchioli2002mechanisms,soldati2009physics}, 
the dependence of the Shannon entropy with the Mach number is probably caused by 
the variation of the characteristics of the streamwise vortices.
In figure~\ref{fig:oxrms} we present the RMS of the streamwise vorticity fluctuations 
$\bar \omega'_x$, normalized by the viscous scales and local viscous scales, respectively.
For the local peaks at $y^+ \approx 10 \sim 20$ that represent the average intensity of
the quasi-streamwise vortices, the former shows weak dependence on the Mach number, 
while the latter manifests a systematic decreasing trend of variation.
From another point of view, the normalization using the viscous scales is more of 
a kinetic description of the fluid motions, while that using the local viscous scales
incorporates the variation of the fluid density and viscosity that influence the motions of
the particles.
Therefore, we conclude that the decreasing $\bar \omega'^*_x$ suggests the weaker impacts of 
the streamwise vortices on the near-wall particle motions, consistent with the observations of 
the less intensified near-wall accumulation with the Mach number.

\begin{figure}
\centering
\begin{overpic}[width=0.7\textwidth,trim={0.2cm 0.2cm 0.2cm 0.2cm},clip]{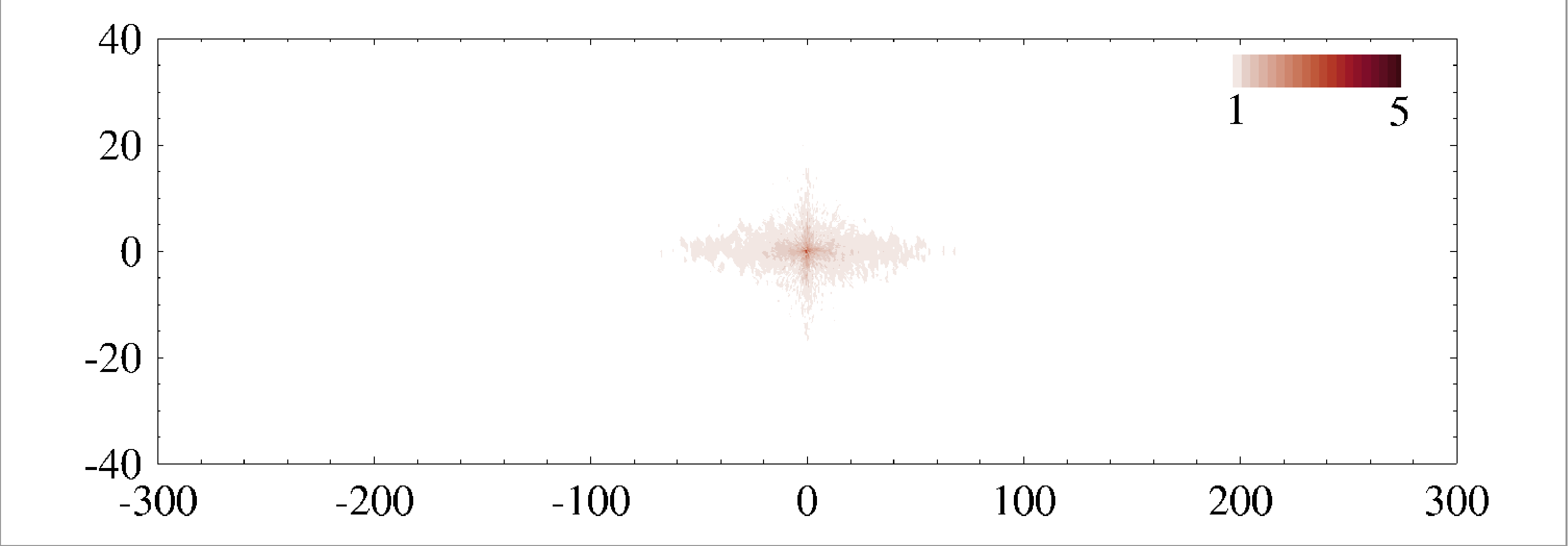}
\put(0,33){(a)}
\put(0,18){\rotatebox{90}{$l^+_z$}}
\put(12,28){$St^+=0.70$}
\end{overpic}\\
\begin{overpic}[width=0.7\textwidth,trim={0.2cm 0.2cm 0.2cm 0.2cm},clip]{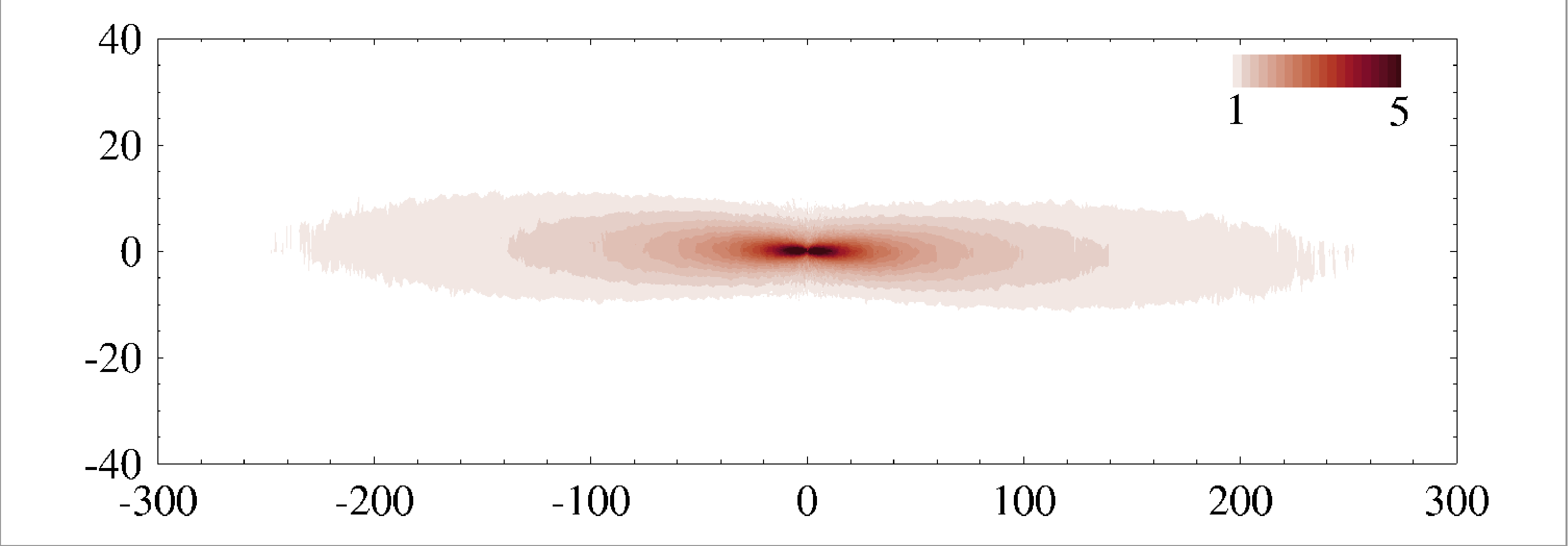}
\put(0,33){(b)}
\put(0,18){\rotatebox{90}{$l^+_z$}}
\put(12,28){$St^+=3.52$}
\end{overpic}\\
\begin{overpic}[width=0.7\textwidth,trim={0.2cm 0.2cm 0.2cm 0.2cm},clip]{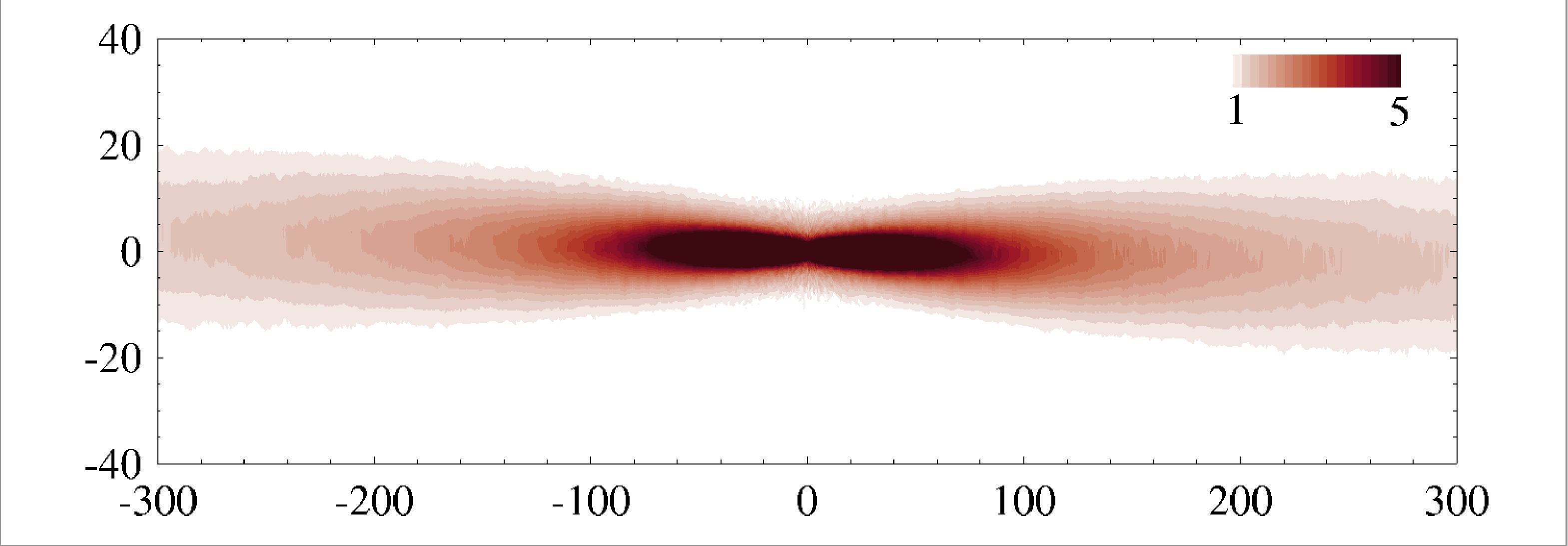}
\put(0,33){(c)}
\put(0,18){\rotatebox{90}{$l^+_z$}}
\put(12,28){$St^+=17.60$}
\end{overpic}\\
\begin{overpic}[width=0.7\textwidth,trim={0.2cm 0.2cm 0.2cm 0.2cm},clip]{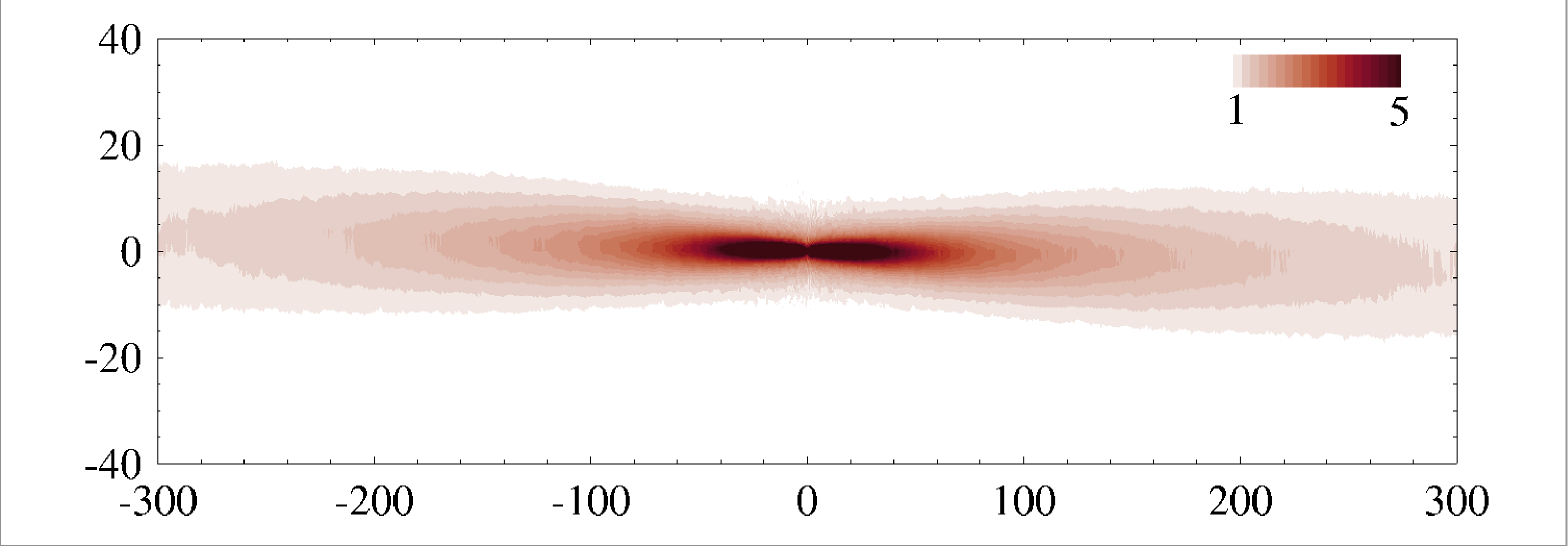}
\put(0,33){(d)}
\put(0,18){\rotatebox{90}{$l^+_z$}}
\put(12,28){$St^+=70.44$}
\end{overpic}\\
\begin{overpic}[width=0.7\textwidth,trim={0.2cm 0.2cm 0.2cm 0.2cm},clip]{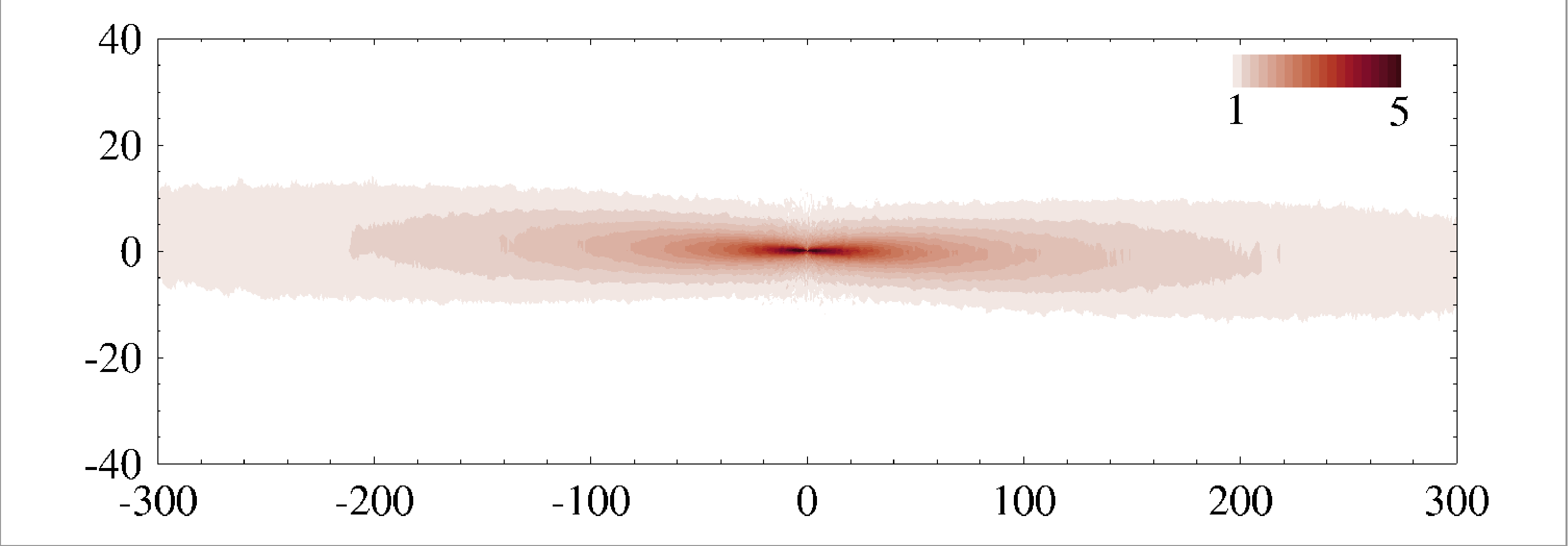}
\put(0,33){(e)}
\put(0,18){\rotatebox{90}{$l^+_z$}}
\put(12,28){$St^+=140.81$}
\end{overpic}\\
\begin{overpic}[width=0.7\textwidth,trim={0.2cm 0.2cm 0.2cm 0.2cm},clip]{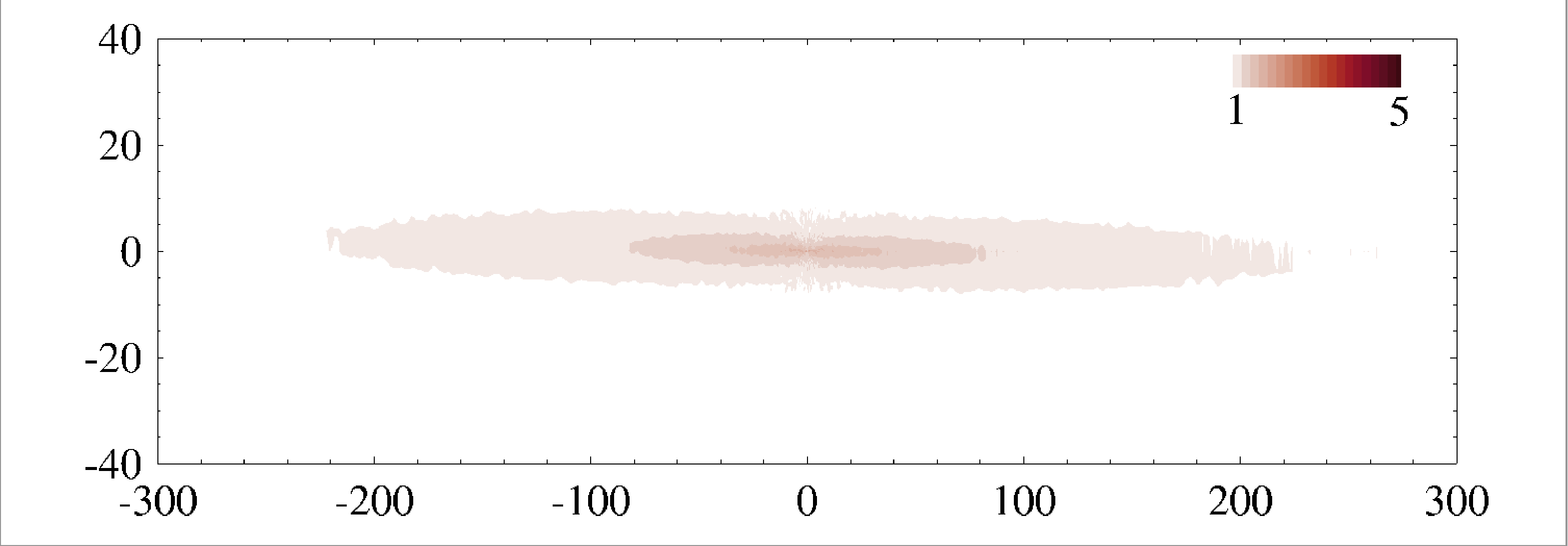}
\put(0,33){(f)}
\put(48,-1){$l^+_x$}
\put(0,18){\rotatebox{90}{$l^+_z$}}
\put(12,28){$St^+=352.21$}
\end{overpic}\\[0.0ex]
\caption{Near-wall particle ADF distribution below $y^+=10$ of case M4, (a) P2, (b) P3, (c) P4, 
(d) P5, (e) P6, (f) P7.}
\label{fig:adf2d}
\end{figure}
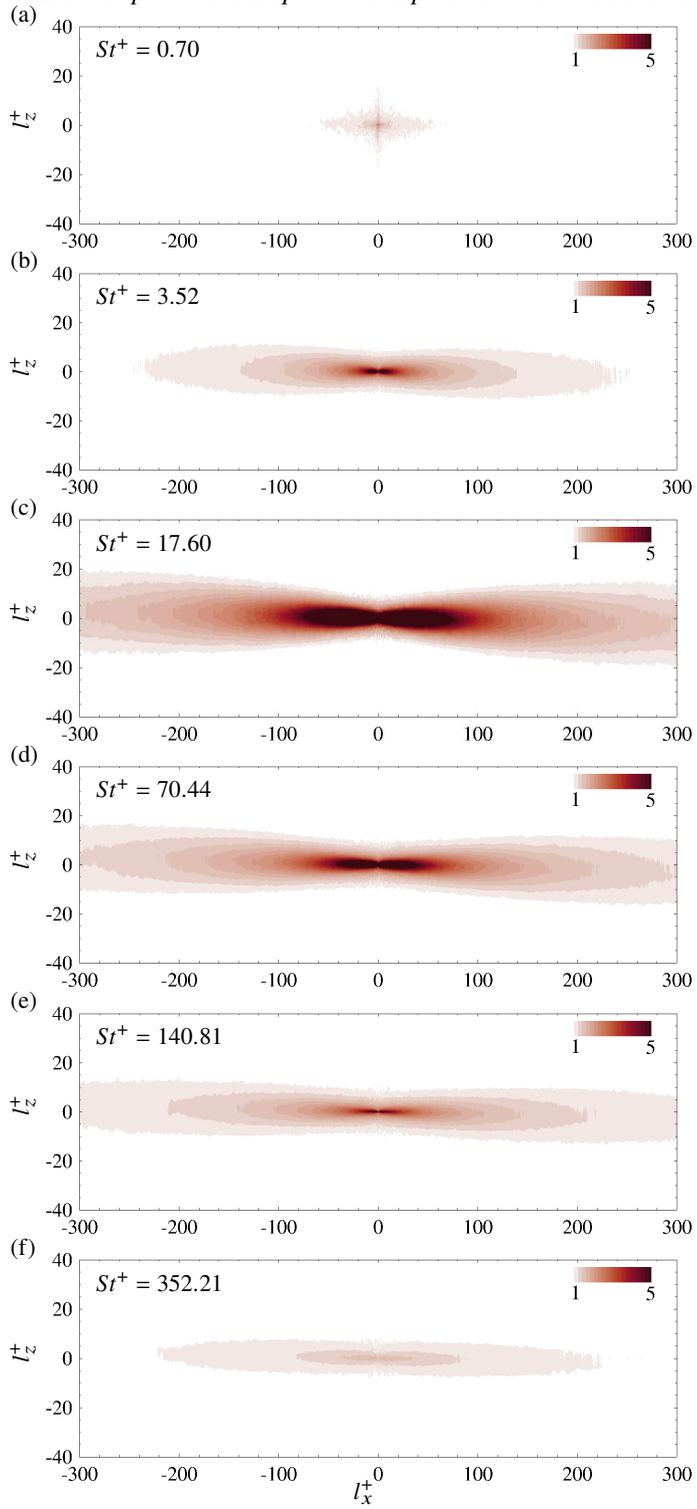

Besides the near-wall accumulation, another crucial aspect of the particle organization
in wall turbulence is the clustering behaviour in the wall-parallel planes
induced by the turbulent structures.
Amongst the multiple methods of quantifying such a phenomenon, such as the maximum deviation
from randomness or the segregation parameter~\citep{fessler1994preferential,
picciotto2005characterization}, the scaling of radial distribution function (RDF)
~\citep{wang1993settling,pumir2016collisional},
the statistics of particle concentration and the Vorono{\"i} tessellation analysis
~\citep{monchaux2010preferential,monchaux2012analyzing},
we adopt the angular distribution function (ADF) proposed by~\citet{gualtieri2009anisotropic}, 
which has been applied to turbulent channel flows in~\citet{sardina2012wall}
for the characterization of the inhomogeneous near-wall particle organization.
Following~\citet{sardina2012wall}, we define the two-dimensional ADF as follows,
\begin{equation}
g(r,\theta)=\frac{1}{r} \frac{{\rm d} \nu_r}{{\rm d} r} \frac{1}{n_0(y)}
\end{equation}
with $r$ the wall-parallel inter-particle distance, $\theta$ the deviation angle from 
the $x$ direction, $\nu_r (r, \theta)$ the averaged particle pair numbers at a certain off-wall 
distance $y$, and $n_0$ the total number of the particle pairs per unit area obtained by
\begin{equation}
n_0(y)=\frac{1}{2A} N_p(y)(N_p(y)-1).
\end{equation}
with $A$ the wall-parallel area where the statistics are performed.

In figure~\ref{fig:adf2d} we present the ADF distributions for particle populations P2 to P7 
in case M4. The trend of variation is qualitatively consistent with that in low-speed 
turbulent channels~\citep{sardina2012wall}. 
Specifically, the particle population P2 with the $St^+$ closest to unity manifest the weakest 
degree of small-scale clustering and the general uniform distribution, as indicated by the merely
slightly higher value of ADF at $r=0$ than those away from it where they remain almost constant.
For particle populations P3 to P5 with the gradually larger $St^+$, 
the probability of finding particles adjacent to them in the streamwise direction is higher, 
corresponding to the formation and the more evident streamwise elongated particle streaks.
At the even larger $St^+$ for particle populations P6 and P7, the particle streaks still exist
but become less evident due to the higher probability away from $l^+_z =0$.
This is consistent with the observations in figure~\ref{fig:instxz4} that the particles
tend to be uniformly distributed when their inertia is large.

\begin{figure}
\centering
\begin{overpic}[width=0.5\textwidth]{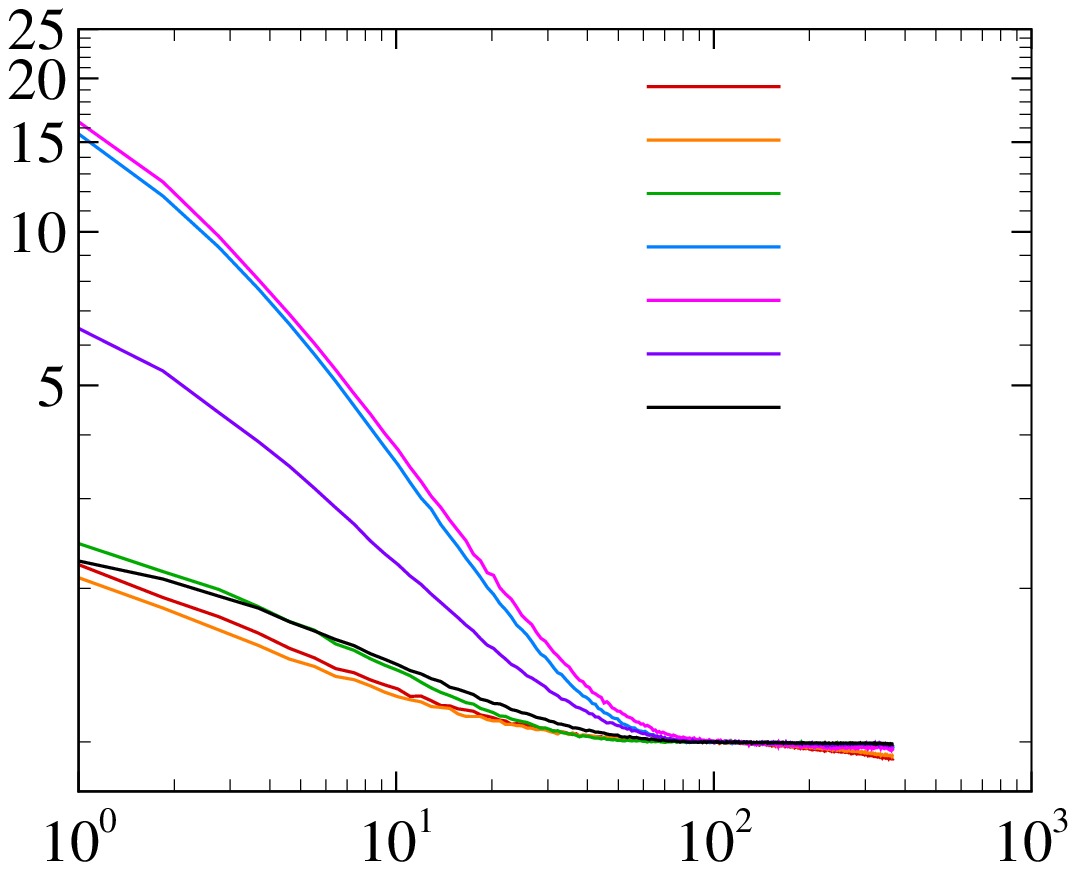}
\put(-1,70){(a)}
\put(50,0){$r^+$}
\put(0,35){\rotatebox{90}{$g_0(r)$}}
\put(71,65.8){\scriptsize P1}
\put(71,61.5){\scriptsize P2}
\put(71,57.8){\scriptsize P3}
\put(71,53.6){\scriptsize P4}
\put(71,49.5){\scriptsize P5}
\put(71,45.0){\scriptsize P6}
\put(71,40.5){\scriptsize P7}
\end{overpic}~
\begin{overpic}[width=0.5\textwidth]{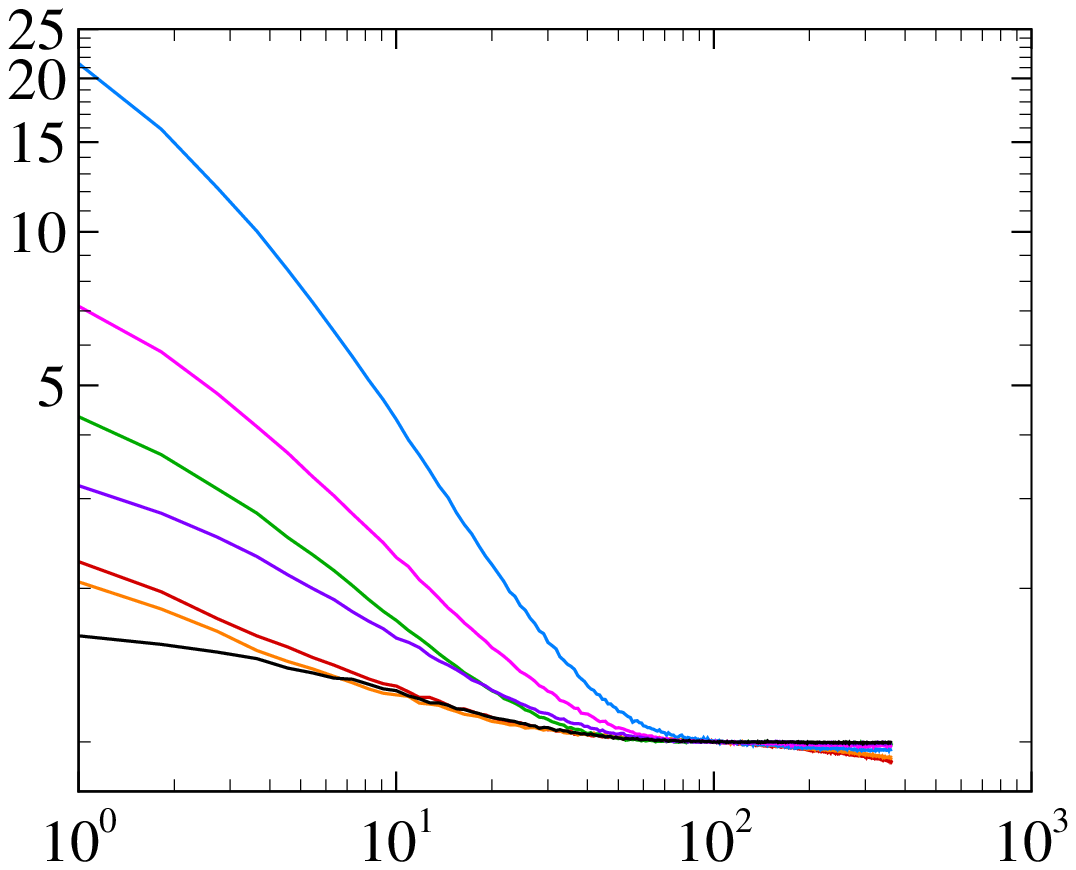}
\put(-1,70){(b)}
\put(50,0){$r^+$}
\put(0,35){\rotatebox{90}{$g_0(r)$}}
\end{overpic}\\[2.0ex]
\begin{overpic}[width=0.5\textwidth]{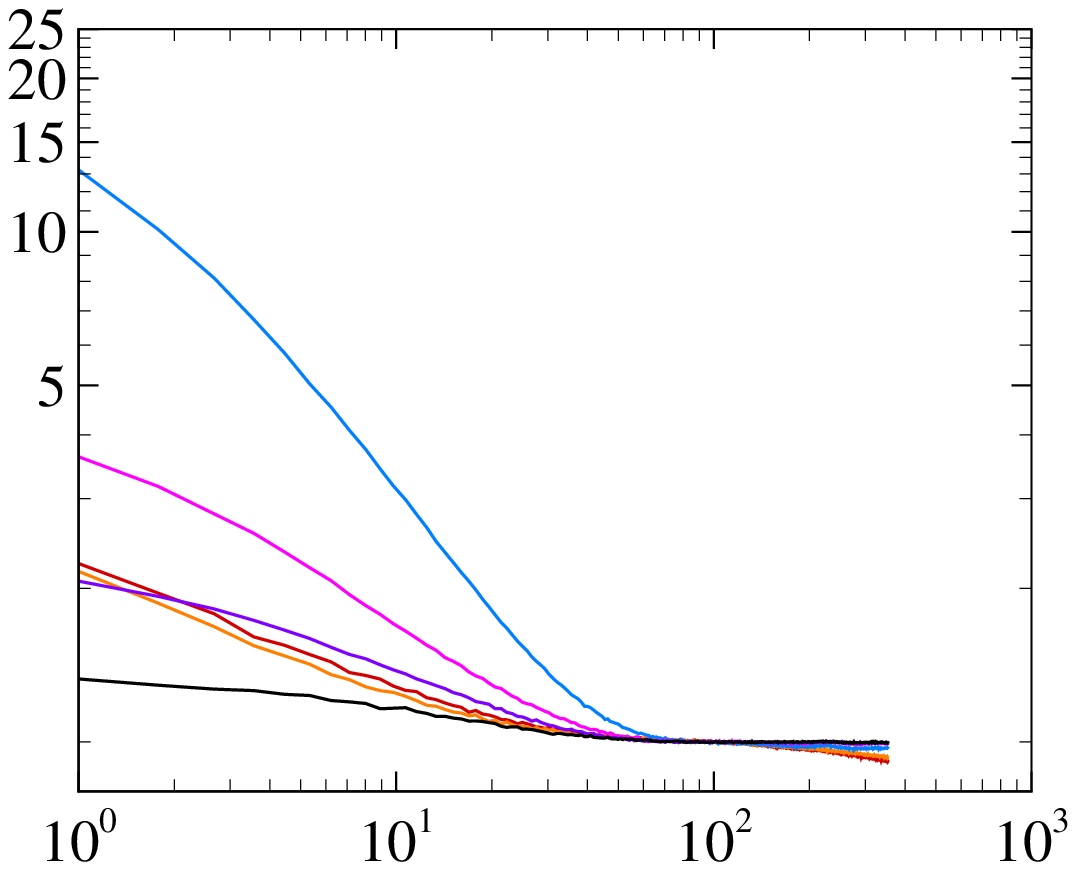}
\put(-1,70){(c)}
\put(50,0){$r^+$}
\put(0,35){\rotatebox{90}{$g_0(r)$}}
\end{overpic}~
\begin{overpic}[width=0.5\textwidth]{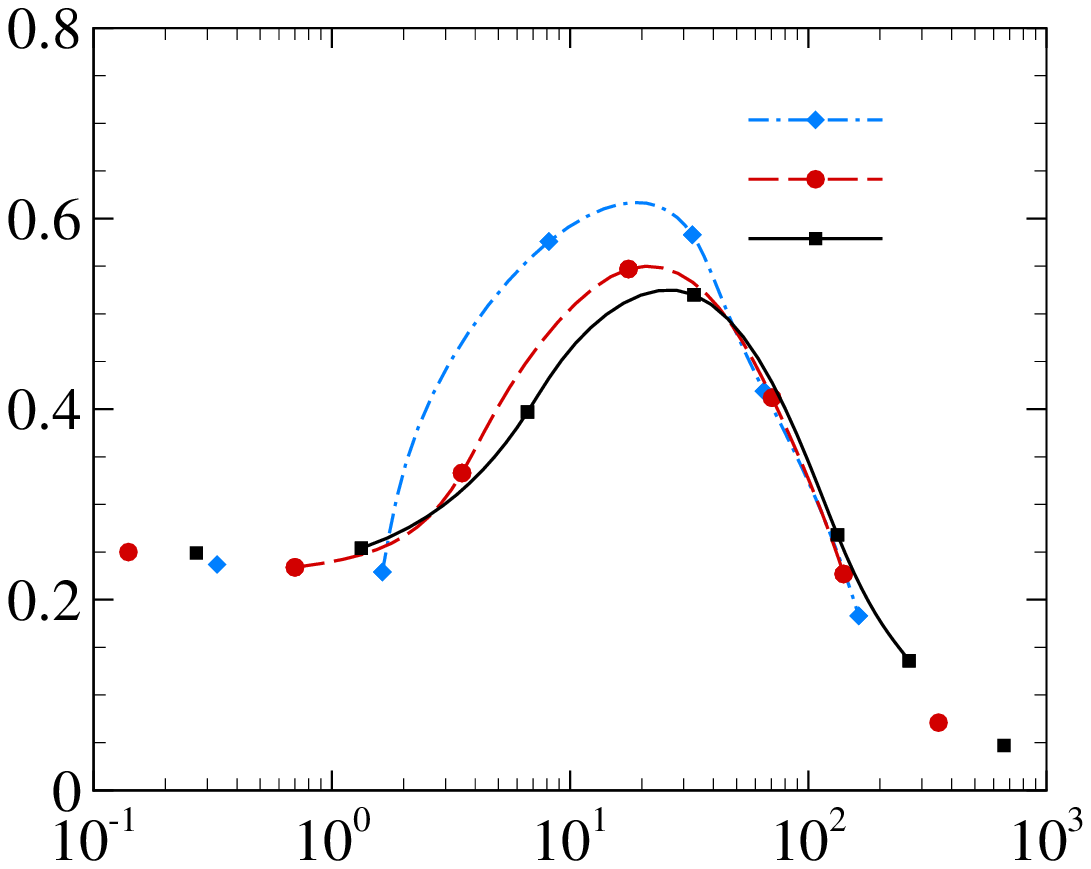}
\put(-1,70){(d)}
\put(50,0){$St^+$}
\put(0,38){\rotatebox{90}{$\alpha$}}
\put(78,64){\scriptsize M2}
\put(78,59){\scriptsize M4}
\put(78,54){\scriptsize M6}
\end{overpic}\\
\caption{Isotropic component of the ADF $g_0(r)$ below $y^+=10$, (a) case M2, (b) case M4, 
(c) case M6, and (d) power law of $g_0(r)$ at $r \rightarrow 0$ against $St^+$.}
\label{fig:adfhom}
\end{figure}

Integrating the ADF in the azimuthal direction, we can get its isotropic component without
incorporating the directional variation, namely the radial distribution function
\begin{equation}
g_0 (r) = \frac{1}{2 \pi} \int^{2 \pi}_0 g(r,\theta) {\rm d} \theta.
\end{equation}
The results of the case M2, M4 and M6 are shown in figures~\ref{fig:adfhom}(a-c).
The $g_0 (r)$ characterizes the probability of finding particle pairs within the circle of 
radius $r$, suggesting the level of radial particle clustering.
For the presently considered cases, particle populations P4 and P5 show the most evident
tendency of clustering, while for the other particle populations such an effect is
alleviated.
Comparing these cases, we found that the particles with the highest value of $g_0(r)$
at the limit of $r \rightarrow 0$ share approximately the same particle inertia under 
wall viscous units $St^+ \approx 15 \sim 50$.

The radial distribution function $g_0 (r)$ is capable of reflecting the small-scale clustering
~\citep{bec2007heavy,saw2008inertial},
as suggested by its association with the number of particle pairs within the radius $r$
\begin{equation}
{\mathscr N}(r) = n_0 \int^{2\pi}_0 \int^r_0 r' g(r',\theta) {\rm d} r {\rm d} \theta.
\end{equation}
For the spatially uniformly distributed particles, the ${\mathscr N}(r)$ can be estimated as
${\mathscr N}(r) = n_0 (y) \pi r^2$, hence the $g_0$ being constant, otherwise the decay of 
$g_0$ at the limit of $r \rightarrow 0$ should follow $g_0 \propto r^{-\alpha}$ with $\alpha>0$.
The exponents in the power law $\alpha$ for all the particle populations in each case are shown
in figure~\ref{fig:adfhom}(d), with cubic spline curve fitting using the data within the range
of $St^+ = 5 \sim 300$.
It can be observed that the peaks of the $\alpha$ are attained at $St^+ \approx 15 \sim 30$,
and the tendency of small-scale clustering is mitigated by the higher Mach number.
This is probably caused by the lower level of the near-wall particle accumulation 
at higher Mach numbers, which reduces the particle numbers close to the wall
(figure~\ref{fig:ccyy}(d)).

\begin{figure}
\centering
\begin{overpic}[width=0.5\textwidth]{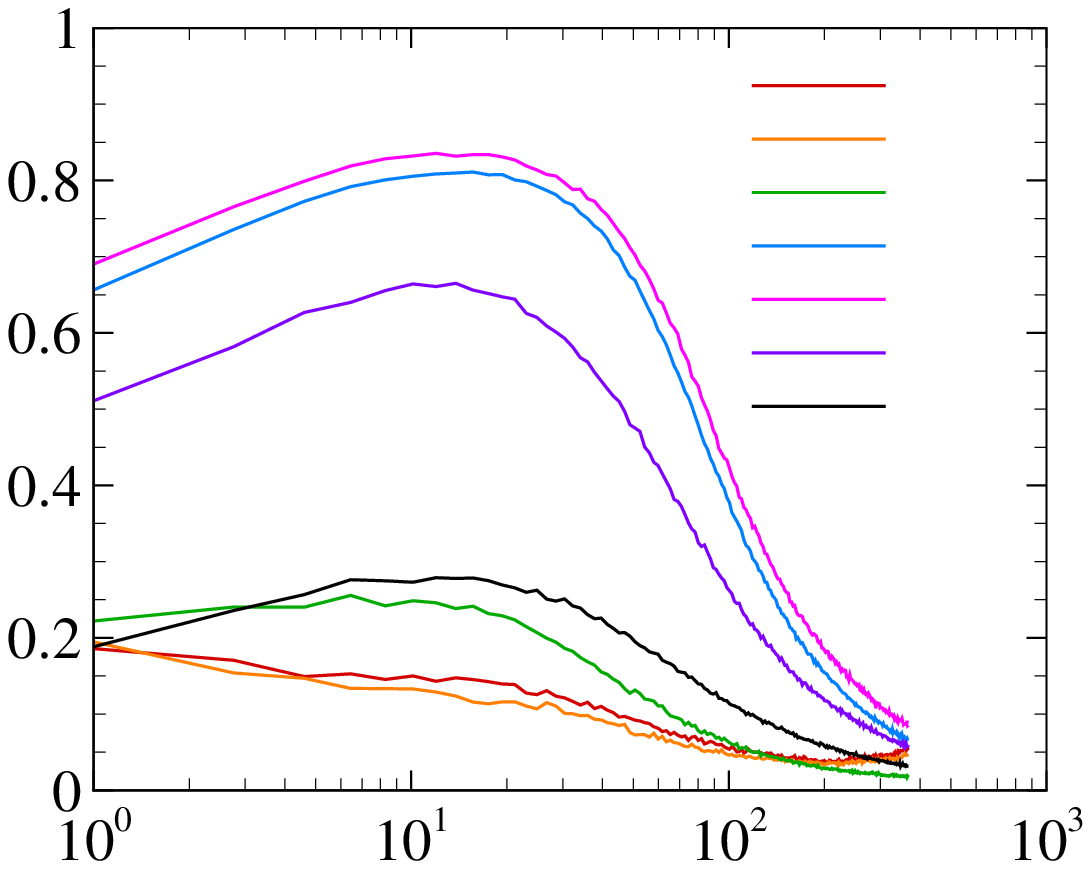}
\put(-1,70){(a)}
\put(50,0){$r^+$}
\put(0,35){\rotatebox{90}{$A(r)$}}
\put(80,65.8){\scriptsize P1}
\put(80,61.5){\scriptsize P2}
\put(80,57.8){\scriptsize P3}
\put(80,53.6){\scriptsize P4}
\put(80,49.5){\scriptsize P5}
\put(80,45.0){\scriptsize P6}
\put(80,40.5){\scriptsize P7}
\end{overpic}~
\begin{overpic}[width=0.5\textwidth]{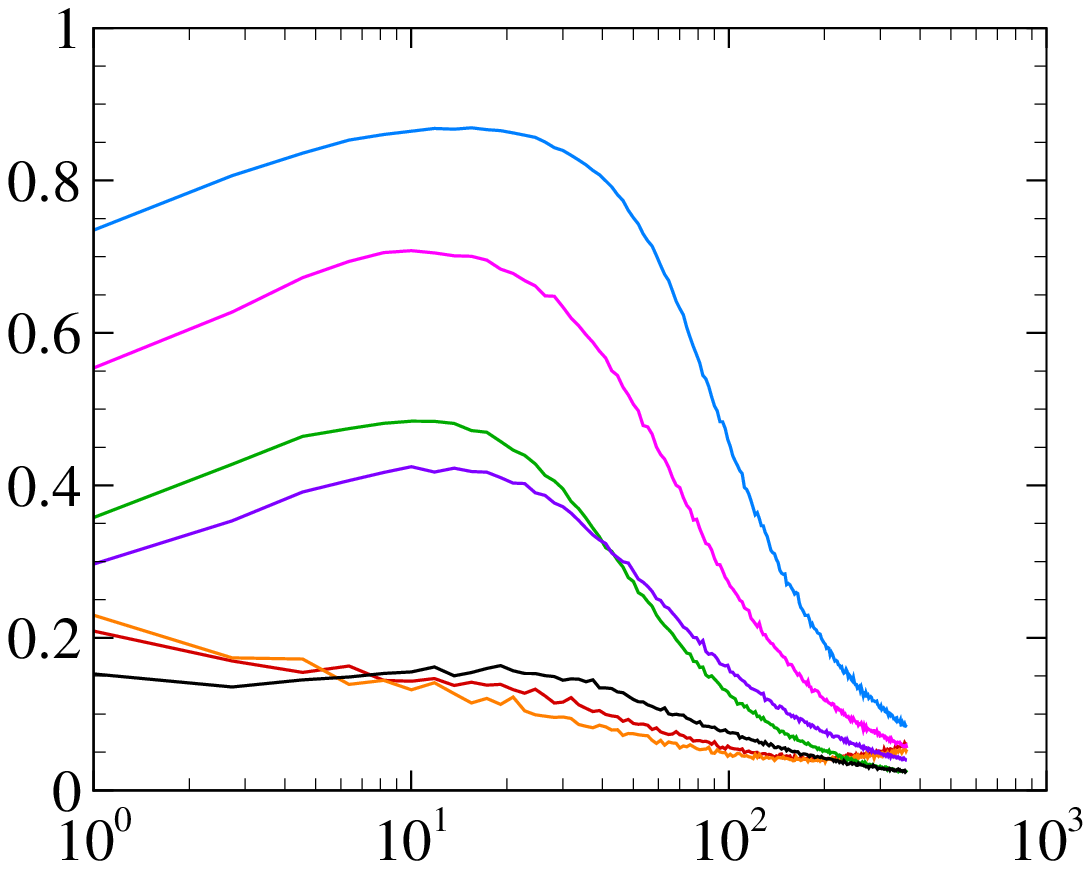}
\put(-1,70){(b)}
\put(50,0){$r^+$}
\put(0,35){\rotatebox{90}{$A(r)$}}
\end{overpic}\\[2.0ex]
\begin{overpic}[width=0.5\textwidth]{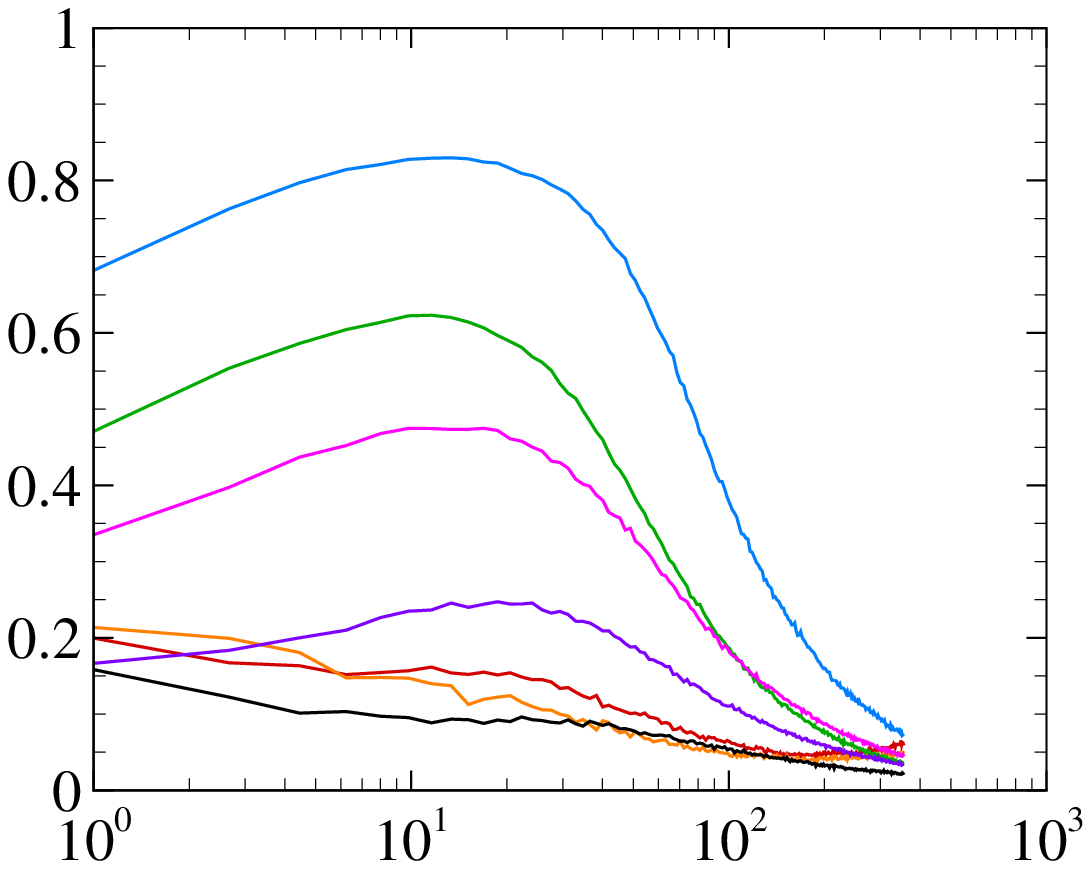}
\put(-1,70){(c)}
\put(50,0){$r^+$}
\put(0,35){\rotatebox{90}{$A(r)$}}
\end{overpic}~
\begin{overpic}[width=0.5\textwidth]{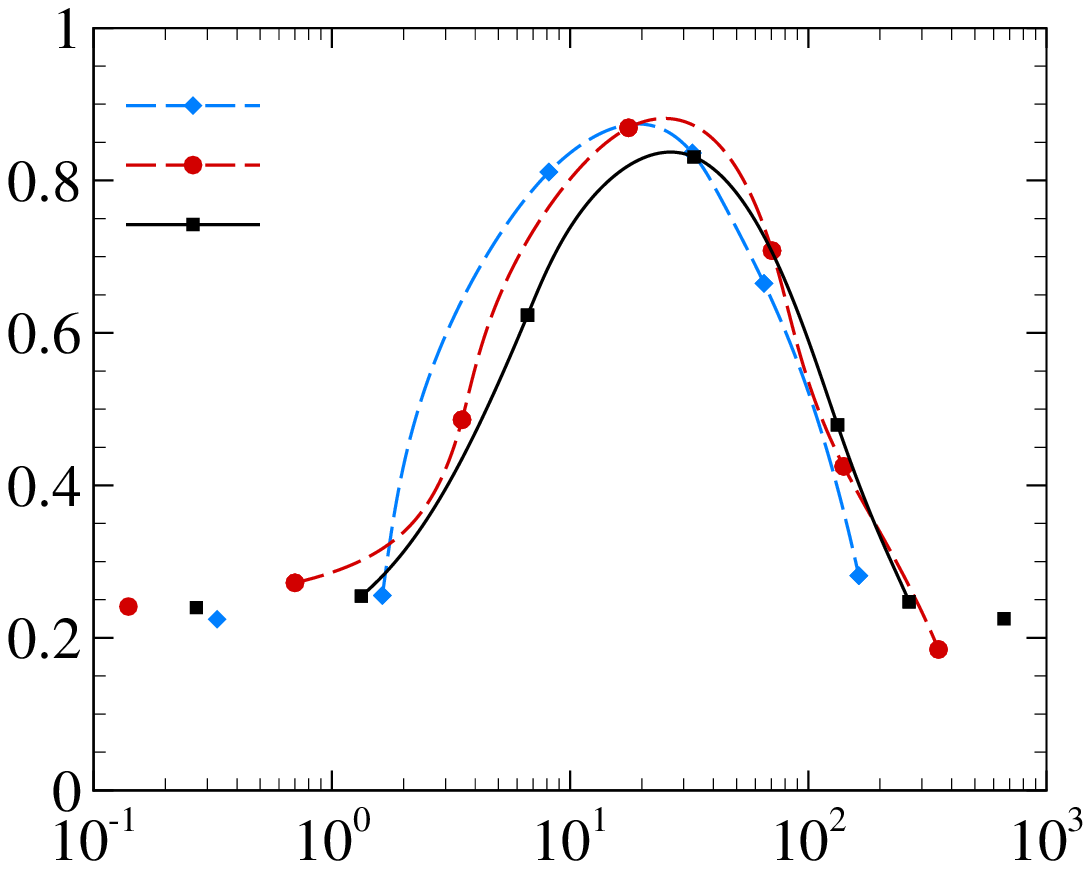}
\put(-1,70){(d)}
\put(50,0){$St^+$}
\put(0,35){\rotatebox{90}{$A(r)_{max}$}}
\put(28,65){\scriptsize M2}
\put(28,60){\scriptsize M4}
\put(28,55){\scriptsize M6}
\end{overpic}\\
\caption{Anisotropic indicator $A(r)$ of the ADF below $y^+=10$, 
(a) case M2, (b) case M4, (c) case M6, (d) the maxima of $A(r)$ against $St^+$.}
\label{fig:adfani}
\end{figure}

The anisotropy indicator of the particle clustering can be obtained by the ADF as
~\citep{casciola2007residual,sardina2012wall}
\begin{equation}
A(r) = \sqrt{\frac{\int (g(r,\theta)-g_0(r))^2 {\rm d} \theta}{\int g^2_0(r) {\rm d} \theta}},
\end{equation}
indicating the directional non-homogeneity of the wall-parallel particle distribution.
The results are shown in figure~\ref{fig:adfani}(a-c).
For particles with small $St^+$ or those with $St^+$ greater than $200$, 
the values of $A(r)$ are small and decay almost monotonically with the increasing $r^+$,
suggesting the weak directional dependence of $g(r,\theta)$ and hence the uniform distribution in
almost all the directions.
For other particle populations with $St^+$ ranging from $5.0$ to $200$, the levels of $A(r)$ are
higher and the peaks are attained not at the limit of $r^+ \rightarrow 0$ but at a finite $r^+$
of $10 \sim 20$, especially for the particle population P4 that displays a high level of 
spatial non-homogeneity for all the cases considered.
This implies that the narrow streaky structures with the width of $(20 \sim 40) \delta_\nu$,
which is in accordance with the spanwise scale of the low-speed streaks
\footnote{The characteristic spanwise length scale $\lambda^+_z \approx 100$ read from 
the spectra~\citep{hwang2013near} is the averaged spanwise intervals between the two streaks.
The low-speed streaks are narrower than the high-speed ones, as suggested by the conditional
average reported by~\citet{wang2015hairpin}.}.
For the purpose of comparison between cases with different Mach numbers, 
in figure~\ref{fig:adfani}(d) we plot the maximum of $A(r)$ against the $St^+$. 
Note that only the maxima of $A(r)$ in particle populations P3 $\sim$ P6 are attained at 
the non-zero $r$.
The level of clustering with the velocity streaks does not manifest significant variation 
in different cases,
further suggesting the weak dependence on the Mach number of the near-wall fluid and particle
dynamics.
Moreover, this trend of variation is similar to the near-wall accumulation discussed 
in figure~\ref{fig:ccyy}.
In fact, it has been previously observed that the near-wall accumulation and the clustering 
are two different aspects of the same process~\citep{sardina2012wall}, 
that the near-wall streamwise vortices bring the high Stokes number particles 
toward the near-wall regions 
where they reside for a rather long period, during which they are shifted horizontally to 
the low-speed streaks~\citep{soldati2009physics}, 
thereby leading to the coexistence of the two phenomena.

In this section, we have presented the instantaneous and statistical organizations of the particles 
with different inertia at various Mach numbers.
We found that the instantaneous distributions of the low Stokes number particles
are similar to those of 
the fluid density, restrained within the turbulent boundary layer, while the high Stokes number 
particles are capable of escaping the turbulent region and reaching the free stream.
The ubiquitous near-wall accumulation and the clustering phenomena in
low-speed flows are observed in the high-speed flows considered herein,
with the former slightly weakened with the increasing Mach number.

\section{Particle dynamics}   \label{sec:pardyn}

In this section, we will further discuss the kinematics and dynamics of particles, 
including their mean and fluctuating velocity and acceleration, 
with the special focus on the effects of the particle inertia at various Mach numbers.

\subsection{Mean and fluctuating particle velocity} \label{subsec:parvel}

\begin{figure}
\centering
\begin{overpic}[width=0.5\textwidth]{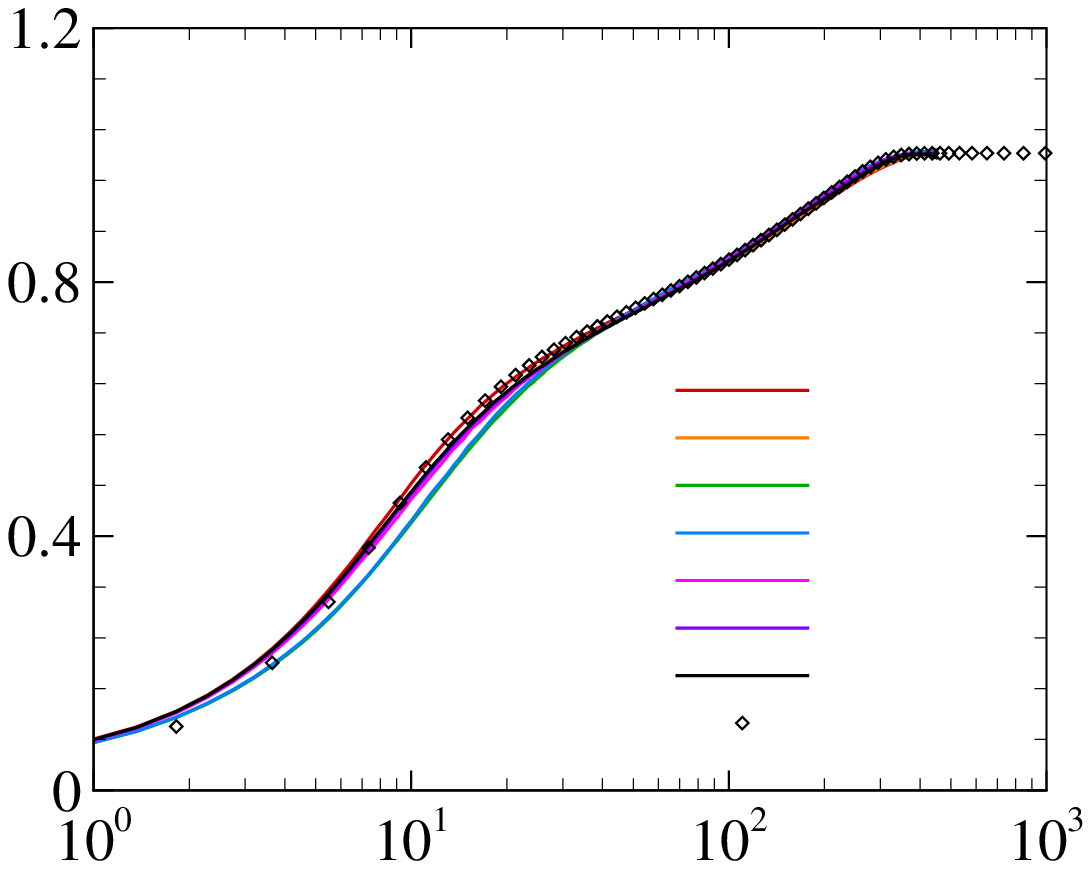}
\put(-2,70){(a)}
\put(50,0){$y^+$}
\put(-3,35){\rotatebox{90}{$\bar u_{1,p}/U_\infty$}}
\put(72,41.5){\scriptsize P1}
\put(72,38){\scriptsize P2}
\put(72,34){\scriptsize P3}
\put(72,30){\scriptsize P4}
\put(72,26.5){\scriptsize P5}
\put(72,22.3){\scriptsize P6}
\put(72,18.3){\scriptsize P7}
\put(72,14.8){\scriptsize $\tilde u_i/U_\infty$}
\end{overpic}~
\begin{overpic}[width=0.5\textwidth]{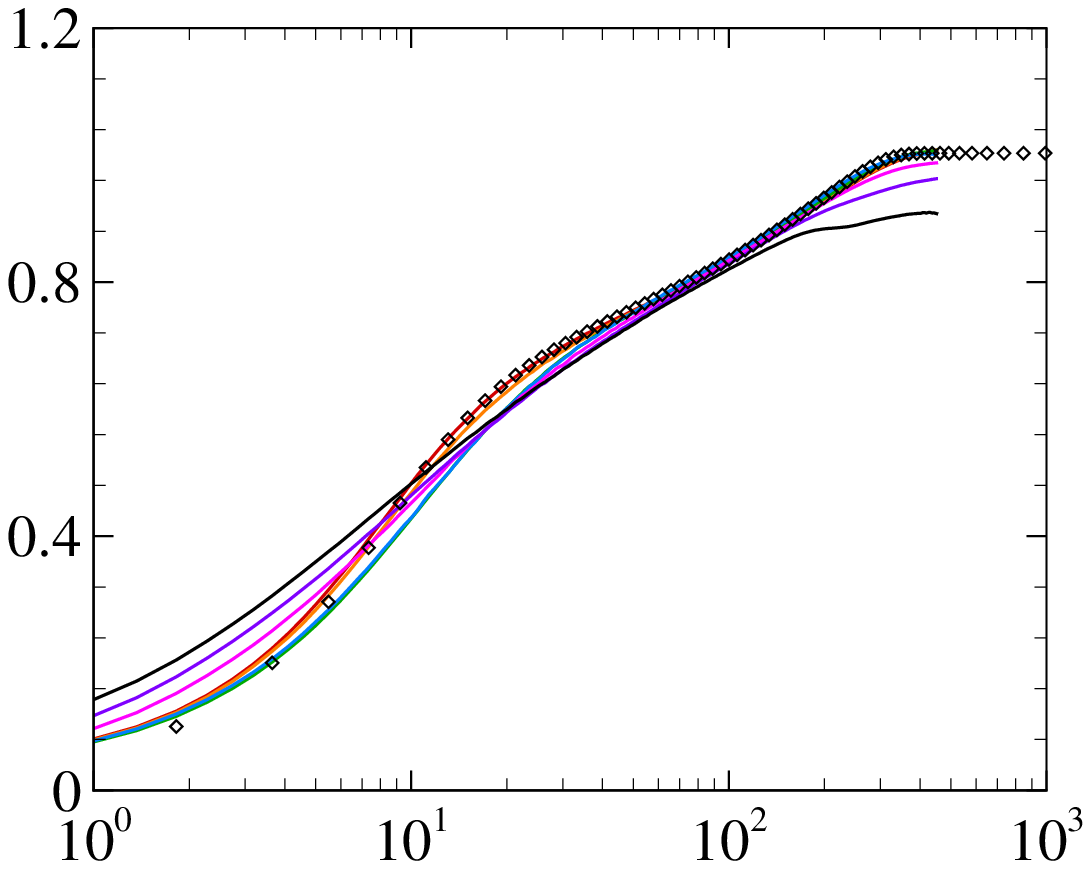}
\put(-2,70){(b)}
\put(50,0){$y^+$}
\put(-3,38){\rotatebox{90}{$\bar v_{1}/U_\infty$}}
\end{overpic}\\[1.0ex]
\begin{overpic}[width=0.5\textwidth]{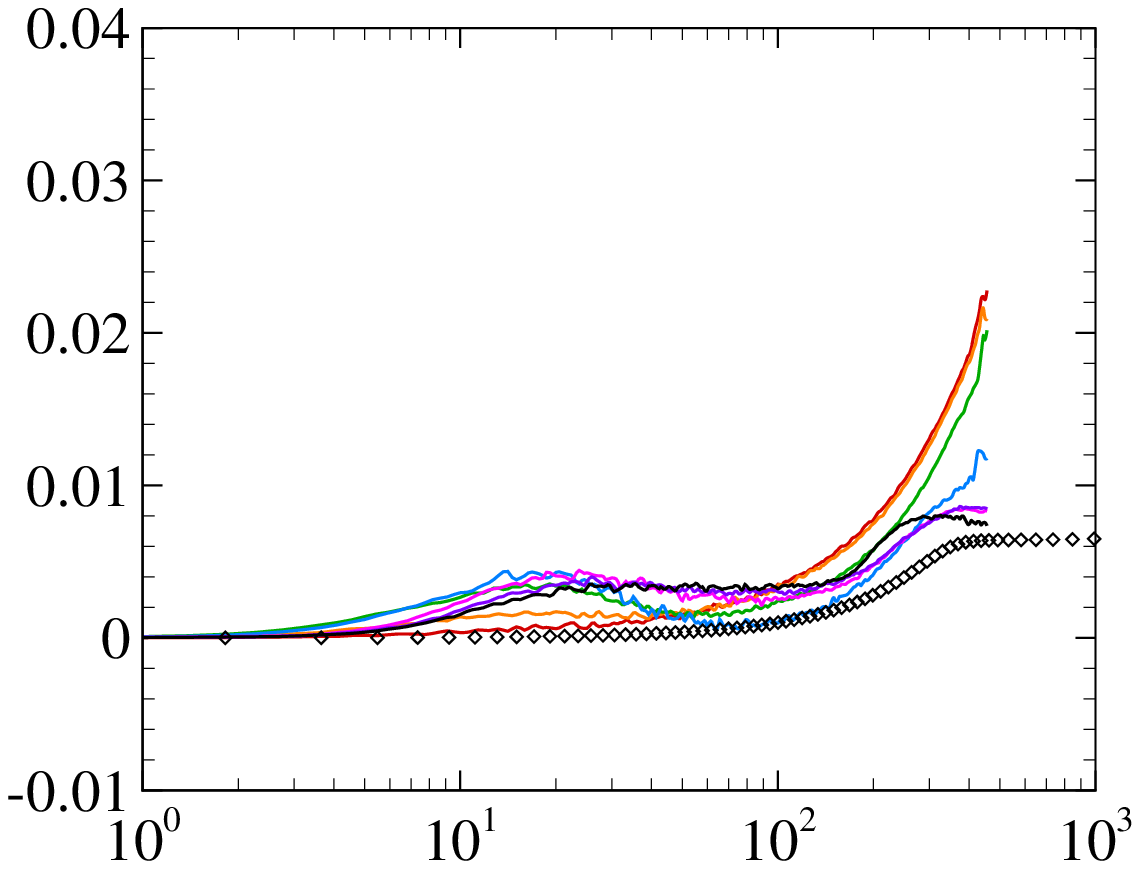}
\put(-2,70){(c)}
\put(50,0){$y^+$}
\put(-3,35){\rotatebox{90}{$\bar u_{2,p}/U_\infty$}}
\end{overpic}~
\begin{overpic}[width=0.5\textwidth]{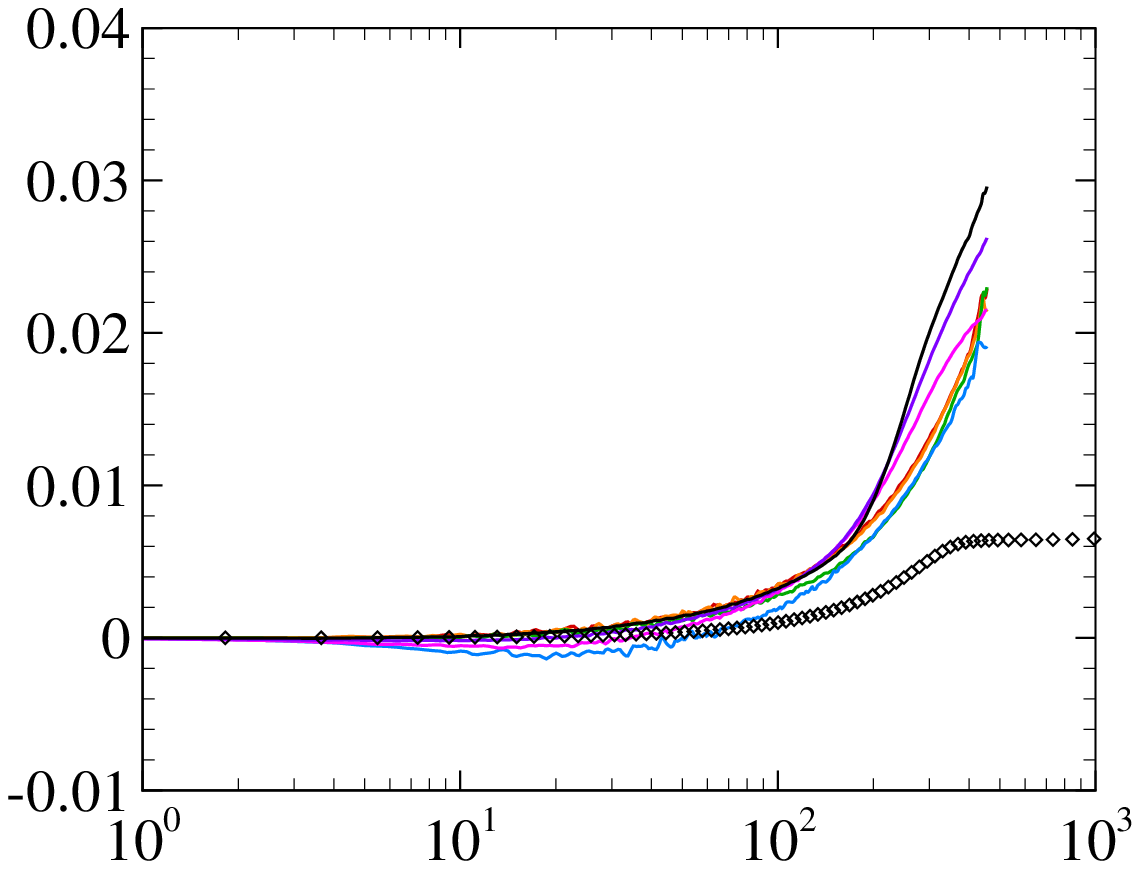}
\put(-2,70){(d)}
\put(50,0){$y^+$}
\put(-3,38){\rotatebox{90}{$\bar v_{2}/U_\infty$}}
\end{overpic}\\
\caption{Mean velocity of (a,c) fluid seen by particles $\bar u_{i,p}$ and of (b,d) particles 
$\bar v_{i}$ in the (a,b) streamwise and (c,d) wall-normal directions in case M4.}
\label{fig:pvelave}
\end{figure}

In figure~\ref{fig:pvelave} we present the wall-normal distributions of the streamwise and 
wall-normal mean fluid velocity seen by particles $\bar u_{i,p}$ and the mean particle velocity 
$\bar v_i$ in case M4, with the other cases omitted here due to the similar conclusions to be drawn.
In the streamwise direction,the mean fluid velocity seen by particles $\bar u_{1,p}$
(figure~\ref{fig:pvelave}(a)) are well-collapsed with the mean velocity $\bar u_1$,
except for the lower values for the particle population P4 that shows the most evident clustering 
beneath the low-speed streaks~\citep{narayanan2003mechanisms,xiao2020eulerian}.
The statistics of particle populations with lower and higher Stokes numbers 
do not show such discrepancies from 
the mean fluid velocity $\bar u_1$ because of their comparatively uniform distributions.
The mean streamwise particle velocity $\bar v_1$ (figure~\ref{fig:pvelave}(b)) behaves differently.
The $\bar v_1$ of the low Stokes number particle populations (P1 and P2) are 
in general consistency with the mean fluid velocity $\bar u_1$.
As the $St^+$ increases to moderate values (P3 and P4), the mean particle velocities $\bar v_1$ 
are lower than that of the fluid $\bar u_1$ in the buffer region but recover as it reaches
$y^+ \gtrsim 60$.
For high Stokes number particle populations P5, P6 and P7, the $\bar v_1$ are
higher than the mean fluid velocity beneath $y^+ \approx 10$ and lower above that location.
These phenomena should be caused by the tendency to their increasing tardiness of 
being accelerated or decelerated because of their comparatively larger inertia,
especially for particle populations P5 $\sim$ P7.
Since they are not bounded by the no-slipping velocity condition at the wall,
these high Stokes number particles either keep rolling on the wall before they are erupted or decelerated to 
zero velocity or reflected away from the wall~\citep{picciotto2005statistics,mortimer2019near}.
As a result, the mean velocity profiles of the particles are consistently flatter than those of
the fluid~\citep{kulick1994particle}.

The wall-normal velocities, on the other hand, show some peculiar phenomena.
Although small in values, the $\bar u_{2,p}$ (figure~\ref{fig:pvelave}(c)) are higher than 
the mean fluid velocity $\bar u_{2}$ across the boundary layer, even for the small particles.
Since the particle populations with moderate and large $St^+$ (P3 $\sim$ P7) tend to accumulate 
within the low-speed regions, which are highly correlated with ejection events,
the fluid velocity seen by particles $\bar u_{2,p}$ are higher within $y^+=50$, 
with the highest values attained for particles P4 and P5 that show the most significant
clustering behaviour~\citep{soldati2009physics,zhao2012stokes,milici2016statistics}.
In the outer region, however, the trend is reversed.
We can attribute such a trend of variation to the fact that the low Stokes number particles are restrained
within the boundary layers so that only the ejections can be reflected in $\bar u_{2,p}$,
whereas the high Stokes number particles are capable of escaping toward the free stream, 
thus alleviating the high $\bar u_{2,p}$ values.
The mean particle velocities $\bar v_2$ (figure~\ref{fig:pvelave}(d)) close to the wall 
are approximately zero, while higher than the mean fluid velocity $\bar u_2$ above $y^+=50$ 
and than that seen by particles $\bar u_{2,p}$ in the outer region for all particle populations,
showing weak dependence on the particle inertia.
Similar results have been reported by~\citet{xiao2020eulerian} at the free stream Mach number $2.0$
for particles with $St^+ = 2.36$ and $24.33$.
This suggests that the mean particle wall-normal velocities are probably merely associated with 
the spatial development of the boundary layer.

\begin{figure}
\centering
\begin{overpic}[width=0.5\textwidth]{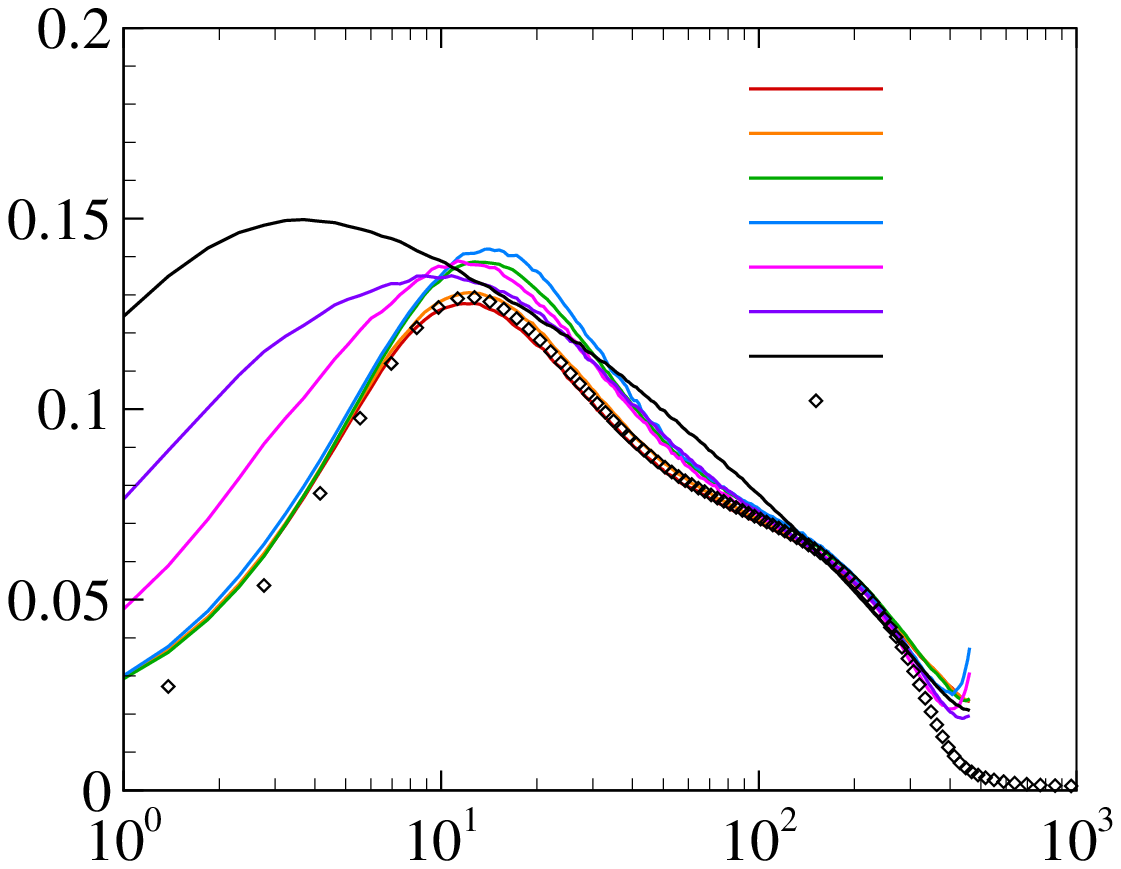}
\put(-2,70){(a)}
\put(50,0){$y^+$}
\put(-3,35){\rotatebox{90}{$\bar v'_{1}/U_\infty$}}
\put(75,66){\scriptsize P1}
\put(75,62){\scriptsize P2}
\put(75,58.4){\scriptsize P3}
\put(75,55){\scriptsize P4}
\put(75,51.5){\scriptsize P5}
\put(75,48){\scriptsize P6}
\put(75,44.3){\scriptsize P7}
\put(75,41.3){\scriptsize $\bar u''_{i}/U_\infty$}
\end{overpic}~
\begin{overpic}[width=0.5\textwidth]{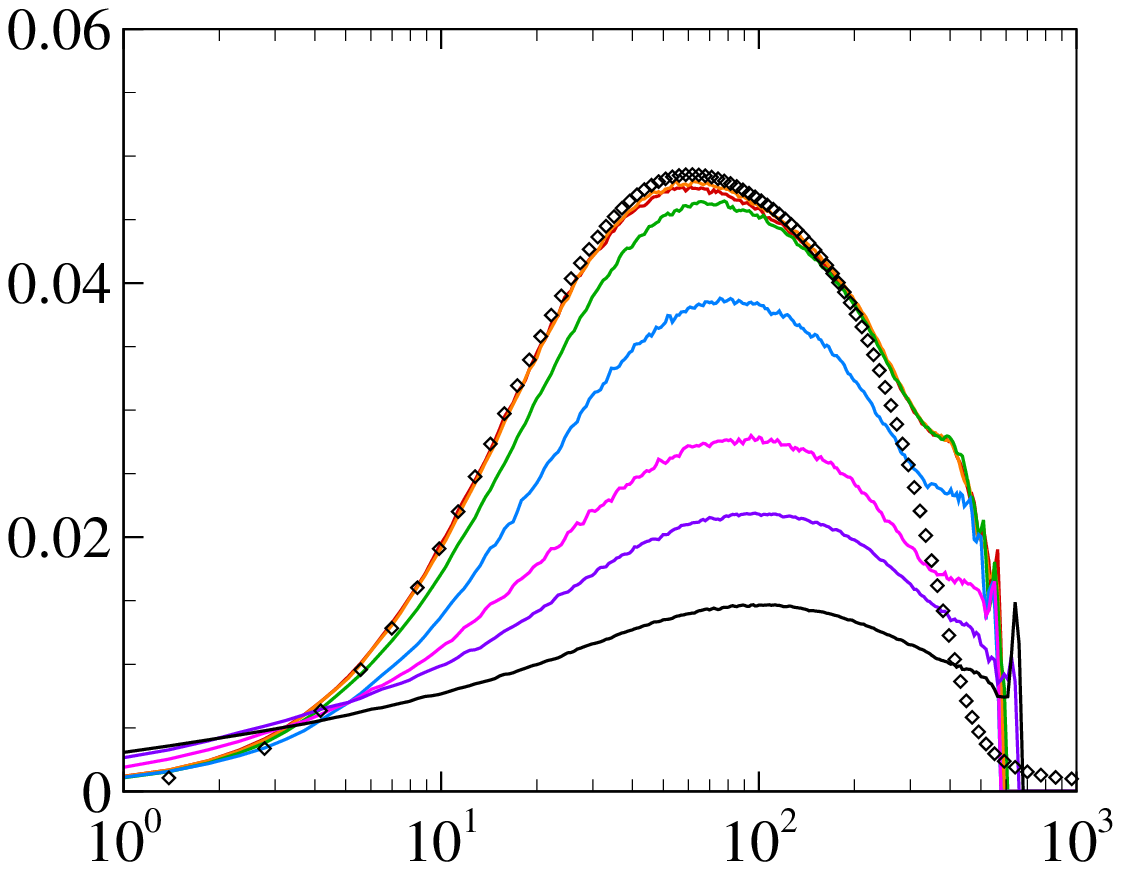}
\put(-2,70){(b)}
\put(50,0){$y^+$}
\put(-3,35){\rotatebox{90}{$\bar v'_{2}/U_\infty$}}
\end{overpic}\\[1.0ex]
\begin{overpic}[width=0.5\textwidth]{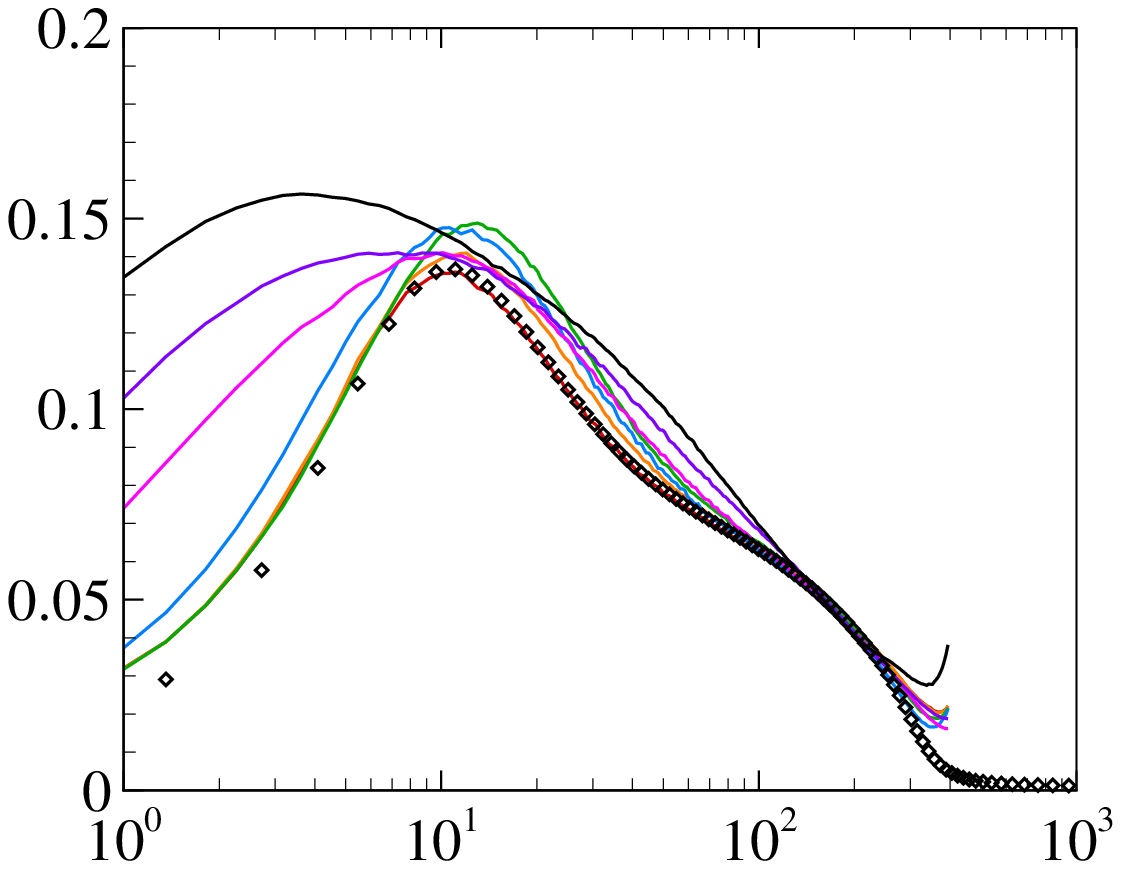}
\put(-2,70){(c)}
\put(50,0){$y^+$}
\put(-3,35){\rotatebox{90}{$\bar v'_{1}/U_\infty$}}
\end{overpic}~
\begin{overpic}[width=0.5\textwidth]{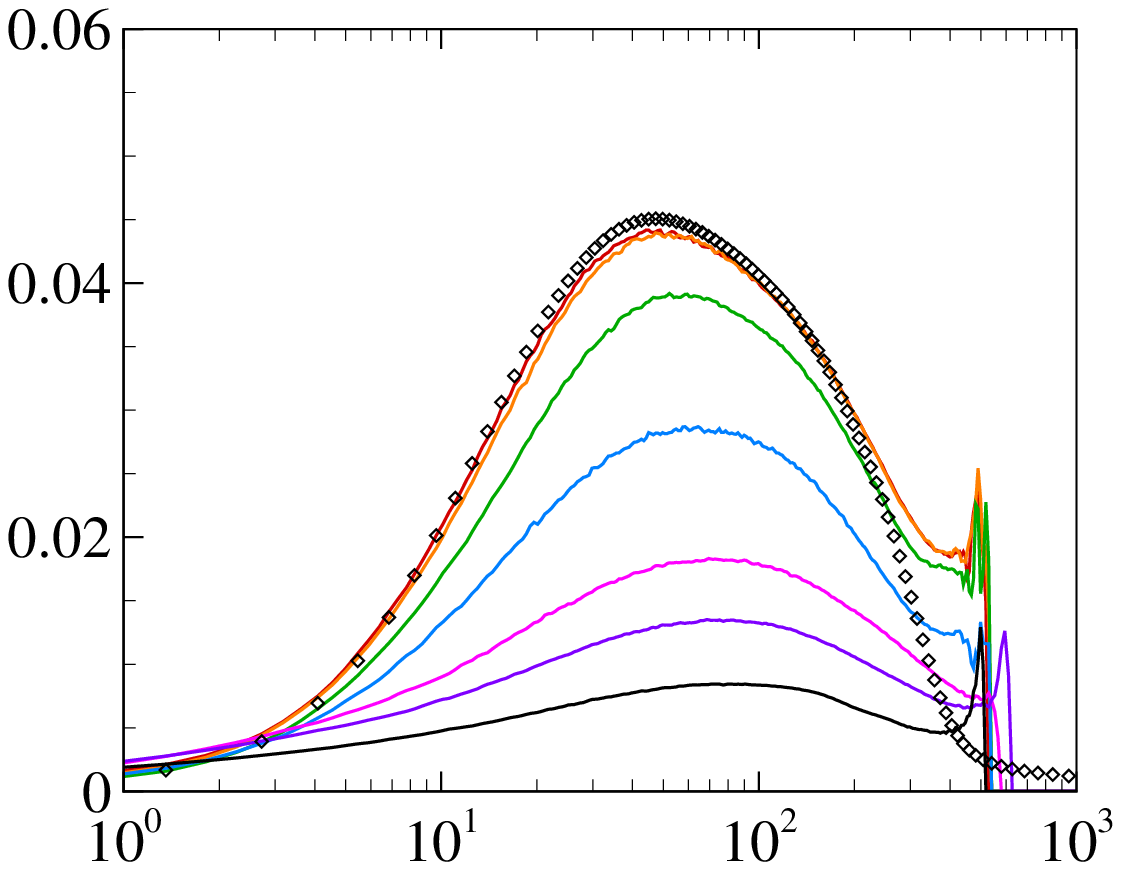}
\put(-2,70){(d)}
\put(50,0){$y^+$}
\put(-3,35){\rotatebox{90}{$\bar v'_{2}/U_\infty$}}
\end{overpic}\\[1.0ex]
\begin{overpic}[width=0.5\textwidth]{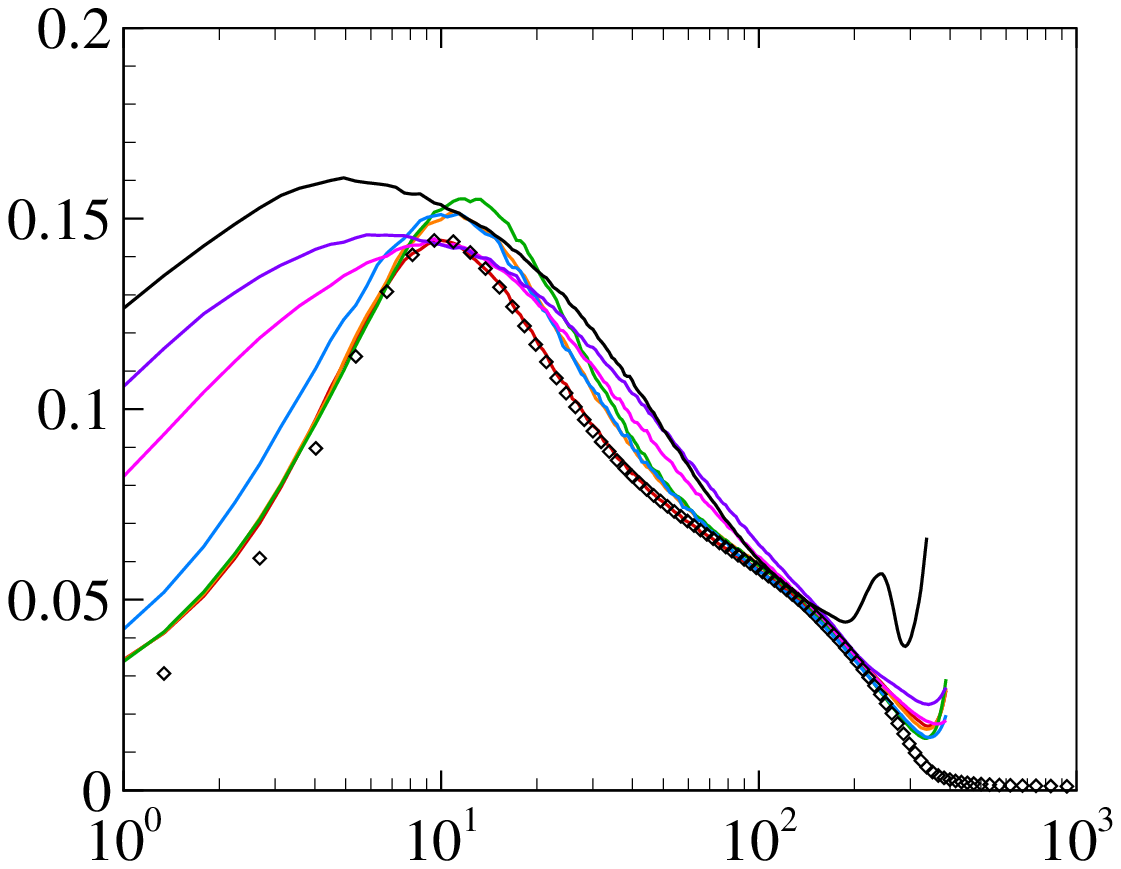}
\put(-2,70){(e)}
\put(50,0){$y^+$}
\put(-3,35){\rotatebox{90}{$\bar v'_{1}/U_\infty$}}
\end{overpic}~
\begin{overpic}[width=0.5\textwidth]{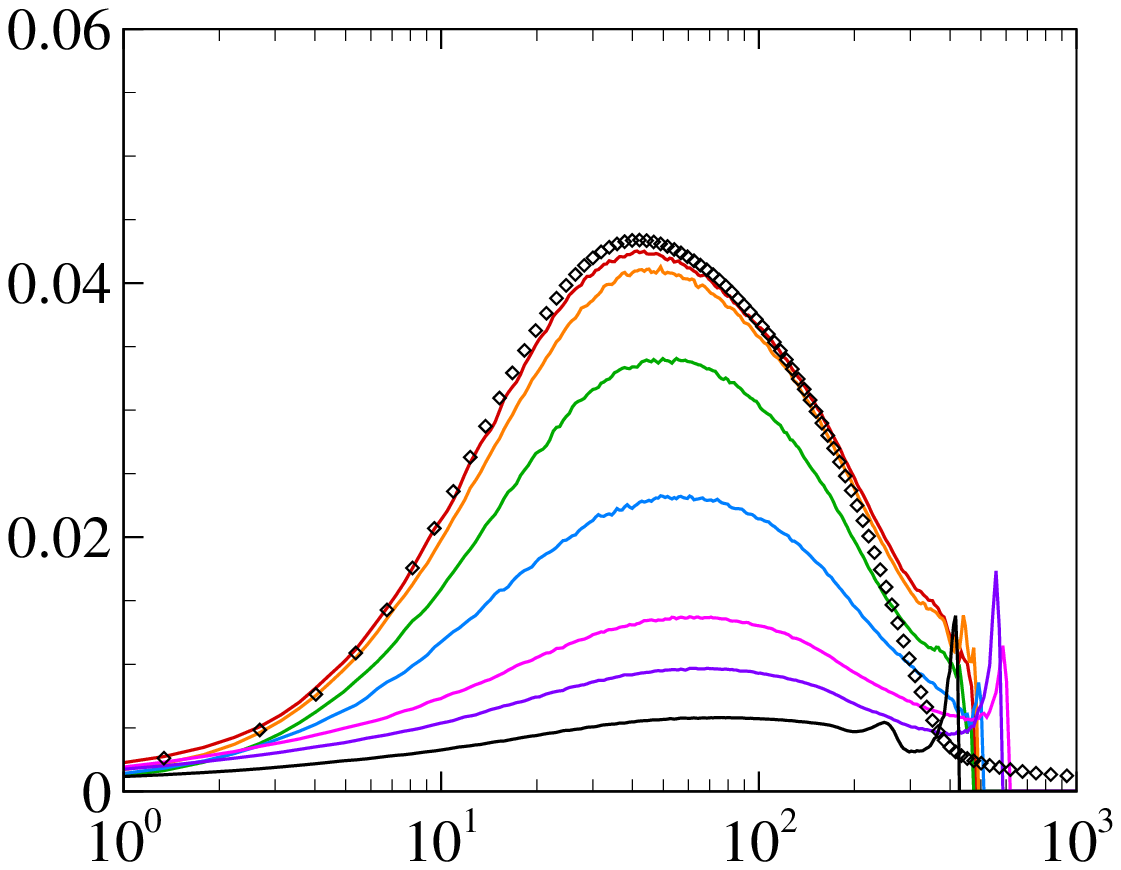}
\put(-2,70){(f)}
\put(50,0){$y^+$}
\put(-3,35){\rotatebox{90}{$\bar v'_{2}/U_\infty$}}
\end{overpic}\\
\caption{RMS of particle velocity fluctuation, (a,c,e) $\bar v'_{1}$, (b,d,f) $\bar v'_{2}$
in (a,b) case M2, (c,d) case M4 and (e,f) case M6.}
\label{fig:pvelrms}
\end{figure}

\begin{figure}
\centering
\begin{overpic}[width=0.5\textwidth]{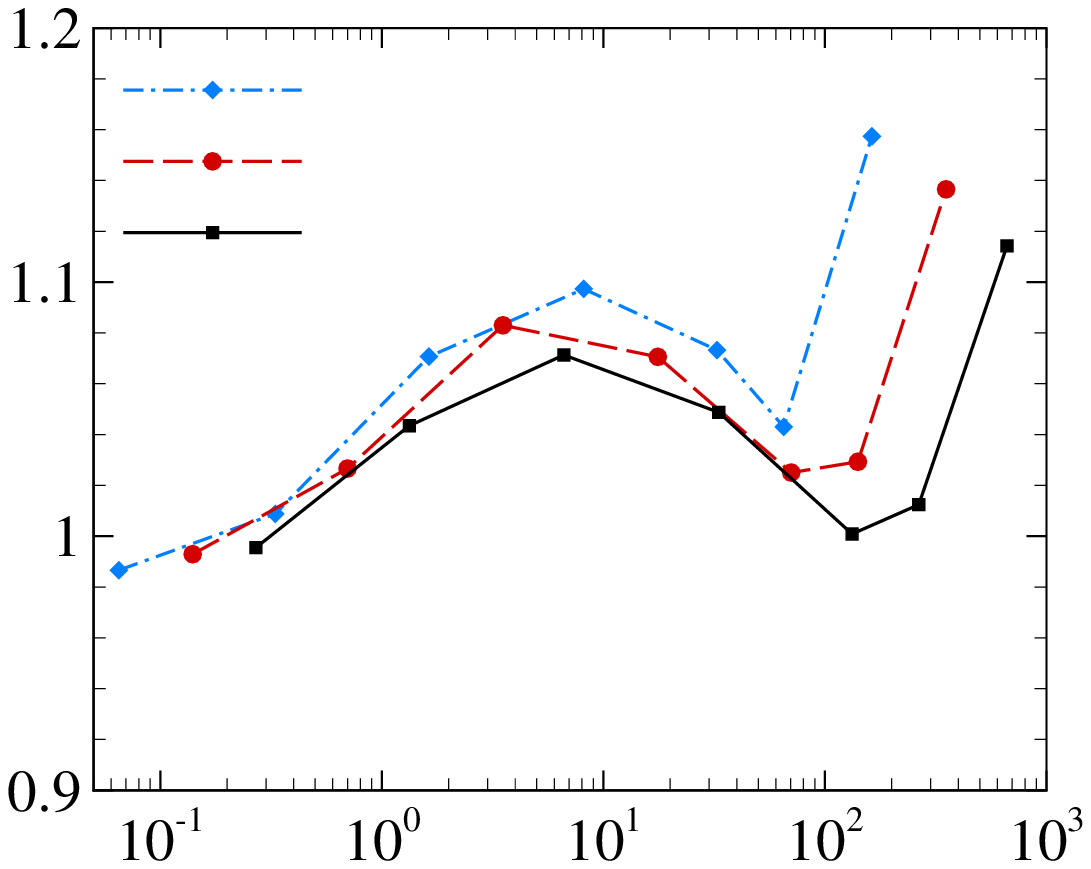}
\put(-2,70){(a)}
\put(50,0){$St^+$}
\put(32,66){\scriptsize M2}
\put(32,60){\scriptsize M4}
\put(32,54.5){\scriptsize M6}
\put(-3,20){\rotatebox{90}{\scriptsize ${\rm max}[\bar v'_{1}]/{\rm max}[\bar u''_{1}]$}}
\end{overpic}~
\begin{overpic}[width=0.5\textwidth]{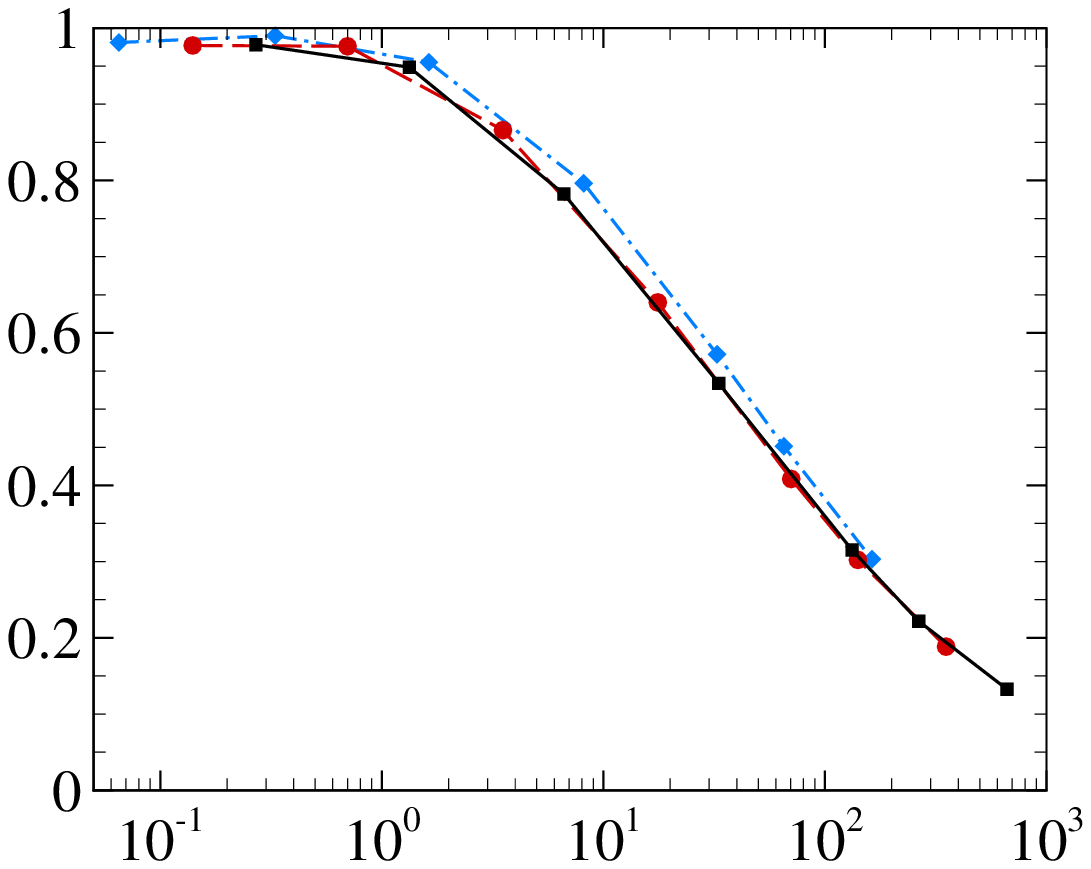}
\put(-2,70){(b)}
\put(50,0){$St^+$}
\put(-3,20){\rotatebox{90}{\scriptsize ${\rm max}[\bar v'_{2}]/{\rm max}[\bar u''_{2}]$}}
\end{overpic}\\
\caption{Ratio between the peaks values of $\bar v'_{i}$ in the near-wall region
and $\bar u''_{i}$ for different cases, components (a) $\bar v'_{1}$ and (b) $\bar v'_{2}$.}
\label{fig:prmsmax}
\end{figure}

In figure~\ref{fig:pvelrms} we present the root-mean-square (RMS) of the streamwise and 
wall-normal components of the particle velocity fluctuations, denoted as $\bar v'_{i}$.
The trend of variation is in general agreement with those in incompressible turbulent channel flows
~\citep{marchioli2008statistics,vinkovic2011direct,li2016direct,fong2019velocity}.
Specifically, within the boundary layer, the RMS of the low Stokes number particles (P1 and P2)
are well-collapsed with those of the fluid.
The peaks of $\bar v'_{1}$ vary non-monotonically with the increasing $St^+$, shifting gradually
toward the wall, showing no tendency of decreasing to zero near the wall. 
In the outer region, the $\bar v'_{1}$ are almost collapsed, manifesting weak dependence on
the Mach number.
The $\bar v'_{2}$, on the other hand, decreases monotonically with $St^+$.
Besides these consistencies, there are two aspects that cannot be covered by the observations
of previous studies.
Firstly, if we plot the near-wall local maximum of $\bar v'_{1}$ against the Stokes number $St^+$,
as displayed in figure~\ref{fig:prmsmax}(a), we found that the former is not varying monotonically
with the latter, with local maxima achieved at $St^+ \approx 10$ for each case but show 
consistent decrement with the increasing Mach number.
The peaks of $\bar v'_{2}$ compared with those of the fluid (figure~\ref{fig:prmsmax}(b)), 
however, decrease systematically, with very weak Mach number effects.
The differences in the $\bar v'_{1}$ in various cases are probably ascribed to
the variation of the particle Mach number that is related to the drag force acting 
on the particles, which will be discussed in the next subsection.
Secondly, both the $\bar v'_{1}$ and $\bar v'_{2}$ deviate from the RMS of fluid velocity 
fluctuation near the edge of the boundary layer, even at the lowest $St^+$.
This is beyond expectation for these low Stokes number particles, which are commonly used in experimental 
measurements to obtain velocity in such methods as particle image velocimetry (PIV), 
should be capable of following the trajectories and replicating the statistics.
The reason is similar to the higher $\bar v_2$ and $\bar u_{2,p}$ compared with $\bar u_2$.
Considering the numerical settings in the present study, the particles with small $St^+$
are primarily restricted beneath the turbulent-non-turbulent interfaces, 
thereby only reflecting the statistics in the turbulent region but
avoiding those in the non-turbulent region and hence the higher intensities.
Such a bias will be amended if the particles at the inlet are statistically uniformly distributed
within a region much higher than the nominal boundary layer thickness.
We have performed extra simulations for the verification of this inference,
details of which can be found in Appendix~\ref{sec:comp}.

\subsection{Particle acceleration}

\begin{figure}
\centering
\begin{overpic}[width=0.5\textwidth]{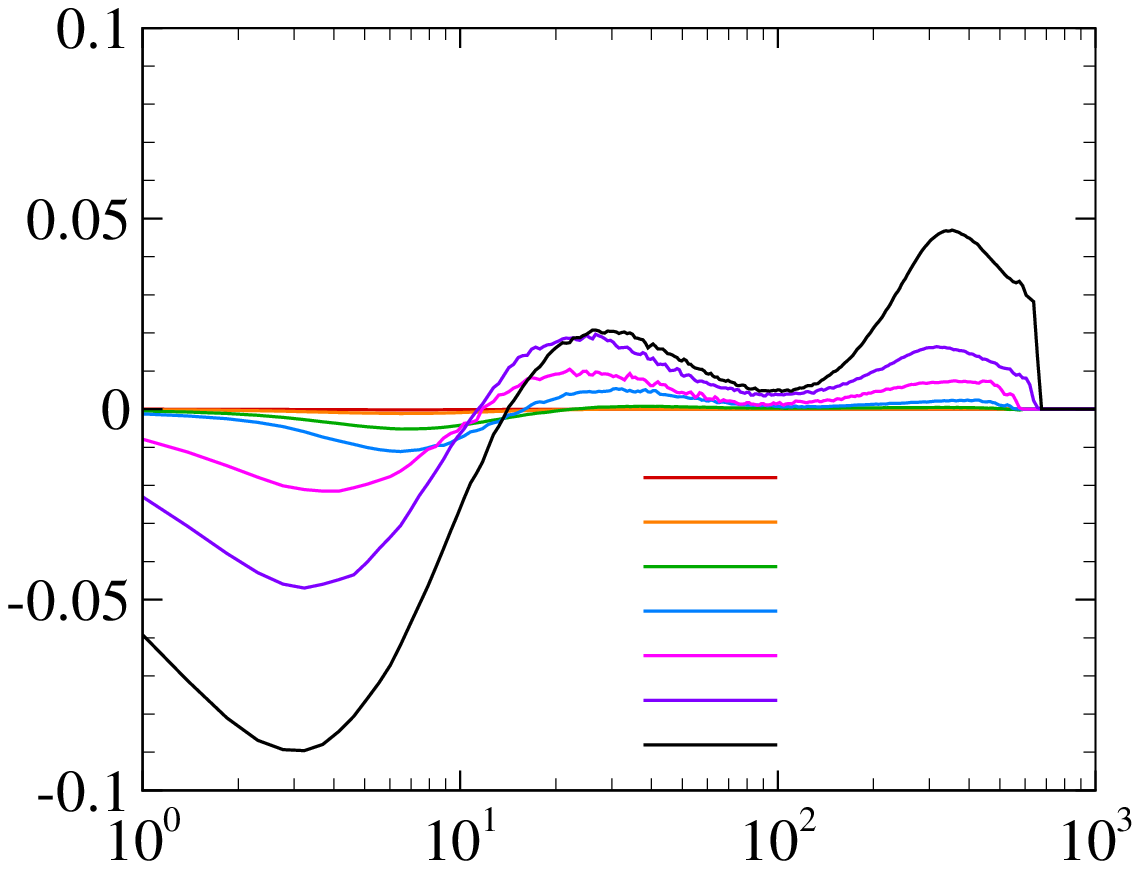}
\put(-2,70){(a)}
\put(48,0){$y^+$}
\put(-3,30){\rotatebox{90}{$\bar v_{1,s}/U_\infty$}}
\put(66,34.5){\scriptsize P1}
\put(66,30.5){\scriptsize P2}
\put(66,27){\scriptsize P3}
\put(66,23.8){\scriptsize P4}
\put(66,20.5){\scriptsize P5}
\put(66,16.5){\scriptsize P6}
\put(66,13){\scriptsize P7}
\end{overpic}~
\begin{overpic}[width=0.5\textwidth]{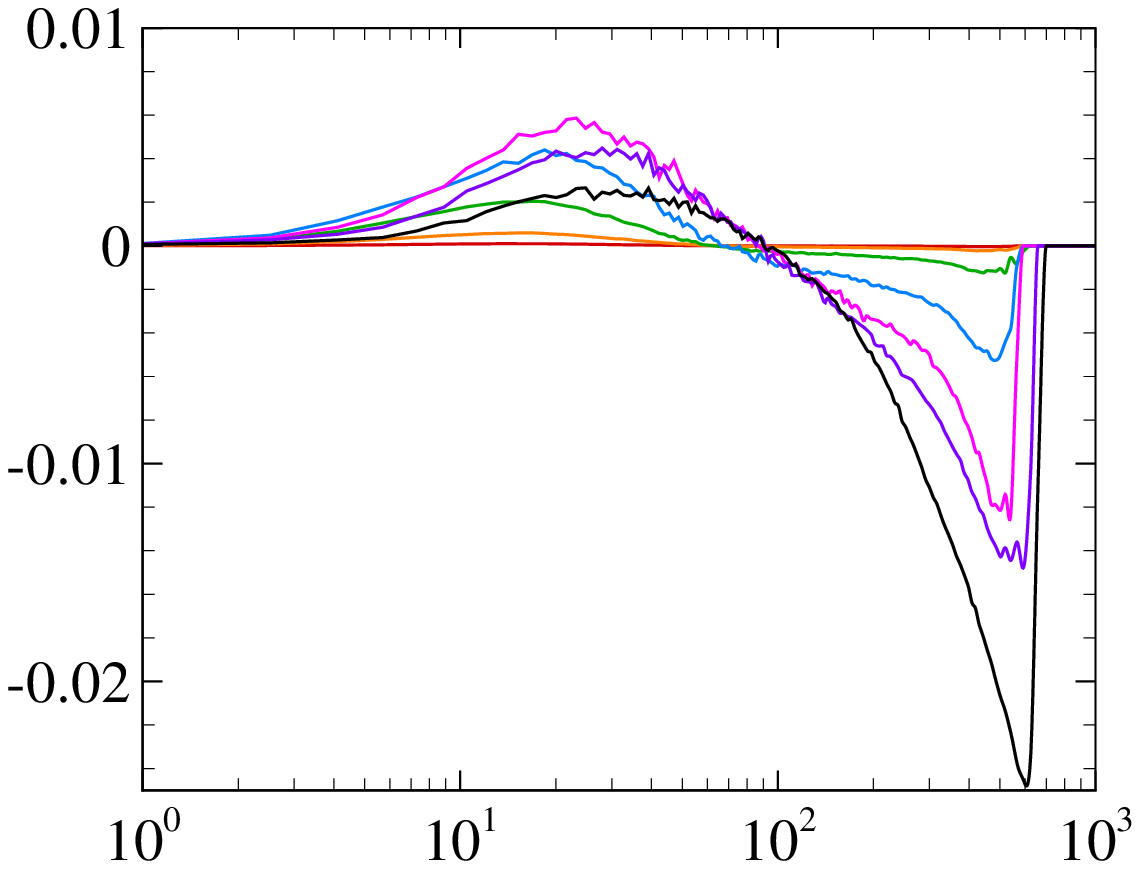}
\put(-2,70){(b)}
\put(48,0){$y^+$}
\put(-3,30){\rotatebox{90}{$\bar v_{2,s}/U_\infty$}}
\end{overpic}\\[1.0ex]
\begin{overpic}[width=0.5\textwidth]{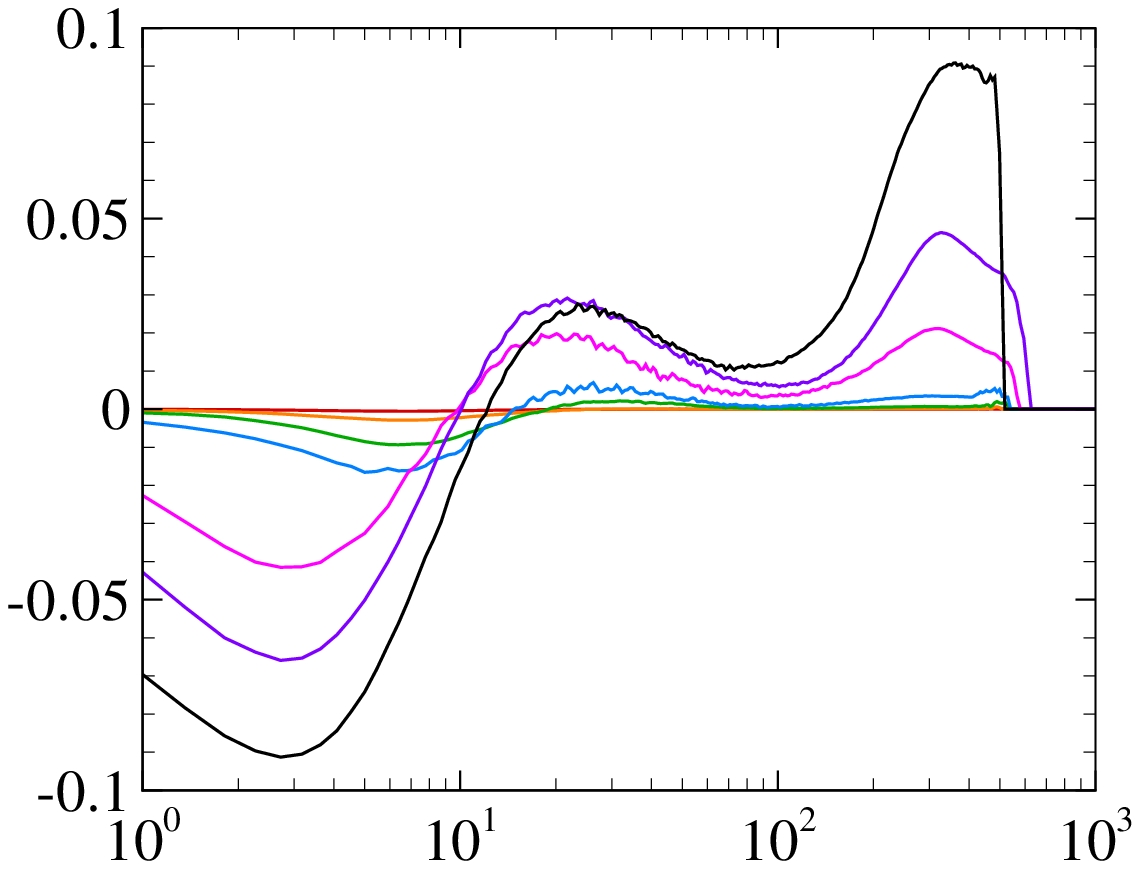}
\put(-2,70){(c)}
\put(48,0){$y^+$}
\put(-3,30){\rotatebox{90}{$\bar v_{1,s}/U_\infty$}}
\end{overpic}~
\begin{overpic}[width=0.5\textwidth]{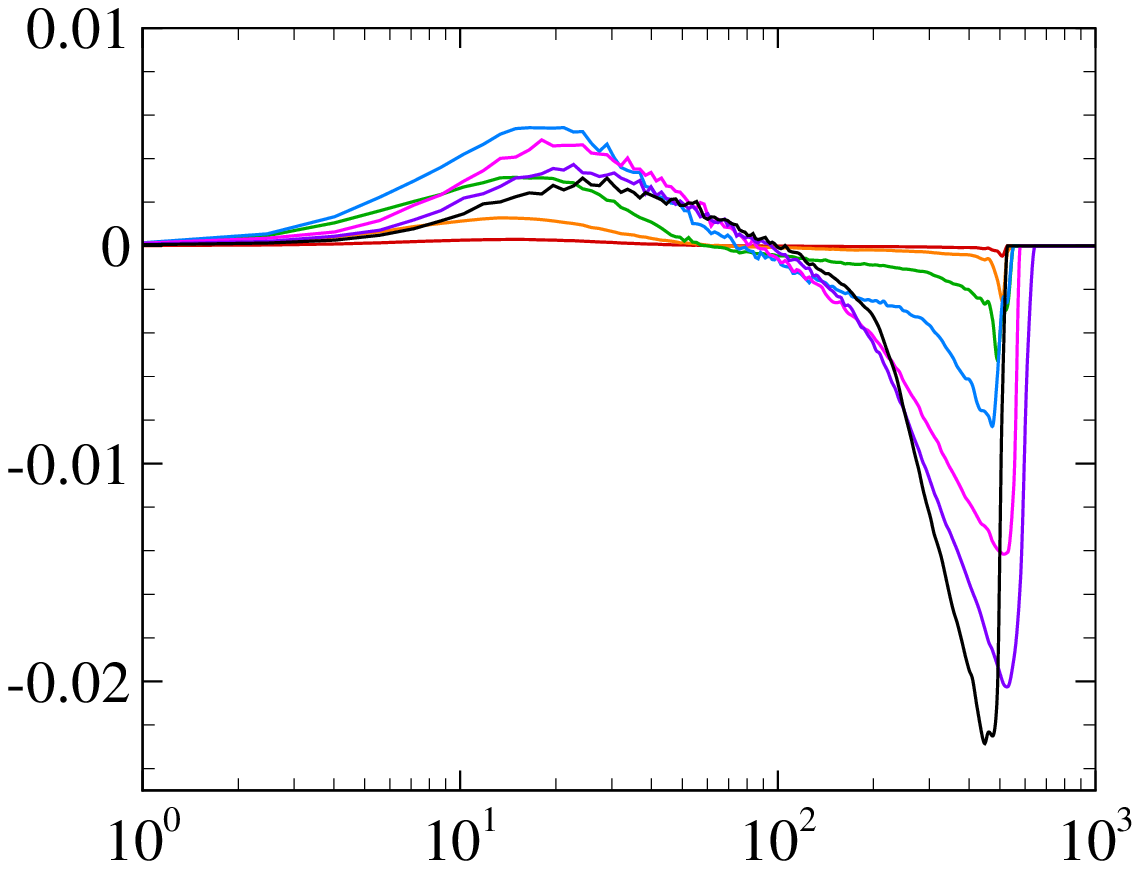}
\put(-2,70){(d)}
\put(48,0){$y^+$}
\put(-3,30){\rotatebox{90}{$\bar v_{2,s}/U_\infty$}}
\end{overpic}\\[1.0ex]
\begin{overpic}[width=0.5\textwidth]{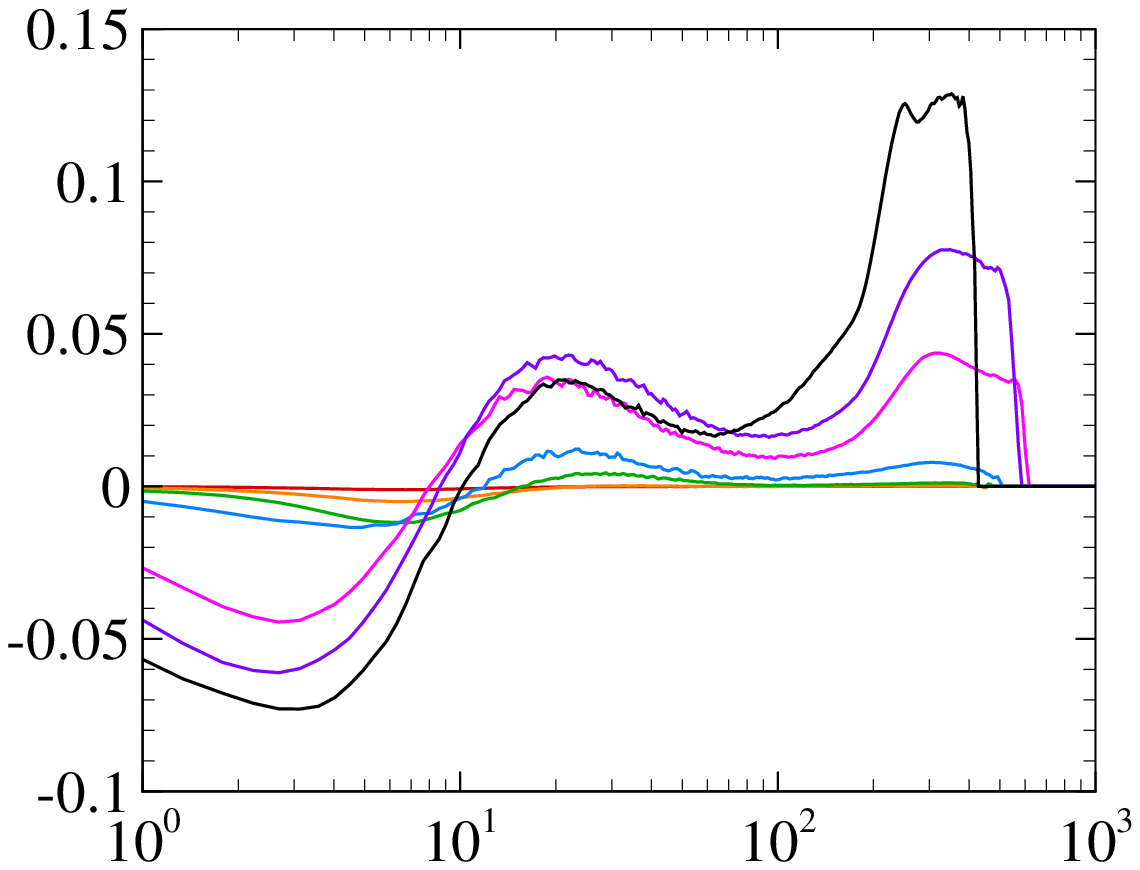}
\put(-2,70){(e)}
\put(48,0){$y^+$}
\put(-3,30){\rotatebox{90}{$\bar v_{1,s}/U_\infty$}}
\end{overpic}~
\begin{overpic}[width=0.5\textwidth]{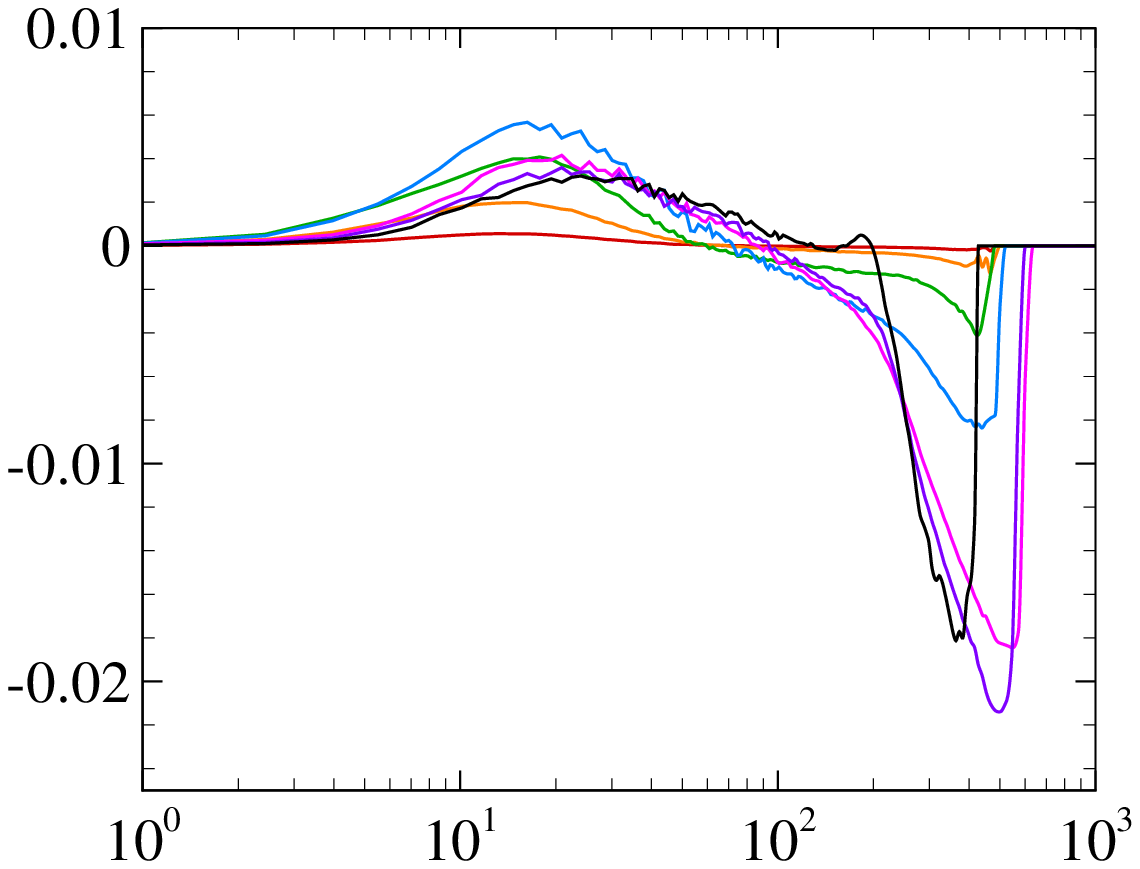}
\put(-2,70){(f)}
\put(48,0){$y^+$}
\put(-3,30){\rotatebox{90}{$\bar v_{2,s}/U_\infty$}}
\end{overpic}\\[0.0ex]
\caption{Mean particle slip velocity, (a,c,e) $\bar v_{1,s}/U_\infty$, 
(b,d,f) $\bar v_{2,s}/U_\infty$ in case (a,b) M2, (c,d) M4 and (e,f) M6.}
\label{fig:slipvelo}
\end{figure}

Equation~\eqref{eqn:pforce} dictates that the forces acted on the particles, or equivalently
the particle accelerations, are related to the slip velocity, particle inertia and
the particle Reynolds and Mach numbers, which will be presented in detail subsequently.

We first consider the slip velocity, namely the difference between the fluid and particle
velocities $u_{i}-v_{i}$.
The mean slip velocity is equivalent to the difference between the mean particle velocity 
$\bar v_i$ and the mean fluid velocity seen by particles $\bar u_{i,p}$. In figure
~\ref{fig:slipvelo} we present the distribution of $\bar v_{i,s}$
across the boundary layer.
The statistics are qualitatively consistent with those in incompressible turbulent channel flows
in the near-wall region~\citep{zhao2012stokes}, but are different in the outer region.
In the streamwise direction, the particles incline to move faster than the fluid 
in the near-wall region below $y^+ \approx 10$ but slower above that location.
The non-zero slip mean velocities close to the wall suggest that the particles 
are capable of sliding horizontally near the wall.
The magnitudes of the peaks and valleys increase monotonically with the Stokes number.
In the wall-normal direction, on the other hand, the slip velocity $\bar v_{2,s}$ is positive below
$y^+=100$ while negative above it. 
The peaks of $\bar v_{2,s}$ are attained at $y^+\approx 20$, manifesting a non-monotonic variation
and reaching a maximum for particle population P4 in case M2 and P5 in cases M4 and M6
that show the utmost near-wall accumulation. 
This indicates that the particles either move much faster toward or slower away from the wall, 
leading to the gradually increasing numbers of particles in the near-wall region.
In the outer region, the particles tend to move faster than the fluid to the free stream or 
slower when re-entrained within the boundary layer, consistent with the observation 
in the instantaneous flow fields (figure~\ref{fig:instxy4}) that the particles with large inertia
are capable of escaping the turbulent region.

\begin{figure}
\centering
\begin{overpic}[width=0.5\textwidth]{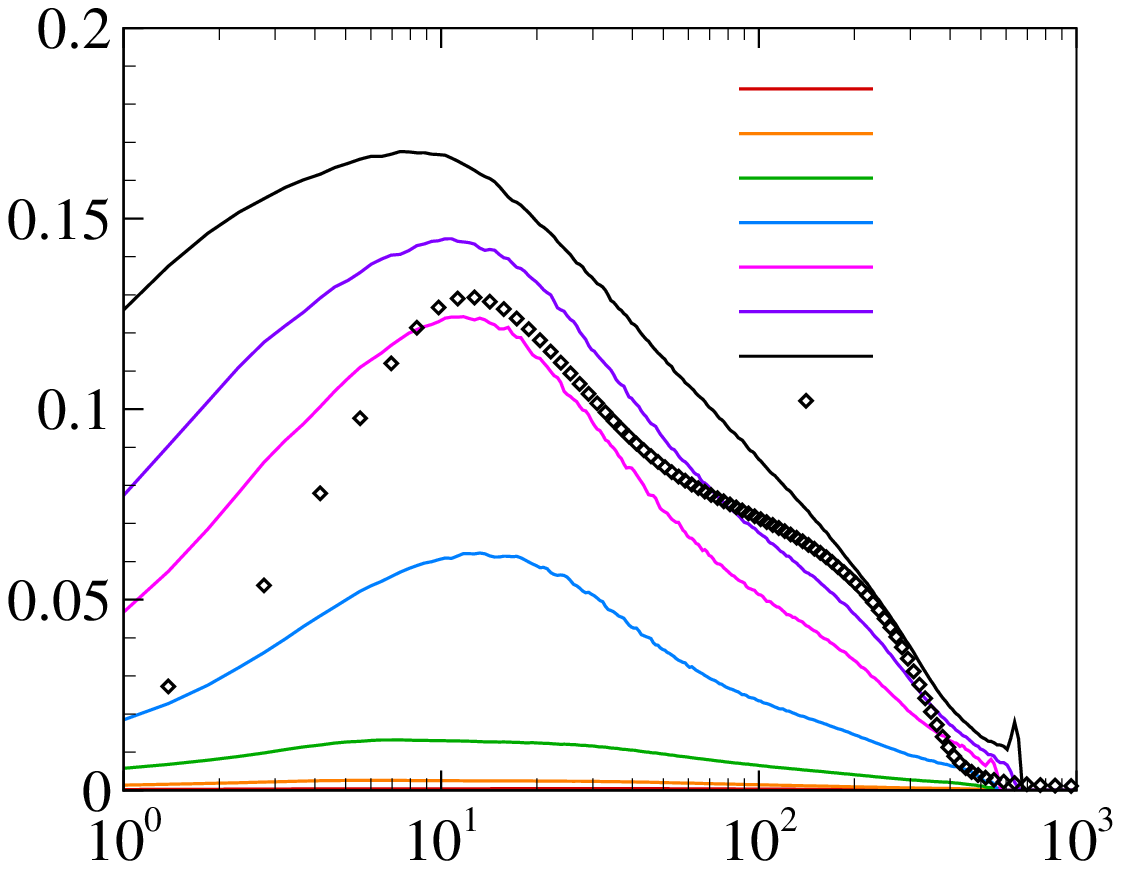}
\put(-2,70){(a)}
\put(50,0){$y^+$}
\put(-3,32){\rotatebox{90}{$\bar v'_{1,s}$}}
\put(75,65.8){\scriptsize P1}
\put(75,62.0){\scriptsize P2}
\put(75,58.5){\scriptsize P3}
\put(75,54.8){\scriptsize P4}
\put(75,51.5){\scriptsize P5}
\put(75,48.0){\scriptsize P6}
\put(75,44.4){\scriptsize P7}
\put(75,40.8){\scriptsize $\bar u''_{i}$}
\end{overpic}~
\begin{overpic}[width=0.5\textwidth]{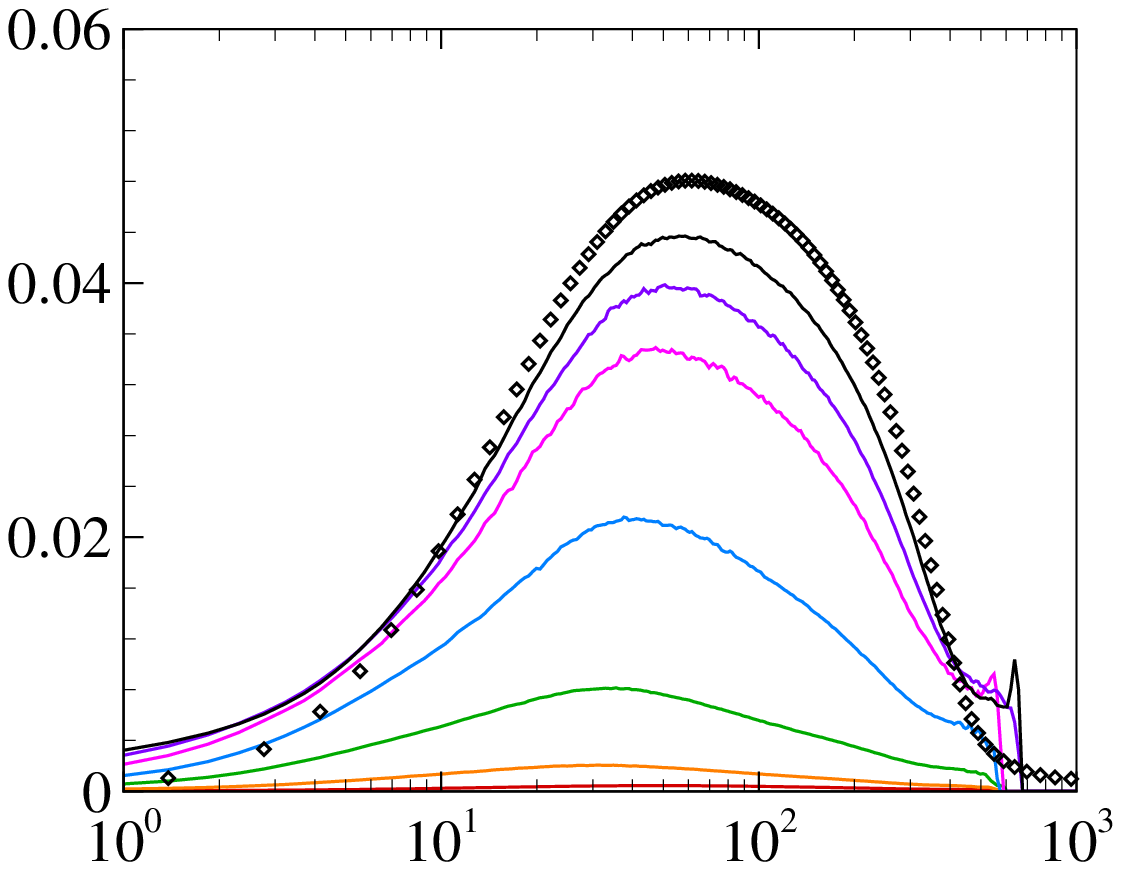}
\put(-2,70){(b)}
\put(50,0){$y^+$}
\put(-3,32){\rotatebox{90}{$\bar v'_{2,s}$}}
\end{overpic}\\[2.0ex]
\begin{overpic}[width=0.5\textwidth]{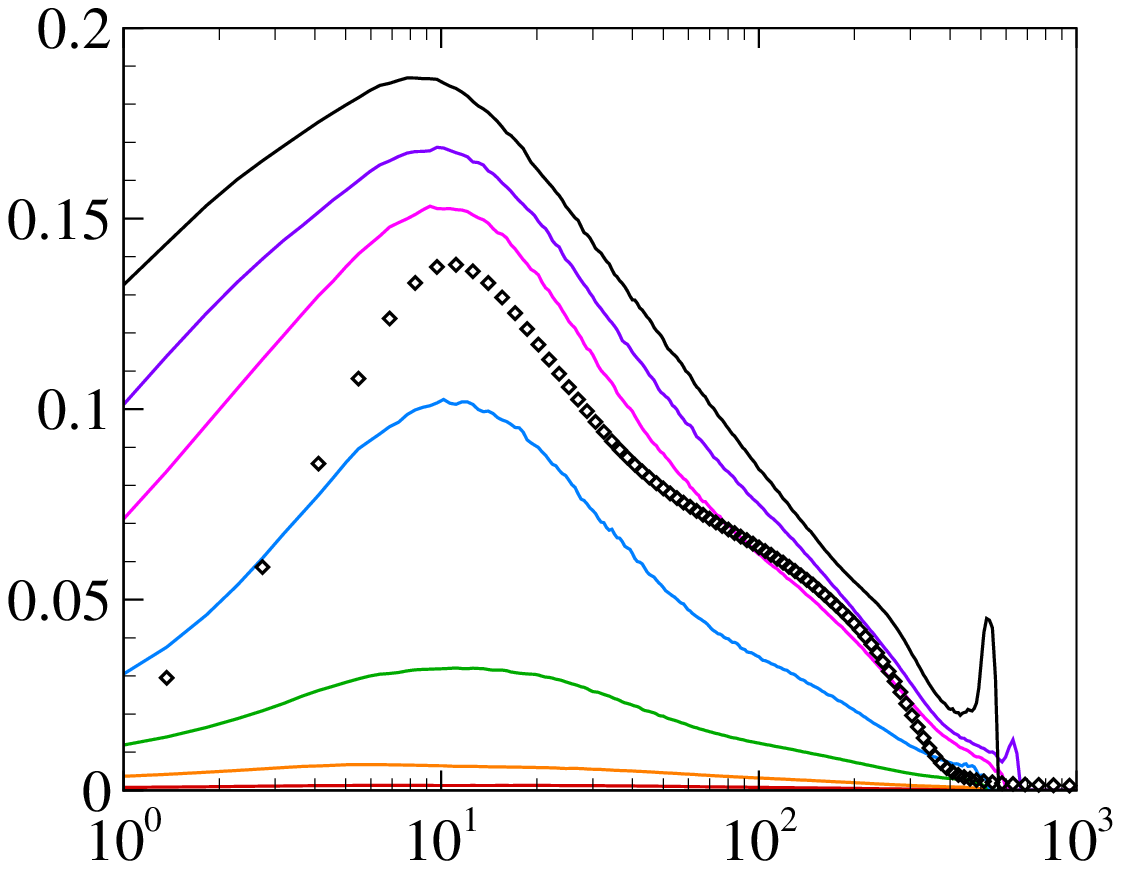}
\put(-2,70){(d)}
\put(50,0){$y^+$}
\put(-3,32){\rotatebox{90}{$\bar v'_{1,s}$}}
\end{overpic}~
\begin{overpic}[width=0.5\textwidth]{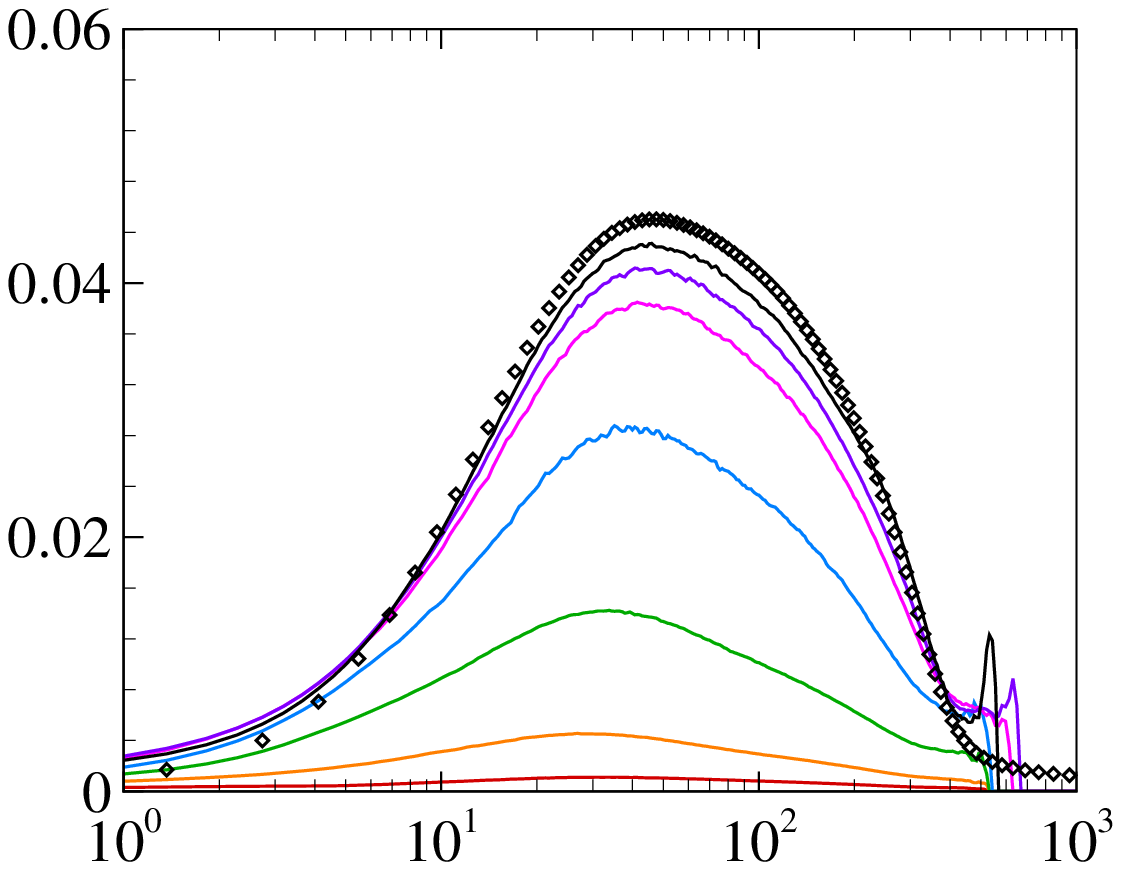}
\put(-2,70){(e)}
\put(50,0){$y^+$}
\put(-3,32){\rotatebox{90}{$\bar v'_{2,s}$}}
\end{overpic}\\[2.0ex]
\begin{overpic}[width=0.5\textwidth]{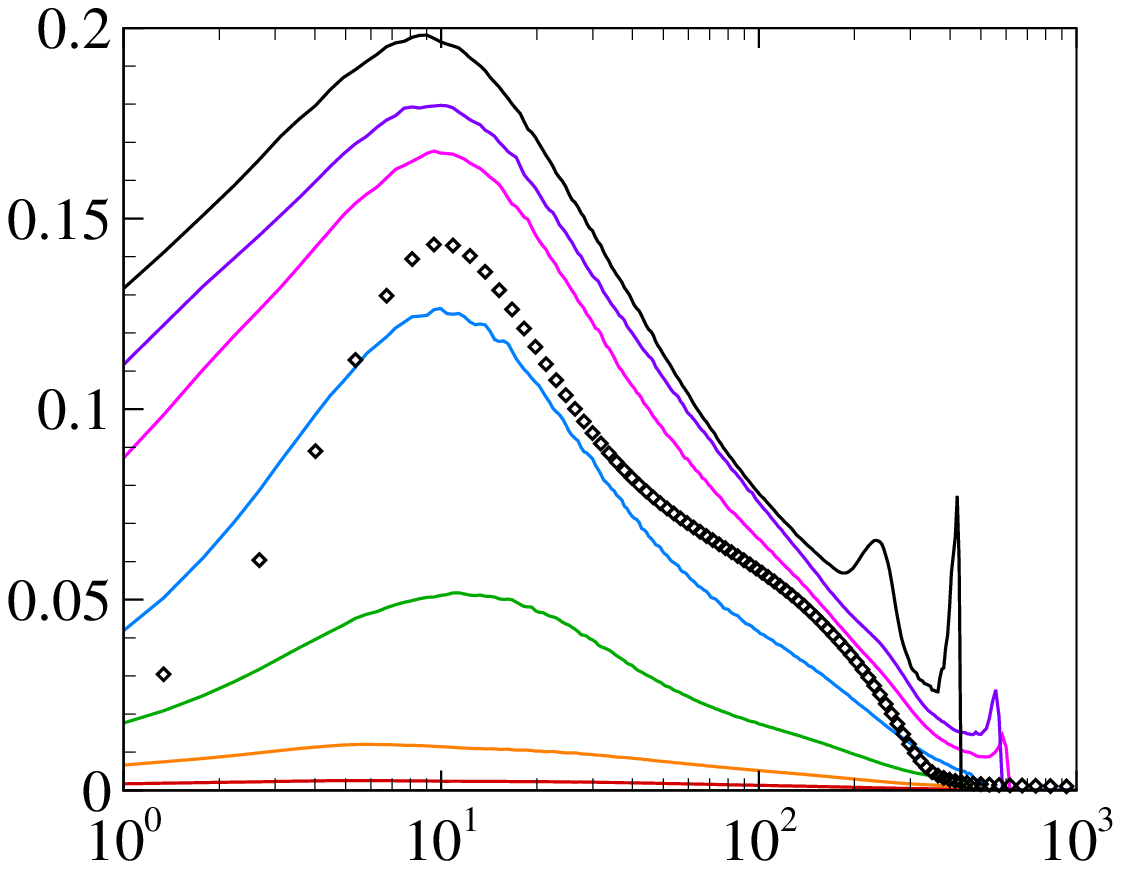}
\put(-2,70){(g)}
\put(50,0){$y^+$}
\put(-3,32){\rotatebox{90}{$\bar v'_{1,s}$}}
\end{overpic}~
\begin{overpic}[width=0.5\textwidth]{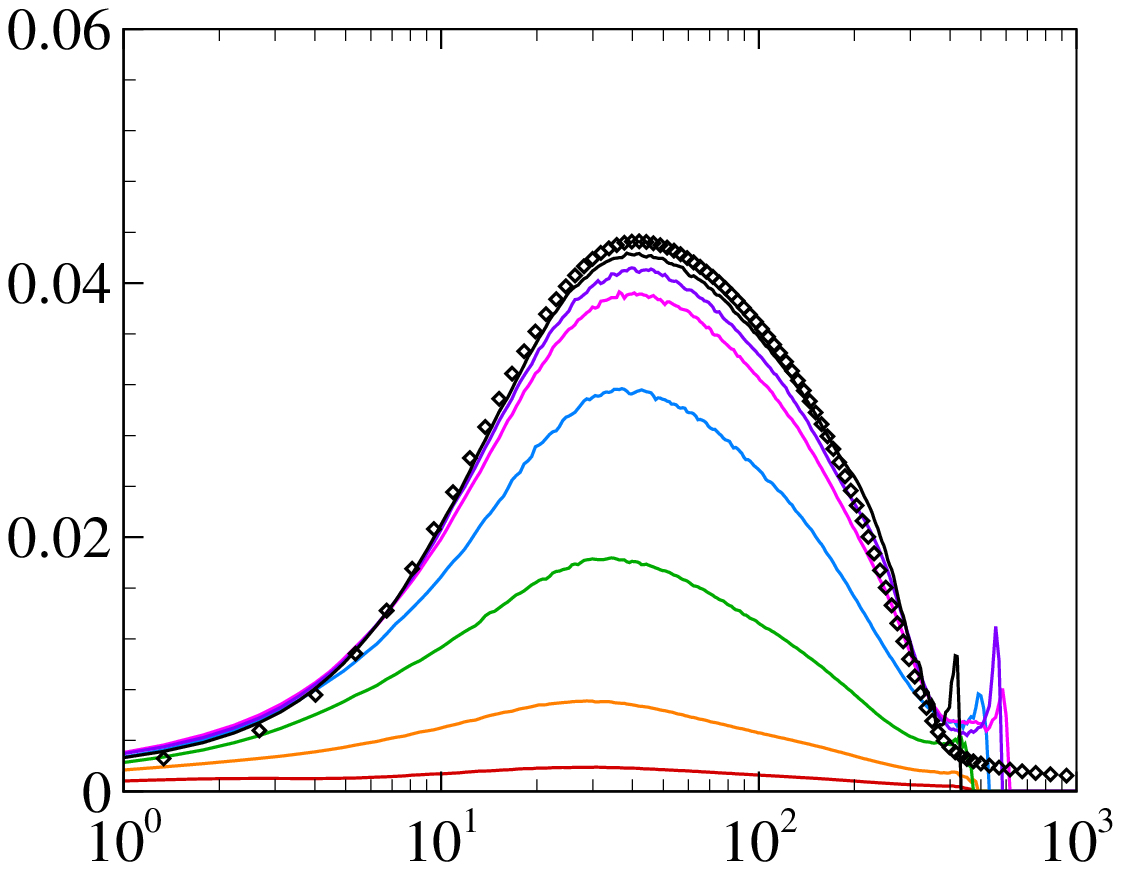}
\put(-2,70){(h)}
\put(50,0){$y^+$}
\put(-3,32){\rotatebox{90}{$\bar v'_{2,s}$}}
\end{overpic}\\
\caption{RMS of particle slip velocity fluctuation, (a,c,e) $\bar v'_{1,s}$, 
(b,d,f) $\bar v'_{2,s}$, (a,b) case M2, (c,d) case M4 and (e,f) case M6.}
\label{fig:slipvelorms}
\end{figure}

In figure~\ref{fig:slipvelorms} we present the RMS of the slip velocity, represented by 
$\bar v'_{i,s}$.
In general, the distributions of $\bar v'_{i,s}$ resemble those of the velocity and become 
more intense with the increasing $St^+$, except that the non-zero values of $\bar v'_{i,s}$ 
close to the wall because of, as explained above, the different boundary conditions implemented 
on the fluid and particles, with the streamwise component $\bar v'_{1,s}$ related to the particle
slipping near the wall and the wall-normal component $\bar v'_{2,s}$ to the particles hitting
and/or bouncing off the wall.
It is interesting to note that the $\bar v'_{1,s}$ for larger inertia particle populations are higher
than the corresponding velocity fluctuations, suggesting that the particle velocity fluctuations
are frequently in the opposite sign of that of the fluid.
The $\bar v'_{2,s}$, on the other hand, is consistently lower than $\bar u'_{2}$, 
except for the small region near the edge of the boundary layer.
Furthermore, we find that the maximum of the $\bar v'_{1,s}$ and $\bar v'_{2,s}$ increase 
monotonically with $St^+$ (not shown here for brevity), 
irrelevant of the different free stream Mach numbers in these cases.

\begin{figure}
\centering
\begin{overpic}[width=0.5\textwidth]{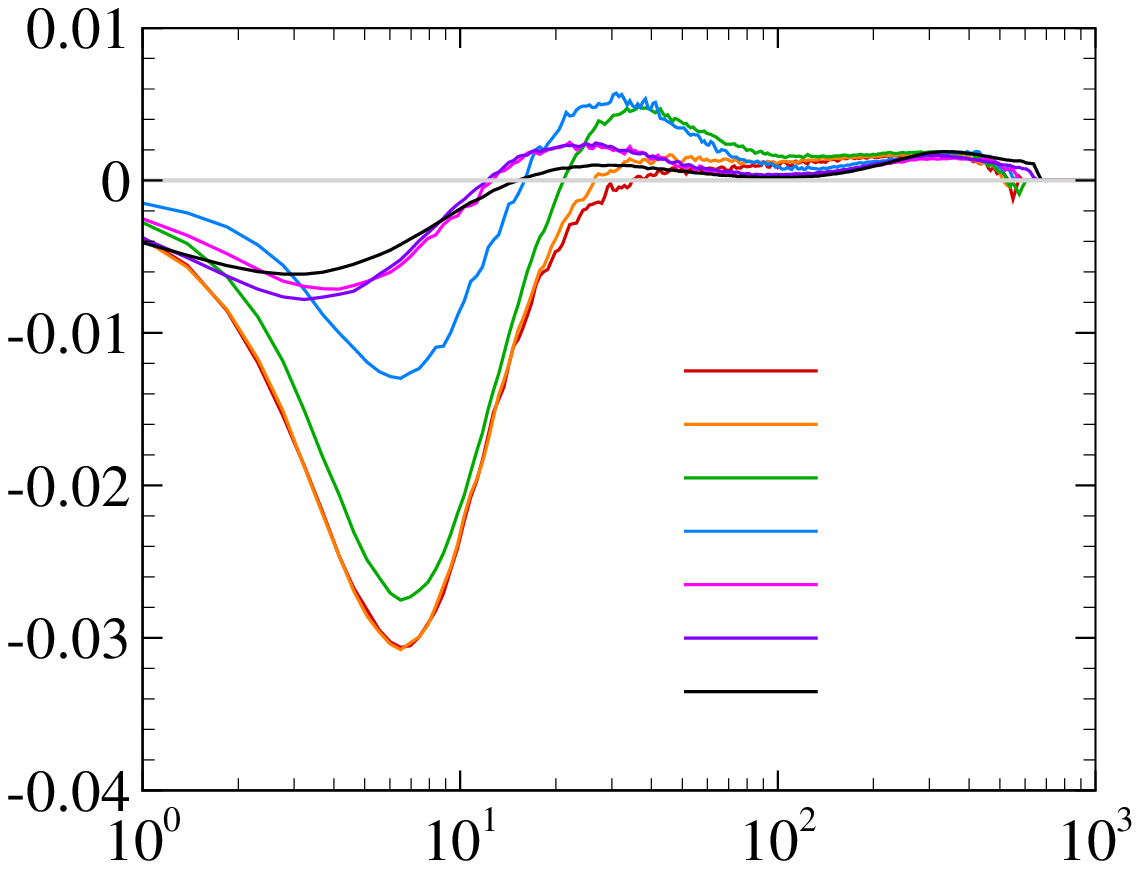}
\put(-2,70){(a)}
\put(50,0){$y^+$}
\put(-2,35){\rotatebox{90}{$\bar a_{1}$}}
\put(68,43){\scriptsize P1}
\put(68,39){\scriptsize P2}
\put(68,35){\scriptsize P3}
\put(68,30.8){\scriptsize P4}
\put(68,26.5){\scriptsize P5}
\put(68,22){\scriptsize P6}
\put(68,17.5){\scriptsize P7}
\end{overpic}~
\begin{overpic}[width=0.5\textwidth]{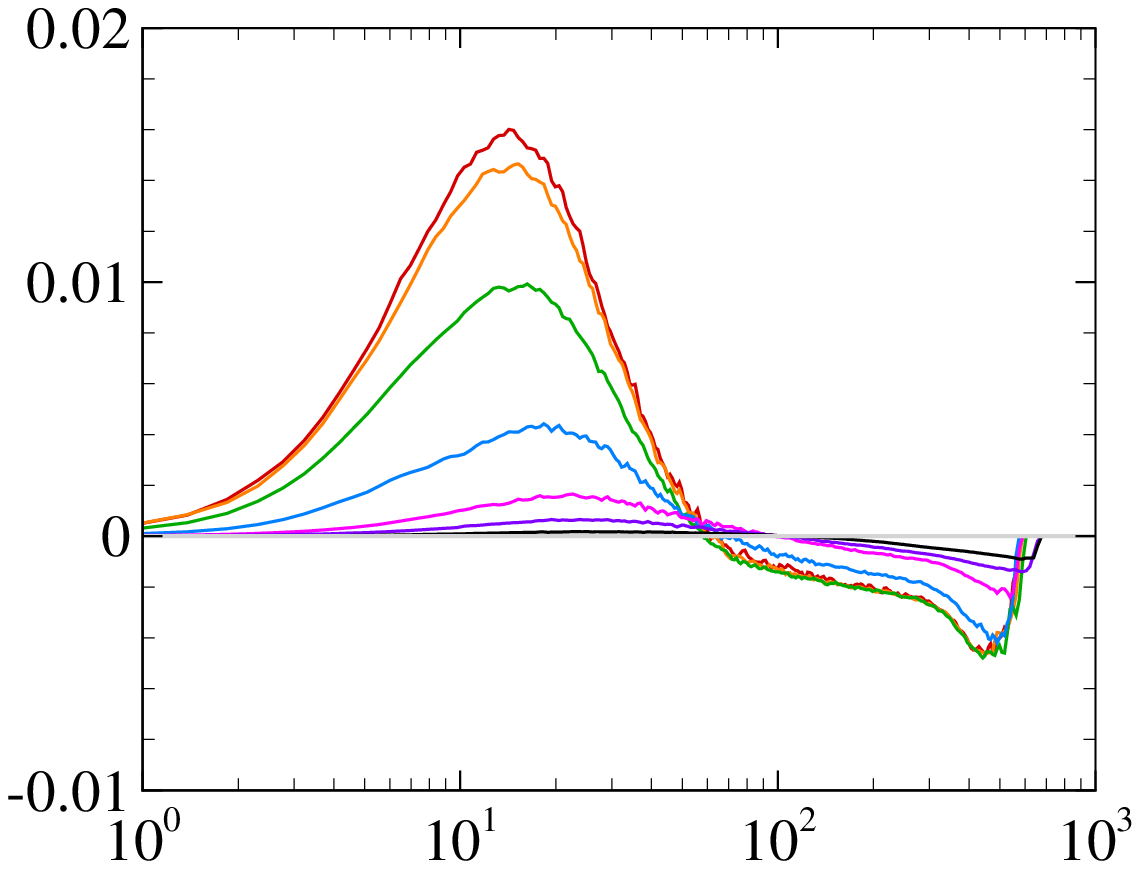}
\put(-2,70){(b)}
\put(50,0){$y^+$}
\put(-2,35){\rotatebox{90}{$\bar a_{2}$}}
\end{overpic}\\[1.0ex]
\begin{overpic}[width=0.5\textwidth]{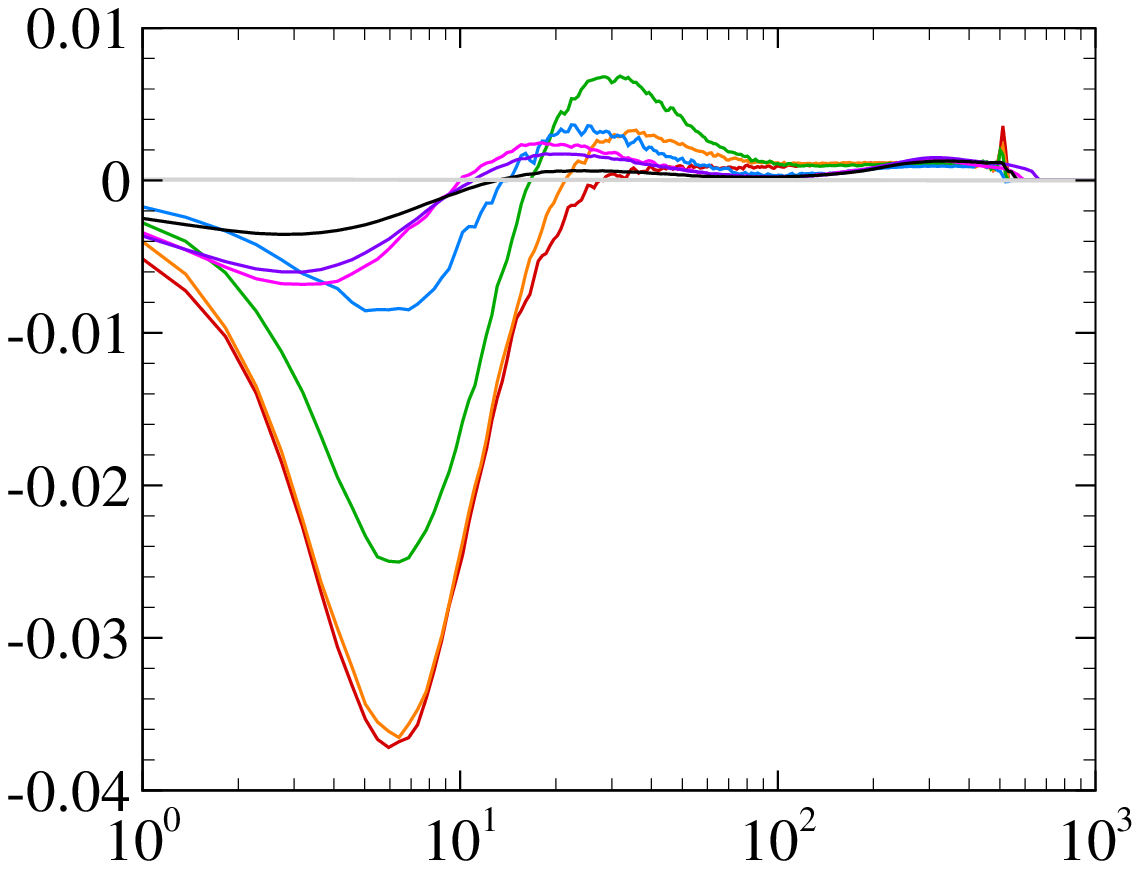}
\put(-2,70){(c)}
\put(50,0){$y^+$}
\put(-2,35){\rotatebox{90}{$\bar a_{1}$}}
\end{overpic}~
\begin{overpic}[width=0.5\textwidth]{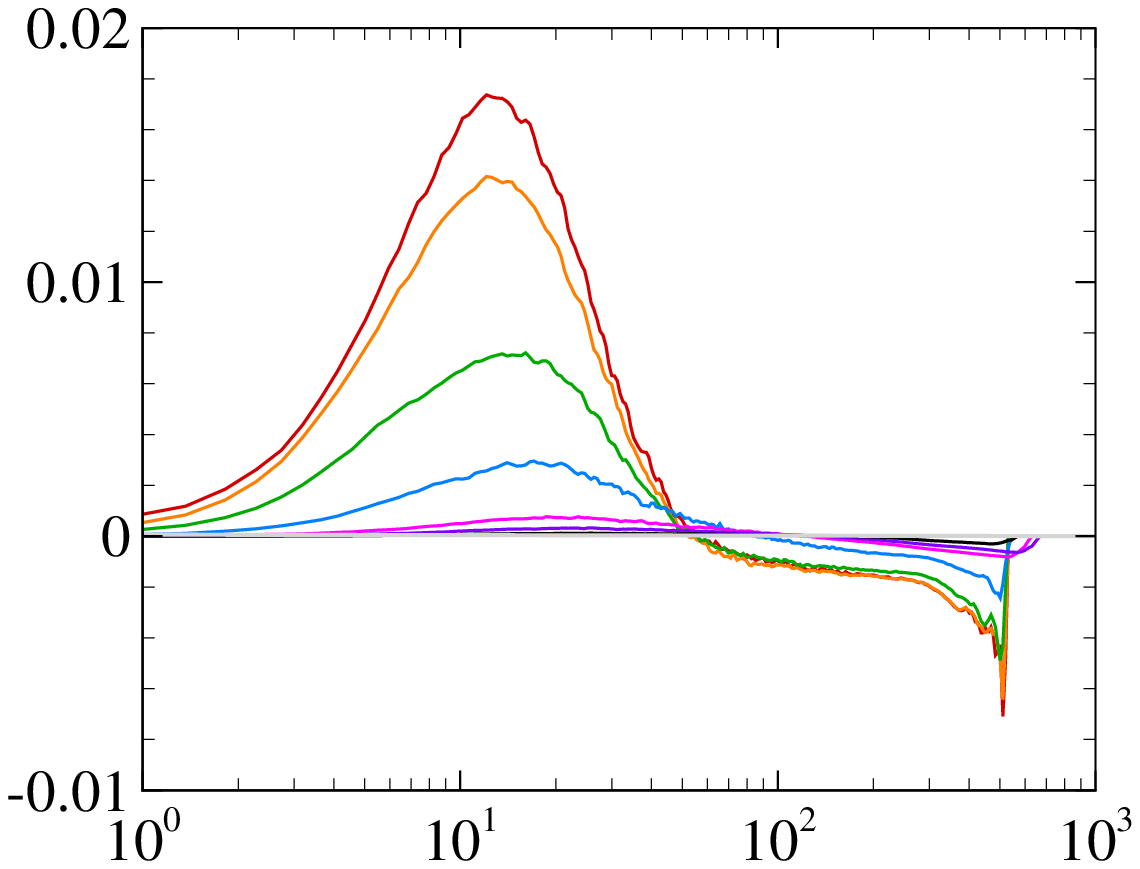}
\put(-2,70){(d)}
\put(50,0){$y^+$}
\put(-2,35){\rotatebox{90}{$\bar a_{2}$}}
\end{overpic}\\[1.0ex]
\begin{overpic}[width=0.5\textwidth]{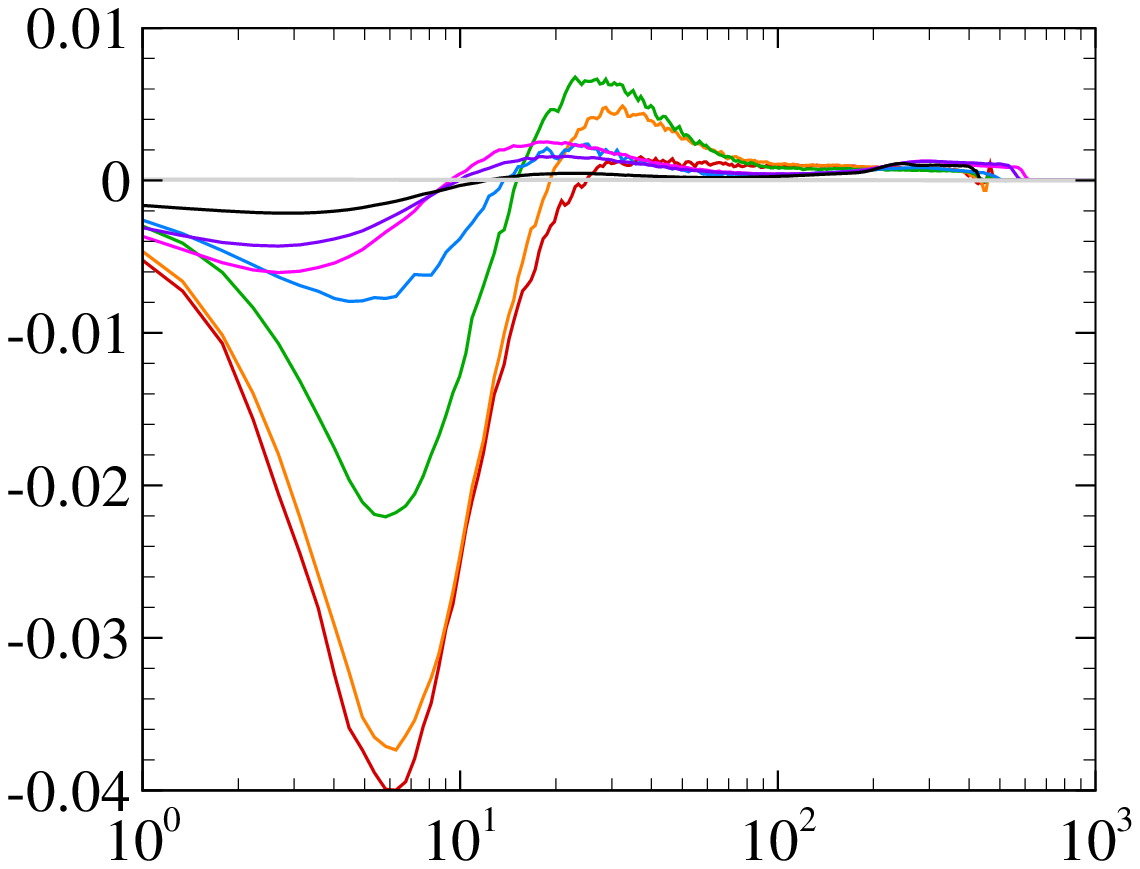}
\put(-2,70){(e)}
\put(50,0){$y^+$}
\put(-2,35){\rotatebox{90}{$\bar a_{1}$}}
\end{overpic}~
\begin{overpic}[width=0.5\textwidth]{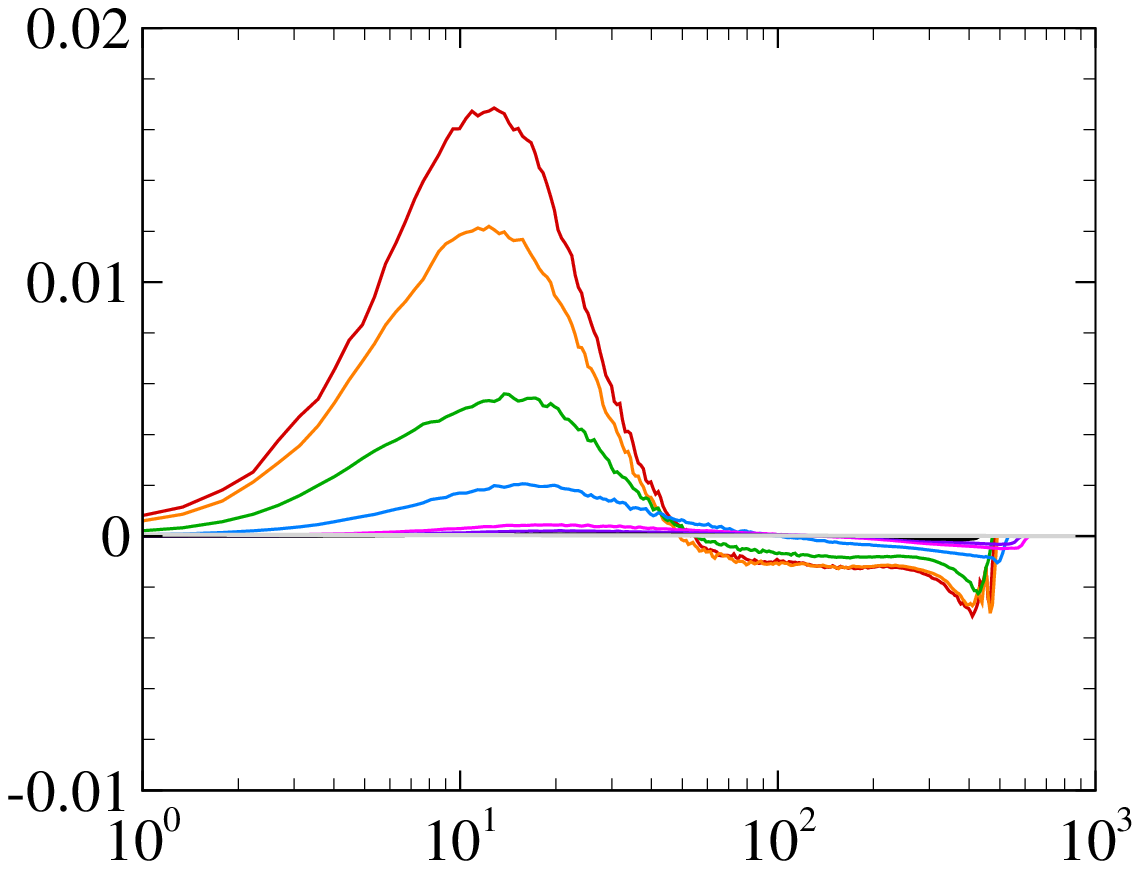}
\put(-2,70){(f)}
\put(50,0){$y^+$}
\put(-2,35){\rotatebox{90}{$\bar a_{2}$}}
\end{overpic}\\
\caption{Mean particle acceleration, normalized by $U_\infty$ and $\delta_0$,
(a,c,e) $\bar a_{1}$ and (b,d,f) $\bar a_{2}$ in (a,b) case M2, (c,d) case M4 and (e,f) case M6.}
\label{fig:accave}
\end{figure}

The slip velocity induces the forces on particles, resulting in their acceleration.
Shown in figure~\ref{fig:accave} are their mean values in the streamwise and wall-normal directions.
By comparing the distribution of the mean particle acceleration $\bar a_i$ for each case,
we firstly find that the streamwise component $\bar a_{1}$ is negative in the near-wall region 
and becomes positive above the buffer region, whereas the wall-normal component $\bar a_{2}$ 
manifests a reverse trend of variation.
Secondly, both the $\bar a_{1}$ and $\bar a_{2}$ possess higher magnitudes in the near-wall region,
which is different from the mean slip velocities $\bar v_{i,s}$ for some of the large inertia 
particle populations.
We can infer from equation~\eqref{eqn:pforce} that the mean acceleration of the particles relies
not only on the mean values of the slip velocity but on their fluctuations and local viscosity
as well. The fluctuations of the slip velocity are more intense in the near-wall region 
and the local viscosity is lower away from the wall, both of these factors
lead to the lower forces acting on the particles in the outer region.
Thirdly, the magnitudes of $\bar a_{1}$ and $\bar a_{2}$ decrease monotonically with the increasing
$St^+$ due to the increasing density of the particle populations, except for a thin layer 
in the buffer layer where the mean slip velocity $\bar v_{1,s}$ changes its sign.
Knowing that the low-inertia particles follow the trajectories of the fluid, the distributions of
the mean particle acceleration indicate that both the dispersed phase particle with low inertia 
and the fluid particles under the Lagrangian frame are decelerated in both the streamwise 
and wall-normal directions when they are swept downwards due to the restriction of the no-slip 
and no-penetration condition imposed on the wall.
As for high Stokes number particles, wall-normal mean acceleration $\bar a_{2}$ is almost zero
while the streamwise mean acceleration $\bar a_{1}$ retain finite values, suggesting that
these particles, either moving toward or away from the wall, are barely or equally affected by
the near-wall sweeping and ejection events, and can be accelerated or decelerated
when they are travelling across the boundary layer due to the comparatively strong shear,
in particular in the near-wall region.

\begin{figure}
\centering
\begin{overpic}[width=0.5\textwidth]{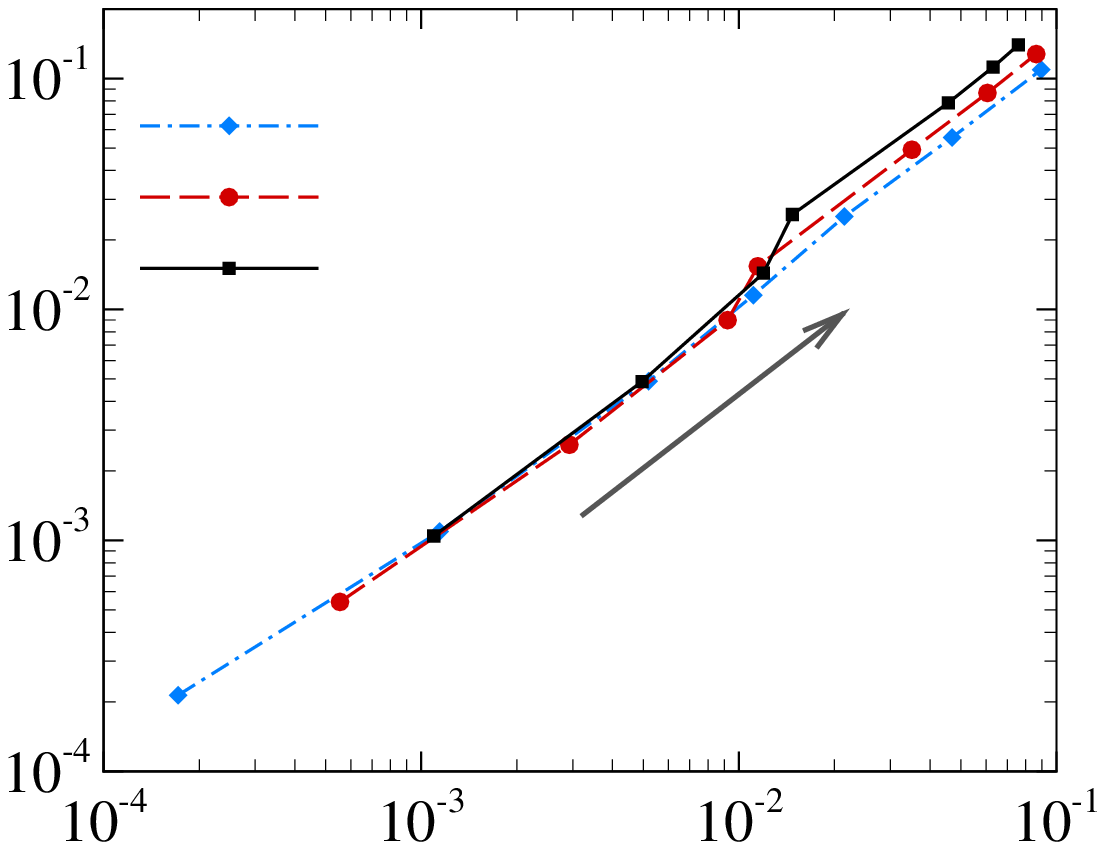}
\put(-2,70){(a)}
\put(48,0){$[\bar v_{1,s}]_{min}$}
\put(-3,32){\rotatebox{90}{$[\bar a_{1} \tau_{p}]_{min}$}}
\put(32,62){\scriptsize M2}
\put(32,56){\scriptsize M4}
\put(32,50){\scriptsize M6}
\put(63,33){$St^+$}
\end{overpic}~
\begin{overpic}[width=0.5\textwidth]{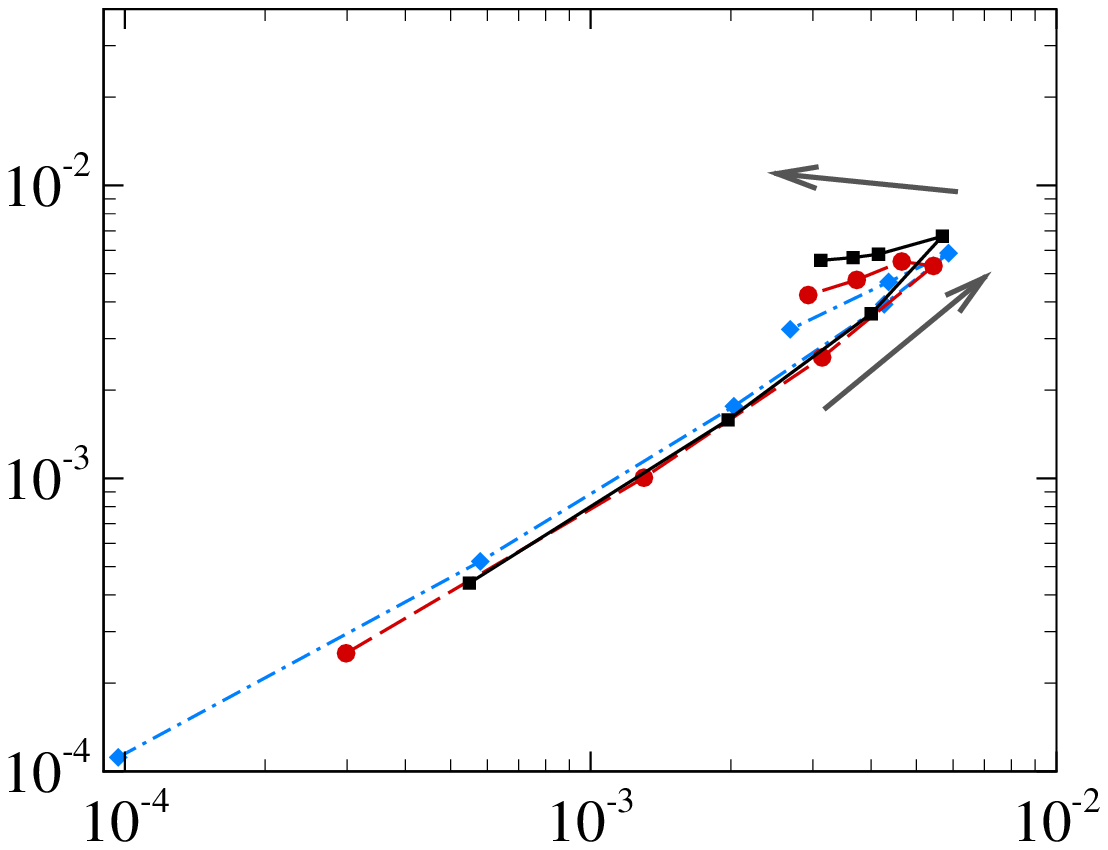}
\put(-2,70){(b)}
\put(48,0){$[\bar v_{2,s}]_{max}$}
\put(-3,32){\rotatebox{90}{$[\bar a_{2} \tau_p]_{max}$}}
\put(75,35){$St^+$}
\end{overpic}\\
\caption{Relation between the mean acceleration and the mean slip velocity,
(a) minimum of $\bar v_{1,s}$ and $\bar a_{1} \tau_{p}$, (b) maximum of $\bar v_{2,s}$ 
and $\bar a_{2} \tau_{p}$.}
\label{fig:forcevel}
\end{figure}

Comparing these cases, however, it appears that the mean particle acceleration $\bar a_{i}$ 
does not obey a monotonic trend of variation with the Stokes number $St^+$,
which is different from the other related flow quantities.
In figure~\ref{fig:forcevel} we plot the minimum of 
$\bar v_{1,s}$ against that of $\bar a_{1}$, and the maximum of $\bar v_{2,s}$ against that of
$\bar a_{2}$, with the acceleration multiplied by the particle relaxation time $\tau_p$ to 
incorporate the particle inertia.
For particles with small $St^+$ (populations P1 $\sim$ P3), the mean particle 
streamwise acceleration $\bar a_{1} \tau_p$ increases with the slip velocity almost 
linearly and manifests no dependence on the free stream Mach number, 
as indicated by the well-collapsed profiles.
The mean streamwise acceleration of high Stokes number particle populations, on the other hand, 
is stronger with the increasing Mach number.
The $\bar a_{2} \tau_p$ does not increase monotonically with the $St^+$ and, more importantly, 
it is greater for high Stokes number particle populations even when the mean slip velocities $\bar v_{2,s}$ 
are the same as those of the low Stokes number particle populations.
According to equation~\eqref{eqn:pforce}, we deduce that the discrepancy in the trend of variation
should be attributed to the function $f_D$ that reflects the non-negligible effects of the 
particle Reynolds and Mach numbers, 
especially the latter due to the manifestation of the Mach number dependence.

\begin{figure}
\centering
\begin{overpic}[width=0.5\textwidth]{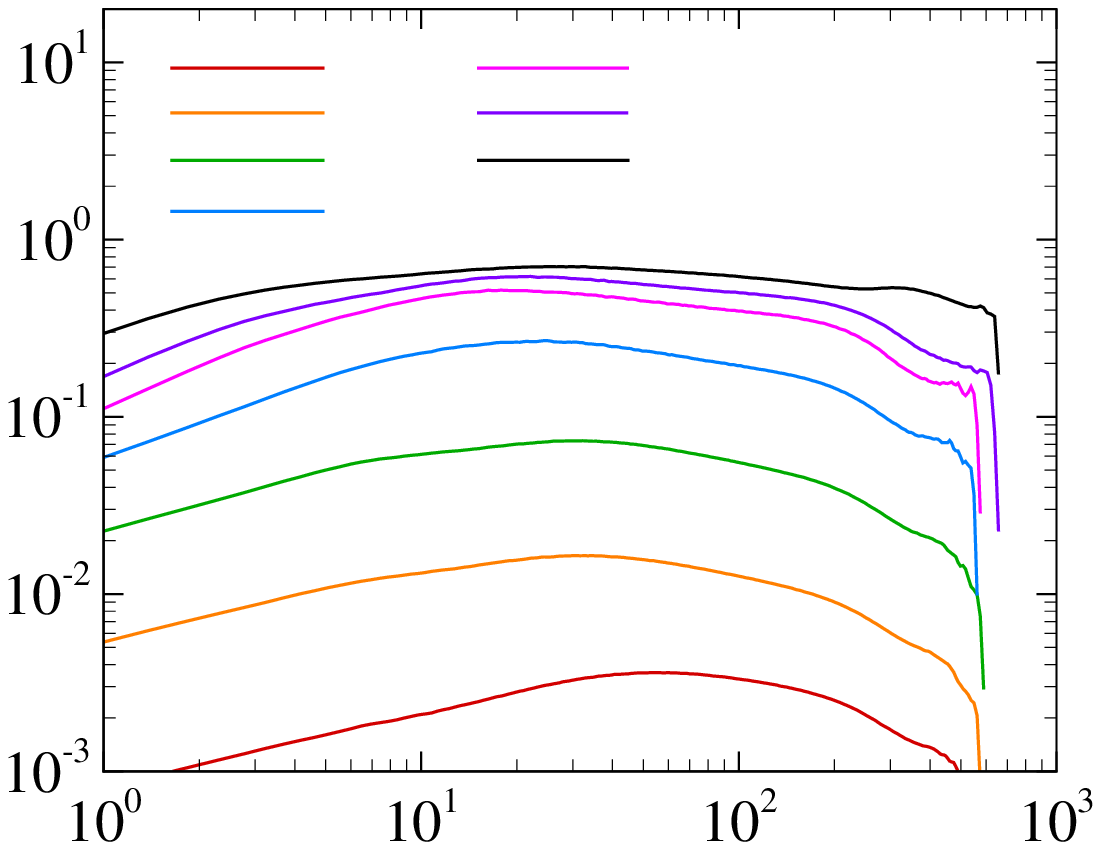}
\put(-2,70){(a)}
\put(50,0){$y^+$}
\put(-2,35){\rotatebox{90}{$Re_p$}}
\put(32,66){\scriptsize P1}
\put(32,62.5){\scriptsize P2}
\put(32,59){\scriptsize P3}
\put(32,55){\scriptsize P4}
\put(57,66){\scriptsize P5}
\put(57,62.5){\scriptsize P6}
\put(57,59){\scriptsize P7}
\end{overpic}~
\begin{overpic}[width=0.5\textwidth]{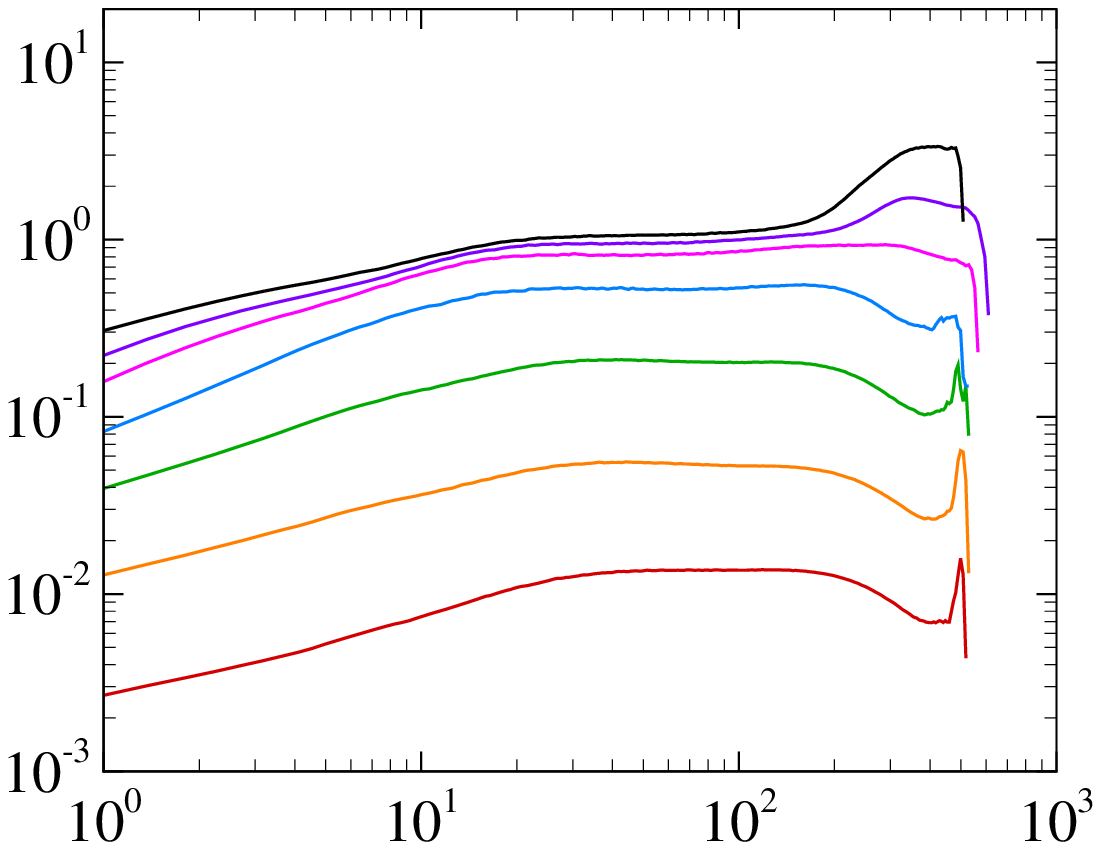}
\put(-2,70){(b)}
\put(50,0){$y^+$}
\put(-2,35){\rotatebox{90}{$Re_p$}}
\end{overpic}\\[2.0ex]
\begin{overpic}[width=0.5\textwidth]{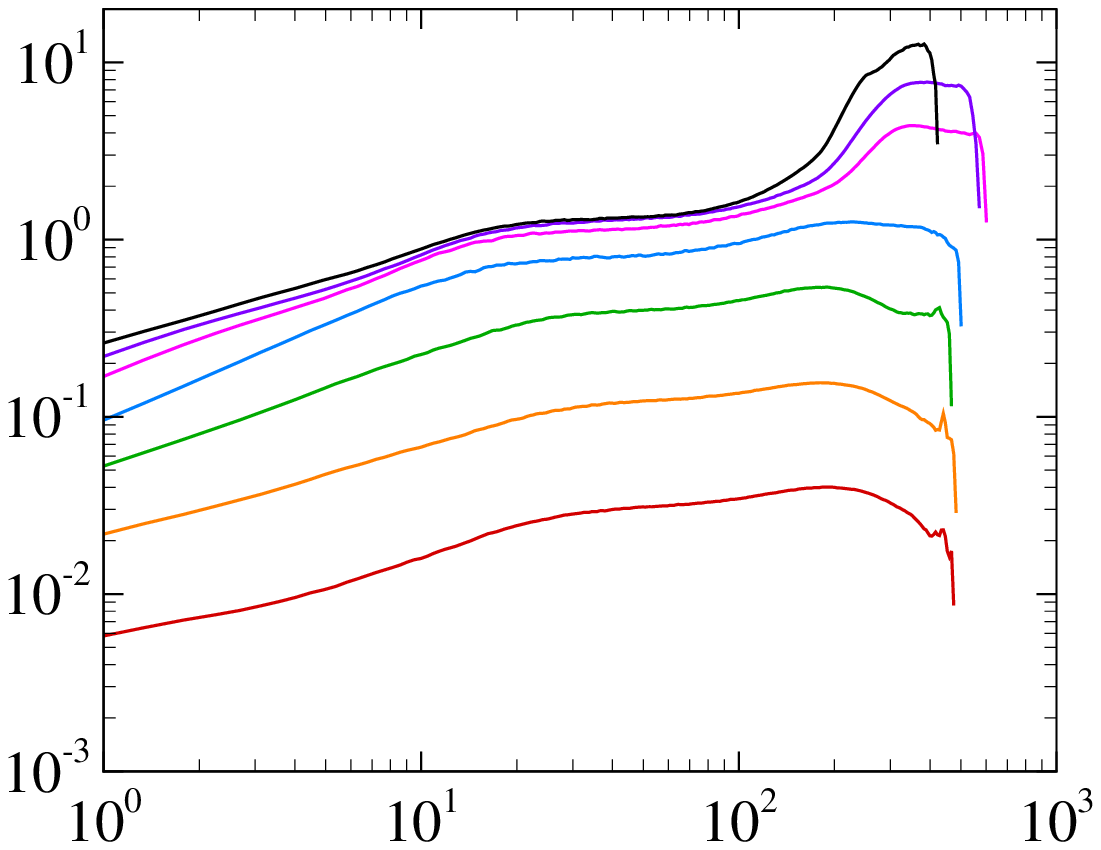}
\put(-2,70){(c)}
\put(50,0){$y^+$}
\put(-2,35){\rotatebox{90}{$Re_p$}}
\end{overpic}~
\begin{overpic}[width=0.5\textwidth]{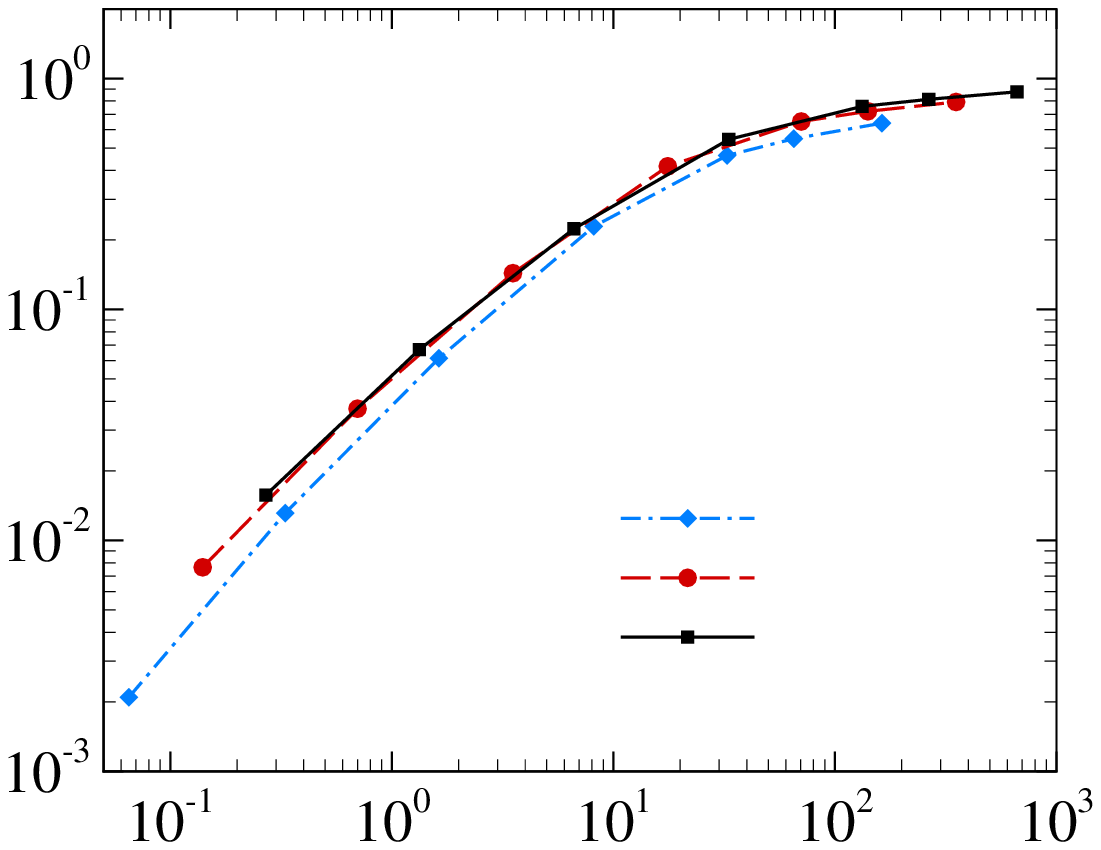}
\put(-2,70){(d)}
\put(50,0){$St^+$}
\put(-2,35){\rotatebox{90}{$Re_p$}}
\put(67,30){\scriptsize M2}
\put(67,25){\scriptsize M4}
\put(67,20.5){\scriptsize M6}
\end{overpic}\\
\caption{Mean particle Reynolds number $\overline{Re}_p$ in (a) case M2, (b) case M4, 
(c) case M6, (d) at $y^+=10$ for the three flow cases.}
\label{fig:rep}
\end{figure}

For the purpose of exploring the cause disparity in the $\bar a_{1} \tau_p$ and $\bar a_{2} \tau_p$
in high Stokes number particles, we further discuss the two factors in the expression of $f_D$,
namely the particle Reynolds numbers $Re_p$ and the particle Mach numbers $M_p$.
In figure~\ref{fig:rep}(a-c) we present the wall-normal distribution of the mean particle
Reynolds number $\overline{Re}_p$ in the cases at different free stream Mach numbers.
As the $St^+$ increases, the $\overline{Re}_p$ increases monotonically across the boundary layer.
Since the diameters are same for different particle populations, the increment of $\overline{Re}_p$ 
should primarily be attributed to the differences in the slip velocity, the local
fluid density and viscosity.
In most of the region, the $\overline{Re}_p$ remains low values no higher than $5.0$, 
a criterion below which the flow surrounding the particle can be regarded as creeping flow 
without shedding vortices.
Even near the edge of the boundary layer, the highest particle Reynolds number is merely $10$.
We have also inspected the probability density function (PDF) of $Re_p$ both in the near-wall 
and the outer region and found that the probability of the $Re_p$ exceeding $20$ is less than 1\%.
The comparatively high values of the $\overline{Re}_p$ near the edge of the boundary layer in
case M6, as shown in figure~\ref{fig:rep}(c), could be caused by the very rare occurrence of
the particles reaching that location, thereby highlighting the comparatively high Reynolds
number ones. 
Figure~\ref{fig:rep}(d) further demonstrates that the $\overline{Re}_p$ at $y^+=10$ are also
weakly dependent on the Mach numbers, showing a slight augmentation with $M_\infty$.
Under such circumstances, the refinement of the drag force regarding the high Reynolds number
effects is expected, at least statistically.

\begin{figure}
\centering
\begin{overpic}[width=0.5\textwidth]{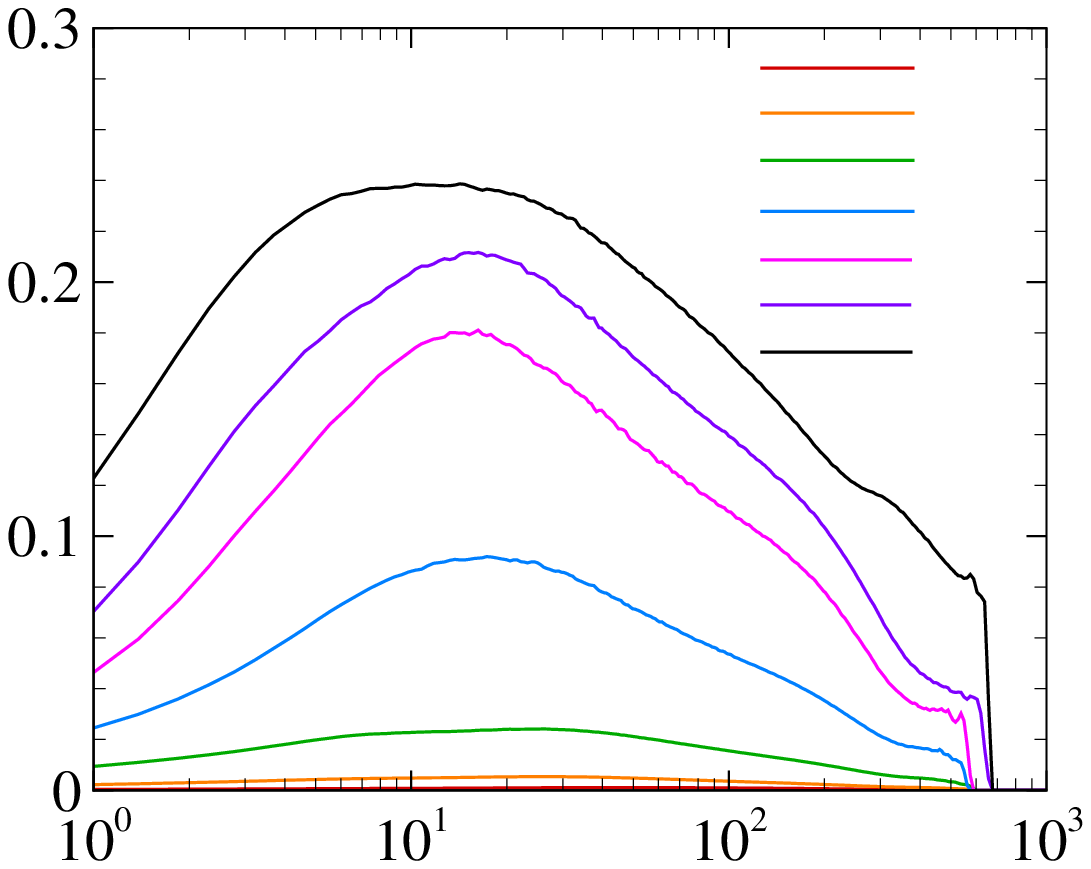}
\put(-2,70){(a)}
\put(50,0){$y^+$}
\put(-2,35){\rotatebox{90}{$\bar M_p$}}
\put(80,68){\scriptsize P1}
\put(80,64){\scriptsize P2}
\put(80,60){\scriptsize P3}
\put(80,56){\scriptsize P4}
\put(80,52.3){\scriptsize P5}
\put(80,48.4){\scriptsize P6}
\put(80,44.5){\scriptsize P7}
\end{overpic}~
\begin{overpic}[width=0.5\textwidth]{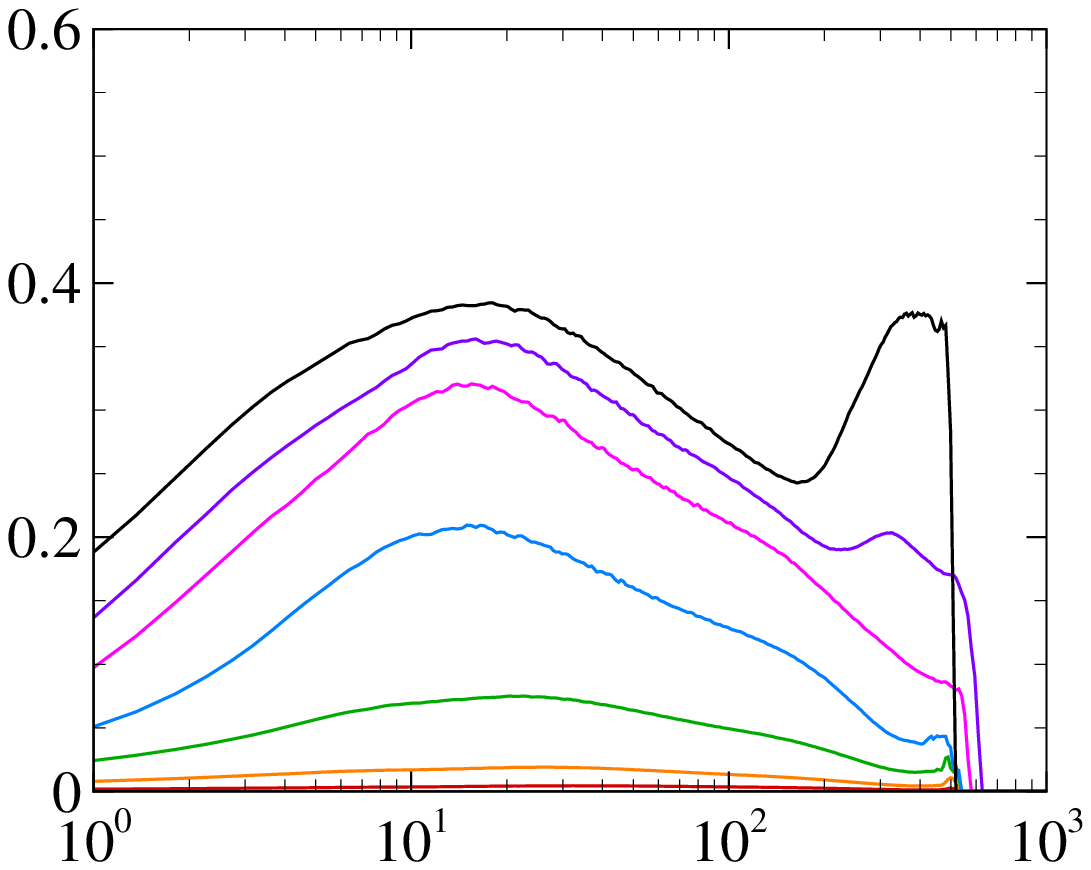}
\put(-2,70){(b)}
\put(50,0){$y^+$}
\put(-2,35){\rotatebox{90}{$\bar M_p$}}
\end{overpic}\\[1.0ex]
\begin{overpic}[width=0.5\textwidth]{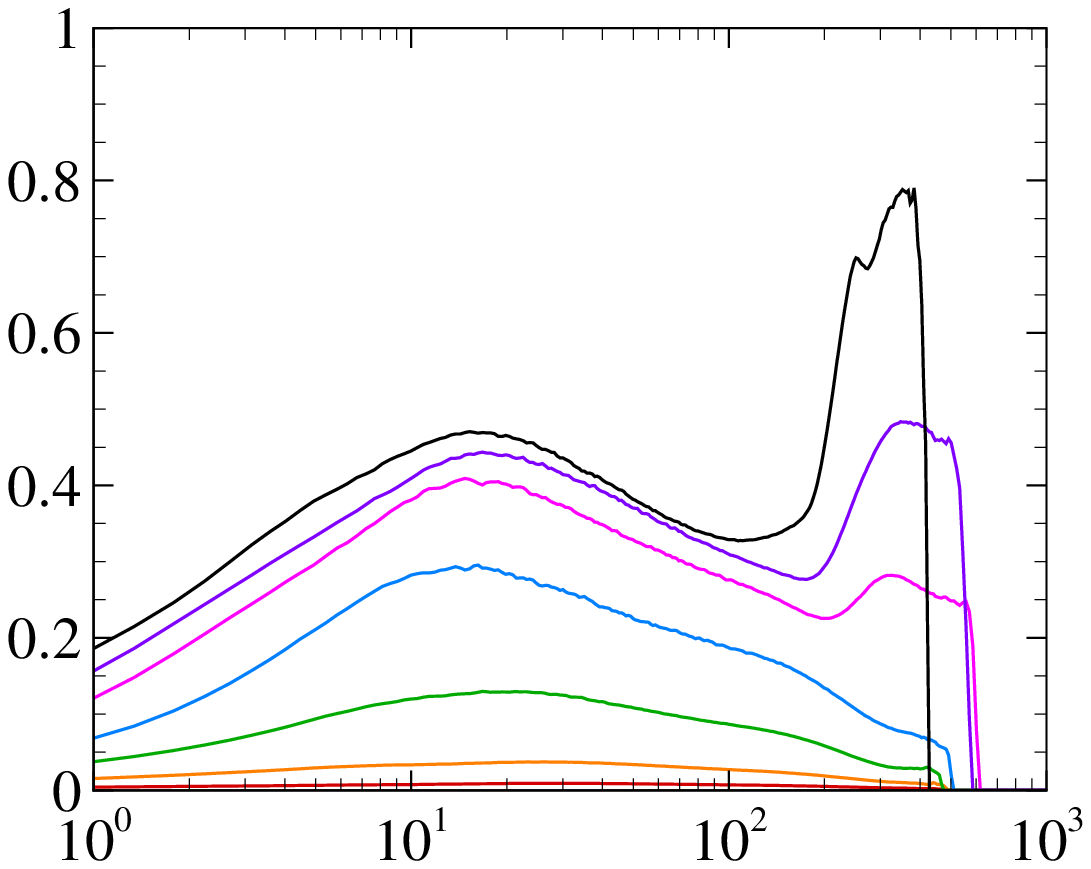}
\put(-2,70){(c)}
\put(50,0){$y^+$}
\put(-2,35){\rotatebox{90}{$\bar M_p$}}
\end{overpic}~
\begin{overpic}[width=0.5\textwidth]{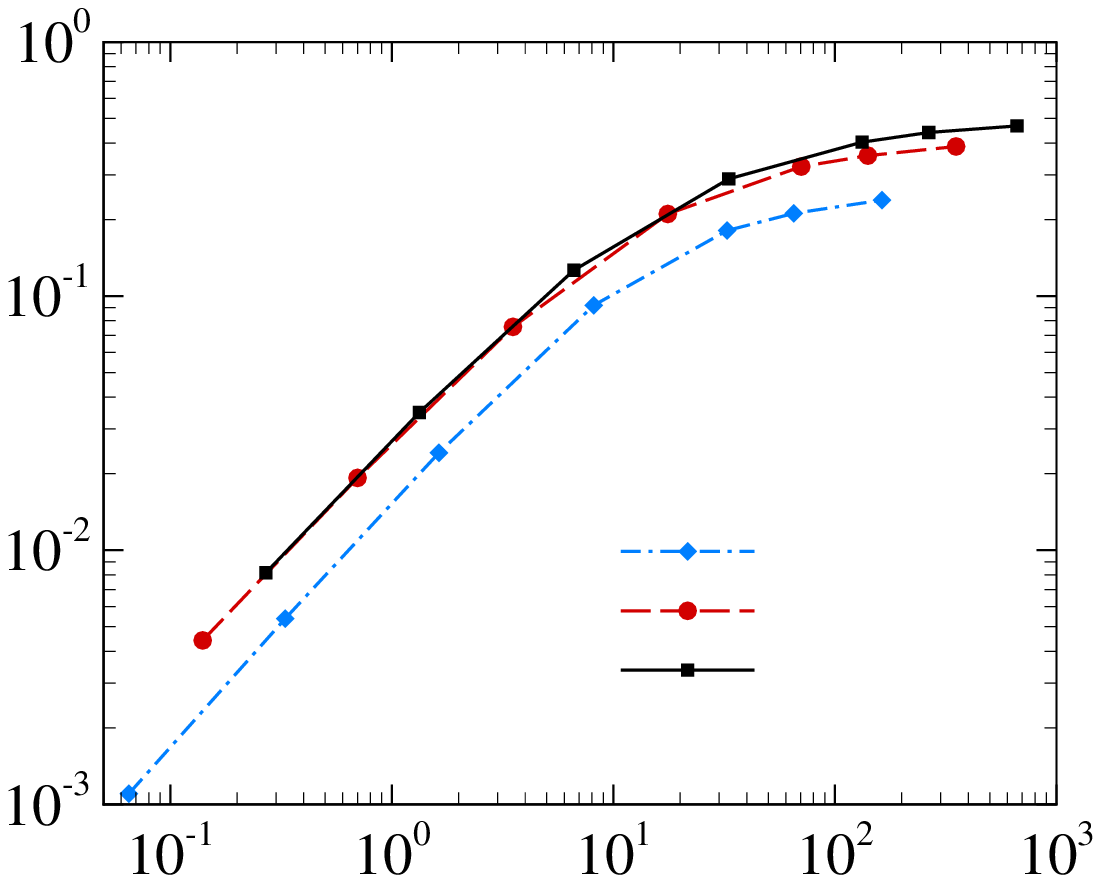}
\put(-2,70){(d)}
\put(50,0){$St^+$}
\put(-2,35){\rotatebox{90}{$\bar M_p$}}
\put(67,30){\scriptsize M2}
\put(67,25){\scriptsize M4}
\put(67,20.5){\scriptsize M6}
\end{overpic}\\
\caption{Mean particle Mach number $\bar M_p$ in (a) case M2, (b) case M4, (c) case M6,
and (d) the near-wall maximum.}
\label{fig:mp}
\end{figure}

Figure~\ref{fig:mp} displays the wall-normal distribution of the mean particle Mach number 
$\bar M_p$.
Consistent with the trend of variation of the flow quantities discussed above, 
the $\bar M_p$ increases with the $St^+$ due to their comparatively high 
mean and fluctuating slip velocities, and also with the free stream Mach number.
At the free stream Mach number $M_\infty = 2$, there only manifests one peak in the profiles of 
$\bar M_p$ for each particle population in the buffer region where the turbulent fluctuations are 
the most intense, whereas at $M_\infty=4$ and $6$, there gradually emerge second peaks in the 
outer region, the values of which even exceed the near-wall peak values for large inertia particles.
Nevertheless, in the presently considered cases, there is no such a region where the particles
are slipping at supersonic speed in the average sense.
The near-wall peaks of $\bar M_p$ (figure~\ref{fig:mp}(d)) increase monotonically with 
the Stokes number $St^+$ but incline to converge to the near-wall maximum of 
the turbulent Mach number, which is defined by the turbulent kinetic energy and 
the local mean acoustic velocity.

\begin{figure}
\centering
\begin{overpic}[width=0.5\textwidth]{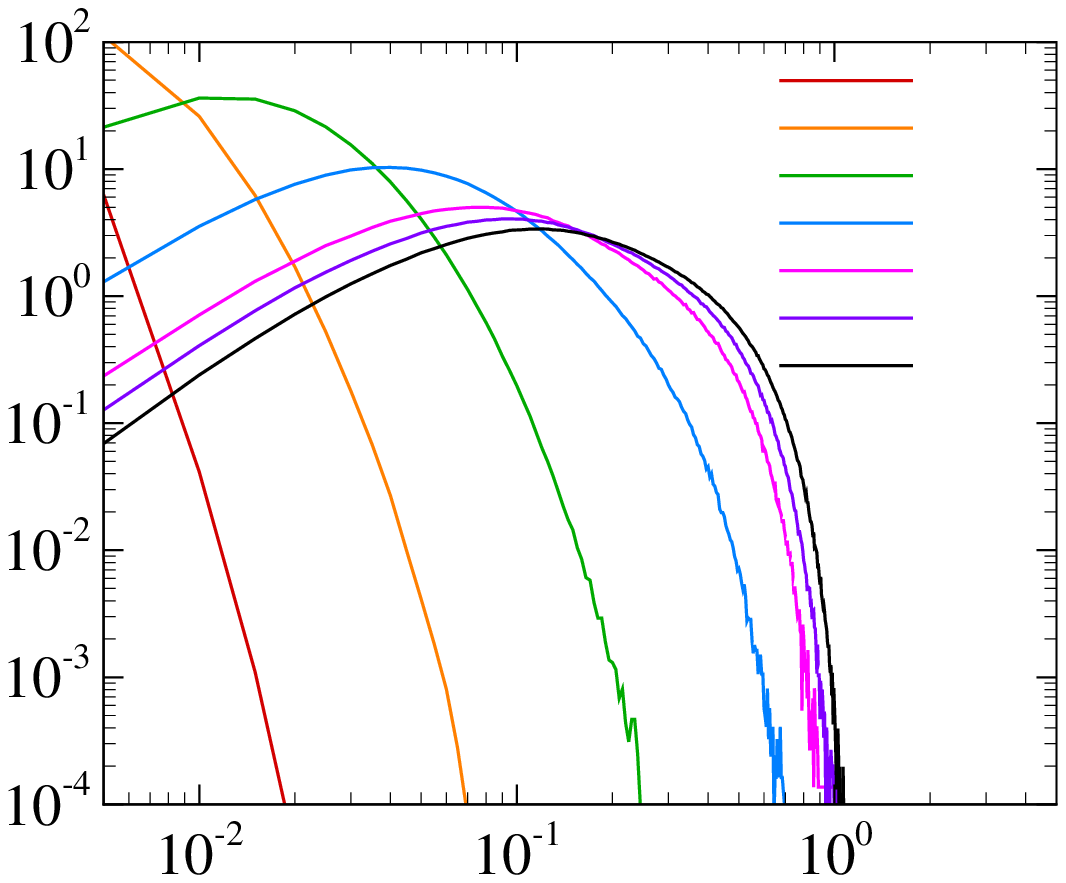}
\put(-2,70){(a)}
\put(50,0){$M_p$}
\put(-2,35){\rotatebox{90}{$P(M_p)$}}
\put(80,68){\scriptsize P1}
\put(80,64){\scriptsize P2}
\put(80,60){\scriptsize P3}
\put(80,56){\scriptsize P4}
\put(80,52.3){\scriptsize P5}
\put(80,48.4){\scriptsize P6}
\put(80,44.5){\scriptsize P7}
\end{overpic}~
\begin{overpic}[width=0.5\textwidth]{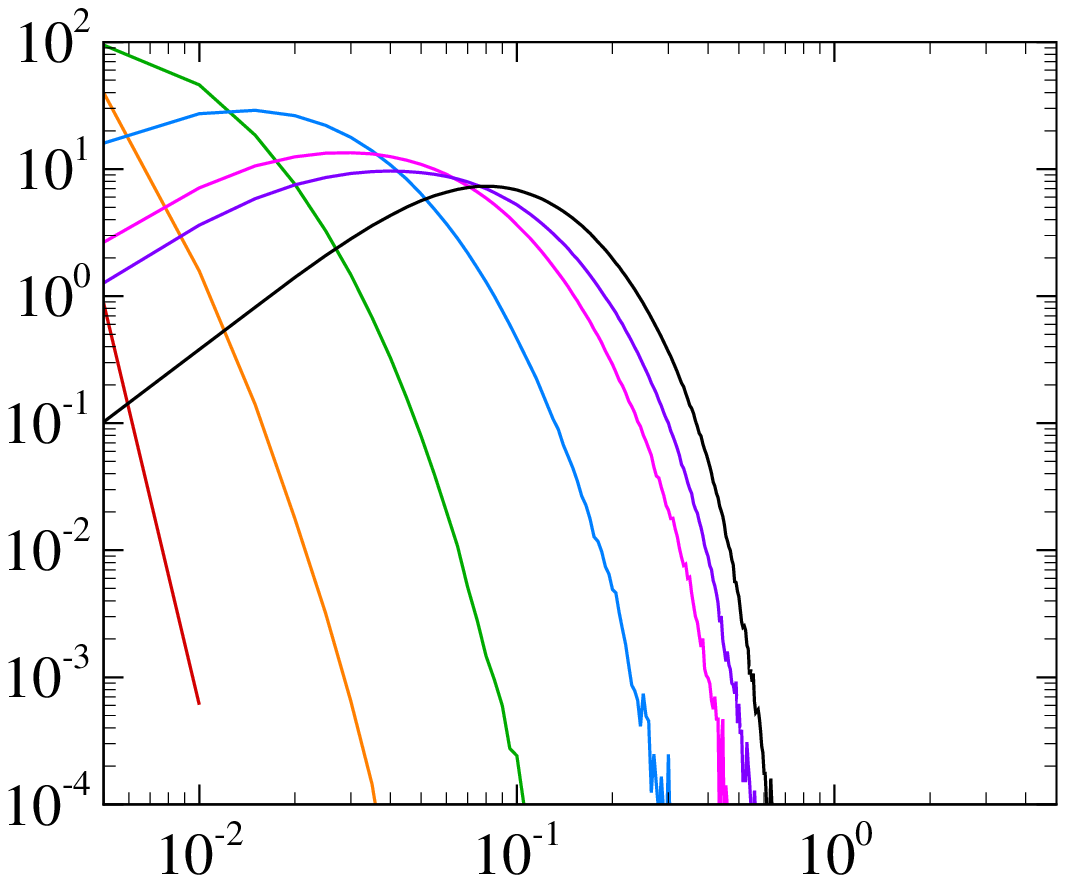}
\put(-2,70){(b)}
\put(50,0){$M_p$}
\put(-2,35){\rotatebox{90}{$P(M_p)$}}
\end{overpic}\\[1.0ex]
\begin{overpic}[width=0.5\textwidth]{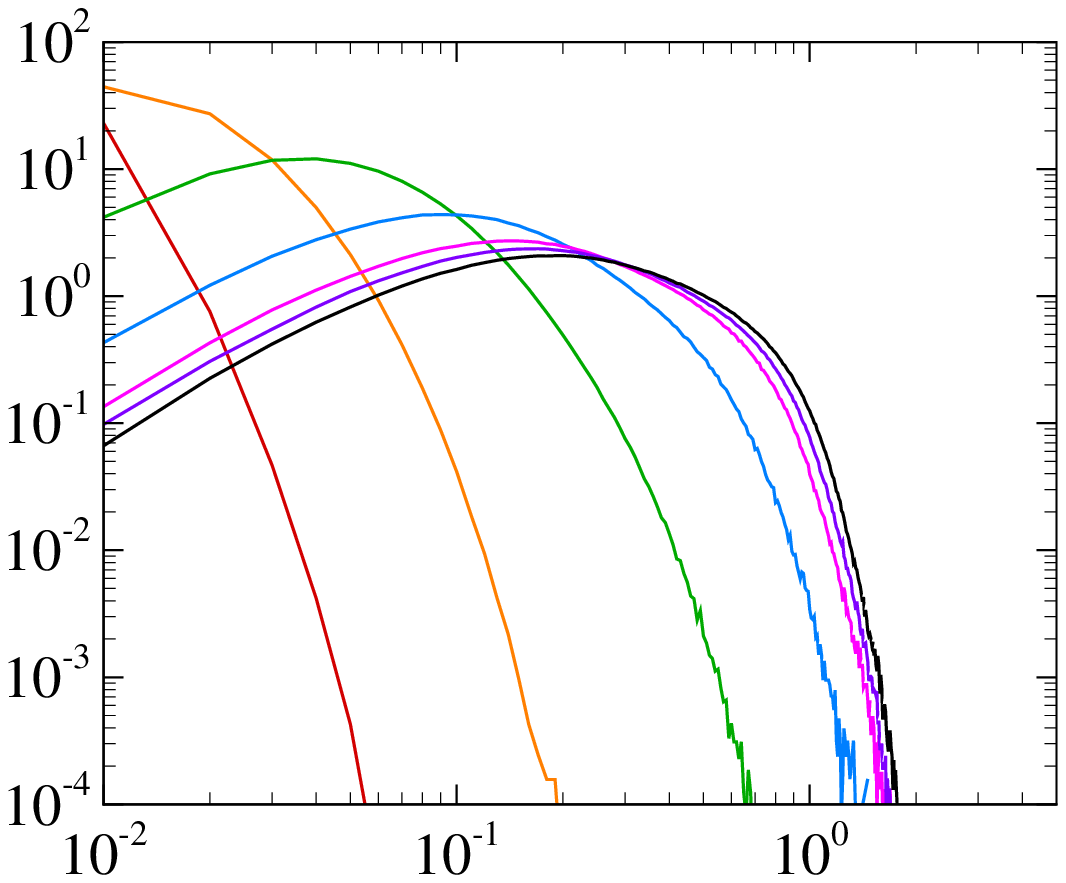}
\put(-2,70){(c)}
\put(50,0){$M_p$}
\put(-2,35){\rotatebox{90}{$P(M_p)$}}
\end{overpic}~
\begin{overpic}[width=0.5\textwidth]{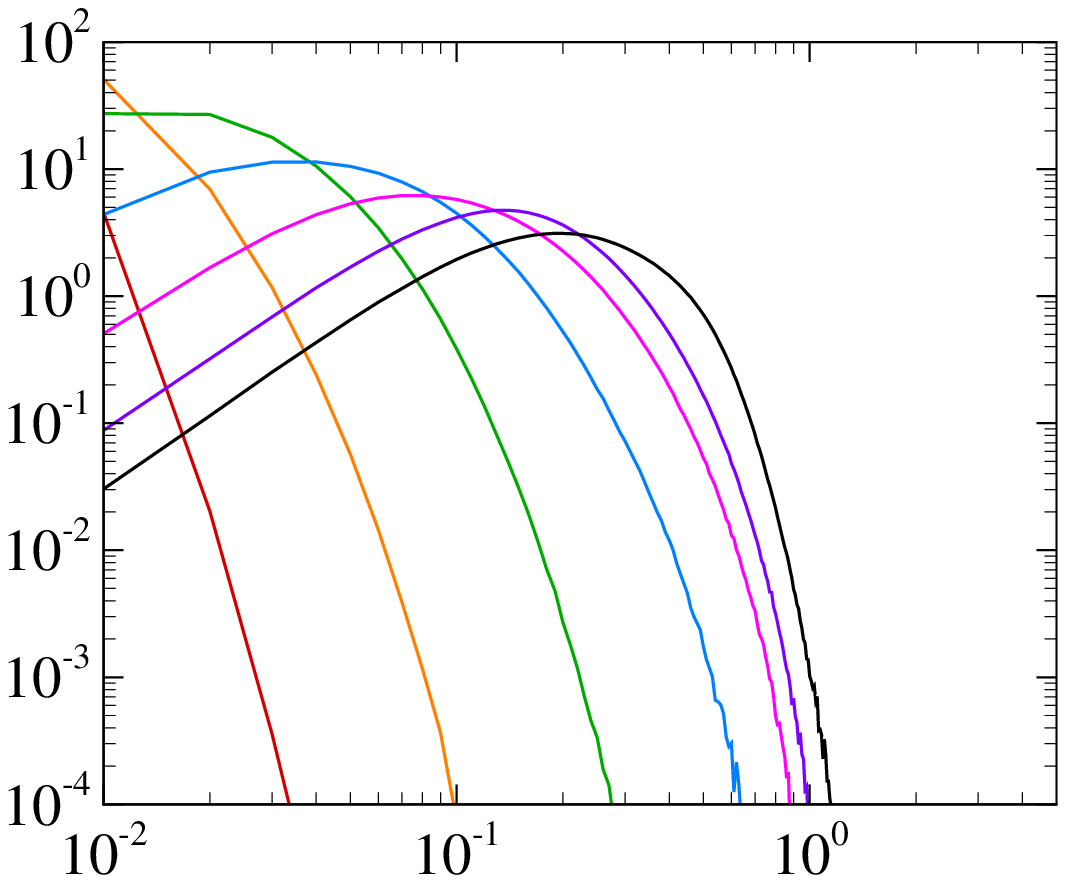}
\put(-2,70){(d)}
\put(50,0){$M_p$}
\put(-2,35){\rotatebox{90}{$P(M_p)$}}
\end{overpic}\\[1.0ex]
\begin{overpic}[width=0.5\textwidth]{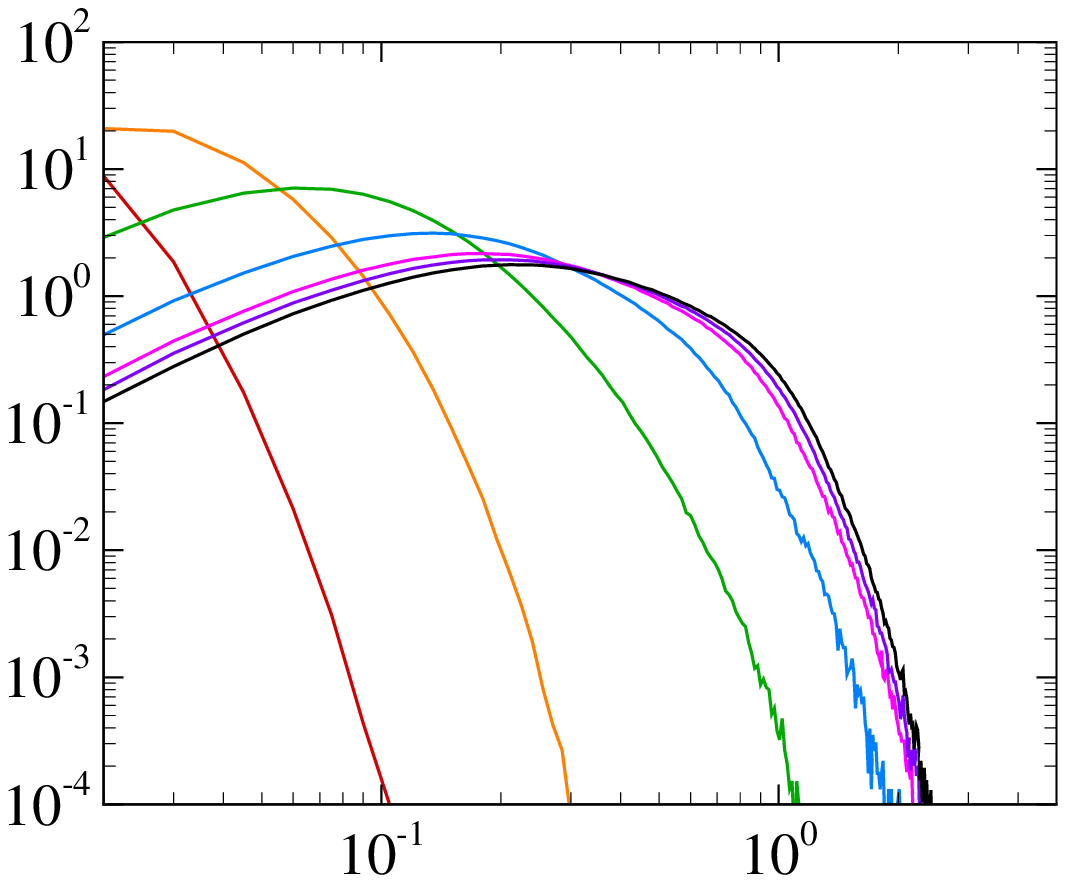}
\put(-2,70){(e)}
\put(50,0){$M_p$}
\put(-2,35){\rotatebox{90}{$P(M_p)$}}
\end{overpic}~
\begin{overpic}[width=0.5\textwidth]{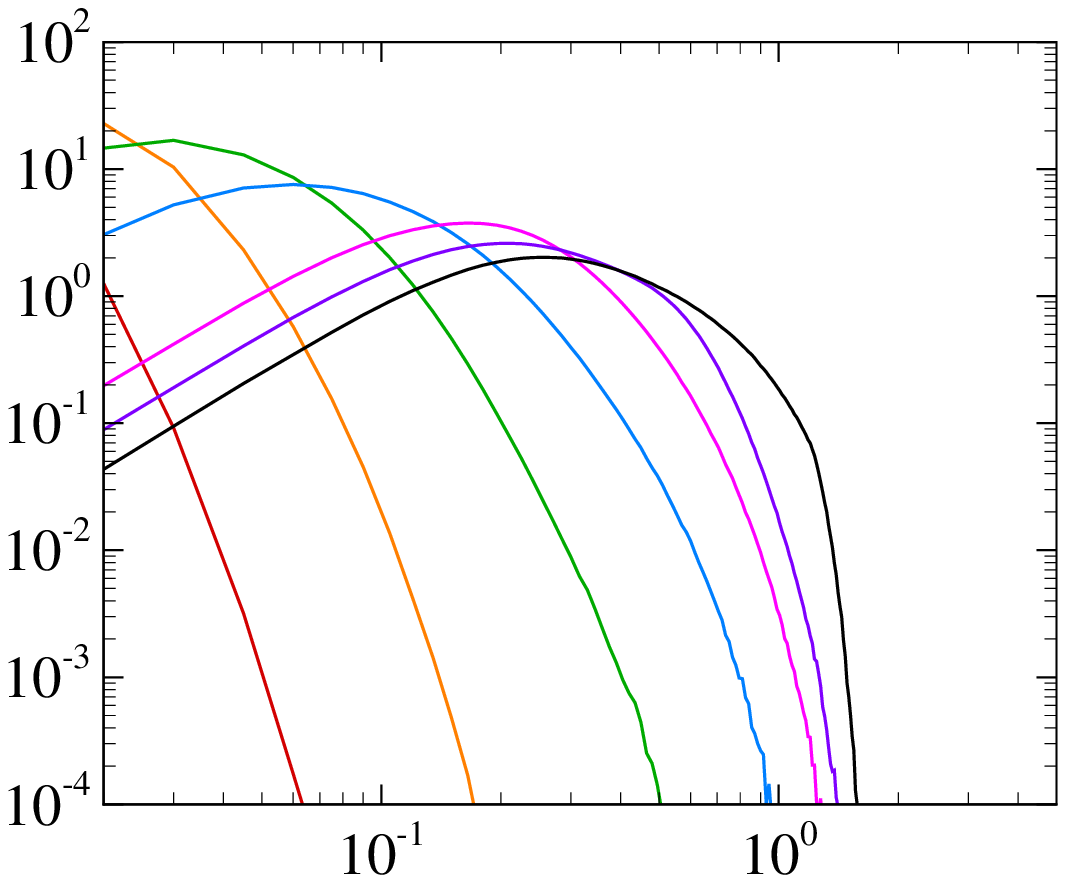}
\put(-2,70){(f)}
\put(50,0){$M_p$}
\put(-2,35){\rotatebox{90}{$P(M_p)$}}
\end{overpic}\\
\caption{PDF of particle Mach number $M_p$  (a,c,e) below $y^+=10$ and
(b,d,f) in the outer region within $y = (0.6 \sim 2.0) \delta$ in (a,b) case M2, (c,d) case M4
and (e,f) case M6.}
\label{fig:mppdf}
\end{figure}

We further report the PDF of the particle Mach number $M_p$ in figure~\ref{fig:mppdf}
in the near-wall region below $y^+=10$ (left column) and in the outer region within
$y = (0.6 \sim 2.0) \delta$ (right column).
It is interesting to note that the probability of the high Stokes number particle slipping
at supersonic speed in cases M4 and M6 is higher in the near-wall region than in the
outer region, as indicated by both their higher values and wider ranges of the PDF distributions.
Such higher particle Mach numbers induce higher drag force, not only their mean values
but also their fluctuations, the integration of which in time will lead to further disparities in
the statistics of the particle velocity.
This serves as a possible explanation of the different behaviour of the particle velocity 
fluctuations with large $St^+$ at various Mach numbers (recall figure~\ref{fig:prmsmax}).

\begin{figure}
\centering
\begin{overpic}[width=0.5\textwidth]{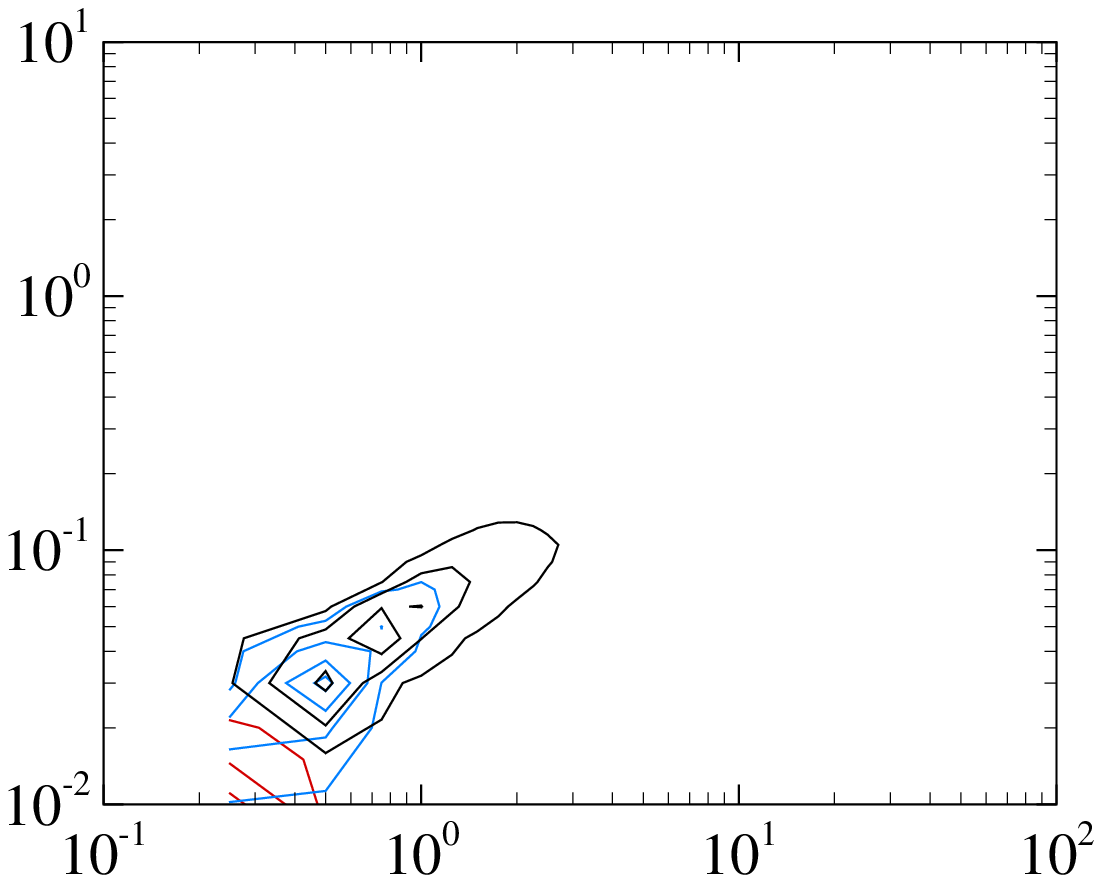}
\put(0,70){(a)}
\put(50,0){$Re_p$}
\put(0,35){\rotatebox{90}{$M_p$}}
\end{overpic}~
\begin{overpic}[width=0.5\textwidth]{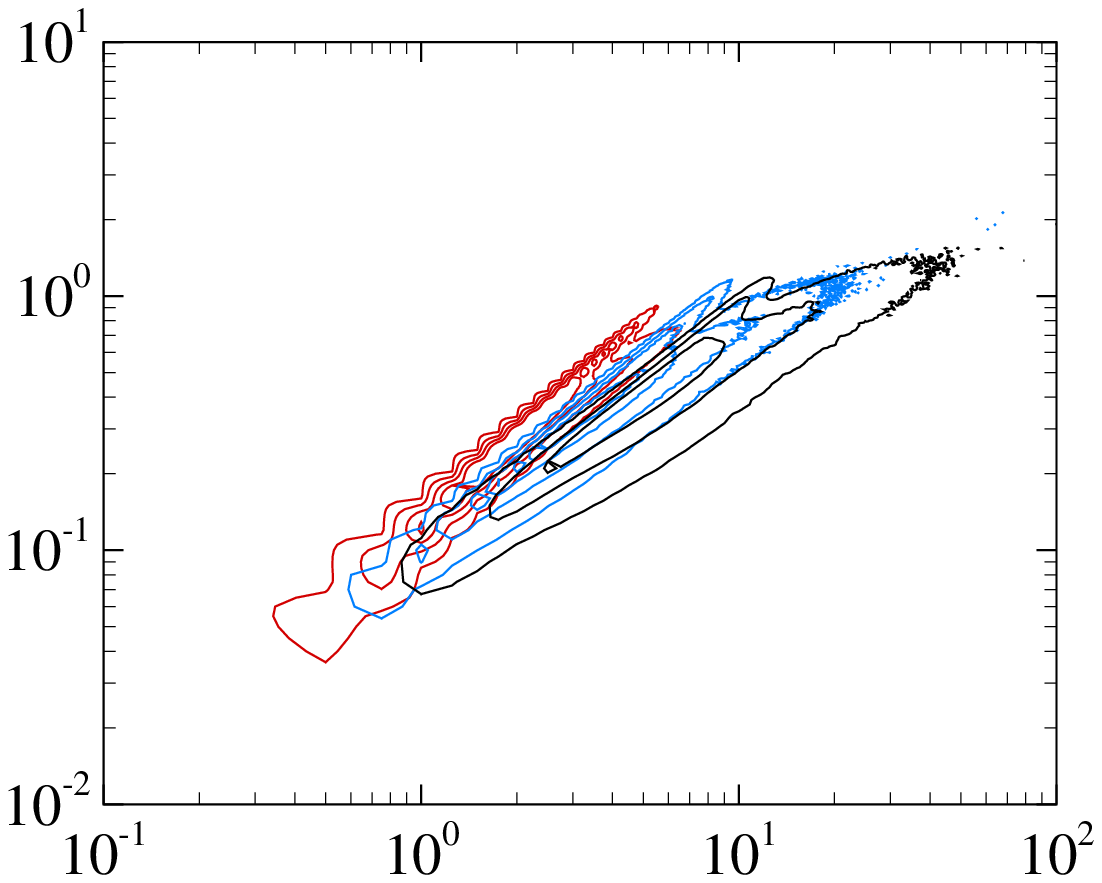}
\put(0,70){(b)}
\put(50,0){$Re_p$}
\put(0,35){\rotatebox{90}{$M_p$}}
\end{overpic}\\[1.0ex]
\caption{ 
Pre-multiplied joint-PDF between $Re_p$ and $M_p$ ($f_D Re_p M_pP(Re_p,M_p)$) of 
particle populations (a) P2, (b) P7 for case M2, P6 for case M4 and P5 for case M6, 
Red lines: M2, blue lines: M4, black lines: M6. Contour levels: (0.1, 0.5, 1.0, 1.5).}
\label{fig:pdfforce}
\end{figure}

To further explore the Reynolds number and Mach number effects on the drag force,
in figure~\ref{fig:pdfforce} we present the joint PDF between $Re_p$ and $M_p$ below $y^+=10$, 
pre-multiplied by $f_D$, $M_p$ and $Re_p$. The integration of the PDF under double-logarithmic
coordinates gives the mean values of $f_D$, corresponding to the modification of the drag forces
by the high $Re_p$ and $M_p$.
For particle population P2 (figure~\ref{fig:pdfforce}(a)), 
the pre-multiplied joint PDFs are mainly concentrated in the low $Re_p$ and $M_p$ region, 
with no significant difference in their distributions at different $M_\infty$.
For higher Stokes number particles, however, such disparity can be clearly visualized,
as demonstrated in figure~\ref{fig:pdfforce}(b), where particle population P7 in case M2,
P6 in case M4 and P5 in case M6 are presented due to their approximately the same $St^+$ 
($\approx 150$).
At higher $M_\infty$, the high values of the pre-multiplied joint PDF extend to the regions of 
both higher $Re_p$ and $M_p$, suggesting that the drag forces acting on the particles
in high Mach number cases are stronger.
This is consistent with the higher acceleration shown in figure~\ref{fig:forcevel} 
but lower velocity fluctuations presented in figures~\ref{fig:prmsmax},
supporting our inferences in the discussion above.

\section{Conclusions}  \label{sec:con}

In the present study, we perform direct numerical simulations of compressible turbulent boundary 
layer flow at the Mach numbers of $2$, $4$ and $6$ that transport the dilute phase of spherical 
particles utilizing the point-particle Eulerian-Lagrangian framework.
We consider seven particle populations, covering a wide range of 
Stokes numbers and examine the instantaneous and statistical distributions and the mean and 
fluctuating particle velocities, with the focus on their consistency and discrepancies 
at different Mach numbers.

In the instantaneous flow fields, it is found that the distributions of the 
low Stokes number particles
resemble those of the density, restricted within the turbulent boundary layers, 
which can be explained by the identical form of the continuity equation and 
the approximation of the transport equation of the particle concentration field.
The high Stokes number particles, on the other hand, are capable of escaping the `restriction' 
of the turbulent-non-turbulent interface and reaching the free stream due to their large inertia.
The particles with moderate Stokes numbers tend to accumulate near the wall,
known as the turbophoresis, and manifest clustering beneath the low-speed region,
consistent with the findings in incompressible canonical wall-bounded turbulence.
Quantitatively, however, both the degrees of near-wall accumulation and the small-scale clustering, 
evaluated by the Shannon entropy and the power-law scaling of the radial density function at 
$r \rightarrow 0$, respectively, are slightly reduced by the higher free stream
Mach number, whereas that of clustering remain unaffected.

The distribution of the particles can be reflected by the mean fluid velocity seen by particles
in that the particle streamwise velocities are lower than the mean fluid velocity
with moderate Stokes numbers, while the wall-normal velocities are higher.
The profiles of the streamwise mean particle velocity are flatter with increasing Stokes number
considering their slow response to the fluid motion when travelling vertically,
whereas the wall-normal mean velocities are roughly independent of the particle inertia,
merely showing the spatial development of the boundary layer.
As for the particle fluctuating velocity, the streamwise component shows non-monotonic variation
against the Stokes number and relevance to the Mach number, while the wall-normal component
decreases monotonically and depends weakly on the Mach number.
By inspecting the statistics of the particle acceleration and the factors by which it is influenced,
we prove that the increasing particle Reynolds and Mach numbers lead to the variation of
the streamwise particle velocity fluctuations for large Stokes number particles.

The present work systematically studies the transport of particles with various Stokes numbers
in the compressible turbulent boundary layers, but only one-way coupling is concerned.
Future work will be dedicated to the investigation of the two-way coupling effects that incorporate
the particle feedback to the phase of fluid turbulence.

\appendix

\section{Validation of numerical solver}  \label{sec:app}

\begin{figure}
\centering
\begin{overpic}[width=0.5\textwidth]{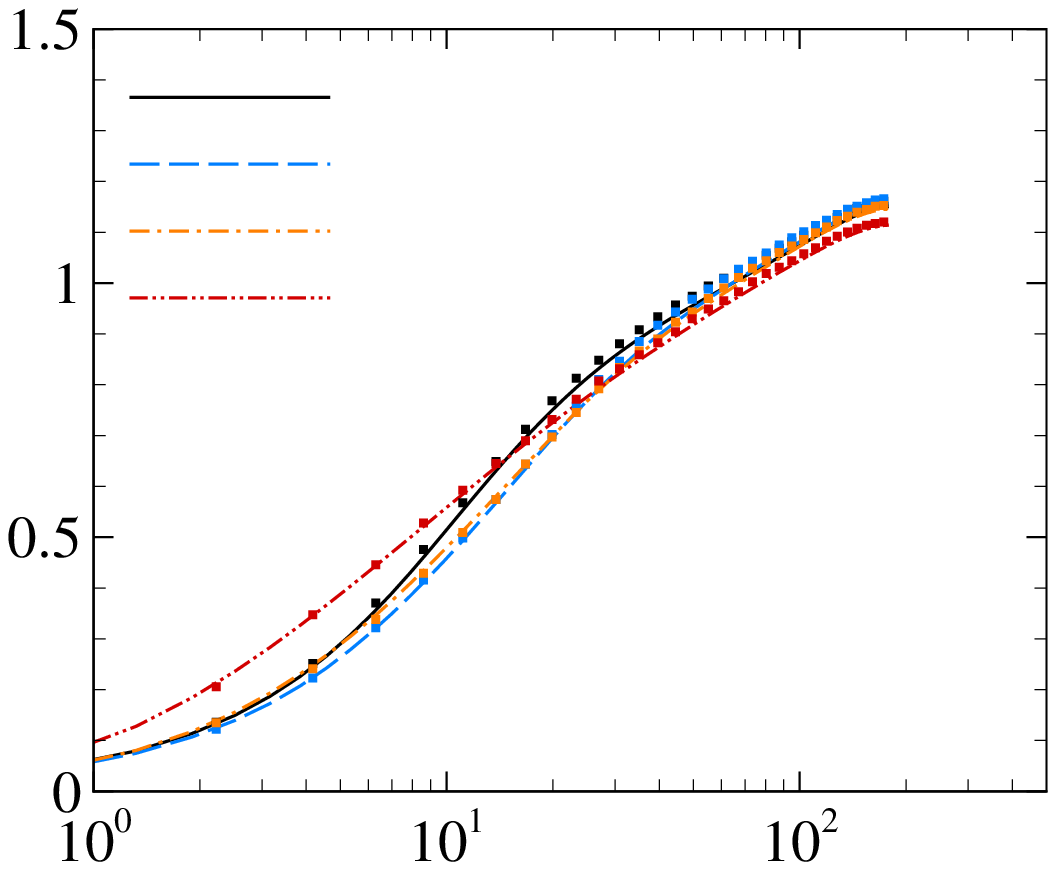}
\put(-2,70){(a)}
\put(50,0){$y^+$}
\put(-2,35){\rotatebox{90}{$\bar v_1$}}
\put(35,65.5){\scriptsize $St^+=1$}
\put(35,60){\scriptsize $St^+=10$}
\put(35,54.5){\scriptsize $St^+=30$}
\put(35,49){\scriptsize $St^+=200$}
\end{overpic}~
\begin{overpic}[width=0.5\textwidth]{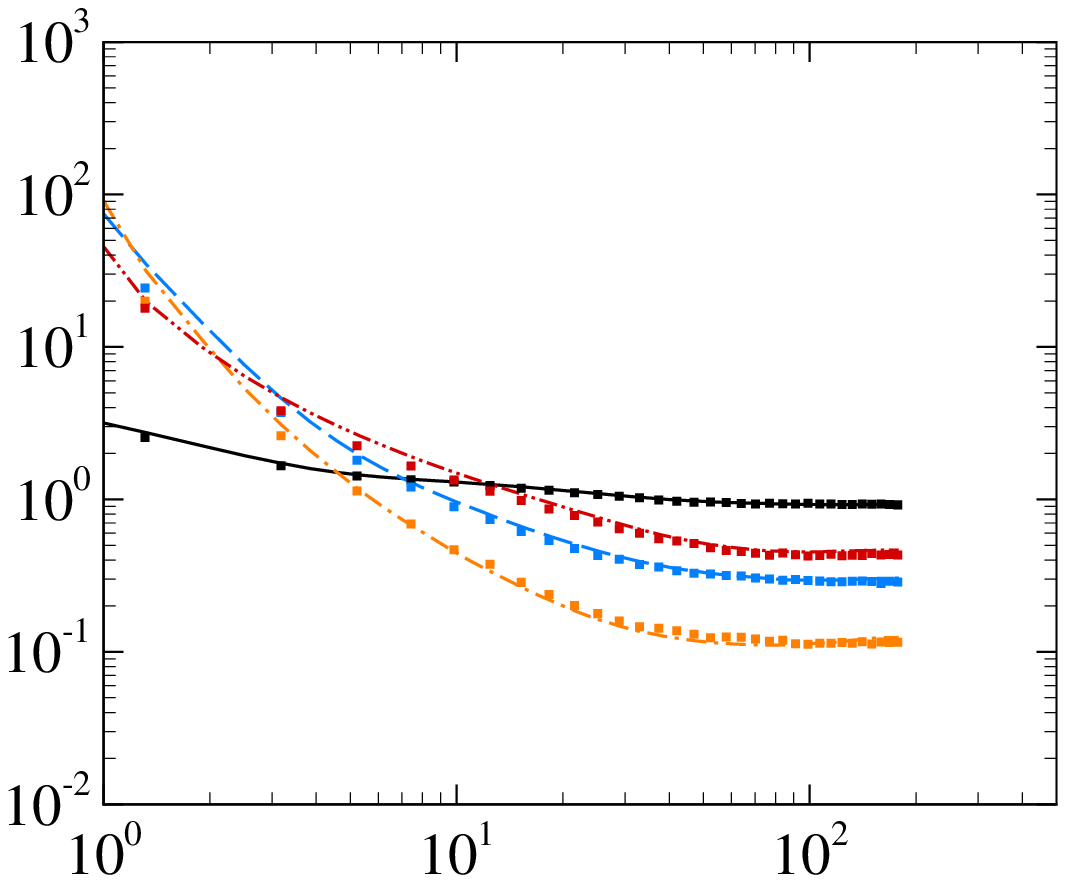}
\put(-2,70){(b)}
\put(50,0){$y^+$}
\put(-2,35){\rotatebox{90}{$\bar c(y)/c_0$}}
\end{overpic}\\[1.0ex]
\begin{overpic}[width=0.5\textwidth]{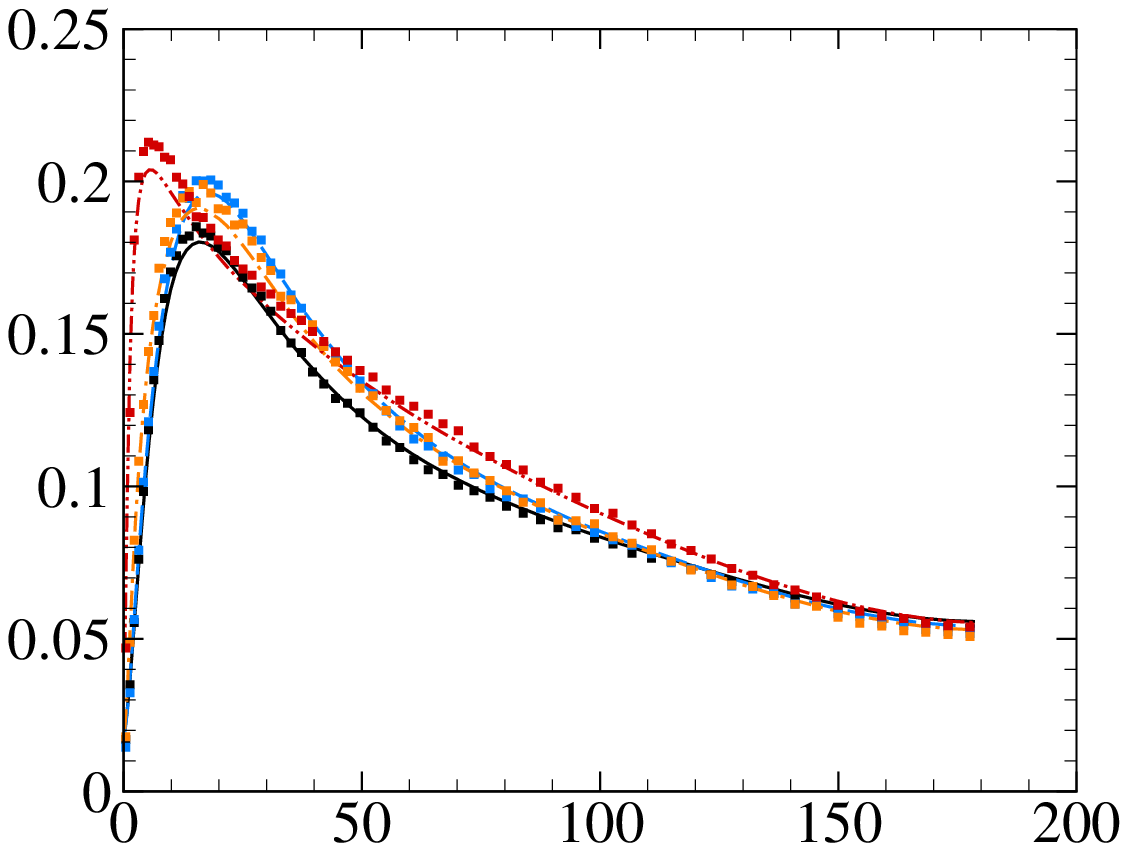}
\put(-2,70){(c)}
\put(50,0){$y^+$}
\put(-2,35){\rotatebox{90}{$\bar v'_{1}/u_b$}}
\end{overpic}~
\begin{overpic}[width=0.5\textwidth]{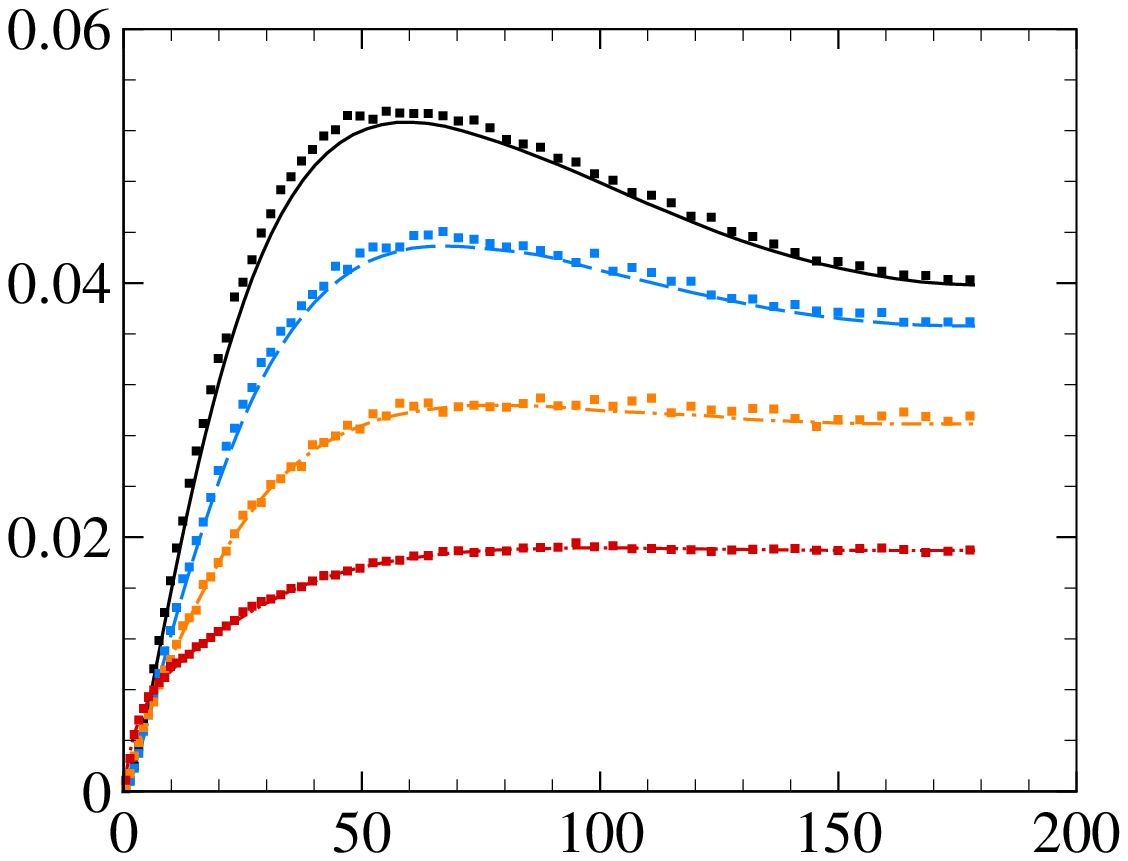}
\put(-2,70){(d)}
\put(50,0){$y^+$}
\put(-2,35){\rotatebox{90}{$\bar v'_{2}/u_b$}}
\end{overpic}\\
\caption{Particle statistics in a turbulent channel flow, (a) $\bar v_1$, (b) $\bar c(y)/c_0$, 
(c) $\bar v'_{1}$, (d) $\bar v'_{2}$. $u_b$ is the bulk velocity of the channel. 
{
Lines: present, symbols: reference data of incompressible flows~\citep{jie2022existence}.}}
\label{fig:valid}
\end{figure}

We perform the direct numerical simulations of the turbulent channel flows 
at the friction Reynolds number of $Re_\tau = \rho_w u_\tau h/\mu_w = 180$ at the bulk Mach number 
of $M_b =u_b/\sqrt{\gamma R T_w}=0.3$
for the validation of the numerical solver utilized in the present study.
The simulation is performed in the computational domain with the streamwise and spanwise sizes
of $2 \pi h$ and $\pi h$, respectively, with $h$ the half channel height.
Four types of particles with the Stokes numbers $St^+=1$, 10, 50 and 200
are uniformly injected into the channel after the turbulence is fully developed.
The statistics are collected within the time span of $7000 \delta_\nu / u_\tau$ after 
the simulation has been run for $7000 \delta_\nu / u_\tau$.
The results are compared with those obtained by the incompressible turbulent channel flow
solvers developed by~\citet{jie2022existence,cui2022shape,cui2021alignment} (private communication)
under the same parameter settings.
As reported in figure~\ref{fig:valid},
the streamwise mean particle velocity $\bar v_1$, mean particle concentration $\bar c(y)$
and the RMS of the streamwise and wall-normal particle velocities obtained by the two distinct
solvers agree well with each other, consolidating the accuracy of the two-phase DNS solver 
utilized in the present study.

\section{Streamwise variation of the Shannon entropy}  \label{sec:stma}

\begin{figure}
\centering
\scriptsize
\begin{overpic}[width=0.33\textwidth]{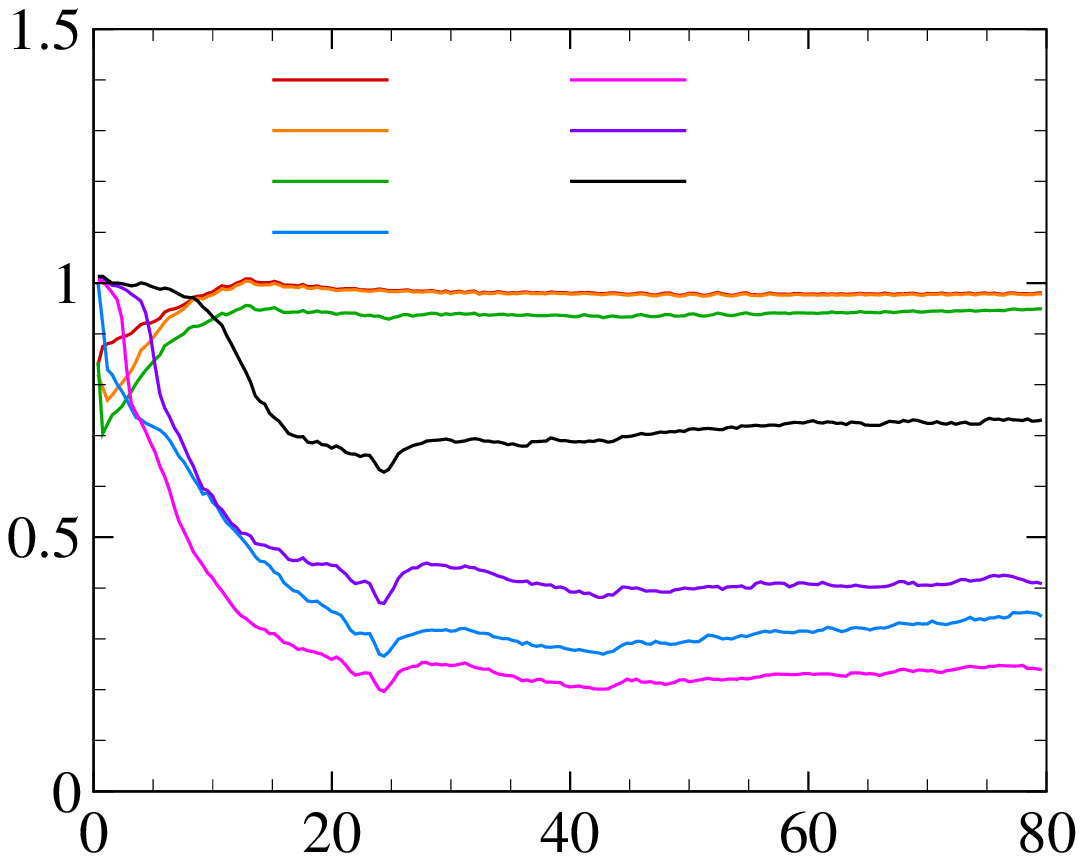}
\put(-2,70){(a)}
\put(50,0){$x/\delta_0$}
\put(-2,35){\rotatebox{90}{$S_p$}}
\put(38,67){\tiny P1}
\put(38,63){\tiny P2}
\put(38,59){\tiny P3}
\put(38,55){\tiny P4}
\put(62,67){\tiny P5}
\put(62,63){\tiny P6}
\put(62,59){\tiny P7}
\end{overpic}~
\begin{overpic}[width=0.33\textwidth]{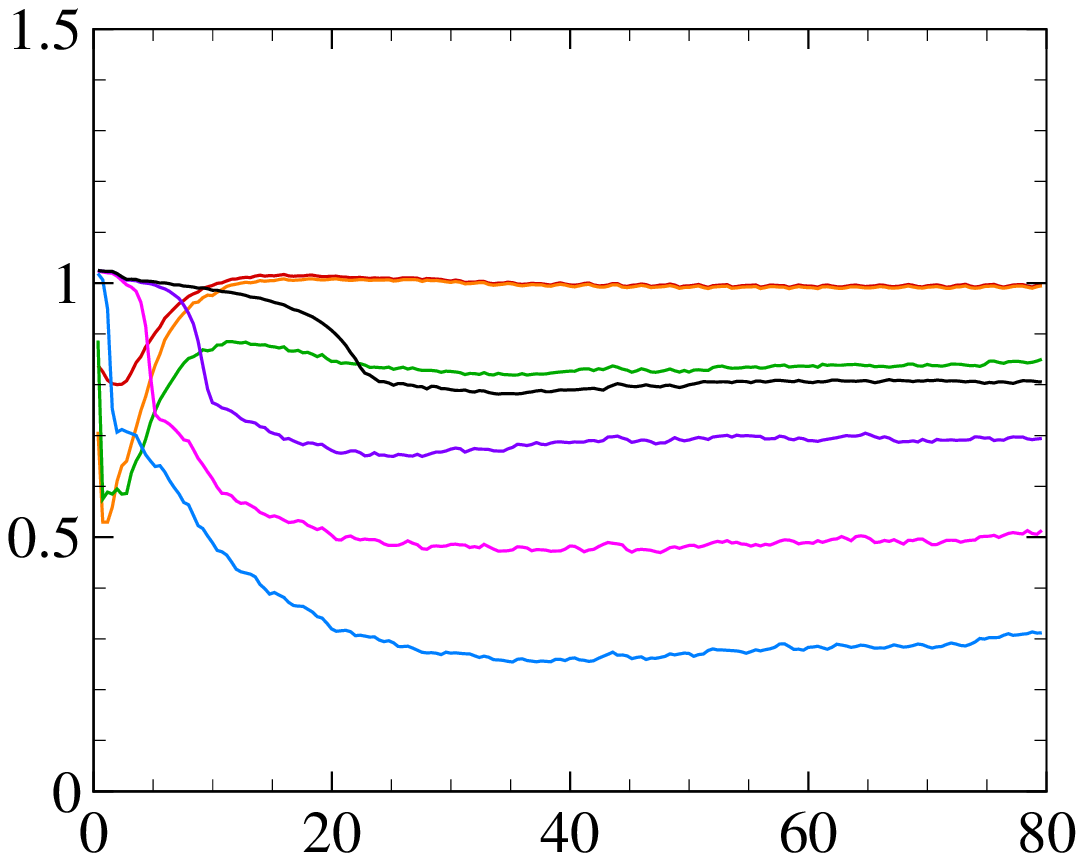}
\put(-2,70){(b)}
\put(50,0){$x/\delta_0$}
\put(-2,35){\rotatebox{90}{$S_p$}}
\end{overpic}~
\begin{overpic}[width=0.33\textwidth]{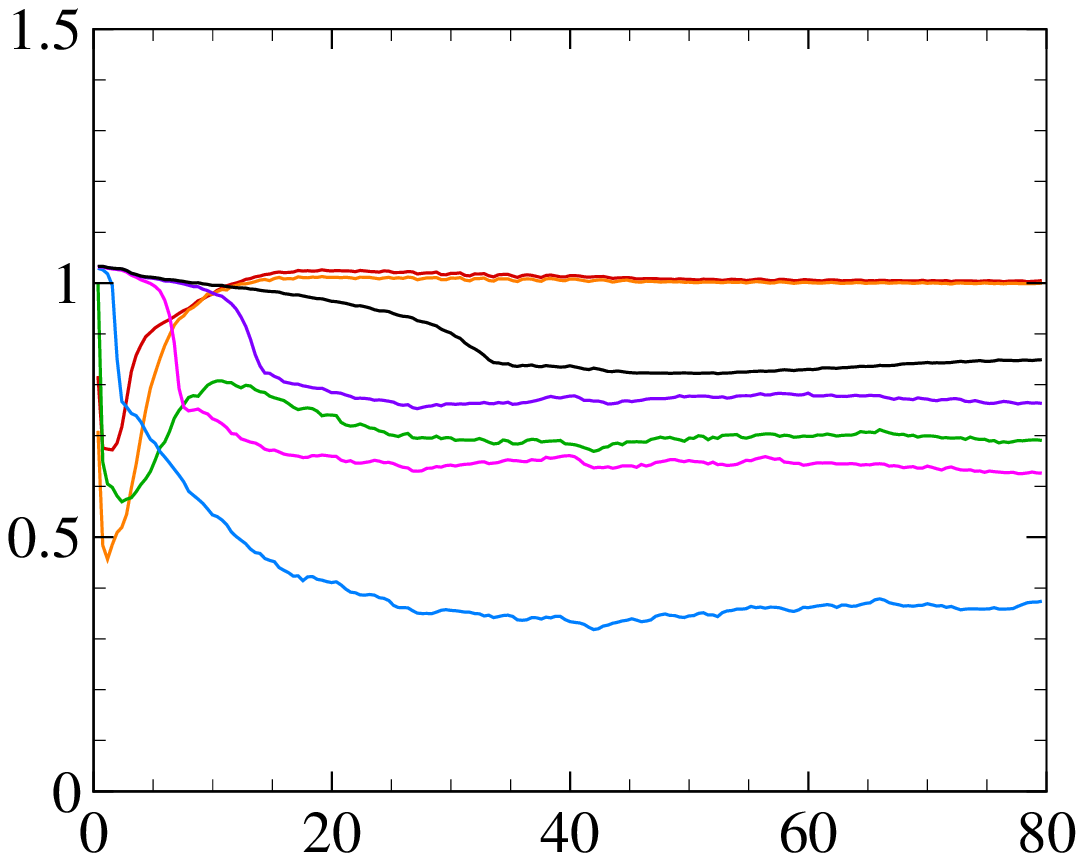}
\put(-2,70){(c)}
\put(50,0){$x/\delta_0$}
\put(-2,35){\rotatebox{90}{$S_p$}}
\end{overpic}
\caption{Streamwise variation of the Shannon entropy $S_p$ in cases (a) M2, (b) M4 and (c) M6.}
\label{fig:stma}
\end{figure}

For the validation of the streamwise statistical invariance within $(60 \sim 70) \delta_0$,
in figure~\ref{fig:stma} we present the variation of the Shannon entropy $S_p$ defined in 
section~\ref{subsec:dist}, which is usually adopted for the purpose of verifying the convergence
and depicting the degree of near-wall accumulation~\citep{picano2009spatial,bernardini2014reynolds}.
For all the cases considered, the Shannon entropy $S_p$ for each particle population remains almost
constant, suggesting that the particle distributions are statistically steady and quasi-homogeneous,
in particular within the domain of interest $(60 \sim 70) \delta_0$ where the statistics are 
obtained.

\section{PDF of wall-normal fluid velocity seen by particles} \label{sec:pdf}

\begin{figure}
\centering
\begin{overpic}[width=0.5\textwidth]{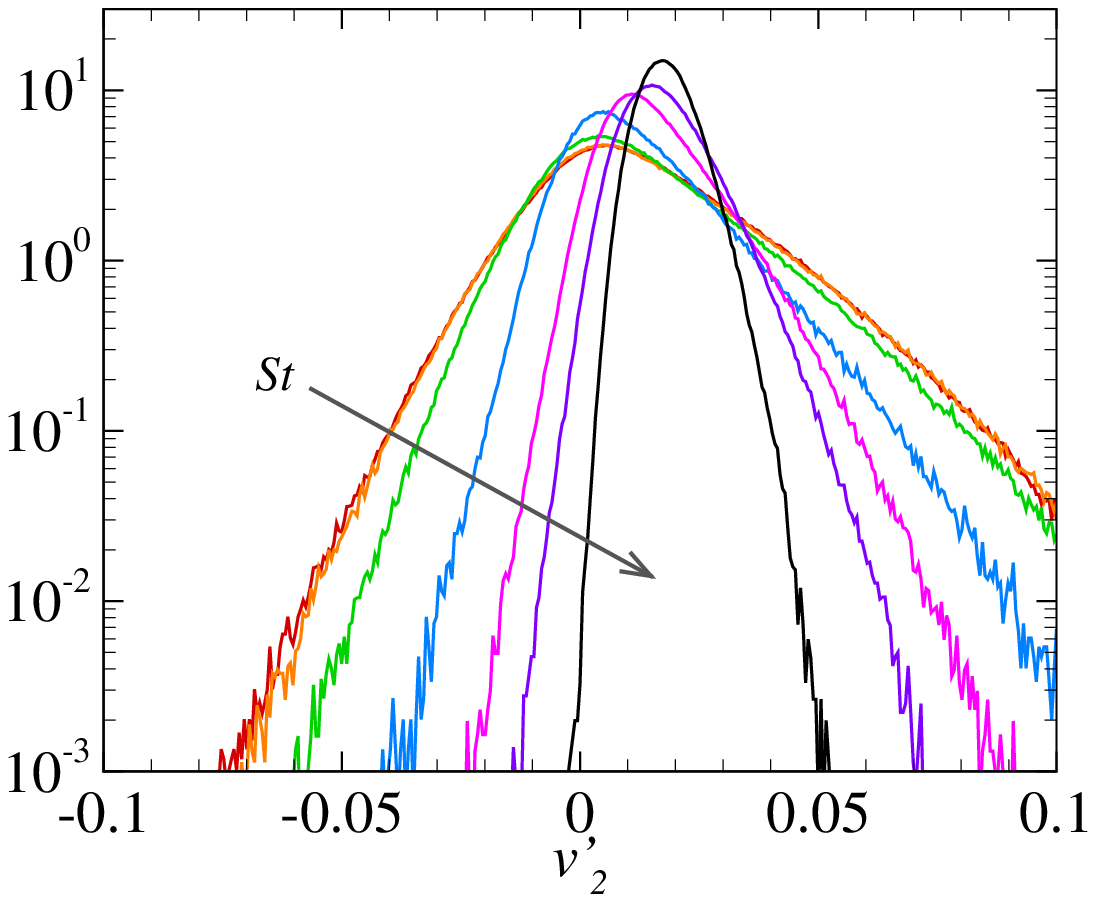}
\put(0,70){(a)}
\put(48,2){\colorbox{white}{$v_2$}}
\end{overpic}~
\begin{overpic}[width=0.5\textwidth]{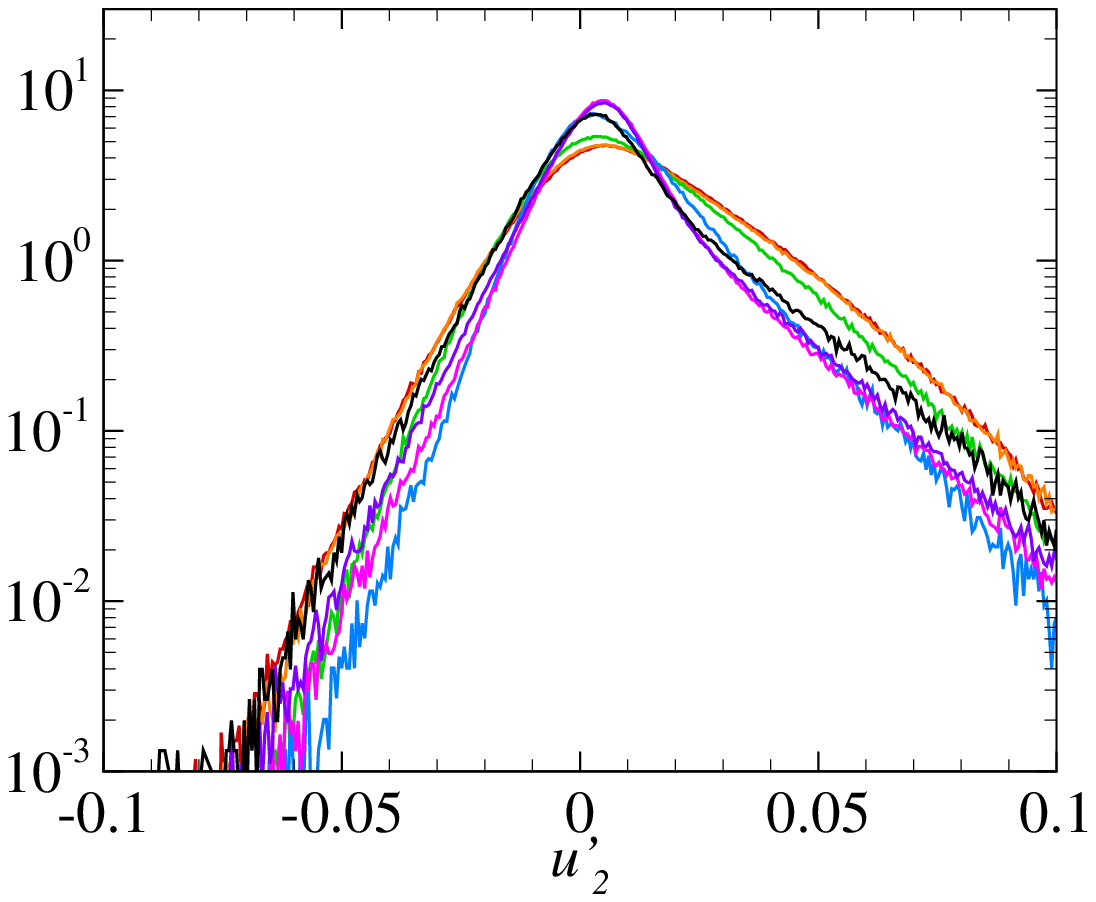}
\put(0,70){(b)}
\put(48,2){\colorbox{white}{$u_{2,p}$}}
\end{overpic}\\
\caption{PDF of (a) $v_2$ and (b) $u_{2,p}$ above $y=0.6\delta$ in case M4, line legends
please refer to figure~\ref{fig:mppdf}.}
\label{fig:addpdf}
\end{figure}

In figure~\ref{fig:addpdf} we present the PDF of the wall-normal particle velocity $v_2$ and
the fluid velocity seen by particles $u_{2,p}$ above $y=0.6\delta$ in case M4.
Obviously, the PDF values are higher in the positive regions than those in the negative regions,
supporting our statement that the particles prefer to accumulate by the strong ejections.

\section{Particle velocity fluctuations with a different flow inlet condition} \label{sec:comp}

\begin{figure}
\centering
\begin{overpic}[width=0.5\textwidth]{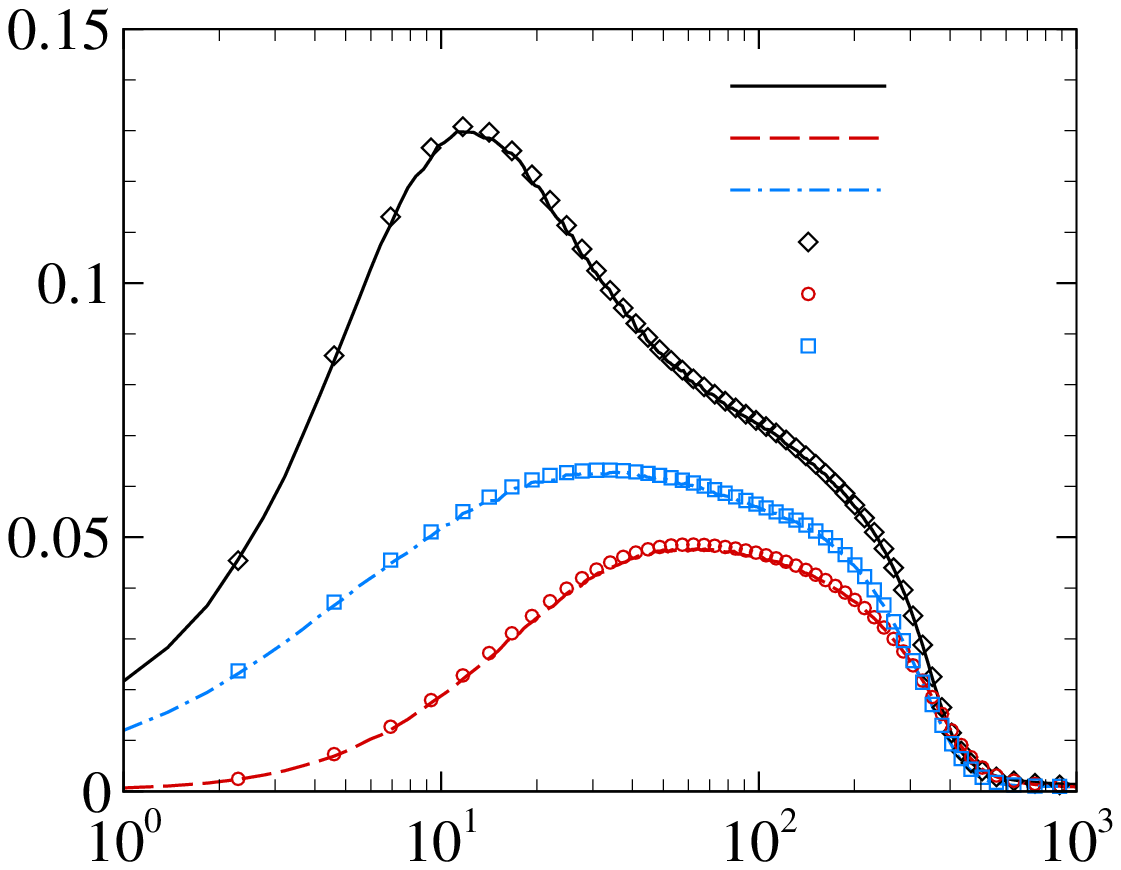}
\put(50,0){$y^+$}
\put(0,38){\rotatebox{90}{$\bar v'_i$}}
\put(76,67){\scriptsize $\bar v'_1$}
\put(76,62.5){\scriptsize $\bar v'_2$}
\put(76,58){\scriptsize $\bar v'_3$}
\put(76,53.5){\scriptsize $\bar u''_1$}
\put(76,49){\scriptsize $\bar u''_2$}
\put(76,44.5){\scriptsize $\bar u''_3$}
\end{overpic}
\caption{RMS of particle (P1) and fluid velocity fluctuations in case M2, with the particles released
randomly at the flow inlet across the whole wall-normal computational domain.}
\label{fig:parrmsexp}
\end{figure}

We have inferred in section~\ref{subsec:parvel} that the discrepancies in the RMS of the velocity
fluctuations between the particle and fluid should be ascribed to the distribution of the particles
within the boundary layer thickness.
In this section, we attempt to verify this inference by performing an extra simulation with the same
computational settings as in case M2, except that the particles are released randomly at the flow
inlet across the whole wall-normal computational domain instead of merely within $\delta_0$.
The obtained RMS of the particle (population P1) and fluid velocity fluctuations are shown in 
figure~\ref{fig:parrmsexp}.
The profiles are well-collapsed, especially near the edge of the boundary layer, validating
the correctness of the inference in section~\ref{subsec:parvel}.

\section*{Funding}
\noindent 
This work is supported by the National Natural Science Foundation of China 
(Grant No. 12202469, 92052301,92252104,12388101).

\section*{Declaration of interests}
\noindent The authors report no conflict of interest.

\section*{Author ORCID}
M. Yu, https://orcid.org/0000-0001-7772-833X ; \\ 
L.H. Zhao, https://orcid.org/0000-0002-3642-3051 ; \\
X.X. Yuan, https://orcid.org/0000-0002-7668-0116 ; \\
C.X. Xu, https://orcid.org/0000-0001-5292-8052

\bibliographystyle{jfm}
\bibliography{bibfile}

\begin{thebibliography}{93}
\expandafter\ifx\csname natexlab\endcsname\relax\def\natexlab#1{#1}\fi
\def\au#1{#1} \def\ed#1{#1} \def\yr#1{#1}\def\at#1{#1}\def\jt#1{\textit{#1}}
  \def\bt#1{#1}\def\bvol#1{\textbf{#1}} \def\vol#1{#1} \def\pg#1{#1}
  \def\publ#1{#1}\def\arxiv#1{#1}\def\org#1{#1}\def\st#1{\textit{#1}}

\bibitem[Armenio \& Fiorotto(2001)]{armenio2001importance}
{\sc \au{Armenio, V.} \& \au{Fiorotto, V.}} \yr{2001}  \at{The importance of
  the forces acting on particles in turbulent flows}.  \jt{Phys.~Fluids}
  \bvol{13}~(8),  \pg{2437--2440}.

\bibitem[Balachandar(2009)]{balachandar2009scaling}
{\sc \au{Balachandar, S}} \yr{2009}  \at{A scaling analysis for point--particle
  approaches to turbulent multiphase flows}.  \jt{Int. J. Multiph. Flow}
  \bvol{35}~(9),  \pg{801--810}.

\bibitem[Balachandar \& Eaton(2010)]{balachandar2010turbulent}
{\sc \au{Balachandar, S.} \& \au{Eaton, J.K.}} \yr{2010}  \at{Turbulent
  dispersed multiphase flow}.  \jt{Annu. Rev. Fluid Mech.}  \bvol{42},
  \pg{111--133}.

\bibitem[Balkovsky {\em et~al.\/}(2001)Balkovsky, Falkovich \&
  Fouxon]{balkovsky2001intermittent}
{\sc \au{Balkovsky, E}, \au{Falkovich, G} \& \au{Fouxon, A}} \yr{2001}
  \at{Intermittent distribution of inertial particles in turbulent flows}.
  \jt{Phys. Rev. Lett.}  \bvol{86}~(13),  \pg{2790}.

\bibitem[Baltussen {\em et~al.\/}(2018)Baltussen, Buist, Peters \&
  Kuipers]{baltussen2018multiscale}
{\sc \au{Baltussen, M.W.}, \au{Buist, K.A.}, \au{Peters, E.} \& \au{Kuipers,
  J.}} \yr{2018}  \at{Multiscale modelling of dense gas--particle flows}.
  \bt{In {\em Adv. Chem. Eng\/}}, ,  \vol{vol.~53},  \pg{pp. 1--52}.
  \publ{Elsevier}.

\bibitem[Bec {\em et~al.\/}(2007)Bec, Biferale, Cencini, Lanotte, Musacchio \&
  Toschi]{bec2007heavy}
{\sc \au{Bec, J.}, \au{Biferale, L.}, \au{Cencini, M.}, \au{Lanotte, A.},
  \au{Musacchio, S.} \& \au{Toschi, F.}} \yr{2007}  \at{Heavy particle
  concentration in turbulence at dissipative and inertial scales}.  \jt{Phys.
  Rev. Lett.}  \bvol{98}~(8),  \pg{084502}.

\bibitem[Berk \& Coletti(2020)]{berk2020transport}
{\sc \au{Berk, T.} \& \au{Coletti, F.}} \yr{2020}  \at{Transport of inertial
  particles in {high-Reynolds-number} turbulent boundary layers}.  \jt{J. Fluid
  Mech.}  \bvol{903},  \pg{A18}.

\bibitem[Bernardini(2014)]{bernardini2014reynolds}
{\sc \au{Bernardini, M.}} \yr{2014}  \at{Reynolds number scaling of inertial
  particle statistics in turbulent channel flows}.  \jt{J. Fluid Mech.}
  \bvol{758},  \pg{R1}.

\bibitem[Bernardini {\em et~al.\/}(2021)Bernardini, Modesti, Salvadore \&
  Pirozzoli]{bernardini2021streams}
{\sc \au{Bernardini, M.}, \au{Modesti, D.}, \au{Salvadore, F.} \&
  \au{Pirozzoli, S.}} \yr{2021}  \at{{STREAmS}: A high-fidelity accelerated
  solver for direct numerical simulation of compressible turbulent flows}.
  \jt{Comput.~Phys.~Commun.}  \bvol{263},  \pg{107906}.

\bibitem[Bernardini \& Pirozzoli(2011)]{bernardini2011wall}
{\sc \au{Bernardini, M.} \& \au{Pirozzoli, S.}} \yr{2011}  \at{Wall pressure
  fluctuations beneath supersonic turbulent boundary layers}.
  \jt{Phys.~Fluids}  \bvol{23}~(8),  \pg{085102}.

\bibitem[Bernardini {\em et~al.\/}(2013)Bernardini, Pirozzoli \&
  Orlandi]{bernardini2013effect}
{\sc \au{Bernardini, M.}, \au{Pirozzoli, S.} \& \au{Orlandi, P.}} \yr{2013}
  \at{The effect of large-scale turbulent structures on particle dispersion in
  wall-bounded flows}.  \jt{Int. J. Multiph. flow}  \bvol{51},  \pg{55--64}.

\bibitem[Bragg \& Collins(2014)]{bragg2014new}
{\sc \au{Bragg, A.D.} \& \au{Collins, L.R.}} \yr{2014}  \at{New insights from
  comparing statistical theories for inertial particles in turbulence: {I.
  Spatial distribution of particles}}.  \jt{New J. Phys.}  \bvol{16}~(5),
  \pg{055013}.

\bibitem[Brandt \& Coletti(2022)]{brandt2022particle}
{\sc \au{Brandt, L.} \& \au{Coletti, F.}} \yr{2022}  \at{Particle-laden
  turbulence: progress and perspectives}.  \jt{Annu. Rev. Fluid Mech.}
  \bvol{54},  \pg{159--189}.

\bibitem[Buchta {\em et~al.\/}(2019)Buchta, Shallcross \&
  Capecelatro]{buchta2019sound}
{\sc \au{Buchta, D.A.}, \au{Shallcross, G.} \& \au{Capecelatro, J.}} \yr{2019}
  \at{Sound and turbulence modulation by particles in high-speed shear flows}.
  \jt{J.~Fluid~Mech.}  \bvol{875},  \pg{254--285}.

\bibitem[Cao {\em et~al.\/}(2014)Cao, Wu \& Xu]{cao2014effects}
{\sc \au{Cao, Y.H.}, \au{Wu, Z.L.} \& \au{Xu, Z.Y.}} \yr{2014}  \at{Effects of
  rainfall on aircraft aerodynamics}.  \jt{Prog. Aerosp. Sci.}  \bvol{71},
  \pg{85--127}.

\bibitem[Capecelatro \& Wagner(2024)]{capecelatro2024gas}
{\sc \au{Capecelatro, J.} \& \au{Wagner, J.L.}} \yr{2024}  \at{Gas--particle
  dynamics in high-speed flows}.  \jt{Annu.~Rev.~Fluid~Mech.}  \bvol{56},
  \pg{379--403}.

\bibitem[Casciola {\em et~al.\/}(2007)Casciola, Gualtieri, Jacob \&
  Piva]{casciola2007residual}
{\sc \au{Casciola, CM}, \au{Gualtieri, P}, \au{Jacob, B} \& \au{Piva, R}}
  \yr{2007}  \at{The residual anisotropy at small scales in high shear
  turbulence}.  \jt{Phys.~Fluids}  \bvol{19}~(10),  \pg{101704}.

\bibitem[Cheng {\em et~al.\/}(2012)Cheng, Zeng \& Hu]{cheng2012stochastic}
{\sc \au{Cheng, X.L.}, \au{Zeng, Q.C.} \& \au{Hu, F.}} \yr{2012}
  \at{Stochastic modeling the effect of wind gust on dust entrainment during
  sand storm}.  \jt{Chin. Sci. Bull}  \bvol{57},  \pg{3595--3602}.

\bibitem[Clift \& Gauvin(1971)]{clift1971motion}
{\sc \au{Clift, R.} \& \au{Gauvin, W.H.}} \yr{1971}  \at{Motion of entrained
  particles in gas streams}.  \jt{Can. J. Chem. Eng.}  \bvol{49}~(4),
  \pg{439--448}.

\bibitem[Cogo {\em et~al.\/}(2022)Cogo, Salvadore, Picano \&
  Bernardini]{cogo2022direct}
{\sc \au{Cogo, M.}, \au{Salvadore, F.}, \au{Picano, F.} \& \au{Bernardini, M.}}
  \yr{2022}  \at{Direct numerical simulation of supersonic and hypersonic
  turbulent boundary layers at moderate-high {Reynolds} numbers and isothermal
  wall condition}.  \jt{J.~Fluid~Mech.}  \bvol{945},  \pg{A30}.

\bibitem[Crowe {\em et~al.\/}(1996)Crowe, Troutt \& Chung]{crowe1996numerical}
{\sc \au{Crowe, C.T.}, \au{Troutt, T.R.} \& \au{Chung, J.N.}} \yr{1996}
  \at{Numerical models for two-phase turbulent flows}.  \jt{Annu. Rev. Fluid
  Mech.}  \bvol{28}~(1),  \pg{11--43}.

\bibitem[Cui {\em et~al.\/}(2021)Cui, Huang, Xu, Andersson \&
  Zhao]{cui2021alignment}
{\sc \au{Cui, Z.W.}, \au{Huang, W.X.}, \au{Xu, C.X.}, \au{Andersson, H.I.} \&
  \au{Zhao, L.H.}} \yr{2021}  \at{Alignment of slender fibers and thin disks
  induced by coherent structures of wall turbulence}.  \jt{Int. J. Multiph.
  Flow}  \bvol{145},  \pg{103837}.

\bibitem[Cui \& Zhao(2022)]{cui2022shape}
{\sc \au{Cui, Z.W.} \& \au{Zhao, L.H.}} \yr{2022}  \at{Shape-dependent regions
  for inertialess spheroids in turbulent channel flow}.  \jt{Phys.~Fluids}
  \bvol{34}~(12),  \pg{123316}.

\bibitem[Dai {\em et~al.\/}(2018)Dai, Jin, Luo \& Fan]{dai2018direct}
{\sc \au{Dai, Q.}, \au{Jin, T.}, \au{Luo, K.} \& \au{Fan, J.R.}} \yr{2018}
  \at{Direct numerical simulation of particle dispersion in a three-dimensional
  spatially developing compressible mixing layer}.  \jt{Phys. Fluids}
  \bvol{30}~(11),  \pg{113301}.

\bibitem[Dai {\em et~al.\/}(2019)Dai, Jin, Luo, Xiao \& Fan]{dai2019direct}
{\sc \au{Dai, Q.}, \au{Jin, T.}, \au{Luo, K.}, \au{Xiao, W.} \& \au{Fan, J.R.}}
  \yr{2019}  \at{Direct numerical simulation of a three-dimensional spatially
  evolving compressible mixing layer laden with particles. {II. Turbulence}
  anisotropy and growth rate}.  \jt{Phys. Fluids}  \bvol{31}~(8),  \pg{083303}.

\bibitem[Dai {\em et~al.\/}(2017)Dai, Luo, Jin \& Fan]{dai2017direct}
{\sc \au{Dai, Q.}, \au{Luo, K.}, \au{Jin, T.} \& \au{Fan, J.R.}} \yr{2017}
  \at{Direct numerical simulation of turbulence modulation by particles in
  compressible isotropic turbulence}.  \jt{J. Fluid Mech.}  \bvol{832},
  \pg{438--482}.

\bibitem[Ducros {\em et~al.\/}(1999)Ducros, Ferrand, Nicoud, Weber, Darracq,
  Gacherieu \& Poinsot]{ducros1999large}
{\sc \au{Ducros, F.}, \au{Ferrand, V.}, \au{Nicoud, F.}, \au{Weber, C.},
  \au{Darracq, D.}, \au{Gacherieu, C.} \& \au{Poinsot, T.}} \yr{1999}
  \at{Large-eddy simulation of the shock/turbulence interaction}.
  \jt{J.~Comput.~Phys.}  \bvol{152}~(2),  \pg{517--549}.

\bibitem[Eaton(2009)]{eaton2009two}
{\sc \au{Eaton, J.K.}} \yr{2009}  \at{Two-way coupled turbulence simulations of
  gas-particle flows using point-particle tracking}.  \jt{Int. J. Multiph.
  Flow}  \bvol{35}~(9),  \pg{792--800}.

\bibitem[Eaton \& Fessler(1994)]{eaton1994preferential}
{\sc \au{Eaton, J.K.} \& \au{Fessler, J.R.}} \yr{1994}  \at{Preferential
  concentration of particles by turbulence}.  \jt{Int. J. Multiph. Flow}
  \bvol{20},  \pg{169--209}.

\bibitem[Elghobashi(1994)]{elghobashi1994predicting}
{\sc \au{Elghobashi, S.}} \yr{1994}  \at{On predicting particle-laden turbulent
  flows}.  \jt{Appl. Sci. Res}  \bvol{52},  \pg{309--329}.

\bibitem[Feng {\em et~al.\/}(2023{\natexlab{{\em a\/}}})Feng, Luo, Song, Xia \&
  Xu]{feng2023numerical}
{\sc \au{Feng, Y.B.}, \au{Luo, S.B.}, \au{Song, J.W.}, \au{Xia, K.X.} \&
  \au{Xu, D.Q.}} \yr{2023{\natexlab{{\em a\/}}}}  \at{Numerical investigation
  on the combustion characteristics of aluminum powder fuel in a supersonic
  cavity-based combustor}.  \jt{Appl. Therm. Eng}  \bvol{221},  \pg{119842}.

\bibitem[Feng {\em et~al.\/}(2023{\natexlab{{\em b\/}}})Feng, Luo, Song \&
  Xu]{feng2023numerical2}
{\sc \au{Feng, Y.}, \au{Luo, S.}, \au{Song, J.} \& \au{Xu, D.}}
  \yr{2023{\natexlab{{\em b\/}}}}  \at{Numerical investigation on flow and
  mixing characteristics of powder fuel under strong shear and shock wave
  interaction}.  \jt{Energy}  \bvol{263},  \pg{126061}.

\bibitem[Ferry \& Balachandar(2001)]{ferry2001fast}
{\sc \au{Ferry, J.} \& \au{Balachandar, S}} \yr{2001}  \at{A fast {Eulerian}
  method for disperse two-phase flow}.  \jt{Int. J. Multiph. Flow}
  \bvol{27}~(7),  \pg{1199--1226}.

\bibitem[Ferry \& Balachandar(2002)]{ferry2002equilibrium}
{\sc \au{Ferry, J.} \& \au{Balachandar, S}} \yr{2002}  \at{Equilibrium
  expansion for the {Eulerian} velocity of small particles}.  \jt{Powder
  Technol.}  \bvol{125}~(2-3),  \pg{131--139}.

\bibitem[Fessler {\em et~al.\/}(1994)Fessler, Kulick \&
  Eaton]{fessler1994preferential}
{\sc \au{Fessler, J.R.}, \au{Kulick, J.D.} \& \au{Eaton, J.K.}} \yr{1994}
  \at{Preferential concentration of heavy particles in a turbulent channel
  flow}.  \jt{Phys.~Fluids}  \bvol{6}~(11),  \pg{3742--3749}.

\bibitem[Fong {\em et~al.\/}(2019)Fong, Amili \& Coletti]{fong2019velocity}
{\sc \au{Fong, K.O.}, \au{Amili, O.} \& \au{Coletti, F.}} \yr{2019}
  \at{Velocity and spatial distribution of inertial particles in a turbulent
  channel flow}.  \jt{J.~Fluid~Mech.}  \bvol{872},  \pg{367--406}.

\bibitem[Gao {\em et~al.\/}(2023)Gao, Samtaney \& Richter]{gao2023direct}
{\sc \au{Gao, W.}, \au{Samtaney, R.} \& \au{Richter, D.H.}} \yr{2023}
  \at{Direct numerical simulation of particle-laden flow in an open channel at
  {$Re_\tau=5186$}}.  \jt{J.~Fluid Mech.}  \bvol{957},  \pg{A3}.

\bibitem[Goto \& Vassilicos(2008)]{goto2008sweep}
{\sc \au{Goto, S.} \& \au{Vassilicos, JC}} \yr{2008}  \at{Sweep-stick mechanism
  of heavy particle clustering in fluid turbulence}.  \jt{Phys. Rev. Lett.}
  \bvol{100}~(5),  \pg{054503}.

\bibitem[Gualtieri {\em et~al.\/}(2009)Gualtieri, Picano \&
  Casciola]{gualtieri2009anisotropic}
{\sc \au{Gualtieri, P.}, \au{Picano, F.} \& \au{Casciola, C.M.}} \yr{2009}
  \at{Anisotropic clustering of inertial particles in homogeneous shear flow}.
  \jt{J. Fluid Mech.}  \bvol{629},  \pg{25--39}.

\bibitem[Hwang(2013)]{hwang2013near}
{\sc \au{Hwang, Y.}} \yr{2013}  \at{Near-wall turbulent fluctuations in the
  absence of wide outer motions}.  \jt{J.~Fluid~Mech.}  \bvol{723},
  \pg{264--288}.

\bibitem[Jie {\em et~al.\/}(2021)Jie, Andersson \& Zhao]{jie2021effects}
{\sc \au{Jie, Y.}, \au{Andersson, H.I.} \& \au{Zhao, L.H.}} \yr{2021}
  \at{Effects of the quiescent core in turbulent channel flow on transport and
  clustering of inertial particles}.  \jt{Int. J. Multiph. Flow}  \bvol{140},
  \pg{103627}.

\bibitem[Jie {\em et~al.\/}(2022)Jie, Cui, Xu \& Zhao]{jie2022existence}
{\sc \au{Jie, Y.C.}, \au{Cui, Z.W.}, \au{Xu, C.X.} \& \au{Zhao, L.H.}}
  \yr{2022}  \at{On the existence and formation of multi-scale particle streaks
  in turbulent channel flows}.  \jt{J. Fluid Mech.}  \bvol{935},  \pg{A18}.

\bibitem[Jim{\'e}nez(2013)]{jimenez2013near}
{\sc \au{Jim{\'e}nez, J.}} \yr{2013}  \at{Near-wall turbulence}.  \jt{Phys.
  Fluids}  \bvol{25}~(10),  \pg{101302}.

\bibitem[Jim{\'e}nez(2018)]{jimenez2018coherent}
{\sc \au{Jim{\'e}nez, J.}} \yr{2018}  \at{Coherent structures in wall-bounded
  turbulence}.  \jt{J.~Fluid~Mech.}  \bvol{842},  \pg{P1}.

\bibitem[Kennedy \& Gruber(2008)]{kennedy2008reduced}
{\sc \au{Kennedy, C.A.} \& \au{Gruber, A.}} \yr{2008}  \at{Reduced aliasing
  formulations of the convective terms within the {Navier--Stokes} equations
  for a compressible fluid}.  \jt{J.~Comput.~Phys.}  \bvol{227}~(3),
  \pg{1676--1700}.

\bibitem[Klein {\em et~al.\/}(2003)Klein, Sadiki \& Janicka]{klein2003digital}
{\sc \au{Klein, M.}, \au{Sadiki, A.} \& \au{Janicka, J.}} \yr{2003}  \at{A
  digital filter based generation of inflow data for spatially developing
  direct numerical or large eddy simulations}.  \jt{J.~Comput.~Phys.}
  \bvol{186}~(2),  \pg{652--665}.

\bibitem[Kuerten(2016)]{m2016point}
{\sc \au{Kuerten, MJG}} \yr{2016}  \at{Point-particle {DNS} and {LES} of
  particle-laden turbulent flow-a state-of-the-art review}.  \jt{Flow Turbul.
  Combust.}  \bvol{97},  \pg{689--713}.

\bibitem[Kulick {\em et~al.\/}(1994)Kulick, Fessler \&
  Eaton]{kulick1994particle}
{\sc \au{Kulick, J.D.}, \au{Fessler, J.R.} \& \au{Eaton, J.K.}} \yr{1994}
  \at{Particle response and turbulence modification in fully developed channel
  flow}.  \jt{J. Fluid Mech.}  \bvol{277},  \pg{109--134}.

\bibitem[Kwon {\em et~al.\/}(2014)Kwon, Philip, De~Silva, Hutchins \&
  Monty]{kwon2014quiescent}
{\sc \au{Kwon, YS}, \au{Philip, J}, \au{De~Silva, CM}, \au{Hutchins, N} \&
  \au{Monty, JP}} \yr{2014}  \at{The quiescent core of turbulent channel flow}.
   \jt{J. Fluid Mech.}  \bvol{751},  \pg{228--254}.

\bibitem[Li {\em et~al.\/}(2023)Li, Cui, Yuan, Zhang, Zhou \&
  Zhao]{li2023particle}
{\sc \au{Li, T.}, \au{Cui, Z.W.}, \au{Yuan, X.X.}, \au{Zhang, Y.}, \au{Zhou,
  Q.} \& \au{Zhao, L.H.}} \yr{2023}  \at{Particle dynamics in compressible
  turbulent vertical channel flows}.  \jt{Phys.~Fluids}  \bvol{35}~(8).

\bibitem[Li {\em et~al.\/}(2016)Li, Wei \& Yu]{li2016direct}
{\sc \au{Li, Z.Z.}, \au{Wei, J.J.} \& \au{Yu, B.}} \yr{2016}  \at{Direct
  numerical study on effect of interparticle collision in particle-laden
  turbulence}.  \jt{AIAA J.}  \bvol{54}~(10),  \pg{3212--3222}.

\bibitem[Liu \& Zheng(2021)]{liu2021large}
{\sc \au{Liu, H.Y.} \& \au{Zheng, X.J.}} \yr{2021}  \at{Large-scale structures
  of wall-bounded turbulence in single-and two-phase flows: advancing
  understanding of the atmospheric surface layer during sandstorms}.  \jt{Flow}
   \bvol{1},  \pg{E5}.

\bibitem[Loth {\em et~al.\/}(2021)Loth, Tyler~Daspit, Jeong, Nagata \&
  Nonomura]{loth2021supersonic}
{\sc \au{Loth, E.}, \au{Tyler~Daspit, J.}, \au{Jeong, M.}, \au{Nagata, T.} \&
  \au{Nonomura, T.}} \yr{2021}  \at{Supersonic and hypersonic drag coefficients
  for a sphere}.  \jt{AIAA J.}  \bvol{59}~(8),  \pg{3261--3274}.

\bibitem[Marchioli \& Soldati(2002)]{marchioli2002mechanisms}
{\sc \au{Marchioli, C.} \& \au{Soldati, A.}} \yr{2002}  \at{Mechanisms for
  particle transfer and segregation in a turbulent boundary layer}.  \jt{J.
  Fluid Mech.}  \bvol{468},  \pg{283--315}.

\bibitem[Marchioli {\em et~al.\/}(2008)Marchioli, Soldati, Kuerten, Arcen,
  Taniere, Goldensoph, Squires, Cargnelutti \&
  Portela]{marchioli2008statistics}
{\sc \au{Marchioli, Ch}, \au{Soldati, A}, \au{Kuerten, JGM}, \au{Arcen, B},
  \au{Taniere, A}, \au{Goldensoph, G}, \au{Squires, KD}, \au{Cargnelutti, MF}
  \& \au{Portela, LM}} \yr{2008}  \at{Statistics of particle dispersion in
  direct numerical simulations of wall-bounded turbulence: Results of an
  international collaborative benchmark test}.  \jt{Int.~J.~Multiph.~Flow}
  \bvol{34}~(9),  \pg{879--893}.

\bibitem[Maxey \& Riley(1983)]{maxey1983equation}
{\sc \au{Maxey, M.~R} \& \au{Riley, J.J.}} \yr{1983}  \at{Equation of motion
  for a small rigid sphere in a nonuniform flow}.  \jt{Phys.~Fluids}
  \bvol{26}~(4),  \pg{883--889}.

\bibitem[Milici \& de~Marchis(2016)]{milici2016statistics}
{\sc \au{Milici, B.} \& \au{de~Marchis, M.}} \yr{2016}  \at{Statistics of
  inertial particle deviation from fluid particle trajectories in horizontal
  rough wall turbulent channel flow}.  \jt{Int. J. Heat Fluid Flow}  \bvol{60},
   \pg{1--11}.

\bibitem[Monchaux {\em et~al.\/}(2010)Monchaux, Bourgoin \&
  Cartellier]{monchaux2010preferential}
{\sc \au{Monchaux, R.}, \au{Bourgoin, M.} \& \au{Cartellier, A.}} \yr{2010}
  \at{Preferential concentration of heavy particles: {A Vorono{\"\i}
  analysis}}.  \jt{Phys. Fluids}  \bvol{22}~(10),  \pg{103304}.

\bibitem[Monchaux {\em et~al.\/}(2012)Monchaux, Bourgoin \&
  Cartellier]{monchaux2012analyzing}
{\sc \au{Monchaux, R.}, \au{Bourgoin, M.} \& \au{Cartellier, A.}} \yr{2012}
  \at{Analyzing preferential concentration and clustering of inertial particles
  in turbulence}.  \jt{Int. J. Multiph. Flow}  \bvol{40},  \pg{1--18}.

\bibitem[Mortimer \& Fairweather(2020)]{mortimer2020density}
{\sc \au{Mortimer, LF} \& \au{Fairweather, M}} \yr{2020}  \at{Density ratio
  effects on the topology of coherent turbulent structures in two-way coupled
  particle-laden channel flows}.  \jt{Phys.~Fluids}  \bvol{32}~(10),
  \pg{103302}.

\bibitem[Mortimer {\em et~al.\/}(2019)Mortimer, Njobuenwu \&
  Fairweather]{mortimer2019near}
{\sc \au{Mortimer, LF}, \au{Njobuenwu, DO} \& \au{Fairweather, M}} \yr{2019}
  \at{Near-wall dynamics of inertial particles in dilute turbulent channel
  flows}.  \jt{Phys. Fluids}  \bvol{31}~(6),  \pg{063302}.

\bibitem[Motoori {\em et~al.\/}(2022)Motoori, Wong \& Goto]{motoori2022role}
{\sc \au{Motoori, Y.}, \au{Wong, C.} \& \au{Goto, S.}} \yr{2022}  \at{Role of
  the hierarchy of coherent structures in the transport of heavy small
  particles in turbulent channel flow}.  \jt{J. Fluid Mech.}  \bvol{942},
  \pg{A3}.

\bibitem[Musker(1979)]{musker1979explicit}
{\sc \au{Musker, A.}} \yr{1979}  \at{Explicit expression for the smooth wall
  velocity distribution in a turbulent boundary layer}.  \jt{AIAA~J.}
  \bvol{17}~(6),  \pg{655--657}.

\bibitem[Narayanan {\em et~al.\/}(2003)Narayanan, Lakehal, Botto \&
  Soldati]{narayanan2003mechanisms}
{\sc \au{Narayanan, C.}, \au{Lakehal, D.}, \au{Botto, L.} \& \au{Soldati, A.}}
  \yr{2003}  \at{Mechanisms of particle deposition in a fully developed
  turbulent open channel flow}.  \jt{Phys. Fluids}  \bvol{15}~(3),
  \pg{763--775}.

\bibitem[Picano {\em et~al.\/}(2009)Picano, Sardina \&
  Casciola]{picano2009spatial}
{\sc \au{Picano, F}, \au{Sardina, G.} \& \au{Casciola, CM}} \yr{2009}
  \at{Spatial development of particle-laden turbulent pipe flow}.
  \jt{Phys.~Fluids}  \bvol{21}~(9),  \pg{093305}.

\bibitem[Picciotto {\em et~al.\/}(2005{\natexlab{{\em a\/}}})Picciotto,
  Marchioli, Reeks \& Soldati]{picciotto2005statistics}
{\sc \au{Picciotto, M.}, \au{Marchioli, C.}, \au{Reeks, M.W.} \& \au{Soldati,
  A.}} \yr{2005{\natexlab{{\em a\/}}}}  \at{Statistics of velocity and
  preferential accumulation of micro-particles in boundary layer turbulence}.
  \jt{Nucl. Eng. Des}  \bvol{235}~(10-12),  \pg{1239--1249}.

\bibitem[Picciotto {\em et~al.\/}(2005{\natexlab{{\em b\/}}})Picciotto,
  Marchioli \& Soldati]{picciotto2005characterization}
{\sc \au{Picciotto, M.}, \au{Marchioli, C.} \& \au{Soldati, A.}}
  \yr{2005{\natexlab{{\em b\/}}}}  \at{Characterization of near-wall
  accumulation regions for inertial particles in turbulent boundary layers}.
  \jt{Phys.~Fluids}  \bvol{17}~(9),  \pg{098101}.

\bibitem[Pirozzoli(2010)]{pirozzoli2010generalized}
{\sc \au{Pirozzoli, S.}} \yr{2010}  \at{Generalized conservative approximations
  of split convective derivative operators}.  \jt{J.~Comput.~Phys.}
  \bvol{229}~(19),  \pg{7180--7190}.

\bibitem[Pirozzoli(2011)]{pirozzoli2011numerical}
{\sc \au{Pirozzoli, S.}} \yr{2011}  \at{Numerical methods for high-speed
  flows}.  \jt{Annu.~Rev.~Fluid Mech.}  \bvol{43},  \pg{163--194}.

\bibitem[Pumir \& Wilkinson(2016)]{pumir2016collisional}
{\sc \au{Pumir, A.} \& \au{Wilkinson, M.}} \yr{2016}  \at{Collisional
  aggregation due to turbulence}.  \jt{Annu. Rev. Condens. Matter Phys.}
  \bvol{7},  \pg{141--170}.

\bibitem[Rashidi {\em et~al.\/}(1990)Rashidi, Hetsroni \&
  Banerjee]{rashidi1990particle}
{\sc \au{Rashidi, Mr}, \au{Hetsroni, G} \& \au{Banerjee, S.}} \yr{1990}
  \at{Particle-turbulence interaction in a boundary layer}.  \jt{Int. J.
  Multiph. Flow}  \bvol{16}~(6),  \pg{935--949}.

\bibitem[Rouson \& Eaton(2001)]{rouson2001preferential}
{\sc \au{Rouson, D.} \& \au{Eaton, J.K.}} \yr{2001}  \at{On the preferential
  concentration of solid particles in turbulent channel flow}.  \jt{J. Fluid
  Mech.}  \bvol{428},  \pg{149--169}.

\bibitem[Rudinger(2012)]{rudinger2012fundamentals}
{\sc \au{Rudinger, G.}} \yr{2012} {\em Fundamentals of gas particle flow\/}, ,
  \vol{vol.~2}.  \publ{Elsevier}.

\bibitem[Salazar {\em et~al.\/}(2008)Salazar, de~Jong, Cao, Woodward, Meng \&
  Collins]{salazar2008experimental}
{\sc \au{Salazar, J.}, \au{de~Jong, J.}, \au{Cao, L.}, \au{Woodward, S.H.},
  \au{Meng, H.} \& \au{Collins, L.}} \yr{2008}  \at{Experimental and numerical
  investigation of inertial particle clustering in isotropic turbulence}.
  \jt{J.~Fluid Mech.}  \bvol{600},  \pg{245--256}.

\bibitem[Sardina {\em et~al.\/}(2012)Sardina, Schlatter, Brandt, Picano \&
  Casciola]{sardina2012wall}
{\sc \au{Sardina, G.}, \au{Schlatter, P.}, \au{Brandt, L.}, \au{Picano, F.} \&
  \au{Casciola, C.~M.}} \yr{2012}  \at{Wall accumulation and spatial
  localization in particle-laden wall flows}.  \jt{J. Fluid Mech.}  \bvol{699},
   \pg{50--78}.

\bibitem[Saw {\em et~al.\/}(2008)Saw, Shaw, Ayyalasomayajula, Chuang \&
  Gylfason]{saw2008inertial}
{\sc \au{Saw, E.W.}, \au{Shaw, R.A.}, \au{Ayyalasomayajula, S.}, \au{Chuang,
  P.Y.} \& \au{Gylfason, A.}} \yr{2008}  \at{Inertial clustering of particles
  in {high-Reynolds-number} turbulence}.  \jt{Phys.~Rev.~Lett.}
  \bvol{100}~(21),  \pg{214501}.

\bibitem[Shu \& Osher(1988)]{shu1988efficient}
{\sc \au{Shu, C.W.} \& \au{Osher, S.}} \yr{1988}  \at{Efficient implementation
  of essentially non-oscillatory shock-capturing schemes}.
  \jt{J.~Comput.~Phys.}  \bvol{77}~(2),  \pg{439--471}.

\bibitem[Soldati(2005)]{soldati2005particles}
{\sc \au{Soldati, A.}} \yr{2005}  \at{Particles turbulence interactions in
  boundary layers}.  \jt{Appl. Math. Mech.}  \bvol{85}~(10),  \pg{683--699}.

\bibitem[Soldati \& Marchioli(2009)]{soldati2009physics}
{\sc \au{Soldati, A.} \& \au{Marchioli, C.}} \yr{2009}  \at{Physics and
  modelling of turbulent particle deposition and entrainment: Review of a
  systematic study}.  \jt{Int. J. Multiph. Flow}  \bvol{35}~(9),
  \pg{827--839}.

\bibitem[Tardu(1995)]{tardu1995characteristics}
{\sc \au{Tardu, S}} \yr{1995}  \at{Characteristics of single and clusters of
  bursting events in the inner layer}.  \jt{Exp. fluids}  \bvol{20}~(2),
  \pg{112--124}.

\bibitem[Vinkovic {\em et~al.\/}(2011)Vinkovic, Doppler, Lelouvetel \&
  Buffat]{vinkovic2011direct}
{\sc \au{Vinkovic, I.}, \au{Doppler, D.}, \au{Lelouvetel, J} \& \au{Buffat,
  M.}} \yr{2011}  \at{Direct numerical simulation of particle interaction with
  ejections in turbulent channel flows}.  \jt{Int. J. Multiph. Flow}
  \bvol{37}~(2),  \pg{187--197}.

\bibitem[Wang \& Maxey(1993)]{wang1993settling}
{\sc \au{Wang, L.P.} \& \au{Maxey, M.R.}} \yr{1993}  \at{Settling velocity and
  concentration distribution of heavy particles in homogeneous isotropic
  turbulence}.  \jt{J. Fluid Mech.}  \bvol{256},  \pg{27--68}.

\bibitem[Wang {\em et~al.\/}(2022)Wang, Wan \& Biferale]{wang2022acceleration}
{\sc \au{Wang, X.}, \au{Wan, M.} \& \au{Biferale, L.}} \yr{2022}
  \at{Acceleration statistics of tracer and light particles in compressible
  homogeneous isotropic turbulence}.  \jt{J.~Fluid Mech.}  \bvol{935},
  \pg{A36}.

\bibitem[Wang {\em et~al.\/}(2015)Wang, Huang \& Xu]{wang2015hairpin}
{\sc \au{Wang, Y.S.}, \au{Huang, W.X.} \& \au{Xu, C.X.}} \yr{2015}  \at{On
  hairpin vortex generation from near-wall streamwise vortices}.  \jt{Acta
  Mech. Sin.}  \bvol{31},  \pg{139--152}.

\bibitem[Wray(1990)]{wray1990minimal}
{\sc \au{Wray, A.}} \yr{1990}  \at{Minimal storage time advancement schemes for
  spectral methods}.  \jt{NASA Ames Research Center, California, Report No. MS}
   \bvol{202}.

\bibitem[Xiao {\em et~al.\/}(2020)Xiao, Jin, Luo, Dai \& Fan]{xiao2020eulerian}
{\sc \au{Xiao, W.}, \au{Jin, T.}, \au{Luo, K.}, \au{Dai, Q.} \& \au{Fan, J.R.}}
  \yr{2020}  \at{{Eulerian--Lagrangian} direct numerical simulation of
  preferential accumulation of inertial particles in a compressible turbulent
  boundary layer}.  \jt{J. Fluid Mech.}  \bvol{903},  \pg{A19}.

\bibitem[Yang {\em et~al.\/}(2014)Yang, Wang, Shi, Xiao, He \&
  Chen]{yang2014interactions}
{\sc \au{Yang, Y.}, \au{Wang, J.}, \au{Shi, Y.}, \au{Xiao, Z.}, \au{He, X.T.}
  \& \au{Chen, S.Y.}} \yr{2014}  \at{Interactions between inertial particles
  and shocklets in compressible turbulent flow}.  \jt{Phys. Fluids}
  \bvol{26}~(9).

\bibitem[Yu \& Xu(2021)]{yu2021compressibility}
{\sc \au{Yu, M.} \& \au{Xu, C.X.}} \yr{2021}  \at{Compressibility effects on
  hypersonic turbulent channel flow with cold walls}.  \jt{Phys.~Fluids}
  \bvol{33}~(7),  \pg{075106}.

\bibitem[Yu {\em et~al.\/}(2019)Yu, Xu \& Pirozzoli]{yu2019genuine}
{\sc \au{Yu, M.}, \au{Xu, C.X.} \& \au{Pirozzoli, S.}} \yr{2019}  \at{Genuine
  compressibility effects in wall-bounded turbulence}.  \jt{Phys.~Rev.~Fluids}
  \bvol{4}~(12),  \pg{123402}.

\bibitem[Yu {\em et~al.\/}(2017)Yu, Lin, Shao \& Wang]{yu2017effects}
{\sc \au{Yu, Z.S.}, \au{Lin, Z.W.}, \au{Shao, X.M.} \& \au{Wang, L.P.}}
  \yr{2017}  \at{Effects of particle-fluid density ratio on the interactions
  between the turbulent channel flow and finite-size particles}.
  \jt{Phys.~Rev.~E}  \bvol{96}~(3),  \pg{033102}.

\bibitem[Zhang {\em et~al.\/}(2016)Zhang, Liu, Ma \&
  Xiao]{zhang2016preferential}
{\sc \au{Zhang, Q.Q.}, \au{Liu, H.}, \au{Ma, Z.Q.} \& \au{Xiao, Z.L.}}
  \yr{2016}  \at{Preferential concentration of heavy particles in compressible
  isotropic turbulence}.  \jt{Phys. Fluids}  \bvol{28}~(5),  \pg{055104}.

\bibitem[Zhang {\em et~al.\/}(2014)Zhang, Bi, Hussain \&
  She]{zhang2014generalized}
{\sc \au{Zhang, Y.}, \au{Bi, W.}, \au{Hussain, F.} \& \au{She, Z.}} \yr{2014}
  \at{A generalized {Reynolds} analogy for compressible wall-bounded turbulent
  flows}.  \jt{J.~Fluid~Mech.}  \bvol{739},  \pg{392--420}.

\bibitem[Zhao {\em et~al.\/}(2012)Zhao, Marchioli \& Andersson]{zhao2012stokes}
{\sc \au{Zhao, L.H.}, \au{Marchioli, C.} \& \au{Andersson, H.I.}} \yr{2012}
  \at{Stokes number effects on particle slip velocity in wall-bounded
  turbulence and implications for dispersion models}.  \jt{Phys. Fluids}
  \bvol{24}~(2),  \pg{021705}.

\end{thebibliography}
\end{document}